\documentclass[aps, prd, reprint, superscriptaddress, amsmath, amssymb,
 floatfix,
 longbibliography, 
 showpacs]{revtex4-2}

\usepackage[utf8]{inputenc}
\usepackage[nowatermark]{fixmetodonotes}
\usepackage[colorlinks, urlcolor=blue, linkcolor=black]{hyperref}
\usepackage{graphicx}
\usepackage{isotope}
\usepackage{xcolor}

\DeclareFontFamily{OT1}{pzc}{}
\DeclareFontShape{OT1}{pzc}{m}{it}{<-> s * [1.10] pzcmi7t}{}
\DeclareMathAlphabet{\mathpzc}{OT1}{pzc}{m}{it}

\makeatletter
\def\LT@LR@e{\LTleft\z@   \LTright\z@}%
\makeatother

\usepackage{siunitx}
\DeclareSIUnit\clight{\text{\ensuremath{c}}}
\DeclareSIUnit\pot{POT}

\usepackage[inline]{enumitem}
\setenumerate[1]{label={(\arabic*)}}

\makeatletter
\protected\def\xvcenter{%
  \hbox\bgroup$\everyvbox{\everyvbox{}\aftergroup\m@th\aftergroup$\aftergroup\egroup}%
  \vcenter
}

\DeclareRobustCommand{\midscript}[1]{
  \mathchoice{\mid@script\scriptstyle{#1}}
    {\mid@script\scriptstyle{#1}}
    {\mid@script\scriptscriptstyle{#1}}
    {\mid@script\scriptscriptstyle{#1}}
}
\newcommand{\mid@script}[2]{
  \vcenter{\hbox{$\m@th#1#2$}}
}

\makeatother

\newcommand{\TotalBinCount}{359}
\newcommand{\TotalDataPOT}{%
  \num[round-mode=places, round-precision=2]{6.7949e20}}
\newcommand{\FOMCutoff}{2.5\%}
\newcommand{\ccnp}{CC$0\pi Np$}
\newcommand{\recoBinIdx}{j}
\newcommand{\trueBinIdx}{\mu}
\newcommand{\secondTrueBinIdx}{\nu}
\newcommand{\secondRecoBinIdx}{\ell}
\newcommand{\PredictedRecoEvtCount}{n}
\newcommand{\PredictedSignalEvtCount}{\phi}
\newcommand{\MeasuredSignalEvtCount}{\hat{\PredictedSignalEvtCount}}
\newcommand{\EXTCount}{O}
\newcommand{\PredictedBkgdCount}{B}
\newcommand{\ResponseMatrix}{\Delta}
\newcommand{\CovMat}{V}
\newcommand{\CovMatFirstIdx}{a}
\newcommand{\CovMatSecondIdx}{b}
\newcommand{\UnivIdx}{u}
\newcommand{\UnivCount}{N_{\mathrm{univ}}}
\newcommand{\MigrationMatrix}{M}
\newcommand{\UnfoldingMatrix}{U}
\newcommand{\ErrPropMatrix}{\mathfrak{E}}
\newcommand{\BkgdSubtractedRecoEvents}{d}
\newcommand{\AllRecoEvents}{D}
\newcommand{\AddSmearMatrix}{A_C}
\newcommand{\IterationIdx}{i}

\newcommand{\BinProb}{P}
\newcommand{\Eff}{\epsilon}
\newcommand{\ConvergenceFOM}{\mathfrak{F}}
\newcommand{\NumTrueBins}{T}
\newcommand{\XsecDimension}{n}
\newcommand{\SecondXsecDimension}{m}
\newcommand{\PhaseSpaceVec}{\mathbf{x}}
\newcommand{\SecondPhaseSpaceVec}{\mathbf{y}}
\newcommand{\PhaseSpaceVecWidths}{\Delta\PhaseSpaceVec}
\newcommand{\SecondPhaseSpaceVecWidths}{\Delta\SecondPhaseSpaceVec}
\newcommand{\IntegratedFlux}{\ensuremath{\Phi}}
\newcommand{\NumTargets}{\mathcal{N}_\mathrm{Ar}}

\newcommand{\TopoScoreCut}{0.1}
\newcommand{\TrackScoreCut}{0.5}

\newcommand{\MuonTrackScoreCut}{0.8}
\newcommand{\MuonVtxDistanceCut}{\SI{4}{\centi\meter}}
\newcommand{\MuonLengthCut}{\SI{10}{\centi\meter}}
\newcommand{\MuonLLRpidCut}{0.2}
\newcommand{\ProtonLLRpidCut}{0.2}

\newcommand{\MuonMinMomentumCut}{\SI{0.1}{\GeV\per\clight}}
\newcommand{\MuonMaxMomentumCut}{\SI{1.2}{\GeV\per\clight}}
\newcommand{\ProtonMinMomentumCut}{\SI{0.25}{\GeV\per\clight}}
\newcommand{\ProtonMaxMomentumCut}{\SI{1.0}{\GeV\per\clight}}

\definecolor{MyDarkBlue}{rgb}{0, 0, 0.3}

\definecolor{MyDarkRed}{rgb}{0.3, 0, 0}

\usepackage{hhline}

\usepackage{booktabs}
\usepackage{multirow}
\usepackage{xcolor}
\usepackage{isotope}
\usepackage{placeins}
\usepackage[caption=false]{subfig}

\usepackage{longtable}
\usepackage{afterpage}

\usepackage{stackengine}

\usepackage{etoolbox}
\makeatletter
\patchcmd\linenumberpar{\@LN@parpgbrk}{\penalty\@LN@parpgpen\relax}{}{}
\makeatother

\robustify\subref

\begin{document}

\title{Measurement of double-differential cross sections for mesonless
charged-current muon neutrino interactions on argon with final-state protons
using the MicroBooNE detector}

%


\newcommand{\ANL}{Argonne National Laboratory (ANL), Lemont, IL, 60439, USA}
\newcommand{\Bern}{Universit{\"a}t Bern, Bern CH-3012, Switzerland}
\newcommand{\BNL}{Brookhaven National Laboratory (BNL), Upton, NY, 11973, USA}
\newcommand{\UCSB}{University of California, Santa Barbara, CA, 93106, USA}
\newcommand{\Cambridge}{University of Cambridge, Cambridge CB3 0HE, United Kingdom}
\newcommand{\CIEMAT}{Centro de Investigaciones Energ\'{e}ticas, Medioambientales y Tecnol\'{o}gicas (CIEMAT), Madrid E-28040, Spain}
\newcommand{\Chicago}{University of Chicago, Chicago, IL, 60637, USA}
\newcommand{\Cincinnati}{University of Cincinnati, Cincinnati, OH, 45221, USA}
\newcommand{\CSU}{Colorado State University, Fort Collins, CO, 80523, USA}
\newcommand{\Columbia}{Columbia University, New York, NY, 10027, USA}
\newcommand{\Edinburgh}{University of Edinburgh, Edinburgh EH9 3FD, United Kingdom}
\newcommand{\FNAL}{Fermi National Accelerator Laboratory (FNAL), Batavia, IL 60510, USA}
\newcommand{\Granada}{Universidad de Granada, Granada E-18071, Spain}
\newcommand{\Harvard}{Harvard University, Cambridge, MA 02138, USA}
\newcommand{\IIT}{Illinois Institute of Technology (IIT), Chicago, IL 60616, USA}
\newcommand{\Indiana}{Indiana University, Bloomington, IN 47405, USA}
\newcommand{\KSU}{Kansas State University (KSU), Manhattan, KS, 66506, USA}
\newcommand{\Lancaster}{Lancaster University, Lancaster LA1 4YW, United Kingdom}
\newcommand{\LANL}{Los Alamos National Laboratory (LANL), Los Alamos, NM, 87545, USA}
\newcommand{\Louisiana}{Louisiana State University, Baton Rouge, LA, 70803, USA}
\newcommand{\Manchester}{The University of Manchester, Manchester M13 9PL, United Kingdom}
\newcommand{\MIT}{Massachusetts Institute of Technology (MIT), Cambridge, MA, 02139, USA}
\newcommand{\Michigan}{University of Michigan, Ann Arbor, MI, 48109, USA}
\newcommand{\MSU}{Michigan State University, East Lansing, MI 48824, USA}
\newcommand{\Minnesota}{University of Minnesota, Minneapolis, MN, 55455, USA}
\newcommand{\Nankai}{Nankai University, Nankai District, Tianjin 300071, China}
\newcommand{\NMSU}{New Mexico State University (NMSU), Las Cruces, NM, 88003, USA}
\newcommand{\Oxford}{University of Oxford, Oxford OX1 3RH, United Kingdom}
\newcommand{\Pitt}{University of Pittsburgh, Pittsburgh, PA, 15260, USA}
\newcommand{\Rutgers}{Rutgers University, Piscataway, NJ, 08854, USA}
\newcommand{\SLAC}{SLAC National Accelerator Laboratory, Menlo Park, CA, 94025, USA}
\newcommand{\SDSMT}{South Dakota School of Mines and Technology (SDSMT), Rapid City, SD, 57701, USA}
\newcommand{\Maine}{University of Southern Maine, Portland, ME, 04104, USA}
\newcommand{\Syracuse}{Syracuse University, Syracuse, NY, 13244, USA}
\newcommand{\TelAviv}{Tel Aviv University, Tel Aviv, Israel, 69978}
\newcommand{\Tennessee}{University of Tennessee, Knoxville, TN, 37996, USA}
\newcommand{\UTA}{University of Texas, Arlington, TX, 76019, USA}
\newcommand{\Tufts}{Tufts University, Medford, MA, 02155, USA}
\newcommand{\UCL}{University College London, London WC1E 6BT, United Kingdom}
\newcommand{\VTech}{Center for Neutrino Physics, Virginia Tech, Blacksburg, VA, 24061, USA}
\newcommand{\Warwick}{University of Warwick, Coventry CV4 7AL, United Kingdom}
\newcommand{\Yale}{Wright Laboratory, Department of Physics, Yale University, New Haven, CT, 06520, USA}

\affiliation{\ANL}
\affiliation{\Bern}
\affiliation{\BNL}
\affiliation{\UCSB}
\affiliation{\Cambridge}
\affiliation{\CIEMAT}
\affiliation{\Chicago}
\affiliation{\Cincinnati}
\affiliation{\CSU}
\affiliation{\Columbia}
\affiliation{\Edinburgh}
\affiliation{\FNAL}
\affiliation{\Granada}
\affiliation{\Harvard}
\affiliation{\IIT}
\affiliation{\Indiana}
\affiliation{\KSU}
\affiliation{\Lancaster}
\affiliation{\LANL}
\affiliation{\Louisiana}
\affiliation{\Manchester}
\affiliation{\MIT}
\affiliation{\Michigan}
\affiliation{\MSU}
\affiliation{\Minnesota}
\affiliation{\Nankai}
\affiliation{\NMSU}
\affiliation{\Oxford}
\affiliation{\Pitt}
\affiliation{\Rutgers}
\affiliation{\SLAC}
\affiliation{\SDSMT}
\affiliation{\Maine}
\affiliation{\Syracuse}
\affiliation{\TelAviv}
\affiliation{\Tennessee}
\affiliation{\UTA}
\affiliation{\Tufts}
\affiliation{\UCL}
\affiliation{\VTech}
\affiliation{\Warwick}
\affiliation{\Yale}

\author{P.~Abratenko} \affiliation{\Tufts}
\author{O.~Alterkait} \affiliation{\Tufts}
\author{D.~Andrade~Aldana} \affiliation{\IIT}
\author{L.~Arellano} \affiliation{\Manchester}
\author{J.~Asaadi} \affiliation{\UTA}
\author{A.~Ashkenazi}\affiliation{\TelAviv}
\author{S.~Balasubramanian}\affiliation{\FNAL}
\author{B.~Baller} \affiliation{\FNAL}
\author{G.~Barr} \affiliation{\Oxford}
\author{D.~Barrow} \affiliation{\Oxford}
% ...

\collaboration{The MicroBooNE Collaboration}
\thanks{microboone\_info@fnal.gov}\noaffiliation



\date{\today}

\begin{abstract}
Charged-current neutrino interactions with final states containing zero mesons
and at least one proton are of high interest for current and future
accelerator-based neutrino oscillation experiments. Using the Booster Neutrino
Beam and the MicroBooNE detector at Fermi National Accelerator Laboratory, we
have obtained the first double-differential cross section measurements of this
channel for muon neutrino scattering on an argon target with a proton momentum
threshold of \SI{0.25}{\GeV\per\clight}. We also report a flux-averaged total
cross section of $\sigma = (\num[round-mode=places, round-precision=1]{11.8278}
\pm \num[round-mode=places, round-precision=1]{1.1627}) \times
\SI{e-38}{\centi\meter\squared} \,/\, \mathrm{Ar}$ and several
single-differential measurements which extend and improve upon previous
results. Statistical and systematic uncertainties are quantified with a full
treatment of correlations across \TotalBinCount~kinematic bins, including
correlations between distributions describing different observables. The
resulting data set provides the most detailed information obtained to date for
testing models of mesonless neutrino-argon scattering.
\end{abstract}

\maketitle

\section{Introduction}
\label{sec:intro}

Accelerator-based measurements of neutrino oscillations will be critical for
definitively answering key open questions in high-energy physics, including
whether charge-parity ($CP$) violation~\cite{cpViolation} occurs in the lepton
sector, whether \textit{sterile} neutrino species~\cite{sterileNeutrinos}
(which do not participate in the weak interaction) exist, and whether $\nu_3$
is the heaviest or lightest of the known neutrino mass
eigenstates~\cite{massOrder}. In the coming years, the Fermilab-based Short
Baseline Neutrino (SBN) program~\cite{SBNprogramReview} and Deep Underground
Neutrino Experiment (DUNE)~\cite{DUNETDRvol2} will both pursue high-precision
neutrino oscillation analyses using liquid argon time projection chambers
(LArTPCs) as the primary detector technology. Maximizing the discovery
potential of these large experimental efforts will require substantial
improvements to the present understanding of \si{\GeV}-scale neutrino-nucleus
scattering physics~\cite{nustec}. An emerging literature of neutrino-argon
cross-section measurements~\cite{DuffyNuAr}, pioneered by the
ArgoNeuT~\cite{ArgoneutDetector} and MicroBooNE~\cite{MicroBooNEDetector}
LArTPC experiments, will provide the most direct constraints for refining
interaction models to the precision needed for SBN and DUNE.

This article contributes to that effort by presenting measurements of
differential cross sections for charged-current (CC) $\nu_\mu$ scattering on
argon leading to final states containing no mesons and one or more protons.
This event topology (hereafter abbreviated as \ccnp) is the dominant neutrino
interaction channel for the SBN program, and its contribution is expected to
remain important for DUNE oscillation analyses. The results presented herein
build upon a previous MicroBooNE investigation~\cite{MCC8ccnpPaper} which
measured single-differential neutrino-argon cross sections in the \ccnp\
channel for the first time. The present work leverages subsequent improvements
to the MicroBooNE simulation~\cite{MicroBooNEGENIETune},
reconstruction~\cite{LLRPID}, and systematic uncertainty
quantification~\cite{DetVarSystematics} to achieve more detailed measurements,
including multiple double-differential distributions. The analysis also
benefits from a significantly larger data set, which was recorded by the
MicroBooNE detector~\cite{MicroBooNEDetector} with a total exposure
of \TotalDataPOT\ protons on target (\si{\pot}) from the Fermilab Booster
Neutrino Beam (BNB)~\cite{MiniBooNEFlux}.

Various kinematic distributions involving the three-momenta of the final-state
muon and leading proton are measured and compared to simulations in this
article. An innovative feature of the data set is the strategy used for its
presentation; covariances between all \TotalBinCount~kinematic bins studied in
the analysis are reported and used to quantify an overall goodness-of-fit of
theoretical predictions. The full group of cross-section measurements may thus
be regarded as a single result with greater combined model discrimination power
(due to the need to describe inter-distribution correlations) than its
constituent parts. Together with the inclusive cross-section measurements
recently reported by MicroBooNE in Refs.~\cite{uBCC0pNpPRL,uBCC0pNpPRD}, this
analysis thus represents a first application of the ``blockwise unfolding''
technique proposed in Ref.~\cite{GardinerXSecExtract}.

The experimental setup and simulations used in the analysis are described in
Sec.~\ref{sec:exp_and_sim}. Section~\ref{sec:selection} presents the signal
definition and the event selection criteria. Section~\ref{sec:binning}
discusses the strategy for reporting the results, including the observables
measured and the general approach to corrections for imperfect event
reconstruction. The procedures used to estimate uncertainties and convert
measured event distributions to cross sections are documented in
Secs.~\ref{sec:unc}~and~\ref{sec:unfold}, respectively.
Section~\ref{sec:results} reports the cross-section results and compares them
to several model predictions. Finally, Sec.~\ref{sec:summary_and_conclusions}
contains a summary and conclusions.

\section{MicroBooNE experiment and simulation}
\label{sec:exp_and_sim}

The MicroBooNE detector~\cite{MicroBooNEDetector} is a LArTPC that operated in
the BNB from 2015--2021. The detector's 85-tonne active mass of liquid argon
was exposed to a \SI{273}{\volt\per\centi\meter} electric field and
instrumented with three wire planes and thirty-two photomultiplier tubes
(PMTs). Acrylic disks coated with tetraphenyl butadiene (TPB) were placed in
front of the PMTs to convert the 128-\si{\nano\meter} argon scintillation light
into visible wavelengths for efficient detection.

Neutrinos from the BNB are generated by bombarding a beryllium target with
8-\si{\GeV} protons. Secondary particles produced by the proton collisions are
focused by a magnetic horn and decay in flight. The decay products
are directed toward a beam stop made of steel and concrete that absorbs
particles other than neutrinos. During operation, the MicroBooNE detector
was located \SI{463}{\meter} downstream from the target along the beam axis.
The BNB neutrino flux at this position is dominated by muon neutrinos with a
mean energy of \SI{0.8}{\GeV}. Minor contributions from $\bar{\nu}_\mu$ (5.8\%)
and a mixture of $\nu_e$ and $\bar{\nu}_e$ (0.5\% combined) are also present.

Interpretation of the data obtained from the MicroBooNE detector is enabled by
a suite of Monte Carlo simulations that provide comprehensive modeling of the
entire experiment. Neutrino beam production is simulated using the Geant4
framework~\cite{Geant4,Allison:2016lfl} and a detailed representation of the
BNB apparatus developed by the MiniBooNE collaboration~\cite{MiniBooNEFlux}.

\subsection{Neutrino interaction model}
\label{sec:uBTune}

Neutrino interactions are modeled in the MicroBooNE simulation chain using
version~3.0.6 of the GENIE neutrino event
generator~\cite{geniev3highlights,GENIEnimA,Andreopoulos:2015wxa} with the
\texttt{G18\_10a\_02\_11a} configuration. This configuration includes a local
Fermi gas (LFG) representation of the nuclear ground
state~\cite{Carrasco:1989vq} and the Valencia model for quasielastic (QE) and
two-particle two-hole (2p2h) interactions in the CC
channel~\cite{NievesQEPaper, NievesQEErratum, Nieves:2012yz,
Gran:2013kda, GENIEValenciaMEC}.
Resonance production (RES) is simulated with the
model of Kuzmin-Lyubushkin-Naumov and Berger-Sehgal
(KLN-BS)~\cite{Nowak:2009se,Kuzmin:2003ji, Berger:2007rq,Graczyk:2007bc}, deep
inelastic scattering~\cite{Paschos:2001np} (DIS) is described using the
structure functions of Bodek and Yang~\cite{Bodek:2004pc,Bodek:2010zz}, and
coherent pion production (COH) follows the Berger-Sehgal
approach~\cite{Berger:2008xs}. Intranuclear hadronic final-state interactions
(FSI) are simulated using the hA2018 model~\cite{hN2018}. Several model
parameters relevant for calculations of neutrino-nucleon cross-sections were
tuned by the GENIE collaboration to data sets from bubble chamber
experiments~\cite{GENIEFreeNucleon}.

The GENIE v3.0.6 \texttt{G18\_10a\_02\_11a} configuration described above was
modified with MicroBooNE-specific tuning~\cite{MicroBooNEGENIETune} of two CCQE
and two CC2p2h model parameters to data from the T2K experiment reported in
Ref.~\cite{T2K2016}. The resulting interaction model is referred to as the
\textit{MicroBooNE Tune} in the remainder of this article.

\subsection{Detector simulation}

Final-state particles produced in GENIE events are transported through the
MicroBooNE detector geometry using version 10.3.3 of
Geant4~\cite{Geant4,Allison:2016lfl}. The response of the detector electronics
to these particles is modeled using a custom simulation implemented within the
\mbox{LArSoft} framework~\cite{Snider2017}. This detector response model
includes data-driven treatments of position dependence in the wire responses,
distortion of the time projection chamber (TPC) electric field due to buildup
of slow-moving positive ions (space-charge effects), and dynamically-induced
charge~\cite{uBSignal1,uBSignal2,uBNoise,uBSpaceCharge,uBLaser}.

Backgrounds generated by cosmic rays and other sources unrelated to the
neutrino beam are directly measured in the analysis using \textit{beam-off}
data samples collected at times when the BNB was not active. To ensure that
these backgrounds are properly treated when reconstructing simulated neutrino
interactions, an \textit{overlay} technique is used; simulated TPC wire and PMT
waveforms induced by GENIE neutrino scattering events are superimposed on
measured waveforms from beam-off data before analysis~\cite{cosmicRejection}.

\section{Signal definition and event selection}
\label{sec:selection}

A neutrino scattering event is considered part of the signal for this analysis
if it fulfills the following criteria:
\begin{enumerate}
\item \label{item:numuCC} A muon neutrino undergoes a CC interaction with an
argon nucleus.
\item \label{item:Np} The final state contains at least one proton.
\item \label{item:muonLimits} The momentum of the outgoing muon
lies within the interval [0.1, 1.2]~\si{\GeV\per\clight}.
\item \label{item:protonLimits} The momentum of the leading final-state
proton (i.e., the proton with the highest momentum) lies within
the interval [0.25, 1.0]~\si{\GeV\per\clight}.
\item \label{item:ZeroMesons} The final state contains zero mesons and zero
\mbox{antimesons}.
\end{enumerate}

The momentum limits on the muon and leading proton in signal
requirements~\ref{item:muonLimits} and \ref{item:protonLimits} are motivated by
considerations of efficiency, resolution, and systematic uncertainties. Apart
from adjustments to these momentum limits, the signal definition given above is
the same as the one adopted in the previous MicroBooNE
\ccnp~analysis~\cite{MCC8ccnpPaper}.

\subsection{Reconstruction workflow}

Identification of candidate \ccnp\ signal events in this article relies upon an
automated event reconstruction workflow implemented within the Pandora
multi-algorithm pattern-recognition toolkit~\cite{uBPandora}. The individual
hits reconstructed from the pulses recorded by each of the TPC wire planes are
clustered in 2D and then correlated to produce 3D particle candidates.
Collections of nearby particle candidates are grouped into \textit{slices}, and
information from both the TPC and PMTs is used to identify a maximum of one
slice per event that contains a candidate neutrino interaction. Algorithms to
calculate additional reconstructed quantities, such as the particle
identification score described in Ref.~\cite{LLRPID} and the muon momentum
estimator described in Ref.~\cite{uBMCS}, are combined with this generic
neutrino selection to perform the analysis.

\subsection{Event categories}
\label{sec:categories}

Plots of reconstructed event distributions in Fig.~\ref{fig:toposcore} and
Figs.~\ref{fig:DataMCblock0}--\ref{fig:DataMCMultipleBlocks2} show the
prediction of the MicroBooNE simulation as a stacked histogram that
distinguishes between the following categories of events:
\begin{description}
\item[Signal] All events which satisfy signal requirements \ref{item:numuCC} to
\ref{item:ZeroMesons} and have a true neutrino vertex position that falls
within the fiducial volume defined in selection
requirement~\ref{item:sel_vtx_in_FV} from Sec.~\ref{sec:preselCCincl}. Signal
events are divided into subcategories based on whether the primary interaction
mode is QE, 2p2h, or any other process. The last of these subcategories is
dominated by resonant pion production followed by intranuclear pion absorption.
\item[Out FV] Events in which the true neutrino vertex falls outside of the
fiducial volume (FV). This category includes all final-state topologies since
it is defined solely in terms of the true vertex location.
\item[CC$N\pi$] Events containing a charged-current $\nu_\mu$ interaction which
produces one or more final-state pions of any charge. These fail to satisfy
signal requirement~\ref{item:ZeroMesons}. Note that $\nu_\mu$ CC events
containing final-state mesons other than pions fall into another event
category.
\item[CC0$\pi$0$p$] Charged-current $\nu_\mu$ events containing zero
final-state pions of any momentum and zero protons within the true momentum
limits imposed by signal requirement~\ref{item:protonLimits}. Events with any
number of mesons other than pions are allowed in this category.
\item[Other CC] Charged-current $\nu_\mu$ events which do not fall into any of
the previous categories. Among these are events which would otherwise be signal
but do not satisfy the muon momentum limits imposed by signal
requirement~\ref{item:muonLimits}. This category also includes events
containing at least one proton within the momentum limits from signal
requirement~\ref{item:protonLimits} together with one or more mesons
other than pions.
\item[$\nu_e$ CC] Events involving a charged-current $\nu_e$ interaction
inside the fiducial volume. These fail to satisfy signal
requirement~\ref{item:numuCC}.
\item[NC] Events in which a neutrino or antineutrino of any flavor undergoes
a neutral-current interaction within the fiducial volume. These fail to
satisfy signal requirement~\ref{item:numuCC}.
\item[Beam-off] Cosmic-ray-induced backgrounds.
\item[Other] All events which do not fall into one of the other categories
defined above. This category is dominated by CC interactions of antineutrinos.
\end{description}

\subsection{Inclusive $\pmb{\nu_\mu}$ CC preselection}
\label{sec:preselCCincl}

Following automated reconstruction using the Pandora framework, twelve
event-selection criteria are applied to distinguish the \ccnp~signal from other
event topologies. These criteria are divided into a preselection that seeks to
identify inclusive $\nu_\mu$ CC events (described in the following paragraphs),
quality checks to eliminate events with a poorly-reconstructed final-state muon
(Sec.~\ref{sec:MuQualityCuts}), and additional requirements designed to isolate
mesonless final states containing one or more protons (Sec.~\ref{sec:selCCNp}).
Overall performance of the selection is then discussed in
Sec.~\ref{sec:overall_selection}, including evolution of the efficiency and
purity as each requirement is applied.

In order of application, the selection criteria used to identify
$\nu_\mu$ CC candidate events are as follows:
\begin{enumerate}[label=(\Roman*)]
\item \label{item:sel_vtx_in_FV} The position of the reconstructed neutrino
vertex must lie within a fiducial volume representing a region in which the
efficiency of the remainder of the selection is appreciable and the detector
response is well-understood. The fiducial volume chosen for this analysis is a
rectangular prism with most boundaries parallel to and \SI{21.5}{\centi\meter}
inward from the edges of the active volume of the MicroBooNE detector. The only
exception is the boundary that is perpendicular to the beam on the downstream
side. This boundary is chosen to be \SI{70}{\centi\meter} inward from the edge
of the active volume to enable better acceptance of forward-going muons.
\item \label{item:sel_containment_vol} Starting points for all reconstructed
primary particles (i.e., those labeled by the reconstruction as direct
progeny of the neutrino) are required to lie within a looser containment
volume offset inward from the active volume edges by \SI{10}{\centi\meter} on
all sides.
\item \label{item:toposcore} The topological score~\cite{WouterThesis} assigned
to the event must be greater than \TopoScoreCut. This score represents the
output of a support-vector machine designed to classify events as either
neutrino-like (scores near 1) or cosmic-ray-like (scores near 0).
Figure~\ref{fig:toposcore} shows the measured distribution of this variable,
obtained at this stage of the selection,
compared to the prediction from the MicroBooNE simulation chain. The
topological score threshold of \TopoScoreCut~is sufficient to remove a large
fraction of beam-off and out-of-fiducial-volume backgrounds.
\begin{figure}
\centering
\includegraphics[width=0.49\textwidth]{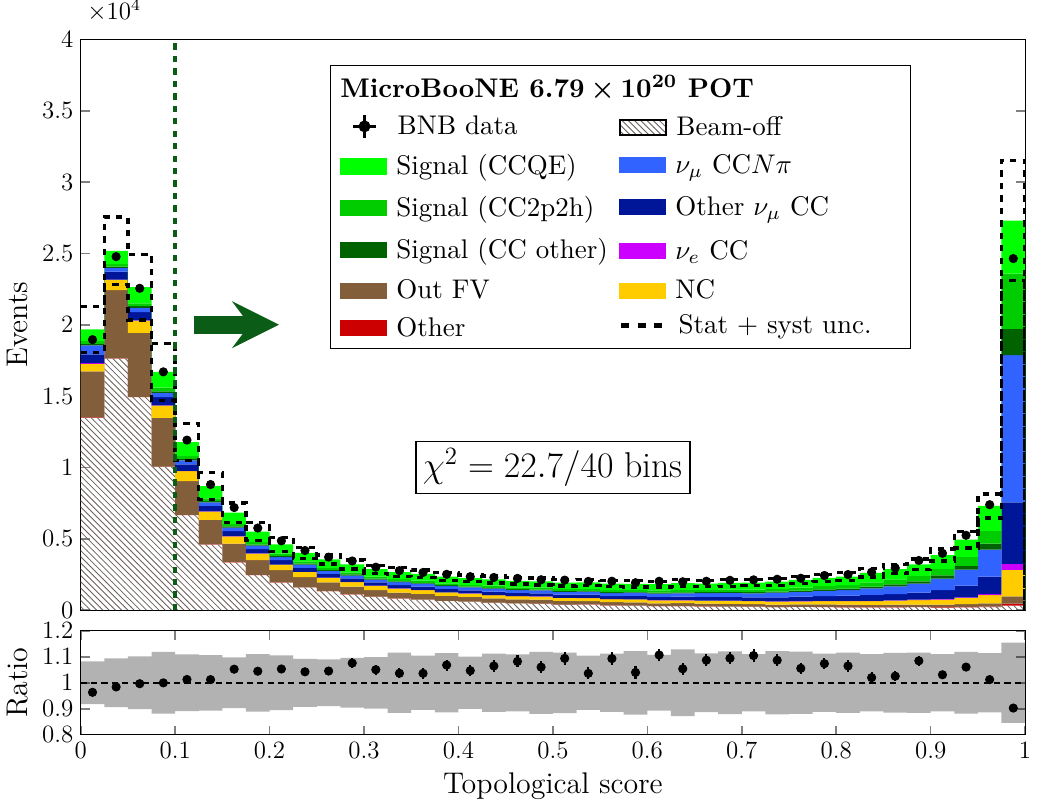}
\caption{Distribution of topological score values obtained when applying the
$\nu_\mu$ CC inclusive preselection. The portion near zero is dominated by
beam-off events as expected. The green arrow points into the region in which
events are accepted by selection requirement~\ref{item:toposcore}. The dashed
lines indicate the total uncertainty on the simulation prediction evaluated
according to the prescription given in Sec.~\ref{sec:unc}. The lower panel
shows the ratio of the data to the simulation prediction. The same total
uncertainty band is represented by the shaded gray region.}
\label{fig:toposcore}
\end{figure}
\item \label{item:has_mu_candidate} The event must contain a muon candidate,
which is defined as a reconstructed primary particle that satisfies the
following criteria:
  \begin{enumerate}[label=\alph*.]
    \item The track score~\cite{WouterThesis,MicroBooNE:2021wad} assigned to
the particle by the reconstruction must be greater than \MuonTrackScoreCut.
This score classifies reconstructed particles as shower-like (values near 0)
and track-like (values near 1). This requirement slightly increases the muon
purity (see Fig.~\ref{fig:trk_score_cut_muon}). The ``cosmic'' (``beam-off'')
category in the legend represents reconstructed particle candidates attributed
to cosmic-ray activity in events that include (do not include) a simulated
neutrino interaction. All other categories represent reconstructed particle
candidates that correspond to a specific kind of true particle produced by a
simulated neutrino interaction.

    \begin{figure}
    \centering
    \includegraphics[width=0.49\textwidth]{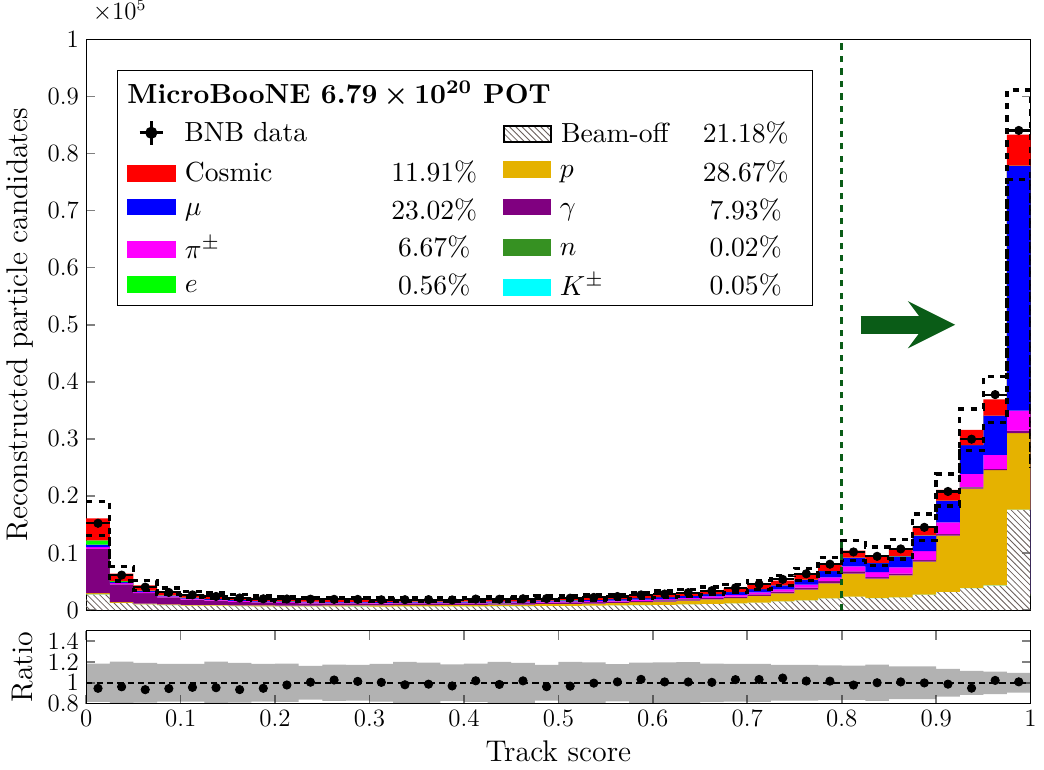}
    \caption{Track score criterion applied to all reconstructed primary
particles when searching for a muon candidate. The green arrow points into the
region in which particle candidates are accepted. The percentages in the legend
indicate the fraction of reconstructed particles of each type before the
illustrated selection is applied. The lower panel shows the ratio of the data
to the simulation prediction. The uncertainty on the simulation prediction is
represented by the dashed lines in the top panel and the gray region in the
bottom panel.}
    \label{fig:trk_score_cut_muon}
    \end{figure}

    \item The Euclidean distance between the particle's reconstructed starting
position and the reconstructed neutrino vertex must be less than
\MuonVtxDistanceCut\ (see Fig~\ref{fig:trk_start_cut_muon}). This is a minor
quality check to ensure that the muon candidate is correctly associated with
the reconstructed neutrino vertex.

    \begin{figure}
    \centering
    \includegraphics[width=0.49\textwidth]{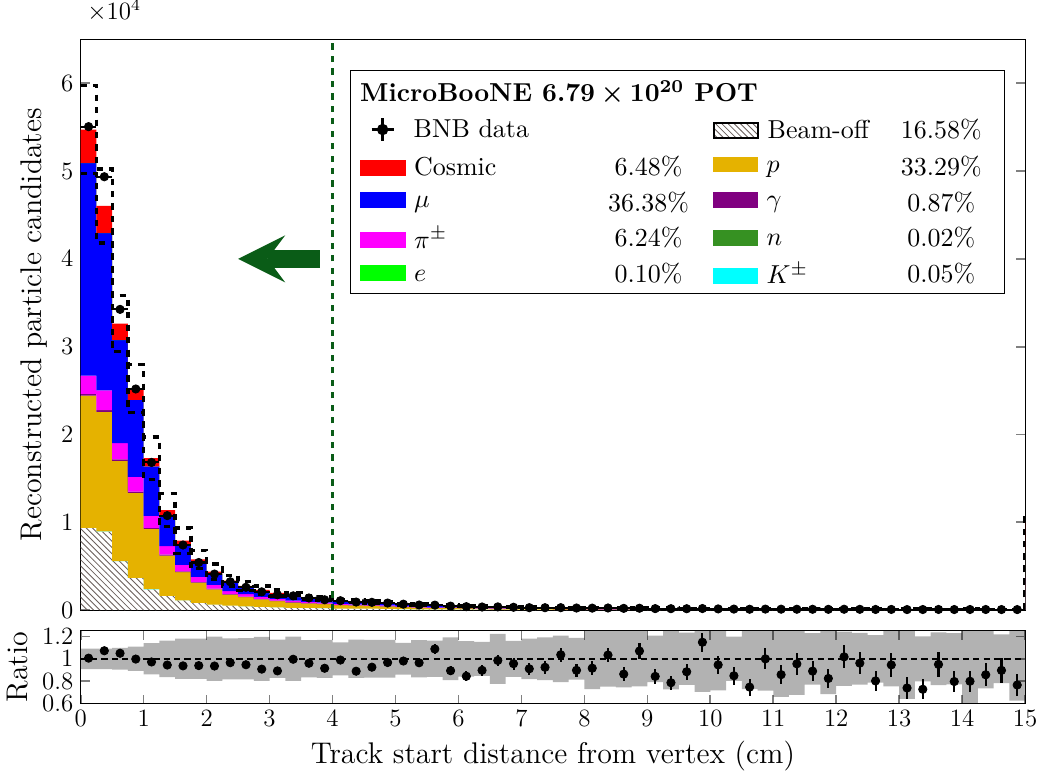}
    \caption{Distribution of distances between the reconstructed neutrino
vertex and reconstructed primary particles. The green arrow points into the
region in which particles are accepted as possible muon candidates. The
percentages in the legend indicate the fraction of reconstructed particles of
each type before the illustrated selection is applied. The lower panel shows
the ratio of the data to the simulation prediction. The uncertainty on the
simulation prediction is represented by the dashed lines in the top panel and
the gray region in the bottom panel.}
    \label{fig:trk_start_cut_muon}
    \end{figure}

    \item The length of the particle track must be greater than \MuonLengthCut\
(see Fig~\ref{fig:trk_length_cut_muon}). Tracks below this length are
overwhelmingly generated by protons and cosmic activity. The lower limit on the
muon momentum applied in signal requirement~\ref{item:muonLimits} helps to
mitigate the impact of this cut on the efficiency for very low-energy muons.

    \begin{figure}
    \centering
    \includegraphics[width=0.49\textwidth]{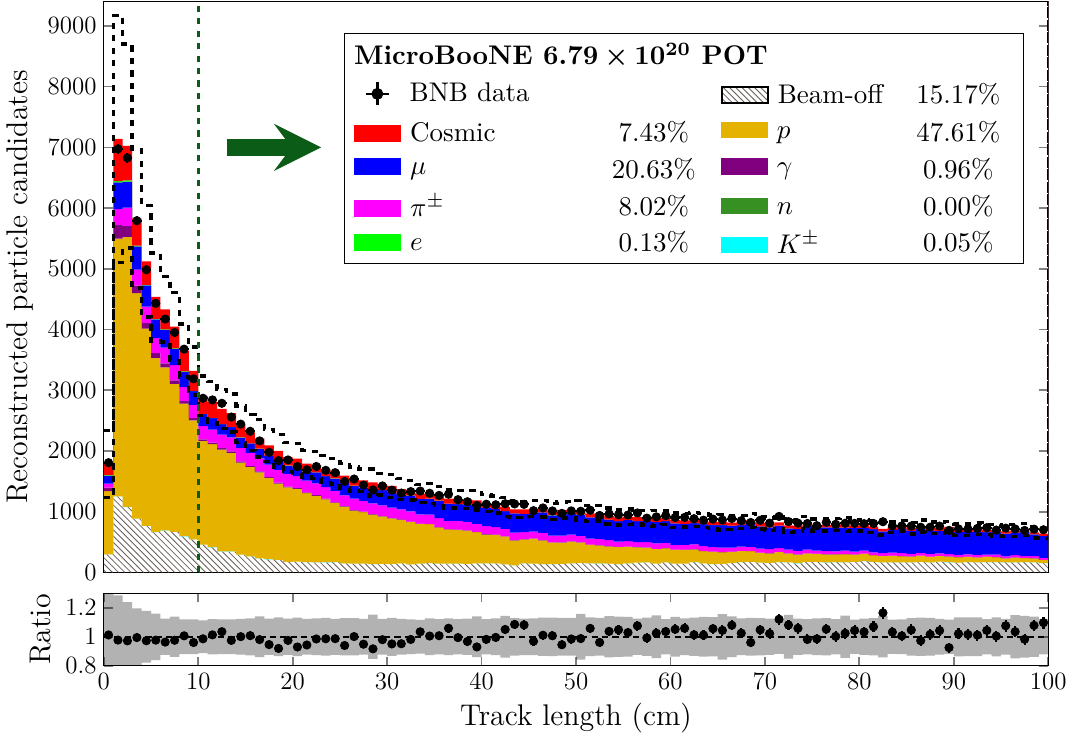}
    \caption{Track length criterion applied to reconstructed primary particles
when searching for a muon candidate. The green arrow points into the region in
which particle candidates are accepted. The percentages in the legend indicate
the fraction of reconstructed particles of each type before the illustrated
selection is applied. The lower panel shows the ratio of the data to the
simulation prediction. The uncertainty on the simulation prediction is
represented by the dashed lines in the top panel and the gray region in the
bottom panel.}
    \label{fig:trk_length_cut_muon}
    \end{figure}

    \item The log-likelihood ratio particle-identification (LLR PID) score
assigned to the particle must exceed \MuonLLRpidCut\ (see
Fig.~\ref{fig:llr_pid_cut_muon}). This score is calculated by comparing track
hit information from all three TPC wire planes to theoretical templates for
muons and protons. The logarithm of a likelihood ratio for these two particle
identification hypotheses is then converted to a score where -1 is most
proton-like and 1 is most muon-like. Reference~\cite{LLRPID} describes this
particle identification technique in greater \mbox{detail}.

    \begin{figure}
    \centering
    \includegraphics[width=0.49\textwidth]{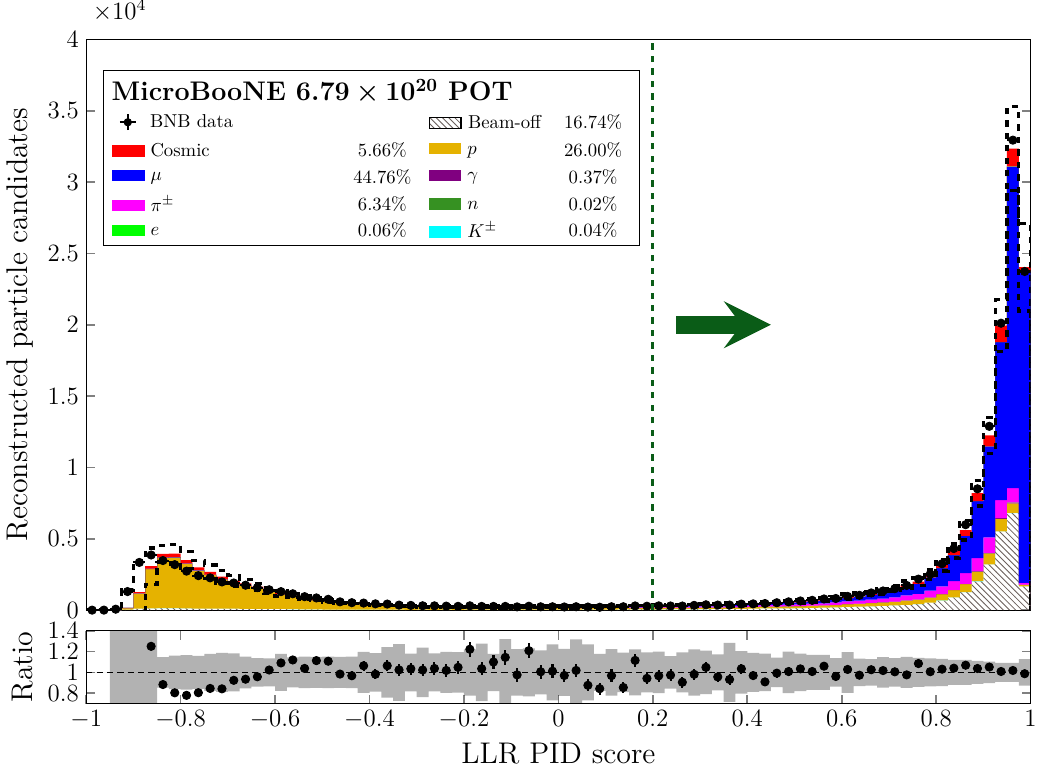}
    \caption{Log-likelihood ratio particle ID score for reconstructed primary
particles under consideration as possible muon candidates. The green arrow
points into the region in which muon candidates are accepted by the selection.
The percentages in the legend indicate the fraction of reconstructed particles
of each type before the illustrated selection is applied. The lower panel shows
the ratio of the data to the simulation prediction. The uncertainty on the
simulation prediction is represented by the dashed lines in the top panel and
the gray region in the bottom panel.}
    \label{fig:llr_pid_cut_muon}
    \end{figure}

  \end{enumerate}

In cases where two or more reconstructed primary
particles satisfy these criteria, the one with the highest LLR PID score (most
muon-like) is considered the muon candidate. \end{enumerate}

\subsection{Muon momentum reconstruction quality checks}
\label{sec:MuQualityCuts}
Three additional selection criteria are applied immediately after the $\nu_\mu$
CC preselection to ensure adequate reconstruction of the outgoing muon
momentum. In order of application, they are as follows:
\begin{enumerate}[label=(\Roman*)]
\setcounter{enumi}{4}
\item \label{item:muon_contained_sel} The muon candidate track must have a
reconstructed end point within the containment volume defined in selection
requirement~\ref{item:sel_containment_vol}. Requiring containment allows for a
reliable track-length-based estimate of the muon momentum. This criterion
substantially improves the resolution when reconstructing the muon momentum and
derived quantities (e.g., $\delta p_T$) at the cost of a notable drop in
efficiency.
\item Muon momentum estimators based on multiple Coulomb
scattering~\cite{uBMCS} ($p_\mu^\mathrm{MCS}$) and based on track
length~\cite{ProtonRangeMomentum} ($p_\mu^\mathrm{range}$) for the muon
candidate track must agree with each other within 25\%, i.e., they must satisfy
the relation
\begin{equation}
\frac{ \big| p_\mu^\mathrm{range} - p_\mu^\mathrm{MCS} \big|}
{ p_\mu^\mathrm{range} } < 0.25 \,.
\end{equation}
This requirement removes events in which only a portion of the full muon track
is successfully reconstructed.
\item \label{item:muon_thresh_sel} The track-length-based estimator for the
reconstructed muon momentum $p_\mu^\mathrm{range}$ must lie within the limits
imposed by requirement~\ref{item:muonLimits} from the signal definition:
$\MuonMinMomentumCut \leq p_\mu^\mathrm{range} \leq \MuonMaxMomentumCut$.
\end{enumerate}

\subsection{$\pmb{0\pi Np}$\ selection}
\label{sec:selCCNp}

The remaining five criteria from the full \ccnp\ selection are intended to
isolate final states containing zero mesons and one or more protons. In order
of application, they are
\begin{enumerate}[label=(\Roman*)]

\setcounter{enumi}{7}

\item All reconstructed primary particles must have a track score higher than
\TrackScoreCut\ (track-like). This requirement eliminates events containing
electromagnetic showers, which are not expected for mesonless $\nu_\mu$
interactions.

\item At least one reconstructed primary particle that is not the muon
candidate must be present in the event. All such particles are considered
proton candidates.

\item All proton candidates must have reconstructed end points that lie within
the containment volume defined in selection
requirement~\ref{item:sel_containment_vol}. This is a quality requirement
intended to ensure that a track-length-based estimator of the proton momentum
will be valid. Containment of the end point for the muon candidate is already
enforced by selection requirement~\ref{item:muon_contained_sel}.

\item \label{item:sel_proton_PID} All proton candidates must have an LLR PID
score less than \ProtonLLRpidCut\ (see Fig.~\ref{fig:llr_pid_cut_proton}). The
chosen cutoff value comes close to optimizing the product of efficiency and
purity while maintaining good acceptance of low-momentum protons.

    \begin{figure}
    \centering
    \includegraphics[width=0.5\textwidth]{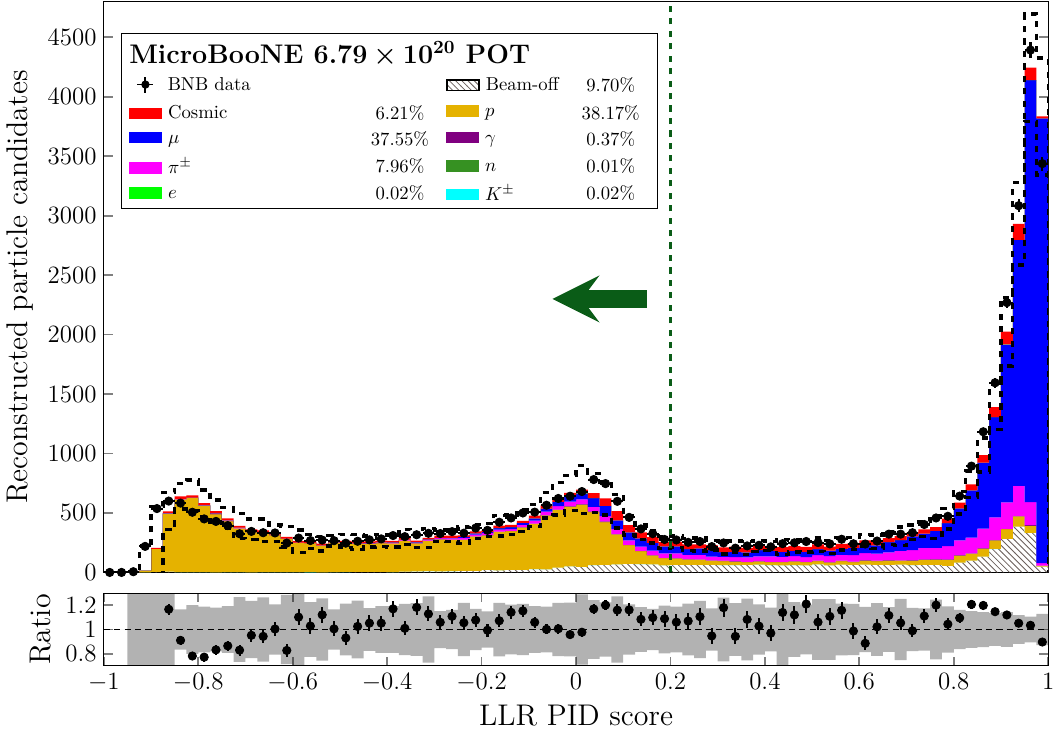}
    \caption{Log-likelihood ratio particle ID requirement applied to proton
candidates in selection requirement~\ref{item:sel_proton_PID}. The green arrow
points into the region in which proton candidates are accepted. The percentages
in the legend indicate the fraction of reconstructed particles of each type
before the illustrated selection is applied. The lower panel shows the ratio of
the data to the simulation prediction. The uncertainty on the simulation
prediction is represented by the dashed lines in the top panel and the gray
region in the bottom panel.}
    \label{fig:llr_pid_cut_proton}
    \end{figure}

\item \label{item:p_limits_sel} The longest proton candidate track
must have a length-based momentum estimator $p_p^\mathrm{range}$
that falls within the limits given in requirement~\ref{item:protonLimits}
from the signal definition: $\ProtonMinMomentumCut \leq p_p^\mathrm{range} \leq
\ProtonMaxMomentumCut$.
\xdef\numSelectionCuts{\theenumi}

\end{enumerate}

\subsection{Overall selection performance}
\label{sec:overall_selection}

Events that satisfy all twelve selection requirements described in the previous
subsections are considered the \ccnp\ candidates of interest for this analysis.
The full \ccnp\ selection achieves an estimated overall efficiency of 12.3\%
and a purity of 78.5\%. The evolution of these quantities as each requirement
is applied is shown in Table~\ref{tab:cuts}, as well as the relative efficiency
(``Rel. eff.''), which is defined as the ratio of the selection efficiency
after the current requirement is applied to the efficiency before it is
applied. Equivalently, it is the efficiency calculated considering only those
signal events accepted after applying all previous criteria.

\begin{table}
\centering
\caption{Evolution of the selection efficiency, purity, and relative efficiency as the various requirements described in the text are applied.}
\label{tab:cuts}
\includegraphics{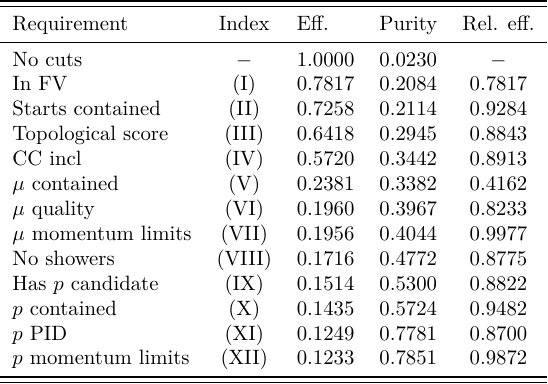}
\end{table}

\section{Data presentation strategy}
\label{sec:binning}

The selection defined in the previous section is used in the remainder of this
article to obtain single- and double-differential cross sections for a variety
of kinematic variables. To maximize the usefulness of these measurements for
the neutrino interaction modeling community, an innovative strategy for
reporting the results has been developed.

In previous MicroBooNE cross-section publications, and typically in the
experimental neutrino scattering literature as a whole, individual kinematic
distributions are treated as entirely separate entities. For example, in the
article describing the previous MicroBooNE \ccnp\
analysis~\cite{MCC8ccnpPaper}, single-differential cross sections are shown for
both the muon momentum and scattering angle (among several other observables),
but the covariance matrices used to report the measurement uncertainties
consider only correlations between bins of the same variable. Readers are thus
obliged to treat each reported differential cross section as if it was obtained
in isolation from all others. This reduces the power of the full multivariable
data set since external users must either ignore correlations between the
observables that are known to be important (e.g., those introduced by limited
data statistics and systematic uncertainties on the neutrino flux prediction),
invent ad hoc correlations that may be inaccurate, or consider only a single
variable at a time when performing model comparisons. This problem has been
encountered and discussed previously in the neutrino interaction
literature~\cite{MINERvAPionTune}, but standard practices for experimental data
releases have not yet been adjusted to address it.

The present analysis overcomes this limitation by providing comprehensive
uncertainties for all final results; cross sections are reported in a binning
scheme that involves multiple variables, and an overall covariance matrix is
computed that accounts for the measurement uncertainties and correlations
between all pairs of bins. To avoid double-counting in the cross-section
extraction procedure, bins describing the same kinematic distribution are
grouped into distinct blocks. As described in the sections that follow, some
mathematical operations are applied to each block independently, but the
blockwise results are later merged together to form the final group of
measurements.

\subsection{Observables measured}

Various observables related to the 3-momenta of the outgoing muon
($\mathbf{p}_\mu$) and leading proton ($\mathbf{p}_p$) are measured in this
analysis. Studies of sub-leading protons in the \ccnp\ channel are reserved for
later work. In addition to the magnitude of the momentum for the muon ($p_\mu$)
and leading proton ($p_p$) and their scattering angles ($\theta_\mu$ and
$\theta_p$, defined with respect to the neutrino direction
$+\hat{\mathbf{z}}$), differential cross sections in several other variables
are reported. These include the opening angle $\theta_{\mu p}$ between the muon
and leading proton
\begin{equation}
\theta_{\mu p} \equiv \arccos\left(\frac{ \mathbf{p}_\mu \cdot \mathbf{p}_p }
{ p_\mu \, p_p }\right) \,,
\end{equation}
as well as the magnitude $\delta p_T$ of the transverse missing momentum
\begin{equation}
\delta \mathbf{p}_T \equiv \mathbf{p}^T_\mu + \mathbf{p}^T_p \,.
\end{equation}
Here, $\mathbf{p}^T_\mu$ is the vector projection of the muon momentum
transverse to the neutrino beam direction and
$\mathbf{p}^T_p$ is the same for the leading proton. These vectors have
magnitudes $p^T_\mu$ and $p^T_p$ respectively. One may also describe the
angular orientation of $\delta \mathbf{p}_T$ with respect to $\mathbf{p}^T_\mu$
using the angle
\begin{equation}
\delta \alpha_T \equiv \arccos\left( \frac{ -\mathbf{p}^T_\mu \cdot
\delta \mathbf{p}_T }{ p^T_\mu \, \delta p_T } \right)
\end{equation}
as well as the vector components
\begin{equation}
\delta p_{T_x} \equiv \frac{ (\hat{\mathbf{z}} \times \mathbf{p}^T_\mu) \cdot
\delta \mathbf{p}_T }{ p^T_\mu } \,,
\end{equation}
and
\begin{equation}
\delta p_{T_y} \equiv \frac{ -\mathbf{p}^T_\mu \cdot
\delta \mathbf{p}_T }{ p^T_\mu } = \delta p_T \cdot \cos\delta \alpha_T \,,
\end{equation}
where the hat on $\hat{\mathbf{z}}$ indicates that it is a unit vector.

Measurements of the transverse kinematic imbalance (TKI) variables $\delta
p_T$, $\delta \alpha_T$ were originally proposed~\cite{TKIidea} as a means of
exploring nuclear effects in neutrino scattering with a reduced dependence on
the neutrino energy. Cross sections as a function of these variables for a
hydrocarbon target have been reported by the T2K~\cite{T2Kcc0piTKI} and
MINERvA~\cite{MinervaCC0piTKI} collaborations, the latter also
extending~\cite{MinervaCC0pipTComponents} the investigation to $\delta p_{T_x}$
and $\delta p_{T_y}$. In response to community
interest~\cite{Bathe-Peters:2022kkj} in measuring these observables using a
LArTPC, a recent MicroBooNE analysis reported first neutrino-argon differential
cross sections in TKI variables for an exclusive one-proton final
state~\cite{uBTKIPRD}, including the first double-differential TKI measurement
for neutrino scattering on any nuclear target~\cite{uBTKIPRL}. The present work
reports similar measurements for the more inclusive \ccnp\ signal event
topology.

Finally, this analysis measures an estimator $p_n$ for the momentum of the
initial struck neutron in the CC interaction. This estimator was proposed
in Ref.~\cite{FurmanskiPn} with similar phenomenological motivations
as the TKI variables, but it is calculated using momentum components
both transverse and longitudinal to the neutrino beam direction.
Specifically, $p_n$ is computed via the expression
\begin{equation}
p_n = \sqrt{ \delta p_L^2 + \delta p_T^2 }
\end{equation}
where the reconstructed longitudinal missing momentum
$\delta p_L$ is given by
\begin{equation}
\delta p_L \equiv \frac{ R }{ 2 } - \frac{ m_f^2 + \delta p_T^2 }{ 2 R } \,.
\end{equation}
Here
\begin{equation}
m_f \equiv m_{\mathrm{Ar40}} - m_n + B
\end{equation}
and
\begin{equation}
R \equiv m_{\mathrm{Ar40}} + p_\mu^z + p_p^z - E_\mu - E_p \,,
\end{equation}
where $m_{\mathrm{Ar40}} = 37.215526$~\si{\GeV\per\clight\squared} is the mass
of the \isotope[40]{Ar} nuclear target (not the atomic mass), $m_n =
0.93956541$~\si{\GeV\per\clight\squared} is the mass of a neutron, and $B =
0.02478$~\si{\GeV} is the assumed average binding energy for a neutron in
\isotope[40]{Ar}. The longitudinal momentum components and total energies of
the muon ($p_\mu^z$, $E_\mu$) and leading proton ($p_p^z$, $E_p$) are also used
above.

The value of $B$ chosen in the present work is a weighted average of the
neutron separation energies $E_\alpha$ given in Table~I of
Ref.~\cite{FurmanskiPn} with the shell occupancies $n_\alpha$ used as the
weights. An equivalent prescription for carbon (using Table~II from the same
reference) was adopted to estimate $B$ in a prior MINERvA analysis that studied
$p_n$~\cite{MinervaCC0piTKI}. The first measured neutrino cross sections in
this observable for an argon target were recently reported by MicroBooNE for a
one-proton final state~\cite{uBGKIPRD} assuming a different value $B =
\SI{0.0309}{\GeV}$ based on a study of electron scattering
data~\cite{Bodek:2018lmc}.

\subsection{Estimation of inefficiency and bin migrations}
\label{sec:bin_migrations}

Corrections for imperfect detector and reconstruction performance in the
analysis are applied using an unfolding procedure documented in
Sec.~\ref{sec:unfold}. The implementation of this procedure requires a
description of the relationship between reconstructed and true values of the
observables of interest. This relationship is estimated quantitatively by
assigning simulated \ccnp~events to two duplicate sets of the bins defined in
Table~\ref{tab:bin_defs} of Appendix~\ref{sec:bin_defs}. The \textit{true bins}
contain all simulated \ccnp~events that have true values of the observables
that fall within the relevant kinematic limits. The \textit{reconstructed bins}
contain those events that pass the full selection and have appropriate
reconstructed values of the observables.

With these definitions, the connection between the reconstructed and true
observables is described using a \textit{response matrix} $\ResponseMatrix$,
which transforms a prediction of signal event counts in true bins $\trueBinIdx$
into a corresponding prediction in reconstructed bins $\recoBinIdx$. Each
element of the response matrix is computed from the MicroBooNE simulation
results according to the relation
\begin{equation}
\label{eq:ResponseMatrixElement}
\ResponseMatrix_{\recoBinIdx\trueBinIdx} \equiv
\frac{ \PredictedSignalEvtCount_{\recoBinIdx\trueBinIdx} }
{ \PredictedSignalEvtCount_{\trueBinIdx} } \,,
\end{equation}
where $\PredictedSignalEvtCount_{\trueBinIdx}$ is the total number of signal
events in true bin $\trueBinIdx$ and
$\PredictedSignalEvtCount_{\recoBinIdx\trueBinIdx}$ is the number of signal
events which fall simultaneously into true bin $\trueBinIdx$ and reconstructed
bin $\recoBinIdx$. Thus, the response matrix quantifies the effects of both
inefficiency and smearing of the observables due to imperfect reconstruction.
This matrix is employed in systematic uncertainty estimation and unfolding as
described in Secs.~\ref{sec:unc}~and~\ref{sec:unfold} respectively.
It is plotted in Sec.~IV of the supplemental materials.

One may also quantify the effect of smearing separately from inefficiency by
calculating the \textit{migration matrix} $\MigrationMatrix$. Each element of
this matrix is estimated from simulation via the expression
\begin{equation}
\label{eq:MigrationMatrixElement}
\MigrationMatrix_{\recoBinIdx\trueBinIdx}
  = \frac{ \PredictedSignalEvtCount_{\recoBinIdx\trueBinIdx} }
{ \sum_{\secondRecoBinIdx}
\PredictedSignalEvtCount_{\secondRecoBinIdx\trueBinIdx} } \,,
\end{equation}
where, to avoid double-counting, the sum over $\secondRecoBinIdx$ includes only
those reconstructed bins in the block of interest. The migration matrix element
$\MigrationMatrix_{\recoBinIdx\trueBinIdx}$ is thus an estimate of the
probability that a selected signal event belonging to true bin $\trueBinIdx$
will be assigned to reconstructed bin $\recoBinIdx$. Plots of the migration
matrices for each of the 14 blocks of bins defined in Table~\ref{tab:bin_defs}
are given in Sec.~II of the supplemental materials.

\section{Uncertainties}
\label{sec:unc}

Systematic uncertainties are estimated by modifying the nominal MicroBooNE
simulation and calculating resultant variations in the expected number of
selected events $\PredictedRecoEvtCount_{\recoBinIdx}$ in each reconstructed
bin $\recoBinIdx$. This expected event count includes contributions from both
simulated neutrino interactions and measured constant-in-time backgrounds from
beam-off data:
\begin{equation}
\label{eq:ExpectedRecoEvents}
\PredictedRecoEvtCount_{\recoBinIdx}
= \PredictedSignalEvtCount_{\recoBinIdx} + \EXTCount_{\recoBinIdx}
+ \PredictedBkgdCount_{\recoBinIdx} \,.
\end{equation}
Here the expected contents of the $\recoBinIdx$-th reconstructed bin are divided
into $\PredictedSignalEvtCount_{\recoBinIdx}$ simulated signal \ccnp\ events,
$\EXTCount_{\recoBinIdx}$ measured beam-off background events, and
$\PredictedBkgdCount_{\recoBinIdx}$ simulated beam-correlated background events.

Uncertainties are quantified using covariance matrices calculated according to
a multiple-universe procedure. Under this approach, the covariance between the
expected event counts $\PredictedRecoEvtCount_{\CovMatFirstIdx}$ and
$\PredictedRecoEvtCount_{\CovMatSecondIdx}$ in the $\CovMatFirstIdx$-th and
$\CovMatSecondIdx$-th reconstructed bins is represented by the matrix element
\begin{equation}
\label{eq:CovMat}
\CovMat_{\CovMatFirstIdx\CovMatSecondIdx} =
\frac{ 1 }{ \UnivCount } \sum_{\UnivIdx = 1}^{\UnivCount} \big(
\PredictedRecoEvtCount_{\CovMatFirstIdx}^\text{CV} -
\PredictedRecoEvtCount_{\CovMatFirstIdx}^{\UnivIdx} \big) \big(
\PredictedRecoEvtCount_{\CovMatSecondIdx}^\text{CV} -
\PredictedRecoEvtCount_{\CovMatSecondIdx}^{\UnivIdx} \big) \,.
\end{equation}
Here $\PredictedRecoEvtCount_{\CovMatFirstIdx}^\text{CV}$ is the total event
count in reconstructed bin $\CovMatFirstIdx$ predicted by the central-value
MicroBooNE simulation. The variable
$\PredictedRecoEvtCount_{\CovMatFirstIdx}^{\UnivIdx}$ is a prediction of the
same quantity computed based on an alternate simulation (i.e., in an alternate
\textit{universe}) in which some aspect of the models used to describe the
neutrino beam, particle interactions, and the detector response has been
changed from the adopted central value. The total number of alternate universes
$\UnivCount$ included when computing the average in Eq.~(\ref{eq:CovMat})
depends on the systematic effect of interest. In some cases, only a single
variation ($\UnivCount = 1$) is used. Covariance matrices
$\CovMat_{\CovMatFirstIdx\CovMatSecondIdx}$ are calculated individually for
each source of uncertainty. A total covariance matrix is then obtained by
summing the individual contributions.

\subsection{Sources of systematic uncertainty}
\label{sec:syst_sources}

Several classes of systematic uncertainties are considered in this analysis.
Those related to the BNB flux prediction, which include variations related to
the horn current and hadron production modeling, follow the treatment developed
by the MiniBooNE collaboration~\cite{MiniBooNEFlux}. Flux shape uncertainties
are handled in a way which allows the final cross-section results to be
reported based upon the predicted BNB $\nu_\mu$ flux (tabulated in the
supplemental materials) rather than the unknown true flux~\cite{KochDolan}.

The GENIE-based neutrino
interaction model and associated uncertainties used by MicroBooNE are
documented in a dedicated publication~\cite{MicroBooNEGENIETune}. Following
Ref.~\cite{uBTKIPRD}, an additional single-universe variation to the neutrino
interaction model is adopted in which an alternative event generator (version
19.02.2 of NuWro~\cite{Golan2012}) replaces GENIE in the MicroBooNE simulation
chain and is used to predict the reconstructed event counts.

Uncertainties related to propagation of the final-state particles emerging from
the neutrino interaction are calculated using the
Geant4Reweight~\cite{Calcutt2021} software package. Systematic variations of
the Geant4~\cite{Geant4,Allison:2016lfl} total cross section model for positive
pions, negative pions, and protons are considered. For protons, variations are
applied to the elastic and reaction (i.e., total inelastic) channels
separately. For pions, the elastic, quasielastic, absorption, single
charge-exchange, double charge-exchange, and pion production channels are
individually varied.

Systematic uncertainties on the detector response model are calculated for both
the MicroBooNE photon detection system and the time projection chamber. For the
former, three alternate universes are constructed in which the scintillation
light yield, attenuation, and Rayleigh scattering length are individually
varied. For the latter, space-charge effects, electron-ion recombination, and
data-driven modifications to the simulated wire
response~\cite{DetVarSystematics} are considered.

Fully-correlated fractional uncertainties are also applied to the beam-related
portion of the reconstructed events ($\PredictedSignalEvtCount_{\recoBinIdx} +
\PredictedBkgdCount_{\recoBinIdx}$) to account for limited precision in the
counting of protons delivered to the beam target (2\%) and the number of argon
atoms in the fiducial volume (1\%).

\subsection{Signal model uncertainty}
\label{sec:syst_signal}

For most of the systematic uncertainties mentioned above, the simulated
reconstructed event counts $\PredictedSignalEvtCount_{\recoBinIdx}$ and
$\PredictedBkgdCount_{\recoBinIdx}$ are varied directly in each alternate
universe. However, since uncertainties related to predicting the \ccnp\ signal
will only affect the final results via their impact on estimating the detection
efficiency and bin migrations, a different treatment is used for neutrino
interaction model variations. In this case, the expected number of selected
signal events $\PredictedSignalEvtCount_{\recoBinIdx}$ is rewritten in the form
\begin{equation}
\PredictedSignalEvtCount_{\recoBinIdx}
= \sum_{\trueBinIdx} \ResponseMatrix_{\recoBinIdx \trueBinIdx} \,
\PredictedSignalEvtCount_{\trueBinIdx}^\mathrm{CV} \,,
\end{equation}
where $\ResponseMatrix_{\recoBinIdx \trueBinIdx}$ is the response matrix element
connecting true bin $\trueBinIdx$ and reconstructed bin $\recoBinIdx$. This
quantity is varied in each alternate universe, while the expected signal event
counts in the $\trueBinIdx$-th true bin
$\PredictedSignalEvtCount_{\trueBinIdx}^\mathrm{CV}$ are held constant at the
central-value prediction (as indicated by the superscript). To avoid
double-counting events, the sum over $\trueBinIdx$ includes only those true bins
that belong to the same block as the $\recoBinIdx$-th reconstructed bin. Details
about the binning scheme and response matrix definition are provided above in
Sec.~\ref{sec:binning}.

\subsection{Statistical correlations}
\label{sec:stat_corr}

The data release strategy outlined in Sec.~\ref{sec:binning} requires a
slightly more complicated treatment of statistical uncertainties than has been
typical for previous neutrino cross-section analyses. When every measured event
belongs to exactly one bin, statistical fluctuations of the bin contents are
described by independent Poisson distributions. In that case, the statistical
covariance matrix is diagonal, and the variance of each bin is estimated by the
number of observed counts.

However, reporting combined measurements over the multiple blocks of bins
defined in Table~\ref{tab:bin_defs} introduces correlations in the statistical
uncertainties because the same event belongs to a unique bin in each block.
Fortunately, the derivation of an estimator for the statistical covariance
between two arbitrary bins is straightforward. As shown in Sec.~III$\,$C$\,$1
of Ref.~\cite{GardinerXSecExtract}, the statistical covariance between any pair
of bins is estimated as simply the number of events that fall simultaneously
into both of them. In the case of weighted Monte Carlo events, the sum of the
squares of the event weights is used rather than the raw event count.

\subsection{Impact on integrated cross section}
\label{sec:unc_integ_xsec}

The flux-averaged total \ccnp\ cross section obtained by this analysis is
$\sigma = (\num[round-mode=places, round-precision=1]{11.8278} \pm
\num[round-mode=places, round-precision=1]{1.1627}) \times
\SI{e-38}{\centi\meter\squared} \,/\, \mathrm{Ar}$. The full fractional
uncertainty of 9.8\% includes contributions from the neutrino flux prediction
(6.6\%), neutrino interaction modeling (5.1\%), detector response modeling
(4.1\%), beam exposure measurements (2.2\%), data statistics (1.4\%),
estimation of the number of argon atoms in the fiducial volume (1.1\%),
modeling of final-state particle propagation (0.9\%), and Monte Carlo
statistics (0.8\%). The data statistical uncertainty includes a contribution
from the beam-off measurements used to estimate cosmic-ray backgrounds. The
Monte Carlo statistical uncertainty includes a contribution from simulated
neutrino-induced backgrounds from the BNB.

For the total cross-section measurement, the additional neutrino interaction
modeling uncertainty estimated by using NuWro as an alternative event generator
(1.7\%) is substantially smaller than the uncertainty arising from GENIE
systematic variations (4.8\%). However, for many of the differential
measurements reported below, the NuWro-based systematic uncertainty is larger.
Full covariance matrices giving the contributions of both of these classes of
interaction model variations to the total measurement uncertainty are reported
in the supplemental materials. For the detector response model, the uncertainty
related to the simulation of electron recombination~\cite{recombArgoneut} has a
dominant impact on the total cross section result (3.3\%).

\subsection{Reconstructed event distributions}
\label{sec:reco_results}

As an intermediate step in the analysis, the event selection presented in
Sec.~\ref{sec:selection} and the uncertainty treatment described above were
applied to measure candidate \ccnp\ event rates in each of the
\TotalBinCount~kinematic bins defined in Table~\ref{tab:bin_defs}. The
measured event rates and associated uncertainties were then used as input
to the unfolding procedure defined in Sec.~\ref{sec:unfold} to obtain the
final cross-section results.

\begin{figure*}
\centering
\includegraphics[page=1, width=0.49\textwidth]{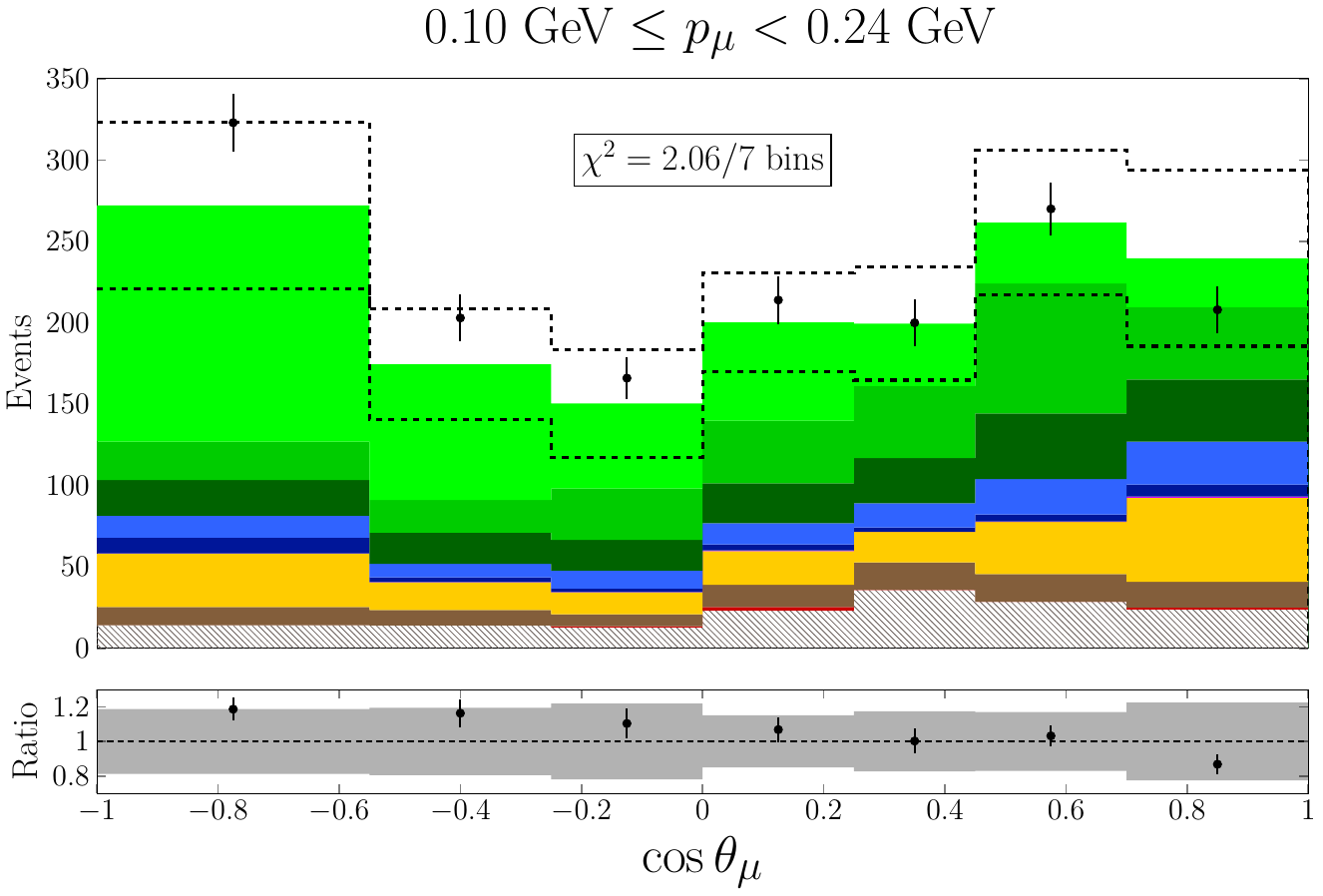}
\hfill
\includegraphics[page=2, width=0.49\textwidth]{figures/reco_results_prd.pdf}
\vspace{0.1cm}

\includegraphics[page=3, width=0.49\textwidth]{figures/reco_results_prd.pdf}
\hfill
\includegraphics[page=4, width=0.49\textwidth]{figures/reco_results_prd.pdf}
\vspace{0.1cm}

\includegraphics[page=5, width=0.49\textwidth]{figures/reco_results_prd.pdf}
\hfill
\includegraphics[page=6, width=0.49\textwidth]{figures/reco_results_prd.pdf}
\vspace{0.1cm}

\includegraphics[page=7, width=0.49\textwidth]{figures/reco_results_prd.pdf}
\hfill
\belowbaseline[-0.23\textheight]{
  \includegraphics[page=8, width=0.35\textwidth]{figures/reco_results_prd.pdf}%
}
\hfill

\caption{Reconstructed event distributions for block \#0
($p_\mu, \cos\theta_\mu$).
The bottom panel of each plot shows the ratio of the data to the MicroBooNE
simulation prediction. The uncertainty on the prediction is represented by the
dashed lines in each top panel and the gray region in each bottom panel.}
\label{fig:DataMCblock0}
\end{figure*}

\begin{figure*}
\centering
\includegraphics[page=9, width=0.49\textwidth]{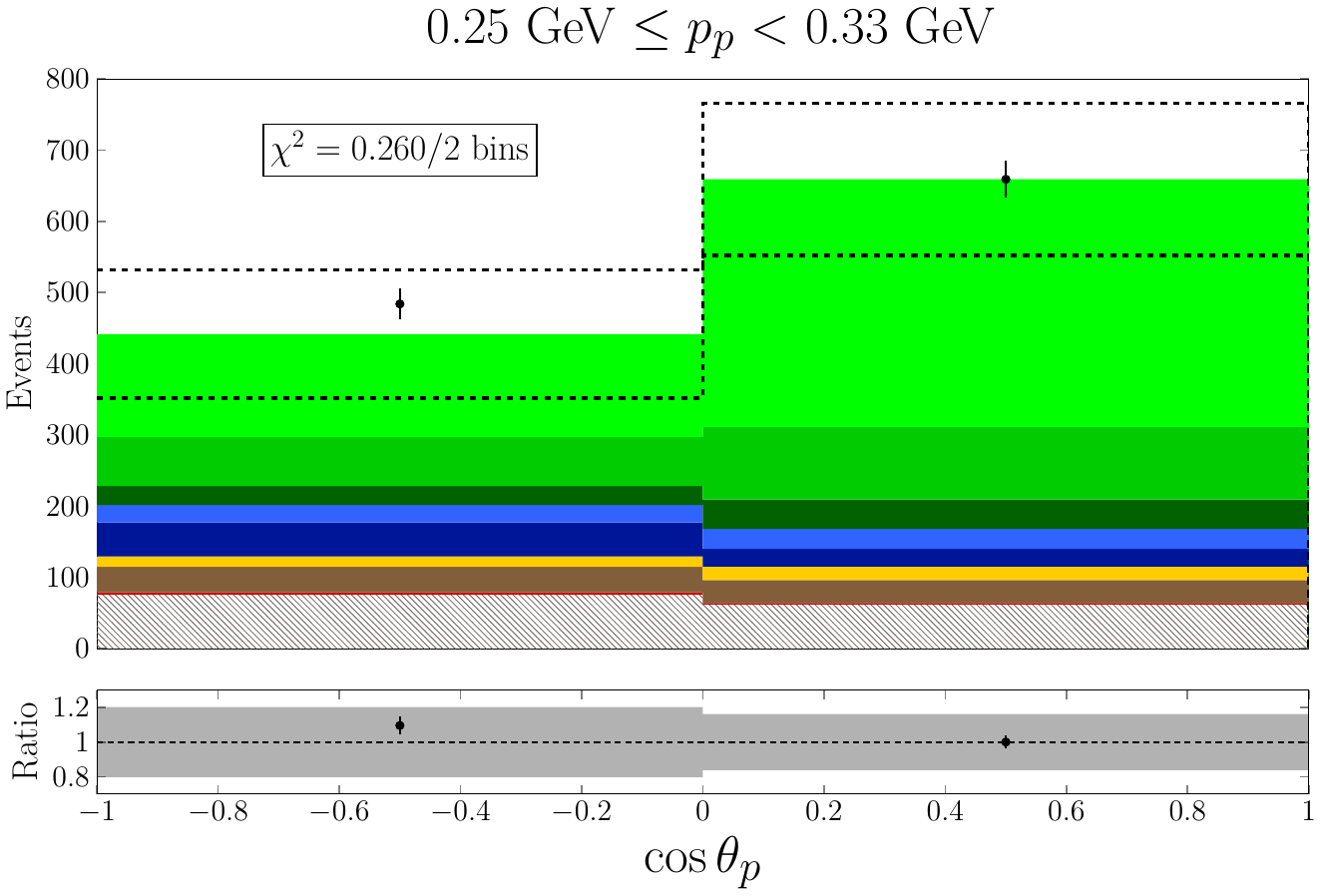}
\hfill
\includegraphics[page=10, width=0.49\textwidth]{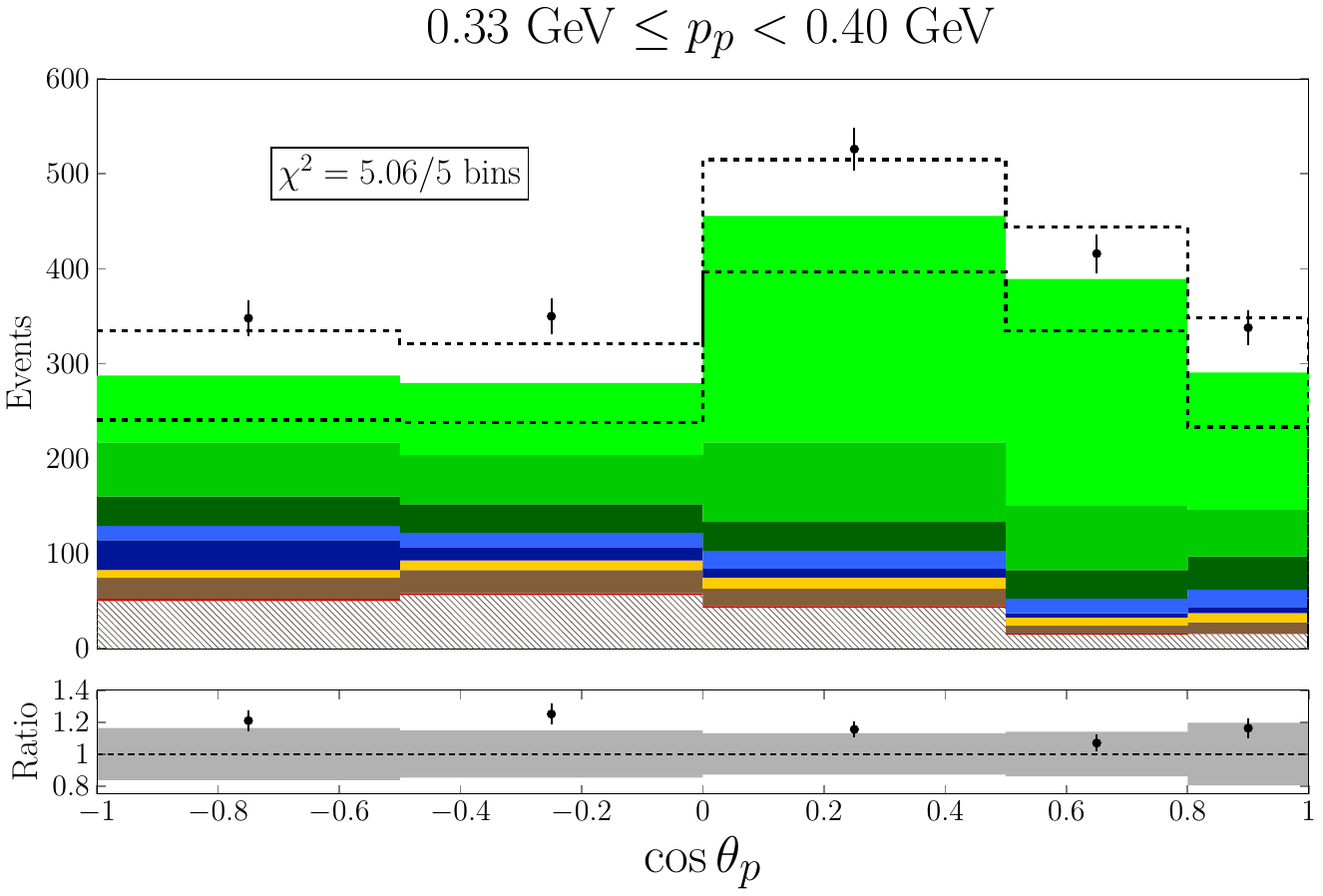}
\vspace{0.2cm}

\includegraphics[page=11, width=0.49\textwidth]{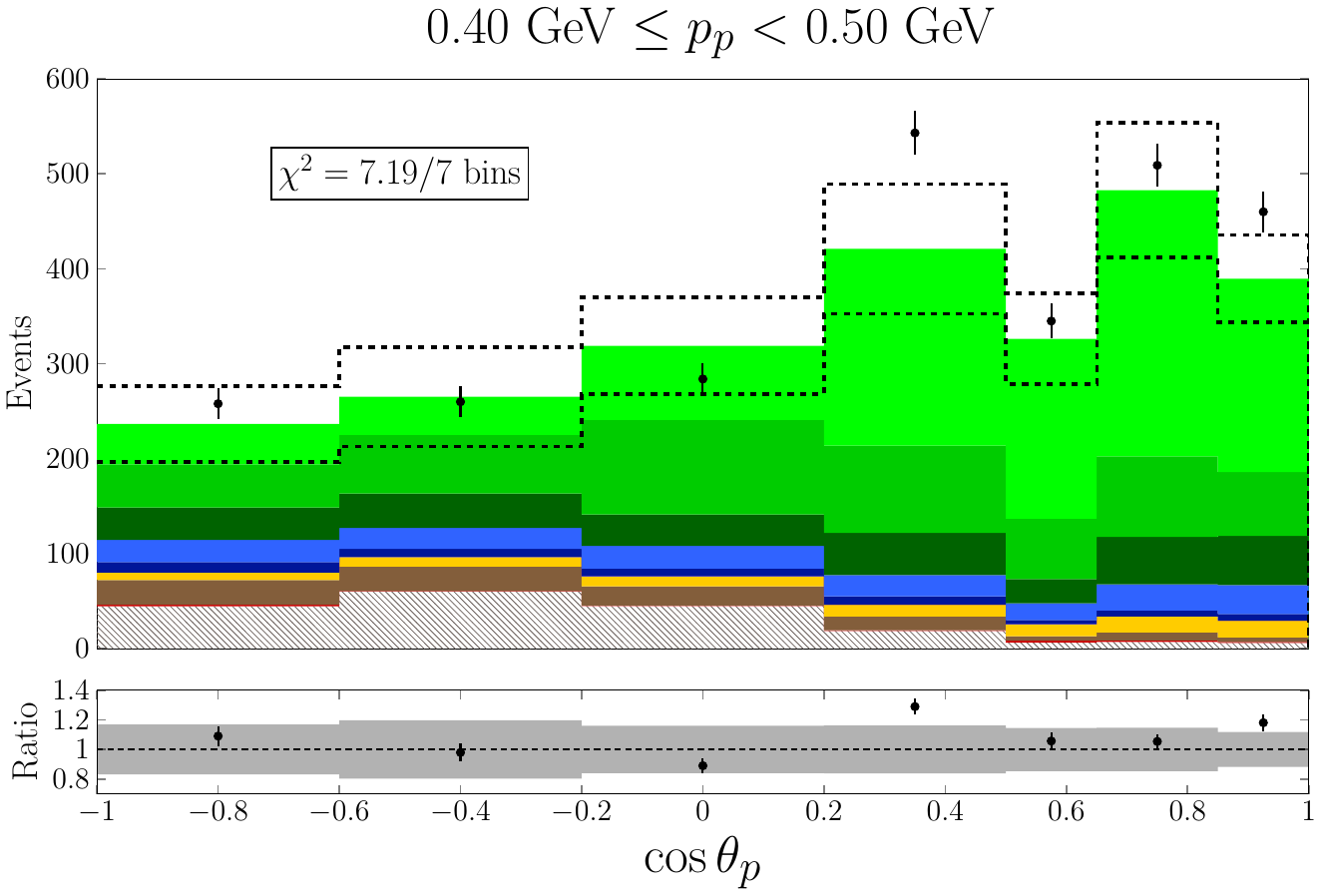}
\hfill
\includegraphics[page=12, width=0.49\textwidth]{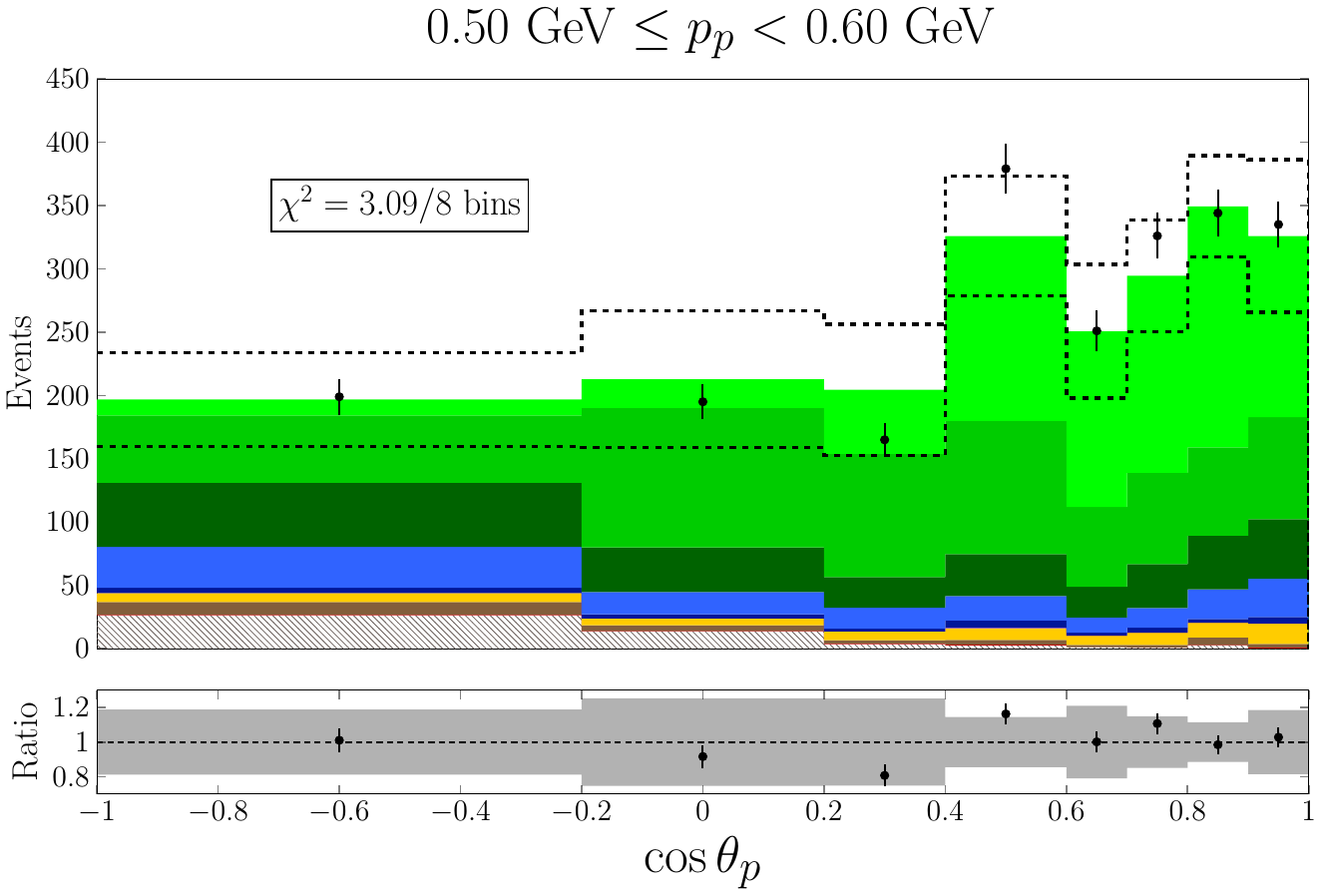}
\vspace{0.2cm}

\includegraphics[page=13, width=0.49\textwidth]{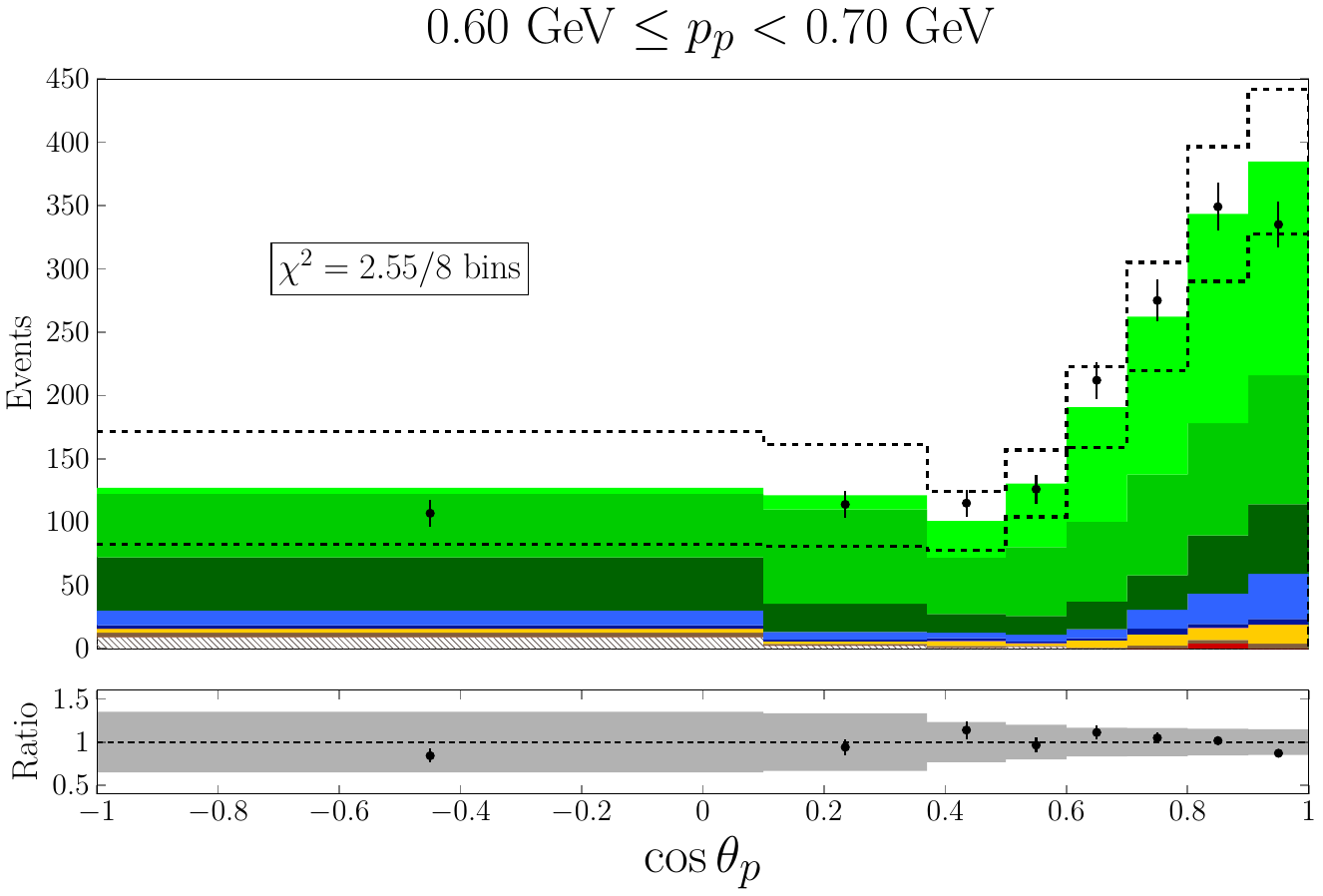}
\hfill
\includegraphics[page=14, width=0.49\textwidth]{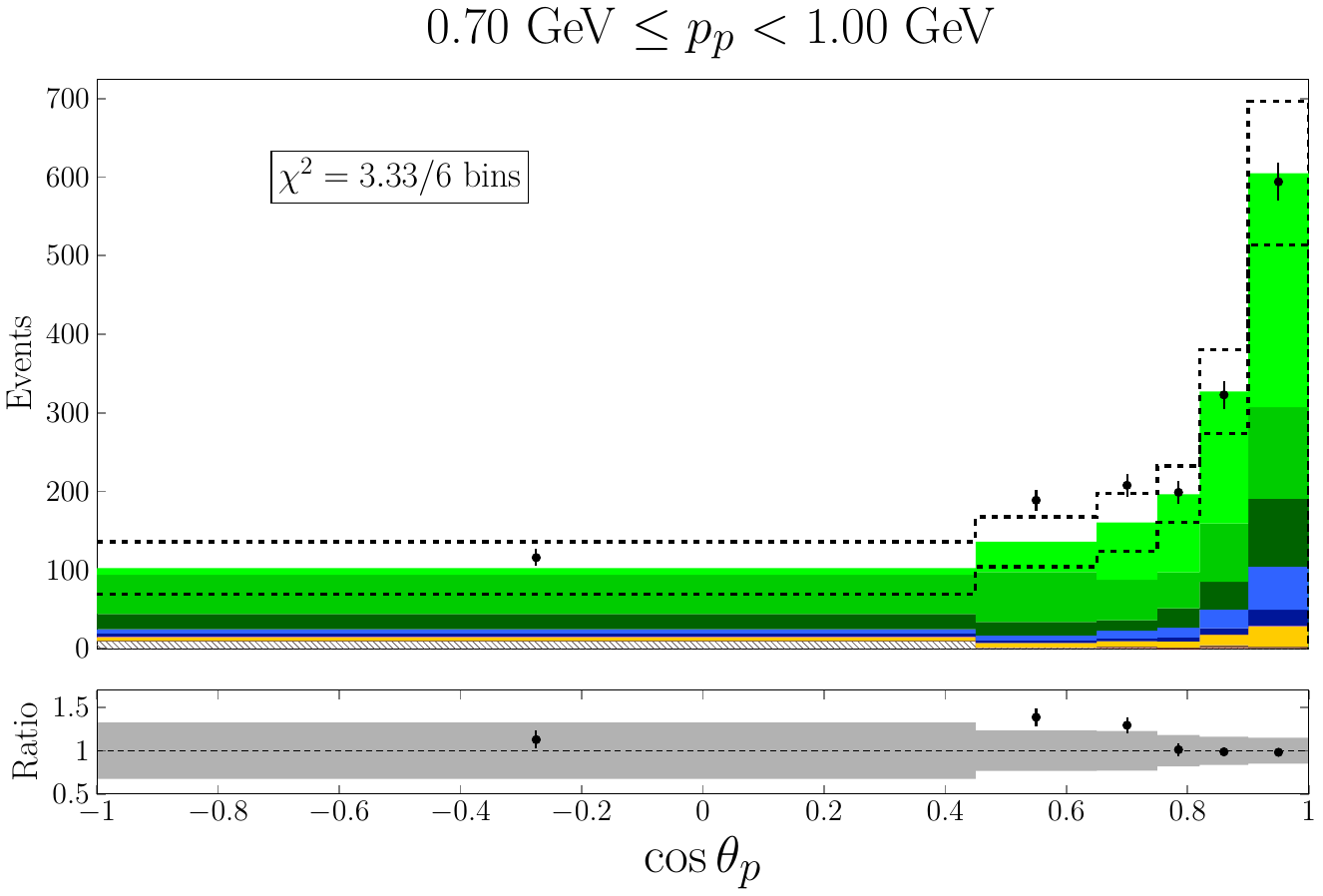}
\vspace{0.2cm}

\includegraphics[page=15, width=0.35\textwidth]{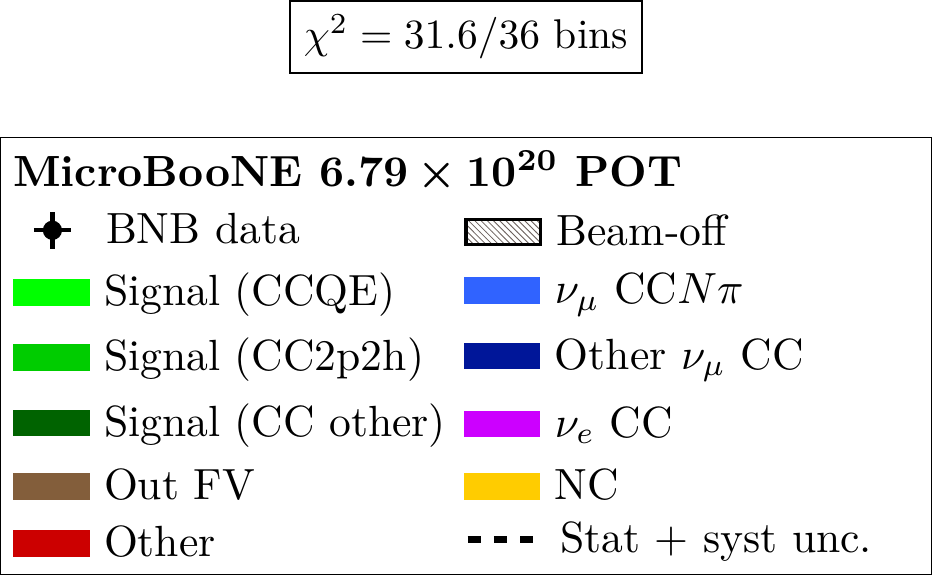}%
\hfill

\caption{Reconstructed event distributions for block \#1
($p_p, \cos\theta_p$).
The bottom panel of each plot shows the ratio of the data to the MicroBooNE
simulation prediction. The uncertainty on the prediction is represented by the
dashed lines in each top panel and the gray region in each bottom panel.}
\label{fig:DataMCblock1}
\end{figure*}

\begin{figure*}
\centering
  \subfloat[\label{subfig:block2}]{%
    \includegraphics[page=16, width=0.49\textwidth]{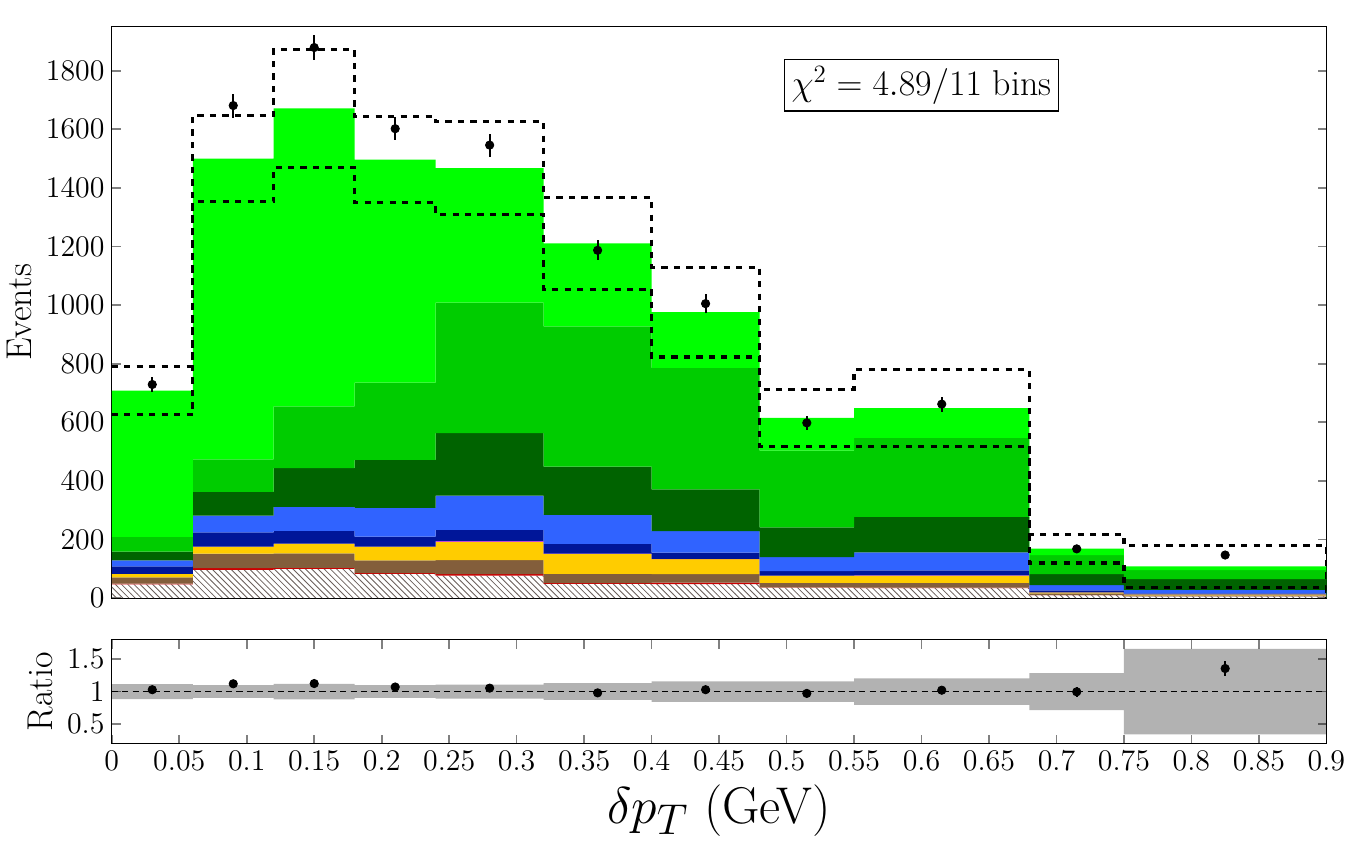}%
}\hfill
  \subfloat[\label{subfig:block5}]{%
    \includegraphics[page=28, width=0.49\textwidth]{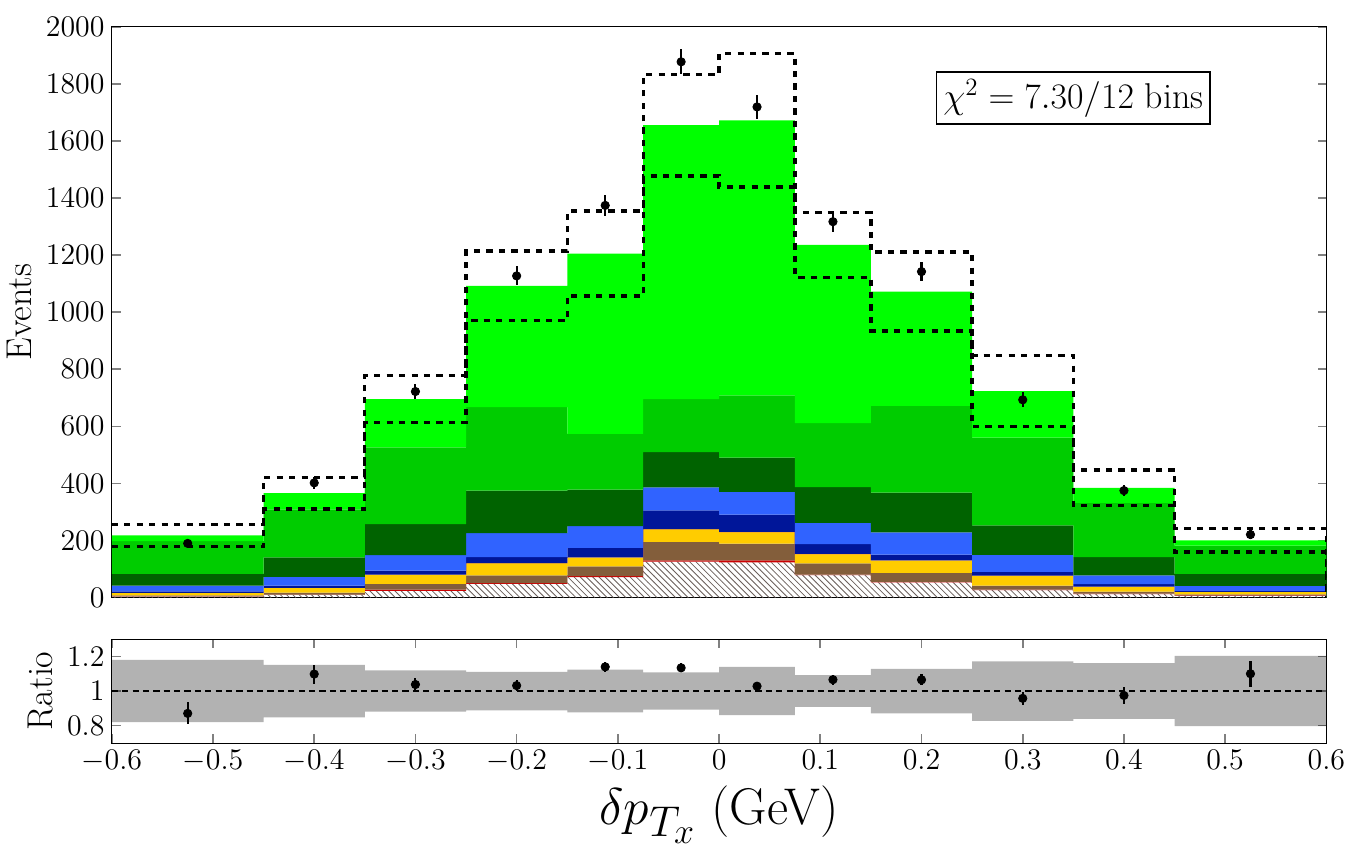}%
}\vspace{0.1em}
  \subfloat[\label{subfig:block7}]{%
    \includegraphics[page=33, width=0.49\textwidth]{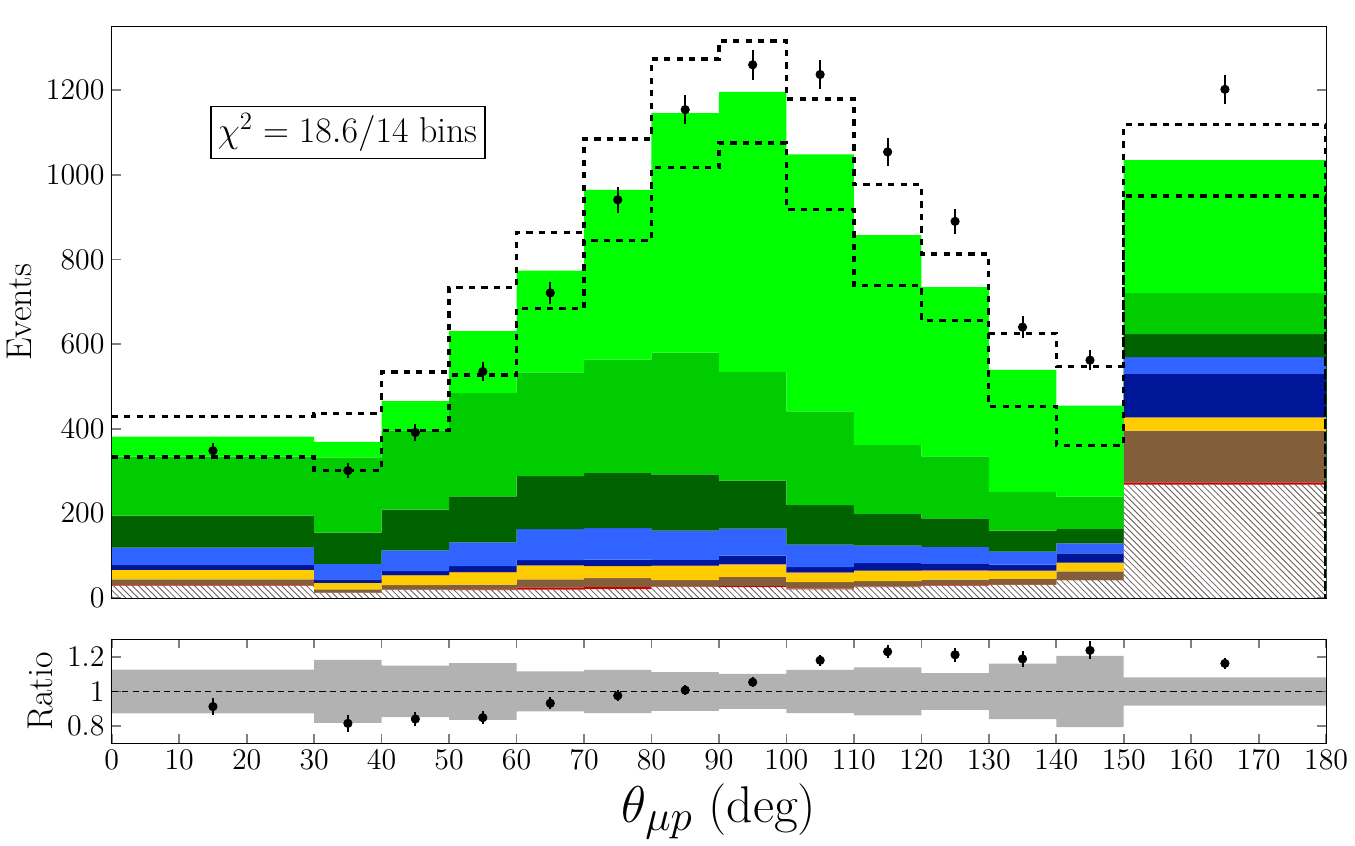}%
}\hfill
  \subfloat[\label{subfig:block8}]{%
    \includegraphics[page=34, width=0.49\textwidth]{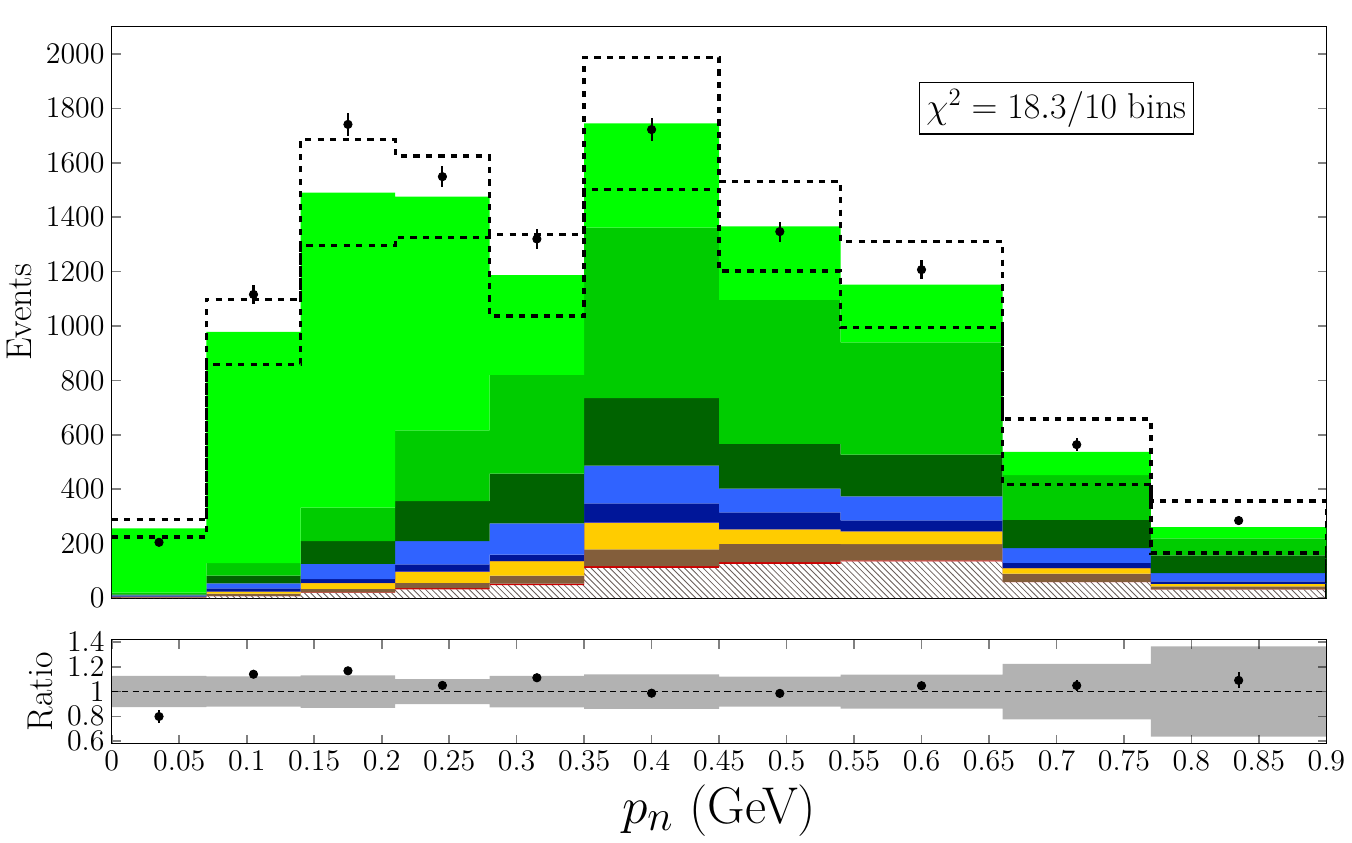}%
}\vspace{0.1em}
\includegraphics[page=17, width=0.35\textwidth]{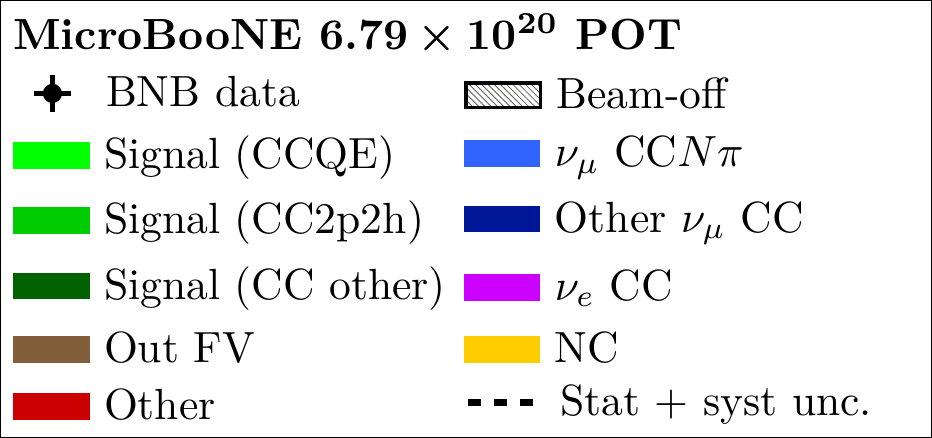}%
\hfill
\caption{Reconstructed event distributions for
(\subref*{subfig:block2})~block~\#2
($\delta p_T$),
(\subref*{subfig:block5})~block~\#5
($\delta p_{T_x}$),
(\subref*{subfig:block7})~block~\#7
($\theta_{\mu p}$), and
(\subref*{subfig:block8})~block~\#8
($p_n$).
The bottom panel of each plot shows the ratio of the data to the MicroBooNE
simulation prediction. The uncertainty on the prediction is represented by the
dashed lines in each top panel and the gray region in each bottom panel.}
\label{fig:DataMCMultipleBlocks1}
\end{figure*}

\begin{figure*}
\centering
\includegraphics[page=18, width=0.49\textwidth]{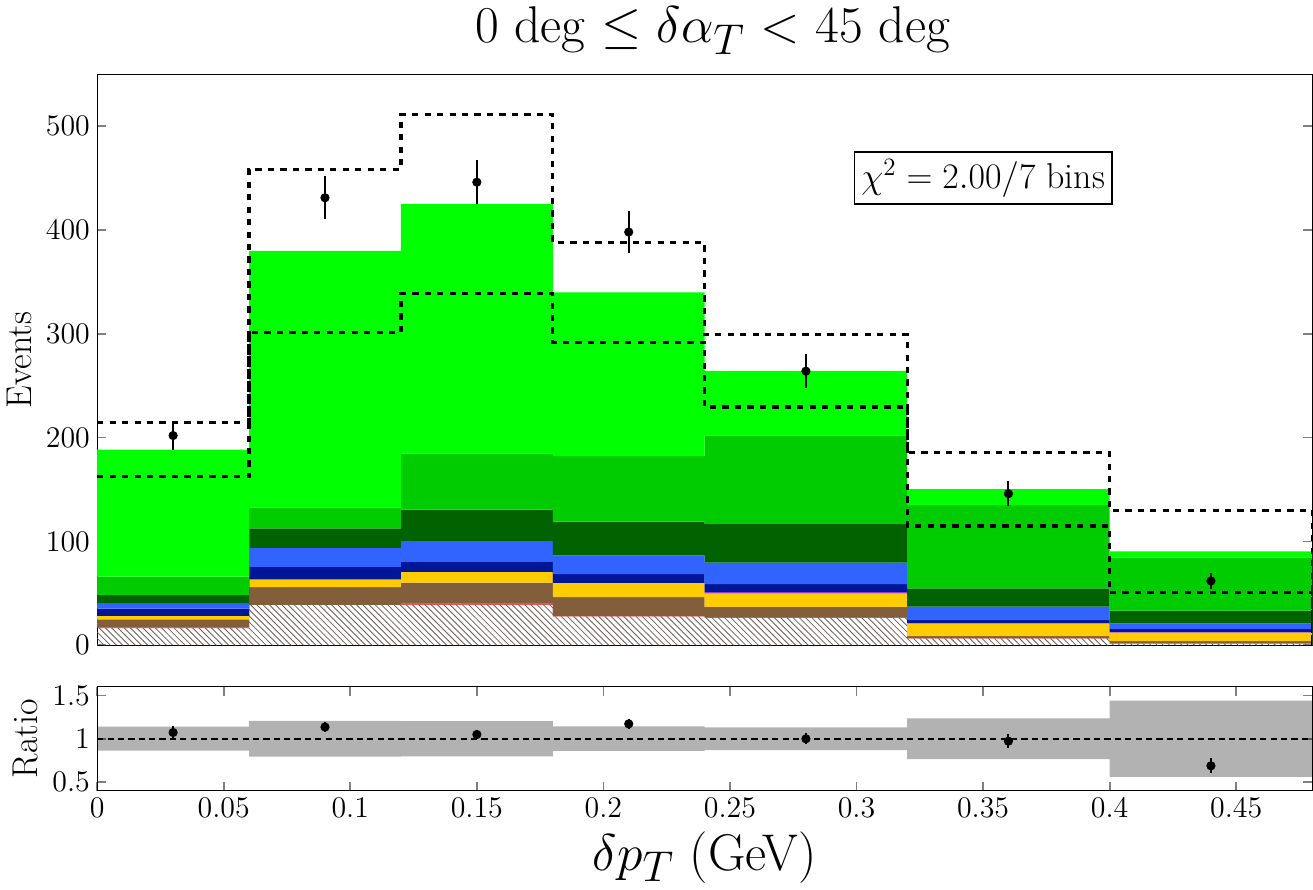}
\hfill
\includegraphics[page=19, width=0.49\textwidth]{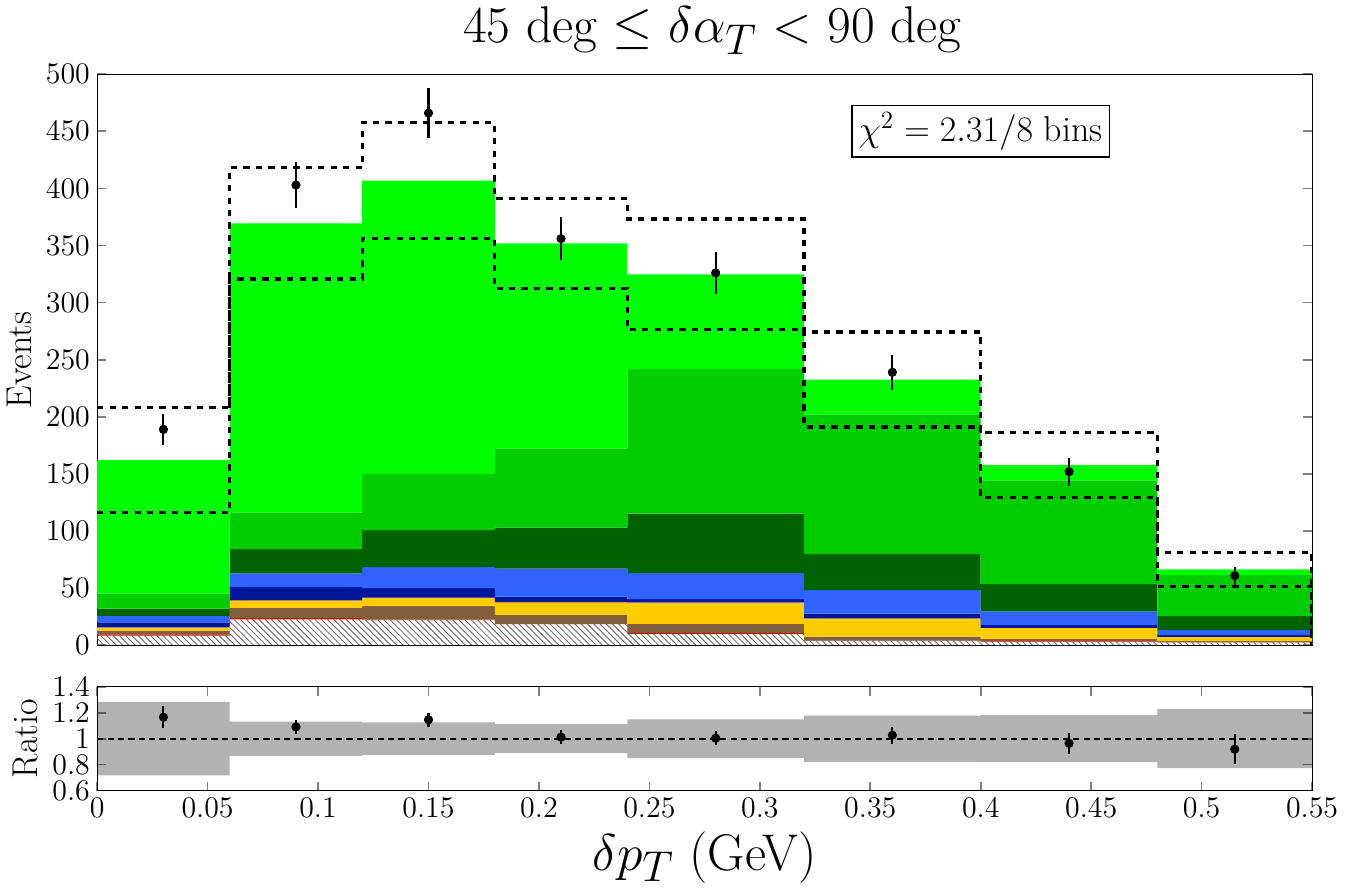}
\vspace{0.2cm}

\includegraphics[page=20, width=0.49\textwidth]{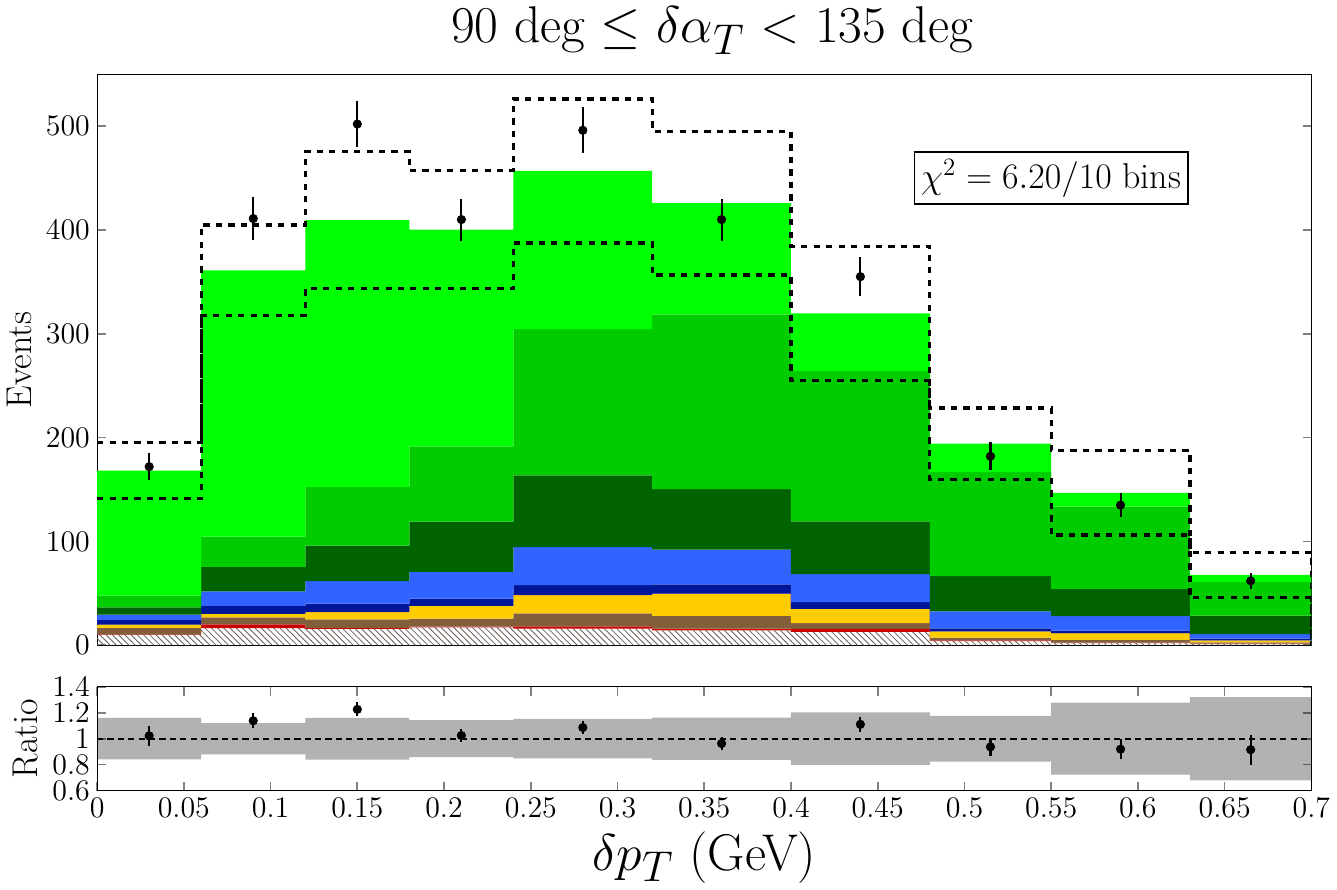}
\hfill
\includegraphics[page=21, width=0.49\textwidth]{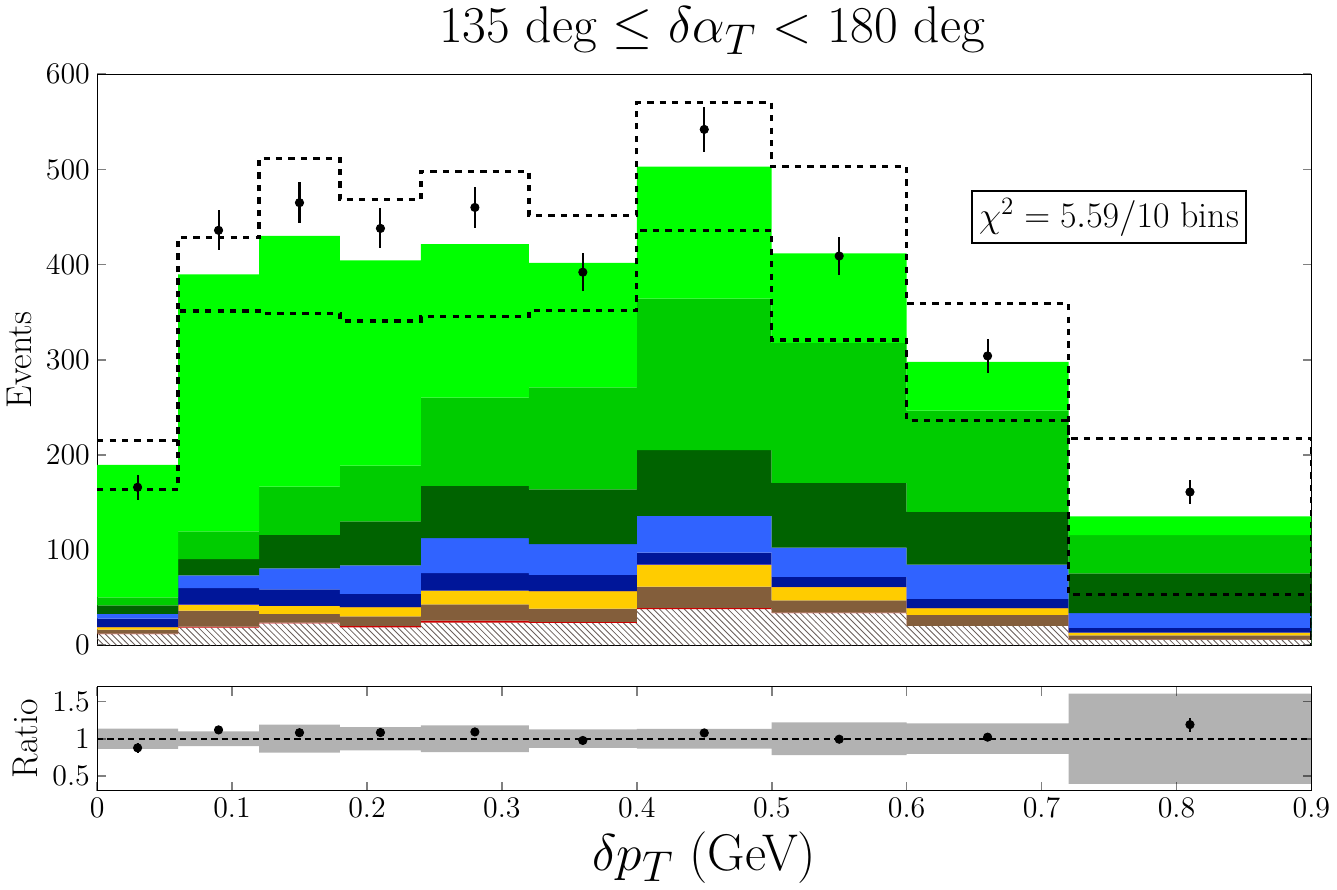}
\vspace{0.2cm}

\includegraphics[page=22, width=0.35\textwidth]{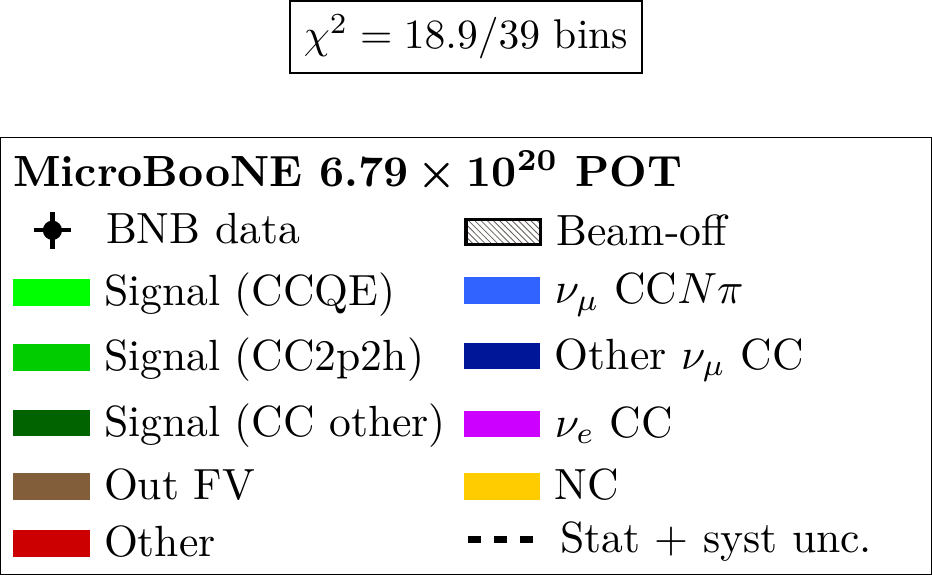}%
\hfill

\caption{Reconstructed event distributions for block \#3
$(\delta \alpha_T, \delta p_T)$. The overall $\chi^2$
value includes contributions from four $\delta p_T$ overflow bins that are not
plotted.
The bottom panel of each plot shows the ratio of the data to the MicroBooNE
simulation prediction. The uncertainty on the prediction is represented by the
dashed lines in each top panel and the gray region in each bottom panel.}

\label{fig:DataMCblock3}
\end{figure*}

\begin{figure*}
\centering
\includegraphics[page=23, width=0.49\textwidth]{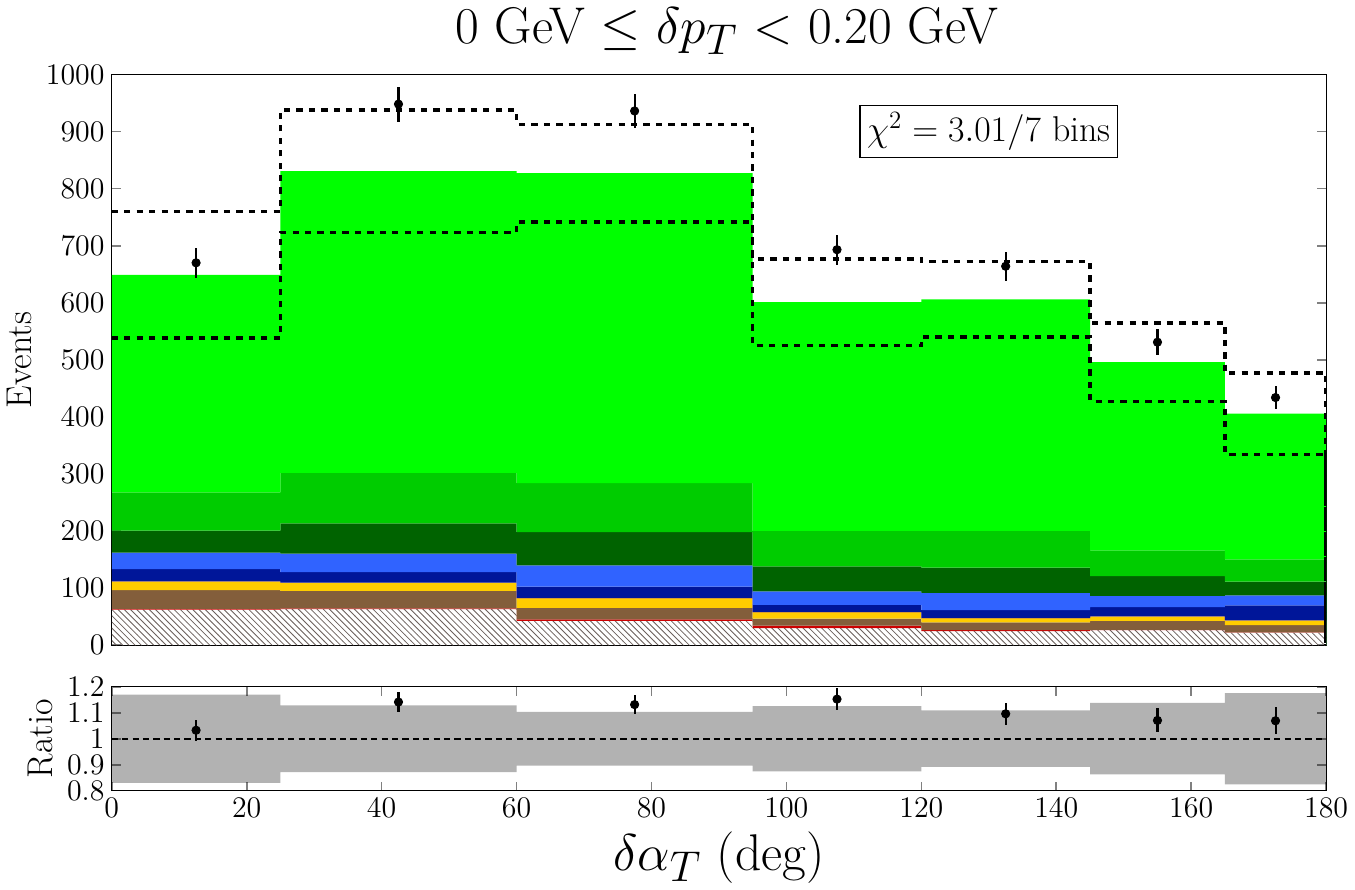}
\hfill
\includegraphics[page=24, width=0.49\textwidth]{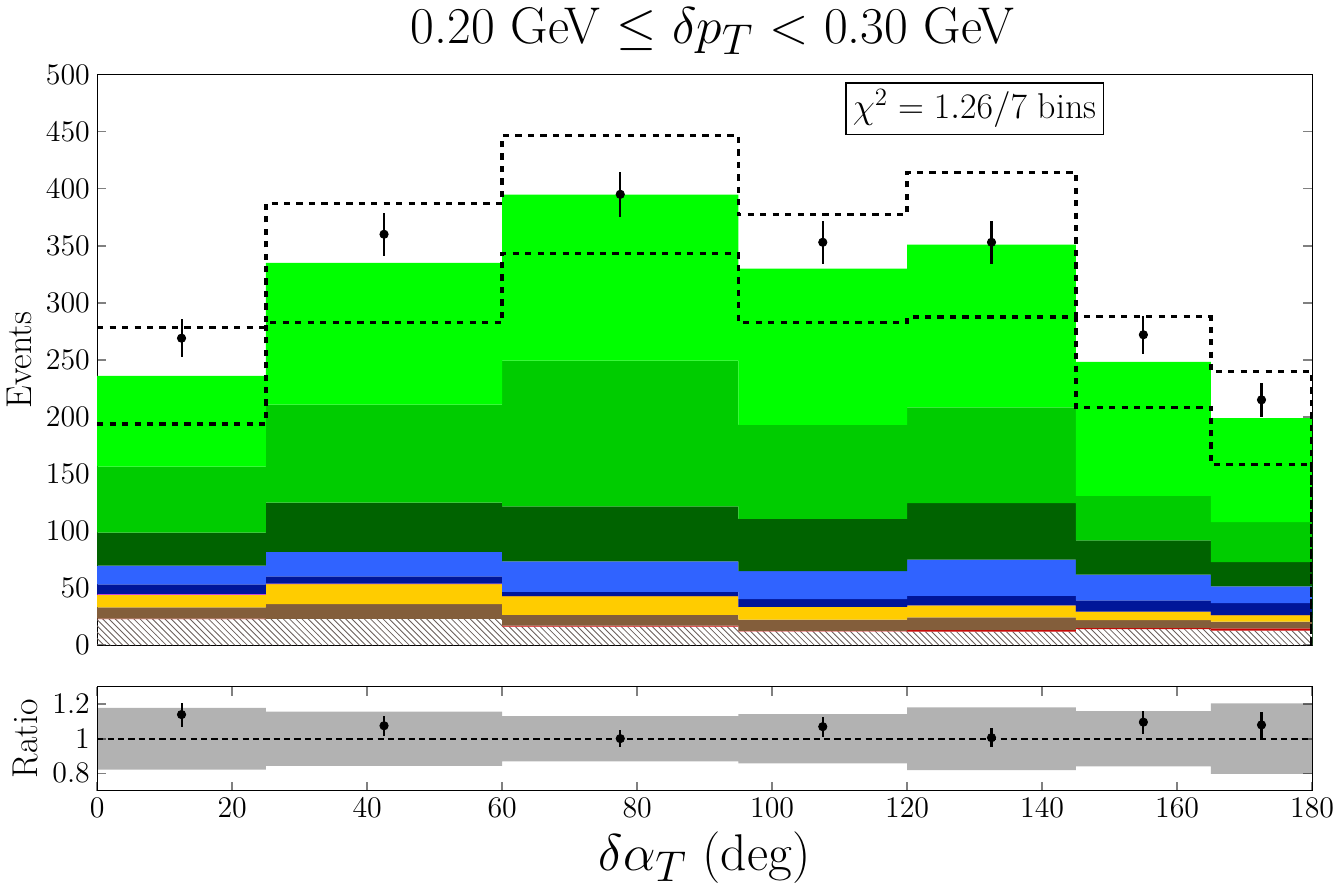}
\vspace{0.2cm}

\includegraphics[page=25, width=0.49\textwidth]{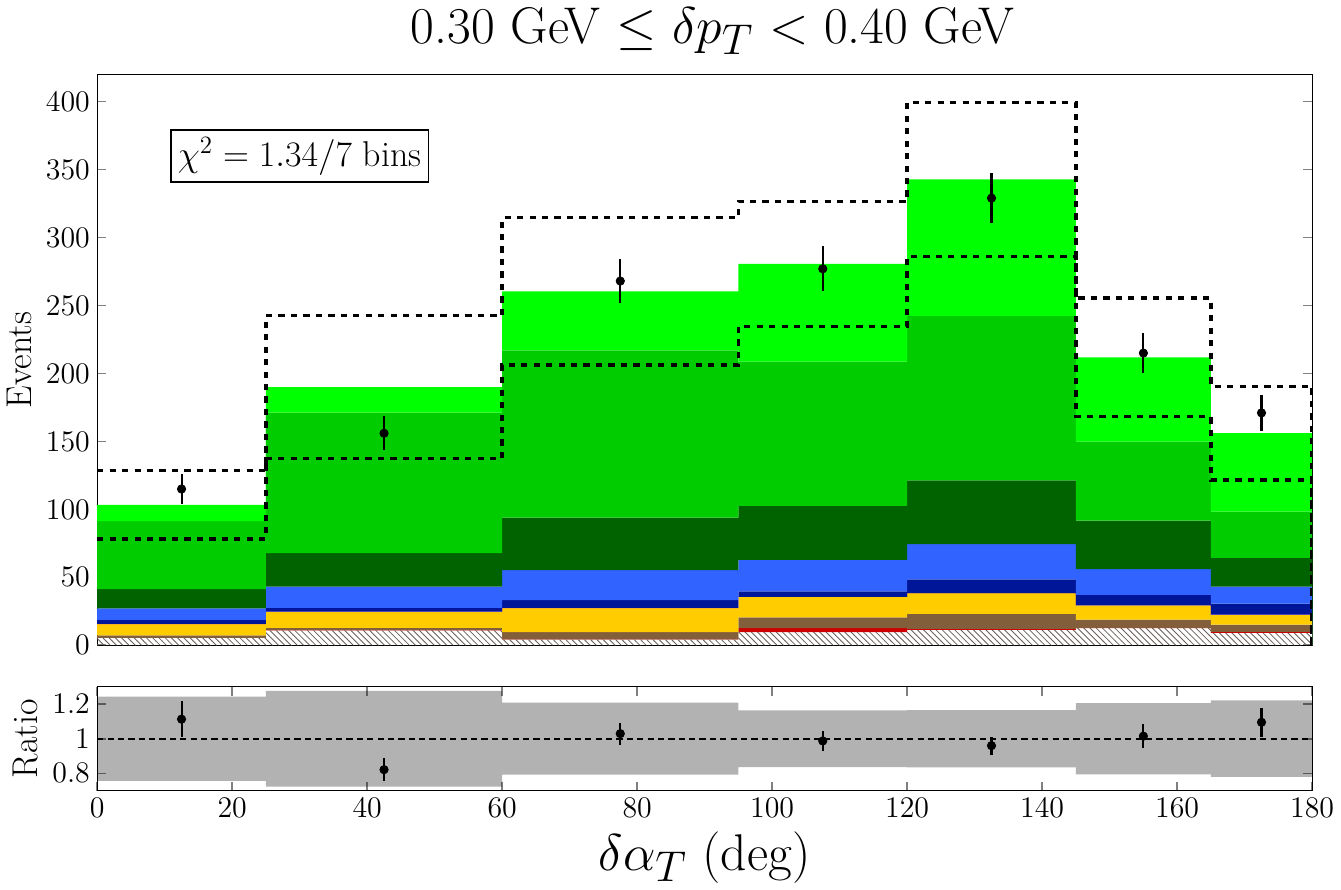}
\hfill
\includegraphics[page=26, width=0.49\textwidth]{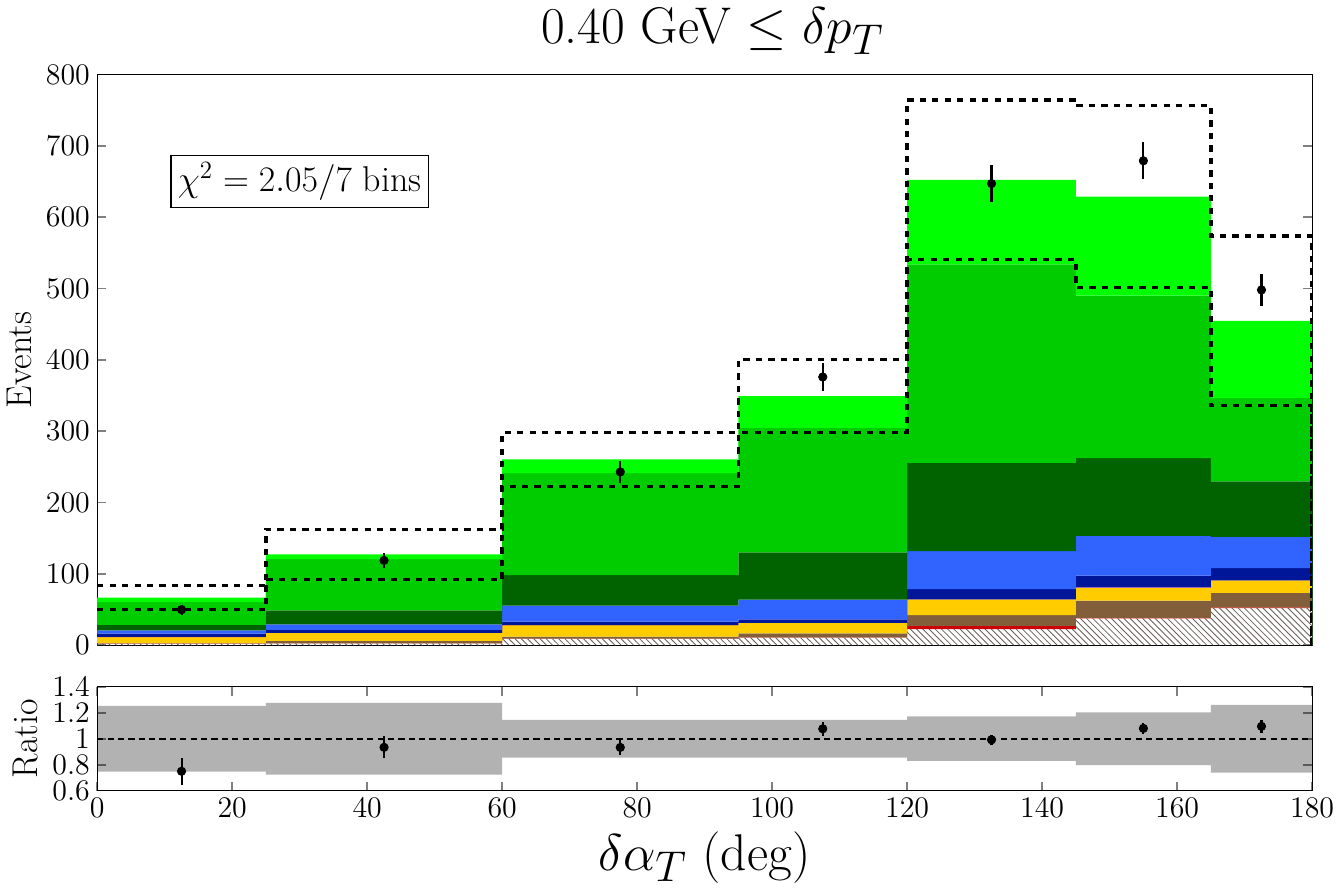}
\vspace{0.2cm}

\includegraphics[page=27, width=0.35\textwidth]{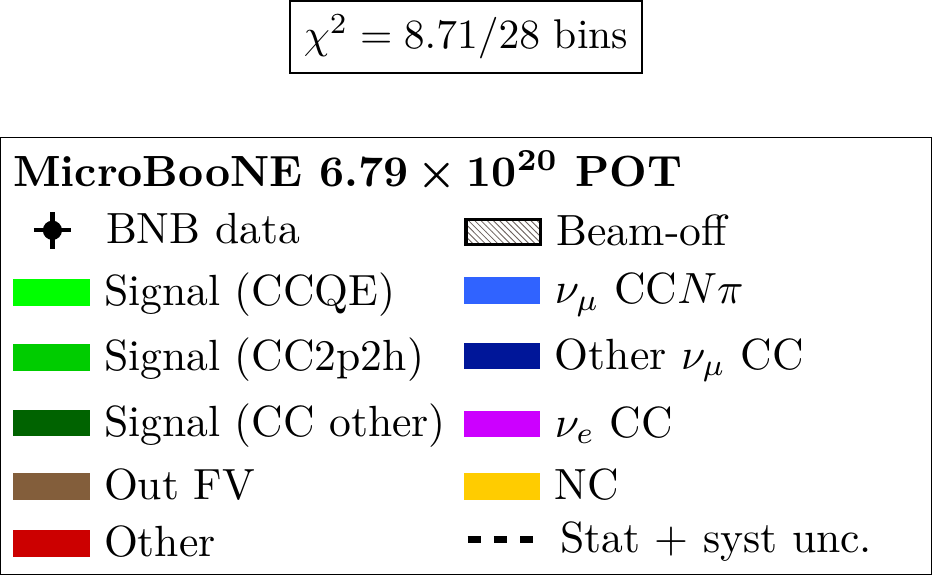}%
\hfill

\caption{Reconstructed event distributions for block \#4
$(\delta p_T, \delta \alpha_T)$.
The bottom panel of each plot shows the ratio of the data to the MicroBooNE
simulation prediction. The uncertainty on the prediction is represented by the
dashed lines in each top panel and the gray region in each bottom panel.}

\label{fig:DataMCblock4}
\end{figure*}

\begin{figure*}
\centering
\includegraphics[page=29, width=0.49\textwidth]{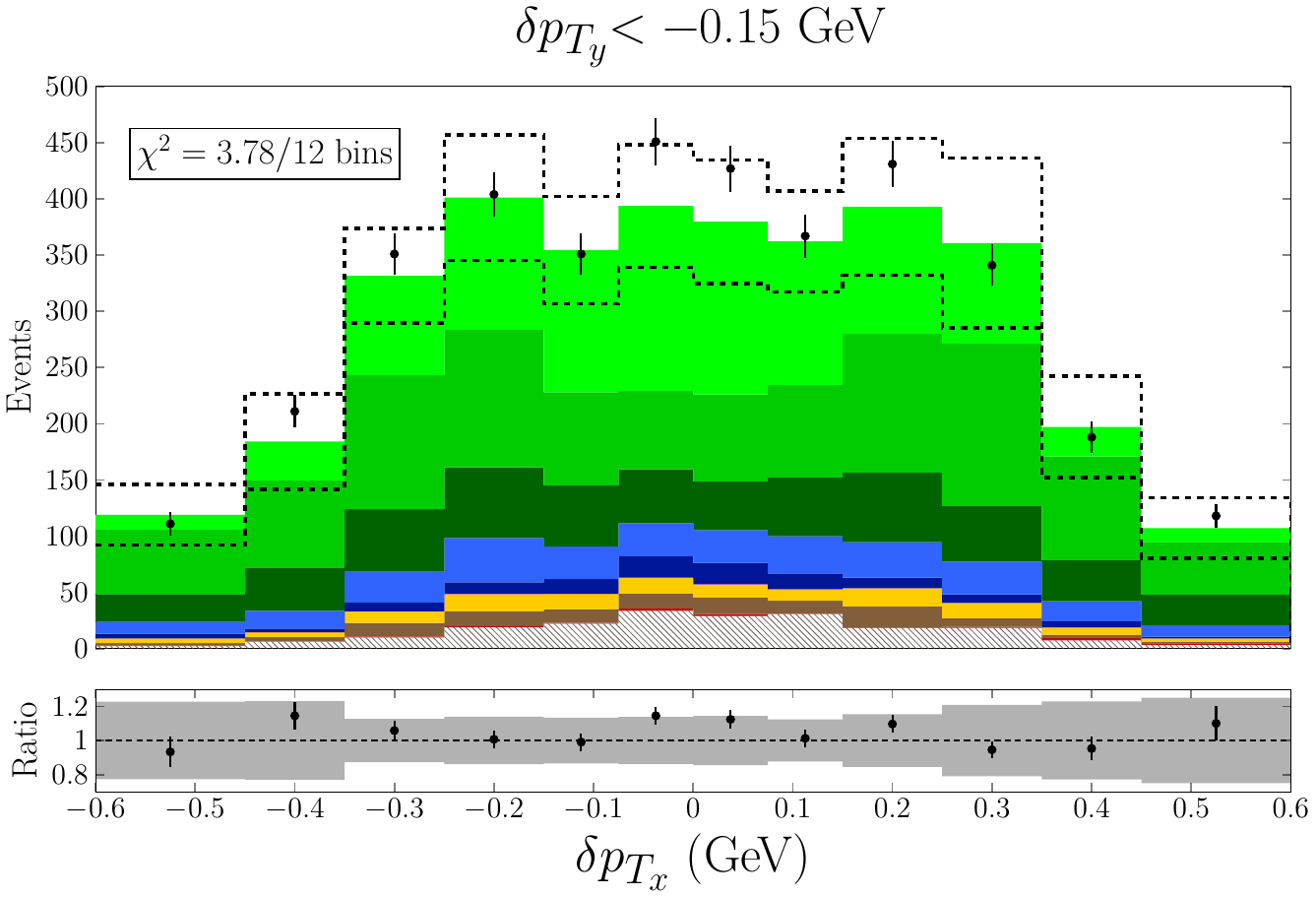}
\hfill
\includegraphics[page=30, width=0.49\textwidth]{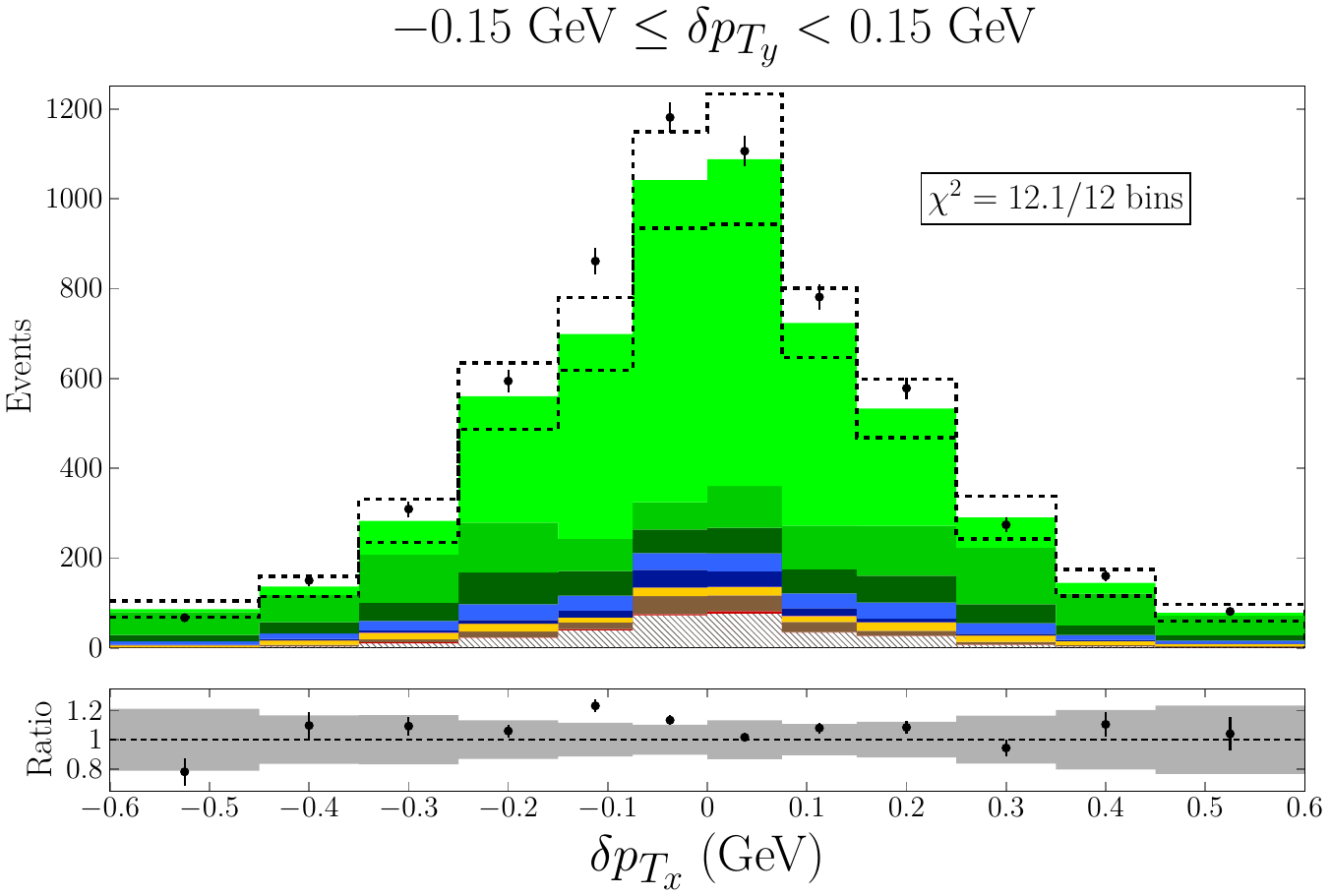}
\vspace{0.2cm}

\includegraphics[page=31, width=0.49\textwidth]{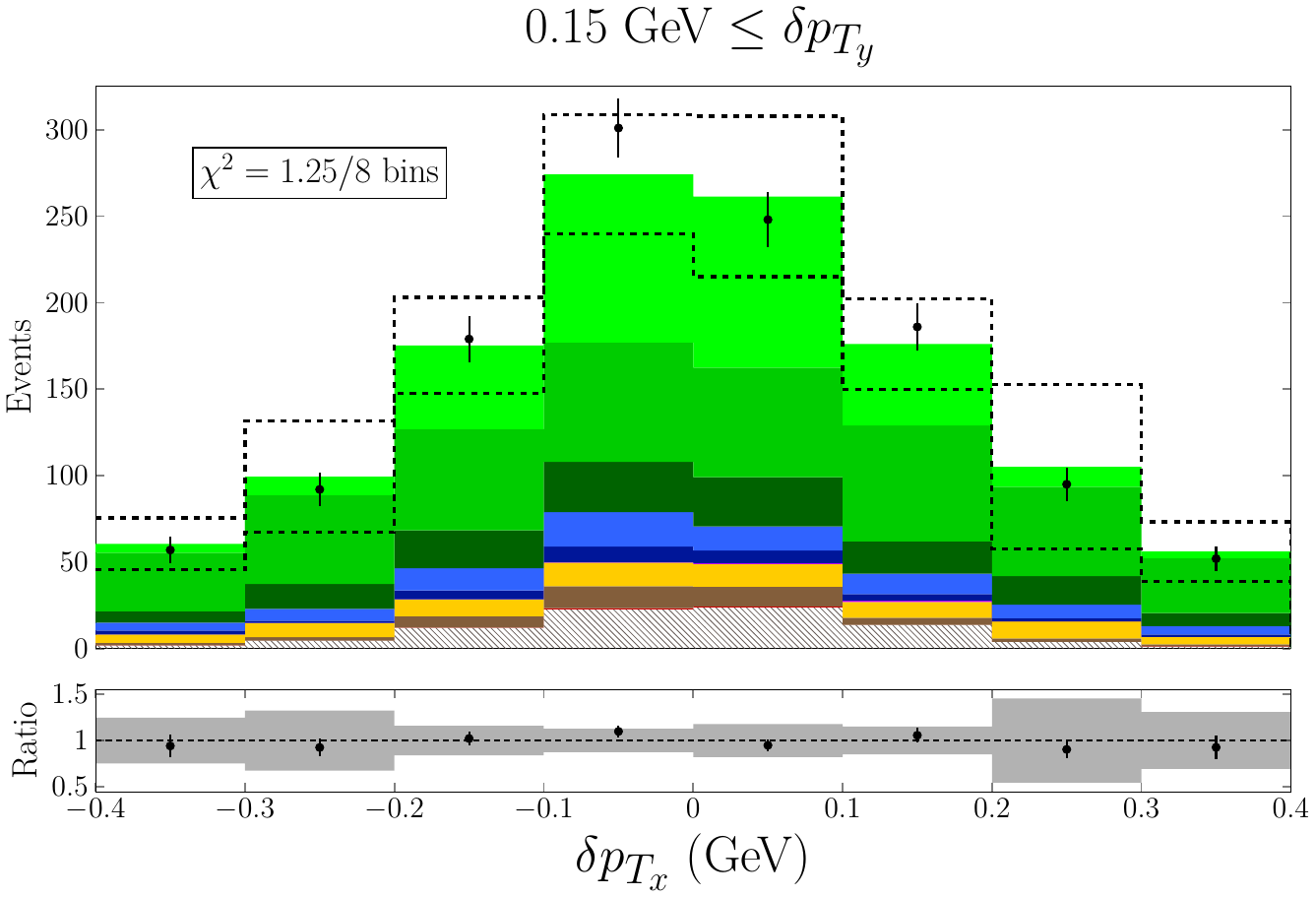}
\hfill
\belowbaseline[-0.23\textheight]{
  \includegraphics[page=32, width=0.35\textwidth]{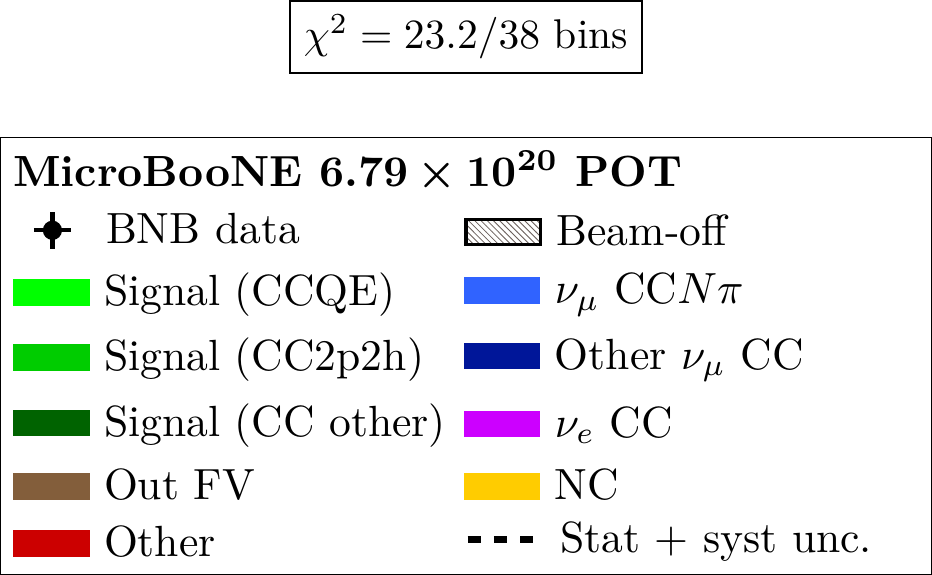}%
}
\hfill

\caption{Reconstructed event distributions for block \#6
$(\delta p_{T_y}, \delta p_{T_x})$. The overall $\chi^2$
value includes contributions from three underflow and three overflow $\delta
p_{T_x}$ bins that are not plotted.
The bottom panel of each plot shows the ratio of the data to the MicroBooNE
simulation prediction. The uncertainty on the prediction is represented by the
dashed lines in each top panel and the gray region in each bottom panel.}
\label{fig:DataMCblock6}
\end{figure*}

\begin{figure*}
\centering
\includegraphics[page=35, width=0.49\textwidth]{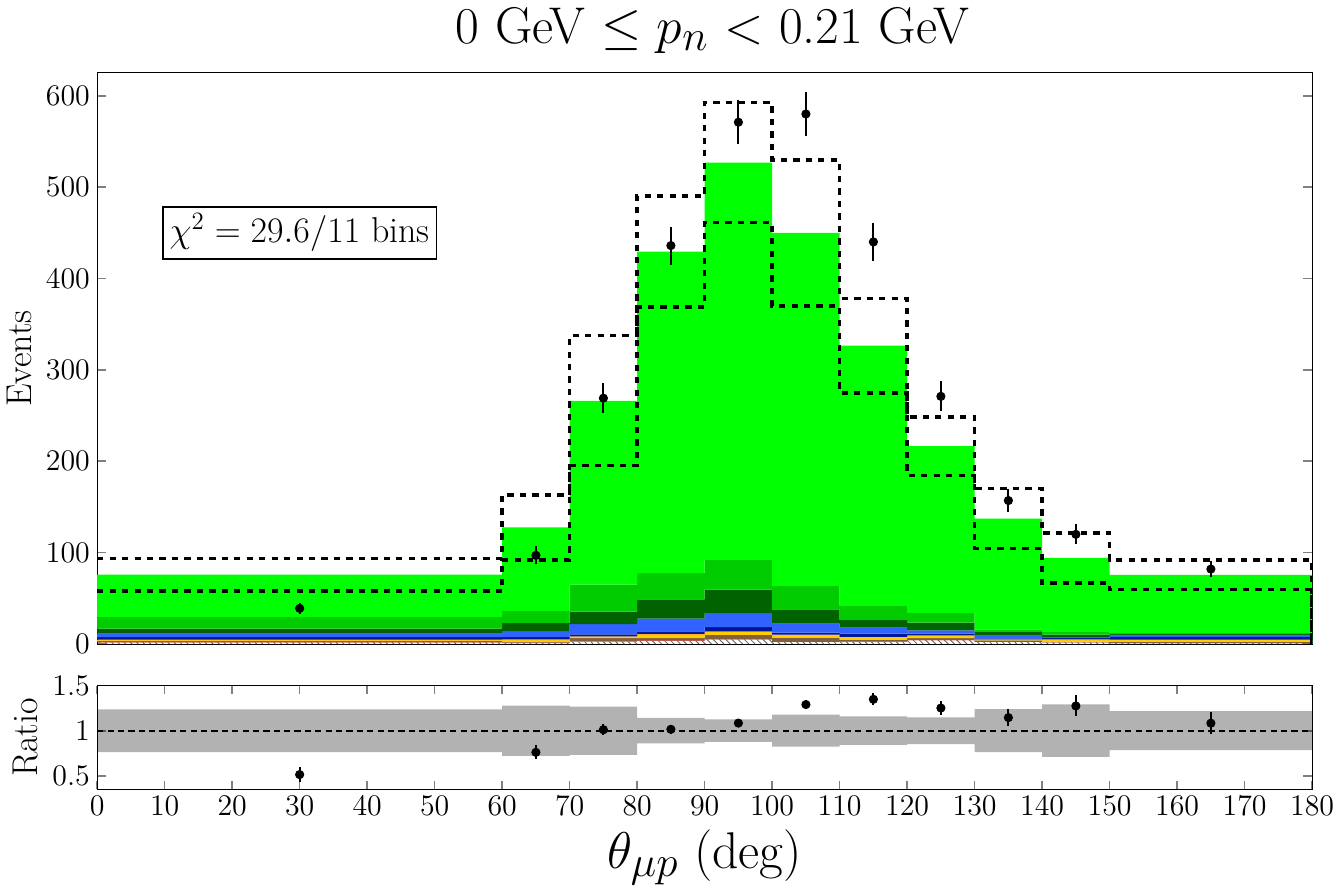}
\hfill
\includegraphics[page=36, width=0.49\textwidth]{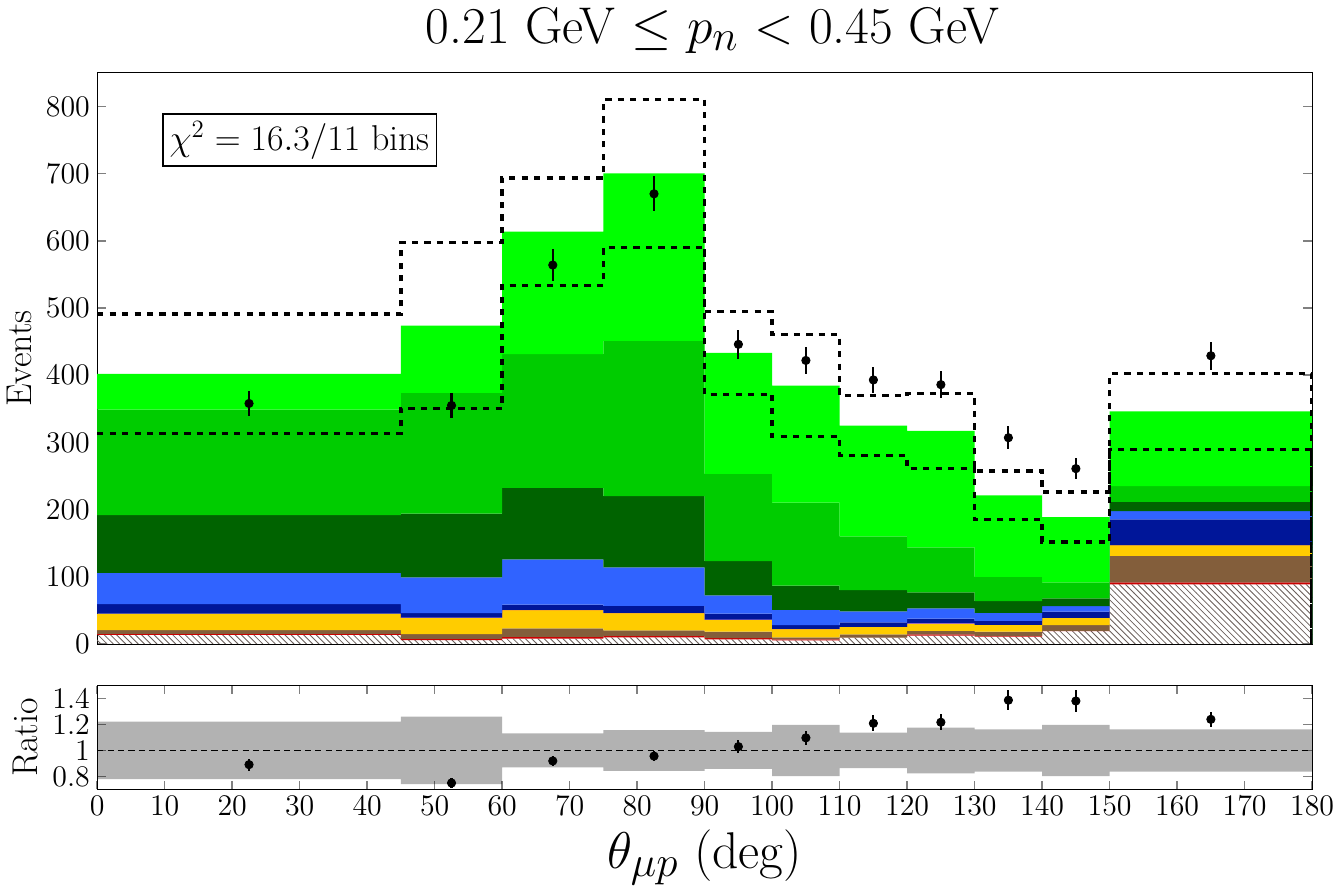}
\vspace{0.2cm}

\includegraphics[page=37, width=0.49\textwidth]{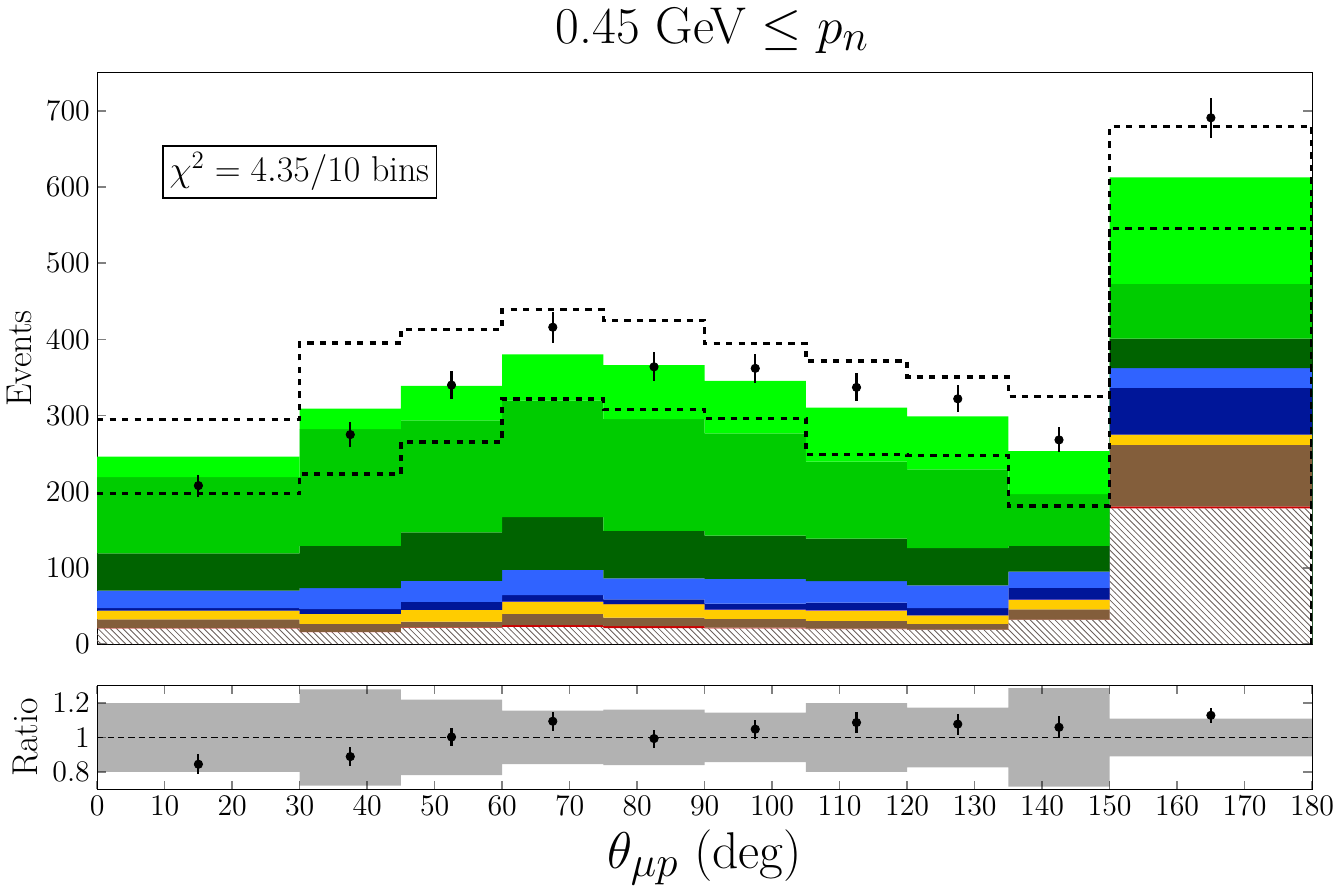}
\hfill
\belowbaseline[-0.23\textheight]{
  \includegraphics[page=38, width=0.35\textwidth]{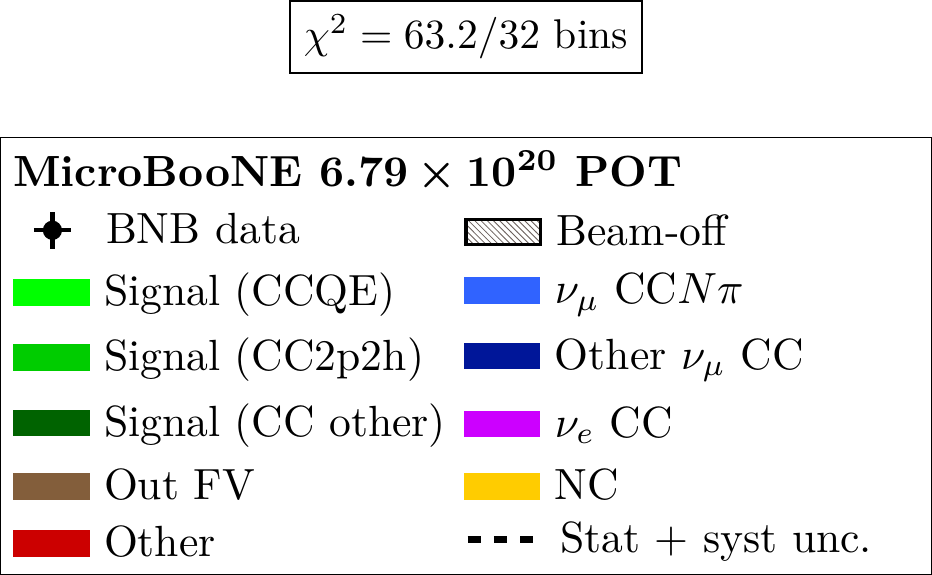}%
}
\hfill

\caption{Reconstructed event distributions for block \#9
$(p_n, \theta_{\mu p})$.
The bottom panel of each plot shows the ratio of the data to the MicroBooNE
simulation prediction. The uncertainty on the prediction is represented by the
dashed lines in each top panel and the gray region in each bottom panel.}
\label{fig:DataMCblock9}
\end{figure*}

\begin{figure*}
\centering
\subfloat[\label{subfig:block10}]{%
  \includegraphics[page=39, width=0.49\textwidth]{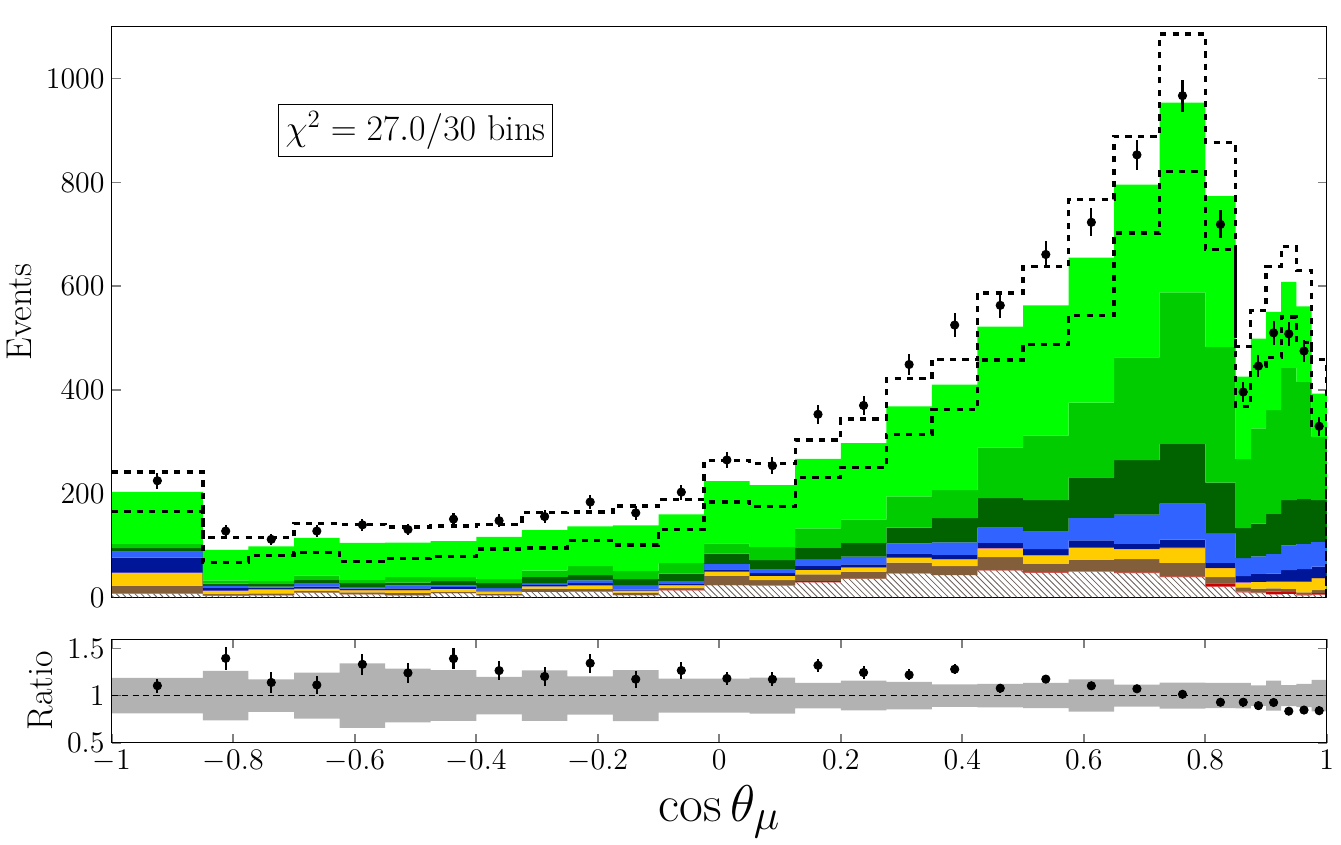}%
}\hfill
\subfloat[\label{subfig:block11}]{%
  \includegraphics[page=40, width=0.49\textwidth]{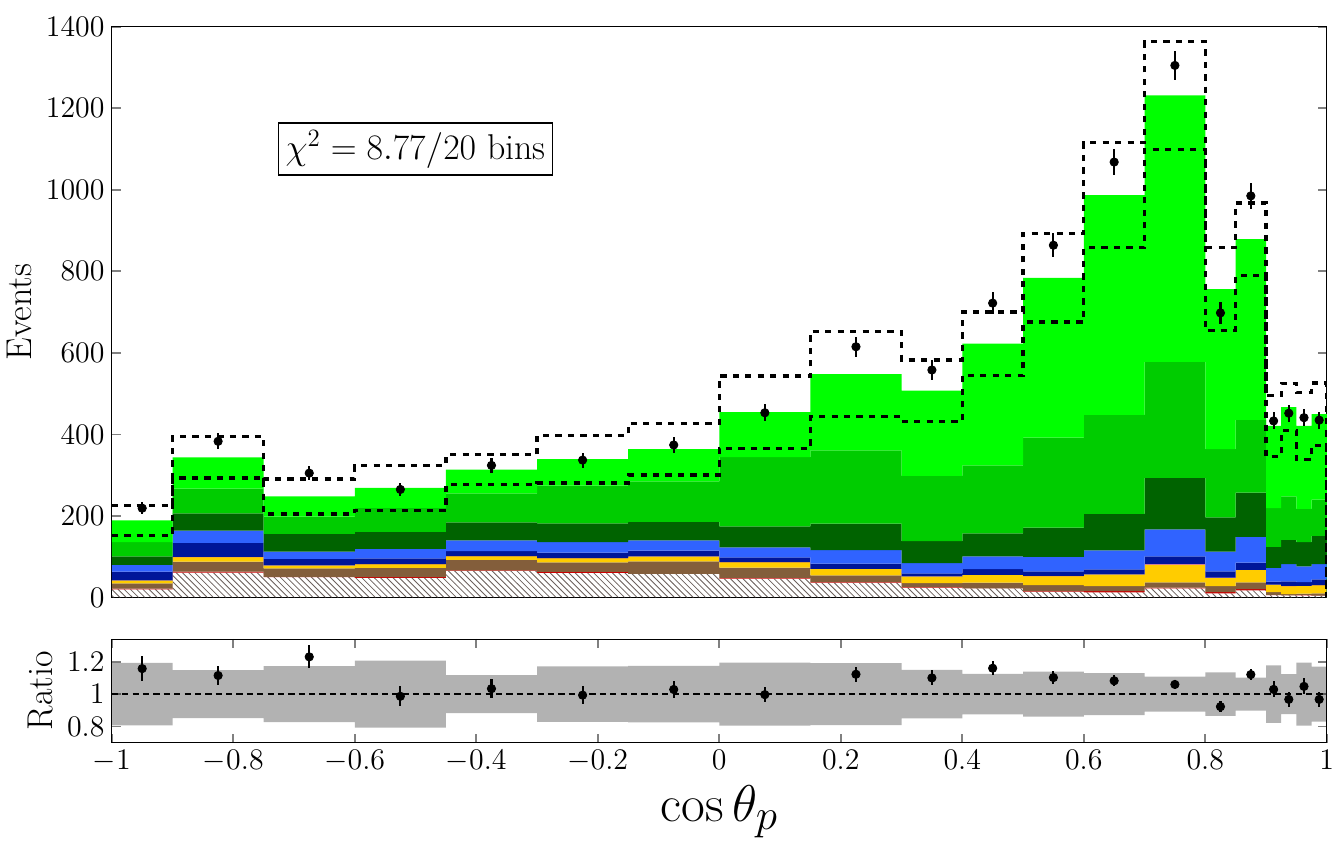}%
}\vspace{0.1em}
\subfloat[\label{subfig:block12}]{%
  \includegraphics[page=41, width=0.49\textwidth]{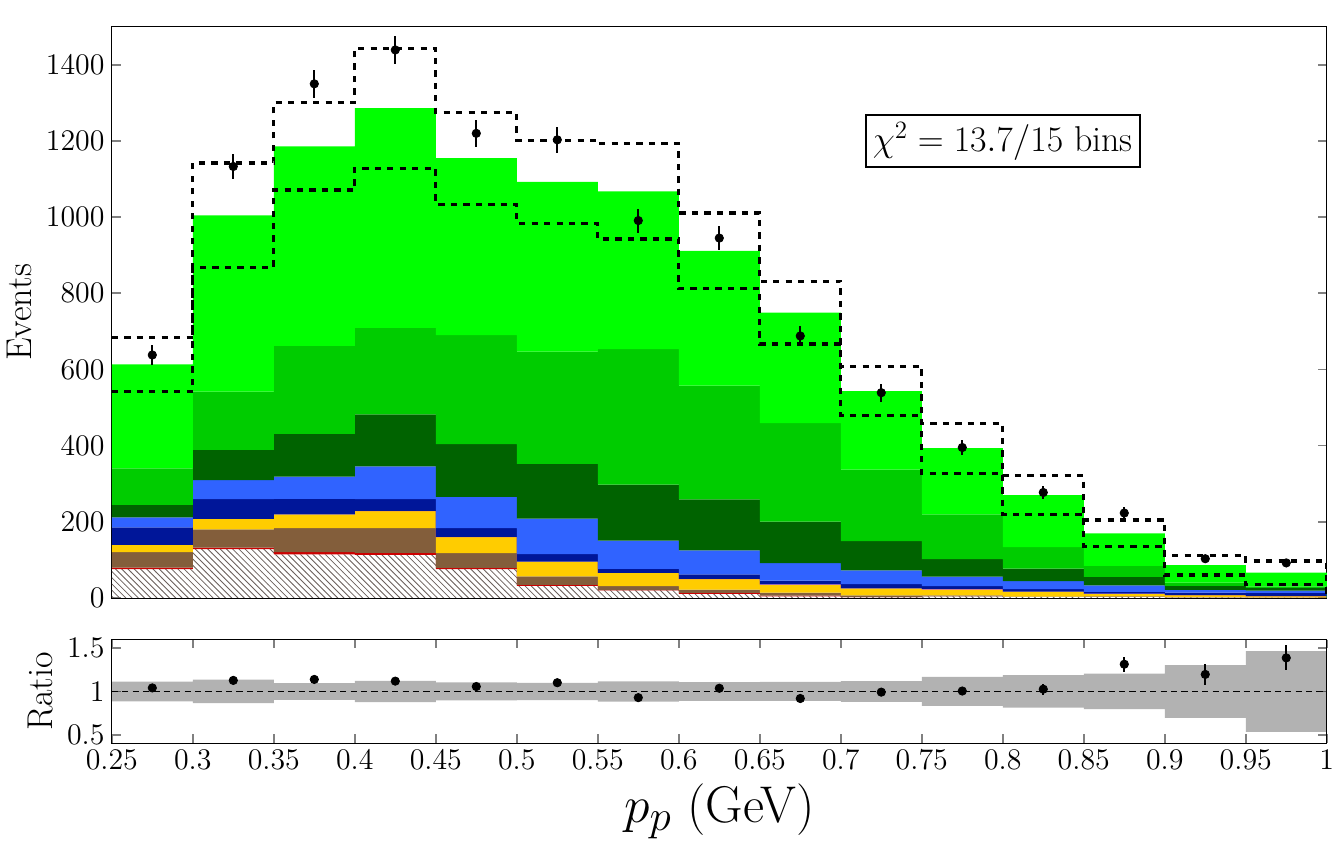}%
}\hfill
\subfloat[\label{subfig:block13}]{%
  \includegraphics[page=42, width=0.49\textwidth]{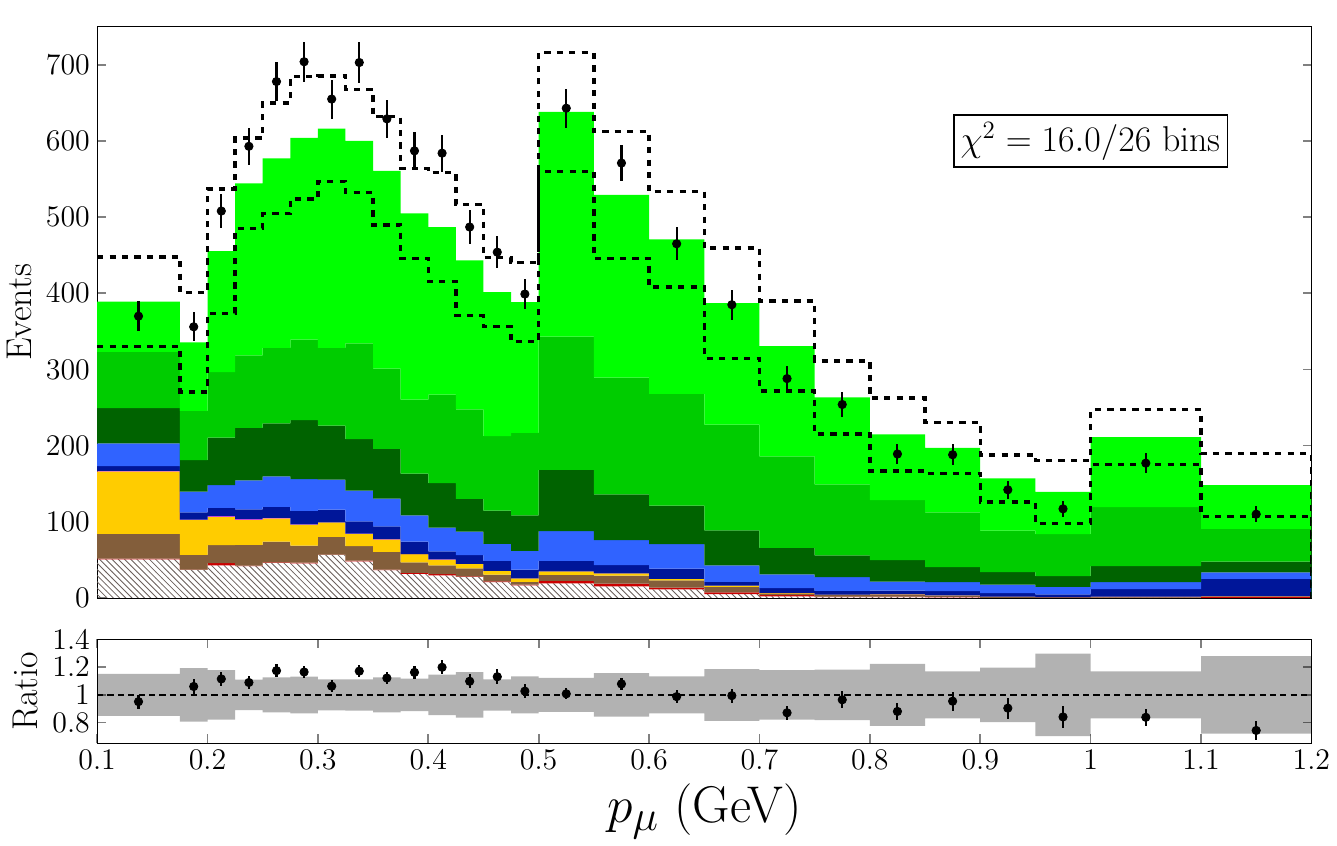}%
}\vspace{0.1em}
\includegraphics[page=17, width=0.35\textwidth]{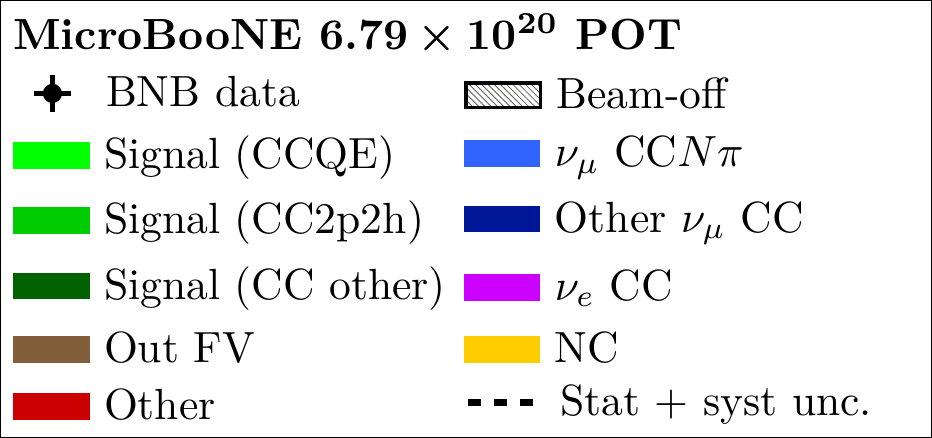}%
\hfill
\caption{Reconstructed event distributions for
(\subref*{subfig:block10})~block~\#10 ($\cos\theta_\mu$),
(\subref*{subfig:block11})~block~\#11 ($\cos\theta_p$),
(\subref*{subfig:block12})~block~\#12 ($p_p$), and
(\subref*{subfig:block13})~block~\#13 ($p_\mu$).
The bottom panel of each plot shows the ratio of the data to the MicroBooNE
simulation prediction. The uncertainty on the prediction is represented by the
dashed lines in each top panel and the gray region in each bottom panel.}
\label{fig:DataMCMultipleBlocks2}
\end{figure*}

The plots in Figs.~\ref{fig:DataMCblock0}~to~\ref{fig:DataMCMultipleBlocks2}
report the measured event rates (black data points) as a function of
reconstructed observables for a total beam exposure of
\TotalDataPOT~\si{\pot}. The error bars on the data points show only the
statistical uncertainty. The stacked histograms in the main panel of each plot
show the central-value prediction from the MicroBooNE simulation chain.
Individual contributions from the event categories defined in
Sec.~\ref{sec:selection} are shown using the same color scheme as in
Fig.~\ref{fig:toposcore}. The total uncertainty (including statistical and
systematic contributions) on the central-value prediction is indicated by the
dashed lines. Below the main panel of each plot, the ratio of the data points
to the central-value prediction is also shown, and the total uncertainty on the
prediction is indicated by the gray band.

Reasonable agreement between the measured event rates and the MicroBooNE
simulation prediction is seen across most of the phase space studied, yielding
an overall $\chi^2$ value of \num[round-mode=places,
round-precision=2]{355.251006} for the \TotalBinCount~reconstructed bins. The
areas of greatest tension are the muon angular distribution at moderate $p_\mu$
(see Fig.~\ref{fig:DataMCblock0}), as well as the individual and
double-differential distributions of $p_n$ and $\theta_{\mu p}$ (see
Figs.~\ref{fig:DataMCMultipleBlocks1}~and~\ref{fig:DataMCblock9}).

\subsection{Sideband-based validation of background model}
\label{sec:sidebands}

Removal of the beam-correlated backgrounds in this analysis relies upon a
simulation-based estimate $\PredictedBkgdCount_{\recoBinIdx}$ of their
contribution to the measured event counts in each reconstructed bin
$\recoBinIdx$. A sideband study, described below, was performed to demonstrate
that the background predictions from the MicroBooNE simulation are sufficiently
precise to obtain the final measurements.

Three sets of alternative selection criteria were developed to enhance
sensitivity to the three dominant backgrounds: events outside of the fiducial
volume (Out FV, see Sec.~\ref{sec:selection}), neutral-current events (NC), and
$\nu_\mu$ charged-current events with one or more final-state pions (CC$N\pi$).
In each case, the full \ccnp\ selection was used as a starting point, and minor
adjustments were made to improve acceptance of the background category of
interest. By construction, all events which pass at least one set of
alternative selection criteria are rejected by the original \ccnp\ selection.
The alternative selection criteria for each class of background are given
below.
\begin{description}
\item[Out FV] The full \ccnp\ selection is unaltered except that the event must
fail requirement~\ref{item:sel_vtx_in_FV} (i.e., the reconstructed
neutrino vertex position must lie outside the fiducial volume).
\item[NC] The event must fail requirement~\ref{item:has_mu_candidate} (there
must not be a muon candidate). The additional requirements
\ref{item:muon_contained_sel}--\ref{item:muon_thresh_sel} applied to the muon
candidate are also removed. All other \ccnp\ selection criteria remain the
same. The event must also contain at least two reconstructed primary particles.
To allow all of the observables of interest to be defined for such events, the
reconstructed primary particle with the longest track length is treated as the
muon candidate, while the second-longest is treated as the leading proton
candidate. Checks of the simulated NC events that passed the original \ccnp\
selection revealed that this ordering by track length occurred in a large
majority (88\%) of cases.
\item[CC$\pmb{N\pi}$] The full \ccnp\ selection is unaltered except that
requirement~\ref{item:sel_proton_PID} is reversed; at least one of the proton
candidates must have an LLR PID score greater than
\ProtonLLRpidCut\ (muon-like).
\end{description}

To assess the adequacy of the background model for this analysis, the logical
OR of these three alternative selections was applied to measure event rates.
That is, an event was accepted by this combined sideband selection if it
satisfied all of the criteria for at least one of the Out-FV-, NC-, or
CC$N\pi$-enhanced selections described above. Good agreement between the
MicroBooNE simulation prediction and the sideband data, within the
uncertainties defined above, was found across the full phase space defined in
Table~\ref{tab:bin_defs}. An overall $\chi^2$ value of \num[round-mode=places,
round-precision=2]{177.935275} was obtained for the full set of
\TotalBinCount~bins. Based on this successful comparison with the sideband data
set, the background predictions of MicroBooNE's simulation are used unaltered
to obtain the final cross-section results. Plots of the measured and predicted
sideband event distributions are provided in Sec.~III of the supplemental
materials.


\section{Cross-section extraction}
\label{sec:unfold}

A neutrino cross-section measurement is ultimately the result of a counting
experiment: estimated backgrounds are subtracted from a measured number of
events in each bin, and scaling factors are then applied to obtain a quantity
with the appropriate units. Corrections to the measured event counts for
detector inefficiency and bin migrations (due to imperfect reconstruction) must
also be applied via a procedure called \textit{unfolding}. Multiple standard
methods for unfolding are described in terms of the matrix transformation
\begin{equation}
\label{eq:unfold}
\MeasuredSignalEvtCount_\trueBinIdx = \sum_\recoBinIdx
\UnfoldingMatrix_{\trueBinIdx \recoBinIdx}
\, \BkgdSubtractedRecoEvents_\recoBinIdx \,,
\end{equation}
where a vector of measured, background-subtracted event counts
$\mathbf{\BkgdSubtractedRecoEvents}$ is
multiplied by an \textit{unfolding matrix}
$\UnfoldingMatrix$ to obtain an estimator
$\pmb{\MeasuredSignalEvtCount}$ for the vector of true signal event counts.
The element of
$\mathbf{\BkgdSubtractedRecoEvents}$ corresponding to the
$\recoBinIdx$-th reconstructed bin is given by
\begin{equation}
\BkgdSubtractedRecoEvents_\recoBinIdx = \AllRecoEvents_\recoBinIdx
- \EXTCount_\recoBinIdx - \PredictedBkgdCount_\recoBinIdx \,,
\end{equation}
where $\AllRecoEvents_\recoBinIdx$ is the total number of measured events in the
bin, and both $\EXTCount_\recoBinIdx$ and $\PredictedBkgdCount_\recoBinIdx$ are
defined as in Eq.~(\ref{eq:ExpectedRecoEvents}). In this context,
$\PredictedBkgdCount_\recoBinIdx$ is evaluated in the central-value universe.
The covariance between the measured event counts in the $\CovMatFirstIdx$-th and
$\CovMatSecondIdx$-th reconstructed bins is used to calculate the covariance
between the unfolded event counts in the $\trueBinIdx$-th and
$\secondTrueBinIdx$-th true bins according to the relation
\begin{equation}
\label{eq:unfold_prop}
\CovMat_{\trueBinIdx \secondTrueBinIdx} =
\mathrm{Cov}(\MeasuredSignalEvtCount_\trueBinIdx,
\MeasuredSignalEvtCount_\secondTrueBinIdx)
= \sum_{\CovMatFirstIdx, \CovMatSecondIdx}
\ErrPropMatrix_{\trueBinIdx \CovMatFirstIdx}
\,
\mathrm{Cov}( \BkgdSubtractedRecoEvents_\CovMatFirstIdx,
\BkgdSubtractedRecoEvents_\CovMatSecondIdx )
\,
\ErrPropMatrix^T_{\CovMatSecondIdx \secondTrueBinIdx} \,,
\end{equation}
where the elements of the error propagation matrix $\ErrPropMatrix$ are
the partial derivatives
\begin{equation}
\ErrPropMatrix_{\trueBinIdx \CovMatFirstIdx} \equiv
\frac{ \partial \MeasuredSignalEvtCount_\trueBinIdx }
{ \partial \BkgdSubtractedRecoEvents_\CovMatFirstIdx } \,.
\end{equation}
An explicit expression for these, appropriate for this analysis, is given below
in Eq.~(\ref{eq:errprop_dagostini}). The contribution of systematic
uncertainties to the covariance on the background-subtracted measured event
counts is estimated from the expected event counts:
\begin{equation}
\mathrm{Cov}( \BkgdSubtractedRecoEvents_\CovMatFirstIdx,
\BkgdSubtractedRecoEvents_\CovMatSecondIdx )
\approx
\mathrm{Cov}( \PredictedRecoEvtCount_\CovMatFirstIdx,
\PredictedRecoEvtCount_\CovMatSecondIdx )
= \CovMat_{\CovMatFirstIdx \CovMatSecondIdx} \,,
\end{equation}
where $\CovMat_{\CovMatFirstIdx \CovMatSecondIdx}$ is calculated according to
the prescription in Eq.~(\ref{eq:CovMat}).

\subsection{The additional smearing matrix $\AddSmearMatrix$}
\label{sec:add_smear}

A simple method of unfolding is to directly invert the response matrix
$\ResponseMatrix$ via
\begin{equation}
\label{eq:unfolding_direct}
\UnfoldingMatrix^\mathrm{direct} = ( \ResponseMatrix^T \ResponseMatrix )^{-1}
\, \ResponseMatrix^T \,.
\end{equation}
However, in most cases of practical interest, this approach leads to strong
anticorrelations between bins and large uncertainties. Overcoming these
difficulties requires introducing new information (specifically, a prior
prediction of the unfolded result) into the otherwise ill-posed unfolding
problem. The process of adding this information is known as
\textit{regularization}. The various unfolding techniques defined in the
literature differ in their prescriptions for how the regularization should be
carried out.

Multiple recent MicroBooNE
analyses~\cite{nueCCDiffXSec,CCinclPRL,uBTKIPRL,uBTKIPRD,uBGKIPRD,uBCC0pNpPRL,uBCC0pNpPRD}
have used the Wiener-SVD~\cite{WienerSVD} unfolding technique, which is based
on an analogy with signal processing: regularization is applied using a
``Wiener filter'' matrix designed to mitigate ``noise'' arising from
uncertainties on the measurement. The Wiener-SVD method involves explicit
construction of an \textit{additional smearing matrix} $\AddSmearMatrix$ which
is related to the unfolding matrix $\UnfoldingMatrix$ via
\begin{equation}
\label{eq:unfolding_and_add_smear}
\UnfoldingMatrix = \AddSmearMatrix \cdot \UnfoldingMatrix^\mathrm{direct} =
\AddSmearMatrix \cdot (\ResponseMatrix^T \ResponseMatrix)^{-1} \,
\ResponseMatrix^T \,.
\end{equation}
The additional smearing matrix thus encapsulates the effect of regularization
on the unfolding procedure. While an expression for $\AddSmearMatrix$ specific
to the Wiener-SVD method is given in Eq.~(3.23) of the original
publication~\cite{WienerSVD}, it follows from the definition in
Eq.~(\ref{eq:unfolding_and_add_smear}) that $\AddSmearMatrix$ may be calculated
for an arbitrary unfolding method via
\begin{equation}
\label{eq:add_smear_general}
\AddSmearMatrix = \UnfoldingMatrix \cdot \ResponseMatrix \,.
\end{equation}

Reference~\cite{WienerSVD} also describes a method to avoid introducing new
measurement uncertainties related to the unfolding procedure itself. By
multiplying theoretical predictions by $\AddSmearMatrix$ before comparisons are
made to the unfolded results, the effects of regularization-related bias on the
measurement are properly taken into account. This prescription is followed
herein: all model predictions are multiplied by $\AddSmearMatrix$ when
comparisons are made to the unfolded data. The $\AddSmearMatrix$ matrix
elements needed to compare the final measurements to new theoretical
predictions are provided in the supplemental materials.

\subsection{D'Agostini iterative unfolding}
\label{sec:dagostini}

The present work adopts an iterative method of unfolding popularized in
high-energy physics by D'Agostini~\cite{Dagostini1995}. An initial estimate for
the unfolded event counts (iteration $i = 0$) is obtained from the
central-value prediction of the MicroBooNE simulation in the true bins:
\begin{equation}
\MeasuredSignalEvtCount_\trueBinIdx^{0}
= \PredictedSignalEvtCount_\trueBinIdx^\mathrm{CV}\,.
\end{equation}
Superscripts on $\MeasuredSignalEvtCount$ and related quantities are
used to denote the number of iterations. Subsequent iterations of the method
are used to refine the initial estimate using the measured event rates in each
reconstructed bin. This is done via the formula
\begin{equation}
\MeasuredSignalEvtCount_\trueBinIdx^{\IterationIdx + 1}
= \sum_\recoBinIdx \UnfoldingMatrix_{\trueBinIdx \recoBinIdx}^{\IterationIdx}
\, \BkgdSubtractedRecoEvents_\recoBinIdx \,.
\end{equation}
where the unfolding matrix element
\begin{equation}
\UnfoldingMatrix_{\trueBinIdx \recoBinIdx}^{\IterationIdx}
= \frac{ \BinProb^{\IterationIdx}_{\recoBinIdx \trueBinIdx} }
{ \Eff_\trueBinIdx }
\end{equation}
is defined in terms of the selection efficiency in the $\trueBinIdx$-th
true bin
\begin{equation}
\label{eq:eff_dagostini}
\Eff_\trueBinIdx = \sum_\secondRecoBinIdx
\ResponseMatrix_{\secondRecoBinIdx \trueBinIdx}
\end{equation}
and the conditional probability that a signal event which belongs to the
$\trueBinIdx$-th true bin will be assigned to the $\recoBinIdx$-th reconstructed
bin
\begin{equation}
\label{eq:binprob_dagostini}
\BinProb^{\IterationIdx}_{\recoBinIdx \trueBinIdx}
= \frac{ \ResponseMatrix_{\recoBinIdx \trueBinIdx}
\, \MeasuredSignalEvtCount_\trueBinIdx^{\IterationIdx} }
{ \sum_\secondTrueBinIdx \ResponseMatrix_{\recoBinIdx \secondTrueBinIdx}
\, \MeasuredSignalEvtCount_\secondTrueBinIdx^{\IterationIdx} } \,.
\end{equation}
Because the unfolding matrix depends on the measured event counts for
iterations $i > 0$, the error propagation matrix elements needed to evaluate
uncertainties on the unfolded result
\begin{equation}
\label{eq:unfold_prop_dagostini}
\CovMat_{\trueBinIdx \secondTrueBinIdx}^\IterationIdx =
\mathrm{Cov}(\MeasuredSignalEvtCount_\trueBinIdx^\IterationIdx,
\MeasuredSignalEvtCount_\secondTrueBinIdx^\IterationIdx)
= \sum_{\CovMatFirstIdx, \CovMatSecondIdx}
\ErrPropMatrix_{\trueBinIdx \CovMatFirstIdx}^\IterationIdx
\,
\mathrm{Cov}( \BkgdSubtractedRecoEvents_\CovMatFirstIdx,
\BkgdSubtractedRecoEvents_\CovMatSecondIdx )
\,
\ErrPropMatrix_{\secondTrueBinIdx \CovMatSecondIdx}^\IterationIdx
\end{equation}
are given by~\cite{Bourbeau2019,Bourbeau2018}
\begin{equation}
\label{eq:errprop_dagostini}
\ErrPropMatrix_{\trueBinIdx \CovMatFirstIdx}^{\IterationIdx + 1} =
\frac{ \partial \MeasuredSignalEvtCount_\trueBinIdx^{\IterationIdx+1} }
{ \partial \BkgdSubtractedRecoEvents_\CovMatFirstIdx } =
\UnfoldingMatrix_{\trueBinIdx \CovMatFirstIdx}^{\IterationIdx}
+ \frac{ \MeasuredSignalEvtCount_\trueBinIdx^{\IterationIdx + 1} }
{ \MeasuredSignalEvtCount^i_\trueBinIdx }
\ErrPropMatrix_{\trueBinIdx \CovMatFirstIdx}^{\IterationIdx}
- \sum_{\secondTrueBinIdx,\CovMatSecondIdx}
\Eff_\secondTrueBinIdx \, \frac{ \BkgdSubtractedRecoEvents_\CovMatSecondIdx }
{ \MeasuredSignalEvtCount^\IterationIdx_\secondTrueBinIdx }
\, \UnfoldingMatrix_{\trueBinIdx \CovMatSecondIdx}^{\IterationIdx}
\, \UnfoldingMatrix_{\secondTrueBinIdx \CovMatSecondIdx}^{\IterationIdx} \,
\ErrPropMatrix_{\secondTrueBinIdx \CovMatFirstIdx}^{\IterationIdx} \,.
\end{equation}
For the first iteration, this reduces to
\begin{equation}
\ErrPropMatrix_{\trueBinIdx \CovMatFirstIdx}^{1} =
\UnfoldingMatrix_{\trueBinIdx \CovMatFirstIdx}^{0}
\end{equation}
since the central-value prediction of the MicroBooNE simulation does not depend
on the data:
\begin{equation}
\ErrPropMatrix_{\trueBinIdx \CovMatFirstIdx}^{0} =
\frac{ \partial \MeasuredSignalEvtCount_\trueBinIdx^{0} }
{ \partial \BkgdSubtractedRecoEvents_\CovMatFirstIdx } =
\frac{ \partial \PredictedSignalEvtCount_\trueBinIdx^{\mathrm{CV}} }
{ \partial \BkgdSubtractedRecoEvents_\CovMatFirstIdx } = 0 \,.
\end{equation}

\subsection{Convergence criterion for unfolding iterations}
\label{sec:unfold_converge}

In the limit of many iterations, the D'Agostini approach converges to the
direct inversion result discussed in Sec.~\ref{sec:add_smear}. Regularization
is thus applied by stopping the method at a finite number of iterations.

Traditionally, the choice of the number of iterations to use in the D'Agostini
unfolding method is a critical step of an analysis: one must iterate enough
times to avoid a strong bias towards the initial estimate
$\MeasuredSignalEvtCount_\trueBinIdx^{0}$ for the true event distribution
without approaching the direct inversion result too closely. These concerns are
addressed in this article by multiplying theoretical predictions
by the additional smearing matrix $\AddSmearMatrix$ before comparing them to
the unfolded data. The implicit regularization in the D'Agostini method is thus
applied to the predictions in a consistent way, avoiding the need to introduce
new unfolding-related systematic uncertainties.

Nevertheless, even while using $\AddSmearMatrix$, one must choose a definite
number of iterations in order to unfold via the D'Agostini approach. In this
analysis, the choice is made by continuing to iterate until the average
fractional deviation $\ConvergenceFOM$ in the $\NumTrueBins$ true bins
\begin{equation}
\label{eq:fom_dagostini}
\ConvergenceFOM \equiv \frac{1}{\NumTrueBins} \sum_\trueBinIdx
\frac{ \big| \MeasuredSignalEvtCount_\trueBinIdx^{\IterationIdx + 1}
 - \MeasuredSignalEvtCount_\trueBinIdx^{\IterationIdx} \big| }
{ \MeasuredSignalEvtCount_\trueBinIdx^{\IterationIdx + 1} }
\end{equation}
between neighboring iterations falls below \FOMCutoff. This cutoff was chosen
empirically based on the unfolding performance in fake data studies and the
observation that fluctuations of the bin counts at this level will be
well-covered by typical systematic uncertainties. The fake data studies used to
validate the unfolding procedure treated the prediction in the NuWro alternate
universe (see Sec.~\ref{sec:syst_sources}) as if it were real data. Recovery of
the known true bin counts $\PredictedSignalEvtCount_{\trueBinIdx}$ predicted by
NuWro was verified within the relevant subset of the uncertainties, i.e., only
the Monte Carlo statistical and neutrino interaction model systematic
uncertainties were included.

\subsection{Blockwise unfolding}
The convergence criterion discussed in Sec.~\ref{sec:unfold_converge} leads to
between two and five iterations being used to unfold each individual block of
bins.

To avoid the double-counting issues mentioned in Sec.~\ref{sec:binning} (e.g.,
an incorrect calculation of the selection efficiency from the elements of the
response matrix $\ResponseMatrix$), the sums in
Eqs.~(\ref{eq:eff_dagostini})--(\ref{eq:errprop_dagostini}) and
Eq.~(\ref{eq:fom_dagostini}) should be understood to include only true and
reconstructed bins within the same block. Also, since
Eq.~(\ref{eq:fom_dagostini}) is intended to be an average, $\NumTrueBins$
should be interpreted as the number of true bins in the block of interest.

An overall unfolding matrix $\UnfoldingMatrix$ used according to
Eq.~(\ref{eq:unfold}) to obtain final results involving all bins is constructed
as the direct sum of the blockwise unfolding matrices:
\begin{equation}
\UnfoldingMatrix = \bigoplus_{b} \UnfoldingMatrix_b
= \UnfoldingMatrix_0 \oplus \UnfoldingMatrix_1 \oplus \dots
= \begin{pmatrix}
U_0 & 0 & 0 & \dots \\
0 & U_1 & 0 & \dots \\
0 & 0 & \ddots & \dots \\
\vdots & \vdots & \vdots & \ddots \\
\end{pmatrix} \,.
\end{equation}
Here $\UnfoldingMatrix_b$ is the unfolding matrix for the $b$-th block of true
and reconstructed bins. A similar direct sum is calculated to build the
combined error propagation matrix needed to evaluate the unfolded covariance
matrix in Eq.~(\ref{eq:unfold_prop}):
\begin{equation}
\ErrPropMatrix = \bigoplus_{b} \ErrPropMatrix_b
= \ErrPropMatrix_0 \oplus \ErrPropMatrix_1 \oplus \dots
\end{equation}

\subsection{Decomposition of the blockwise covariance matrices}
\label{sec:cov_mat_decomp}

The covariance matrix $\CovMat_{\trueBinIdx\secondTrueBinIdx}$ describing the
unfolded event counts $\MeasuredSignalEvtCount_\trueBinIdx$ and
$\MeasuredSignalEvtCount_\secondTrueBinIdx$ (see Eq.~(\ref{eq:unfold_prop})) for
an individual block of bins can be decomposed into components representing
normalization-only, shape-only, and mixed uncertainties according to the
relations~\cite{NormShape}
\begin{equation}
\CovMat_{\trueBinIdx\secondTrueBinIdx} =
 \CovMat_{\trueBinIdx\secondTrueBinIdx}^\mathrm{norm}
+ \CovMat_{\trueBinIdx\secondTrueBinIdx}^\mathrm{shape}
+ \CovMat_{\trueBinIdx\secondTrueBinIdx}^\mathrm{mixed}
\end{equation}
\begin{equation}
\CovMat_{\trueBinIdx\secondTrueBinIdx}^\mathrm{norm}
=
\frac{ \MeasuredSignalEvtCount_\trueBinIdx \,
\MeasuredSignalEvtCount_\secondTrueBinIdx }
{ \MeasuredSignalEvtCount_\mathrm{tot}^2 }
\sum_{\alpha\beta} \CovMat_{\alpha\beta}
\end{equation}
%
%
\begin{equation}
\CovMat_{\trueBinIdx\secondTrueBinIdx}^\mathrm{mixed}
=
\frac{ \MeasuredSignalEvtCount_\trueBinIdx }
{ \MeasuredSignalEvtCount_\mathrm{tot} }
\sum_{\alpha} \CovMat_{\alpha\secondTrueBinIdx}
+ \frac{ \MeasuredSignalEvtCount_\secondTrueBinIdx }
{ \MeasuredSignalEvtCount_\mathrm{tot} }
\sum_{\alpha} \CovMat_{\trueBinIdx\alpha}
- 2\CovMat_{\trueBinIdx\secondTrueBinIdx}^\mathrm{norm}
\end{equation}
where
\begin{equation}
\label{eq:total_signal_events_in_block}
\MeasuredSignalEvtCount_\mathrm{tot} \equiv \sum_{\alpha}
  \MeasuredSignalEvtCount_{\alpha} \,,
\end{equation}
and the sums are understood to include all true bins belonging to the block of
interest and no others. Note that the same formulas cannot be meaningfully
applied to the multi-block covariance matrix due to double-counting when
computing the total number of events in
Eq.~(\ref{eq:total_signal_events_in_block}). This decomposition of the
covariance matrix is used when presenting cross-section results for individual
blocks of bins in Sec.~\ref{sec:diff_xsec}.

\subsection{Calculation of differential cross sections}

The unfolded event counts $\MeasuredSignalEvtCount_\trueBinIdx$ from the final
iteration provide an estimator for the true number of signal events in the
$\trueBinIdx$-th bin. They may be converted to a flux-averaged differential
cross section according to the formula
\begin{equation}
\Big< \frac{d^\XsecDimension\sigma}{d\PhaseSpaceVec} \Big>_\trueBinIdx
= \frac{ \MeasuredSignalEvtCount_\trueBinIdx }
{ \IntegratedFlux \, \NumTargets \, \PhaseSpaceVecWidths_\trueBinIdx }
\end{equation}
where $\IntegratedFlux = \num{5.00846e+11}\text{ }
\nu_\mu\,/\si{\centi\meter\squared}$ is the integrated flux of muon neutrinos
calculated for the beam exposure (\TotalDataPOT~\si{\pot}) used in the
analysis, $\NumTargets = \num{7.99249e+29}$ is the number of argon nuclei in
the fiducial volume, and $\PhaseSpaceVecWidths_\trueBinIdx$ is the product of
the $\XsecDimension$ widths of the $\trueBinIdx$-th bin of the
$\XsecDimension$-dimensional measurement.

The covariance between the $\XsecDimension$- and
$\SecondXsecDimension$-dimensional differential cross sections in the
$\trueBinIdx$-th and $\secondTrueBinIdx$-th true bins is likewise given by
\begin{equation}
\mathrm{Cov}\left(
\Big< \frac{d^\XsecDimension\sigma}{d\PhaseSpaceVec} \Big>_\trueBinIdx,
\Big< \frac{d^\SecondXsecDimension\sigma}{d\SecondPhaseSpaceVec}
\Big>_\secondTrueBinIdx \right)
= \frac{ \mathrm{Cov}(\MeasuredSignalEvtCount_\trueBinIdx,
\MeasuredSignalEvtCount_\secondTrueBinIdx) }
{ \IntegratedFlux^2 \, \NumTargets^2 \, \PhaseSpaceVecWidths_\trueBinIdx
\, \SecondPhaseSpaceVecWidths_\secondTrueBinIdx } \,.
\end{equation}

\section{Results}
\label{sec:results}

The measurements shown in the remainder of this article are flux-averaged
differential \ccnp\ cross sections obtained via the unfolding procedure
presented above. Several event generator predictions are compared to the data
using NUISANCE~\cite{Stowell2017}, and goodness-of-fit is quantified for the
data set as a whole and for the individual blocks of kinematic bins listed in
Table~\ref{tab:bin_defs}.

In Sec.~I of the supplemental materials, the full \TotalBinCount-bin data set
is presented in terms of flux-averaged total cross sections integrated over
each bin (trivially obtainable by multiplying by the bin widths
$\PhaseSpaceVecWidths_\trueBinIdx$). This allows the measured cross sections
and the covariance matrix elements for all observables to be expressed using
consistent units.

\subsection{Interaction models}
\label{sec:models}

In addition to the MicroBooNE Tune model described in Sec.~\ref{sec:uBTune} and
used to execute the analysis, the predictions of several other simulation-based
neutrino interaction models are compared to the measured cross-section results.
These models include multiple configurations of GENIE as well as
three alternative neutrino event generators.

The GENIE-based prediction labeled \textit{GENIE 3.0.6} is produced using the
same code version and model set (\texttt{G18\_10a\_02\_11a}) as the MicroBooNE
Tune, but the custom MicroBooNE-specific modifications described in
Ref.~\cite{MicroBooNEGENIETune} are omitted. The \textit{GENIE 2.12.10}
prediction uses a near-default configuration of this older version of the code,
which includes the Bodek-Ritchie Fermi gas description of the nuclear ground
state~\cite{PhysRevD.23.1070}, the Llewellyn Smith CCQE
calculation~\cite{LlewellynSmith:1971uhs}, an empirical model for 2p2h
interactions~\cite{Katori:2013eoa}, the Rein-Sehgal treatment of RES and COH
scattering~\cite{Rein:1980wg}, and the hA FSI model~\cite{hAOld1,hAOld2}. The
modeling ingredients in \textit{GENIE 3.2.0 G18\_02a}, which uses the default
configuration (\textit{G18\_02a\_00\_000}) of this recent GENIE release, are
largely similar to GENIE~2.12.10. In addition to code updates, however, the
KLN-BS RES model~\cite{Nowak:2009se,Kuzmin:2003ji,
Berger:2007rq,Graczyk:2007bc}, Berger-Sehgal COH model~\cite{Berger:2008xs},
and hA2018 FSI model~\cite{hN2018} replace their prior counterparts. A final
configuration, \textit{GENIE 3.2.0 G21\_11b}, uses the more recently-added
\texttt{G21\_11b\_00\_000} model set, which adopts the SuSAv2
calculation~\cite{PhysRevD.101.033003, SuSAv2QE, RuizSimo2016, Simo2017} of
CCQE and 2p2h cross sections as well as the hN2018 FSI model~\cite{hN2018}. The
remaining components of the model set are similar to GENIE 3.0.6.

The first prediction of an alternate neutrino event generator, \textit{NuWro
19.02.2}, uses the NuWro~\cite{Golan2012} implementations of the LFG nuclear
ground state~\cite{Carrasco:1989vq}, the Llewellyn Smith CCQE
model~\cite{LlewellynSmith:1971uhs}, the Valencia CC2p2h
model~\cite{Gran:2013kda, GENIEValenciaMEC}, the Adler-Rarita-Schwinger
treatment of $\Delta$ resonance production~\cite{Graczyk:2007bc}, the
Berger-Sehgal COH model~\cite{Berger:2008xs} and an intranuclear cascade
approach to FSI. A comparison to \textit{NEUT 5.6.0} is also provided, in which
the NEUT event generator~\cite{hayato2021, hayato2009} is configured to use an
LFG nuclear model~\cite{Carrasco:1989vq}, the Valencia model for CCQE and
2p2h~\cite{NievesQEPaper, NievesQEErratum, Nieves:2012yz, Gran:2013kda,
GENIEValenciaMEC}, the KLN-BS RES
calculation~\cite{Nowak:2009se,Kuzmin:2003ji,Berger:2007rq,Graczyk:2007bc}, the
Berger-Sehgal COH model~\cite{Berger:2008xs}, and an FSI cascade treatment with
nuclear medium corrections for pions~\cite{Salcedo1988}.

Finally, a prediction from the GiBUU event generator~\cite{Buss2012} (labeled
\textit{GiBUU 2021.1}) is studied using the 2021 ``patch 1'' release of the
code announced to users on 5 November 2021. Ingredients of the GiBUU physics
model include an LFG representation of the nuclear ground
state~\cite{Carrasco:1989vq}, a standard expression for the neutrino-nucleon
CCQE cross section~\cite{Leitner:2006ww}, an empirical 2p2h model, a treatment
of RES based upon the MAID analysis~\cite{Mosel:2019vhx}, and a DIS model from
PYTHIA~\cite{Sjostrand:2006za}. A unique feature of the code is its dynamical
model of intranuclear hadron transport based upon numerical solution of the
Boltzmann-Uehling-Uhlenbeck equation. A consistent nuclear potential is used in
both the description of the target nucleus and FSI.

The $\chi^{2}$ metric used to assess goodness-of-fit when comparing these
models to data accounts for the total covariance matrix associated with the
measurements but neglects any theoretical uncertainties on the predictions
themselves. Documentation needed to enable new comparisons beyond those shown
here is provided in Sec.~I of the supplemental materials. The data set from
this article has also recently been incorporated into the
NUISANCE~\cite{Stowell2017} software framework for convenient use by the
neutrino interaction modeling community.

\subsection{Differential cross-section results}
\label{sec:diff_xsec}

\begin{figure*}
\centering
\includegraphics[page=1, width=0.49\textwidth]{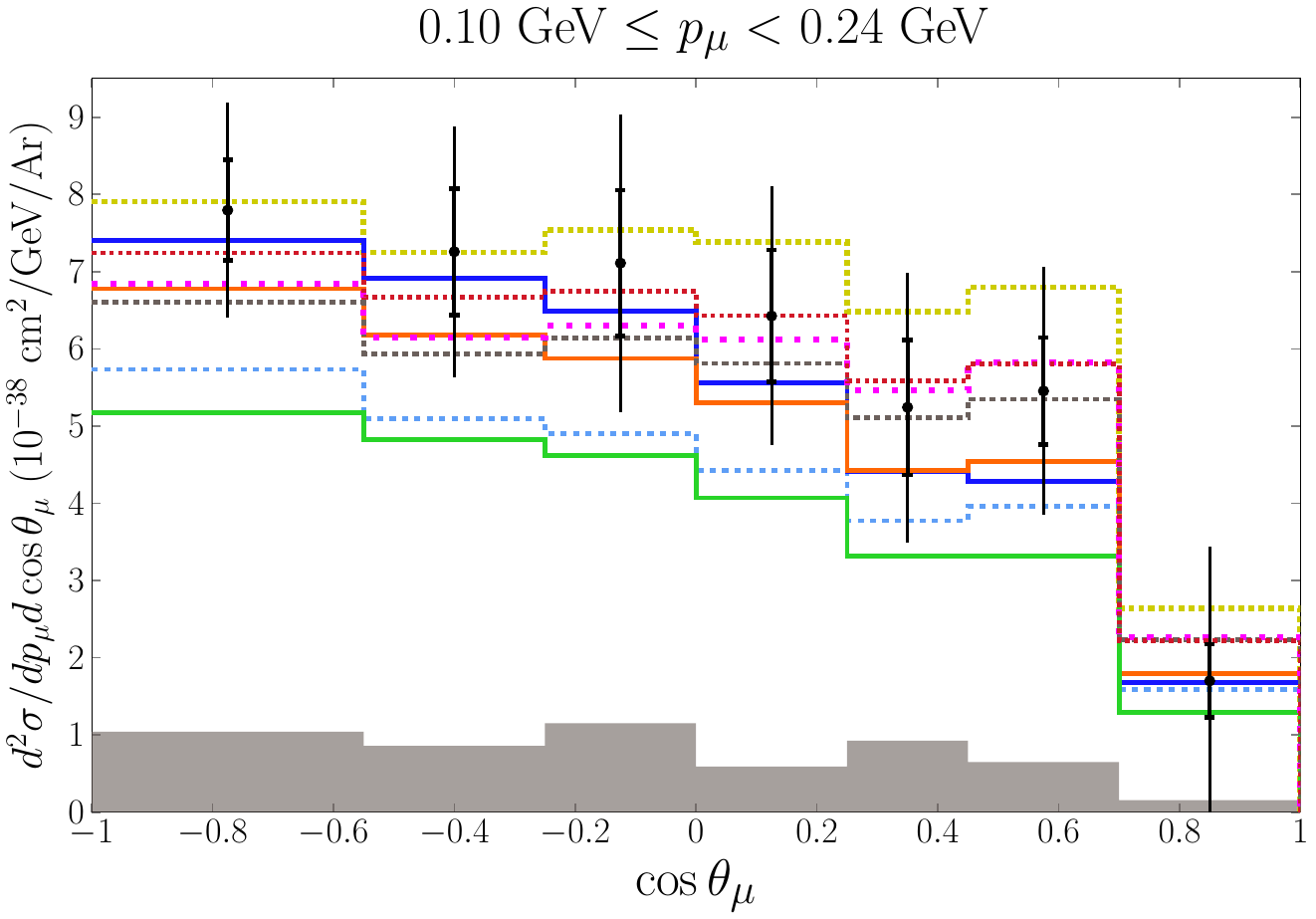}
\hfill
\includegraphics[page=2, width=0.49\textwidth]{figures/real_data_prd.pdf}
\vspace{0.1cm}

\includegraphics[page=3, width=0.49\textwidth]{figures/real_data_prd.pdf}
\hfill
\includegraphics[page=4, width=0.49\textwidth]{figures/real_data_prd.pdf}
\vspace{0.1cm}

\includegraphics[page=5, width=0.49\textwidth]{figures/real_data_prd.pdf}
\hfill
\includegraphics[page=6, width=0.49\textwidth]{figures/real_data_prd.pdf}
\vspace{0.1cm}

\includegraphics[page=7, width=0.49\textwidth]{figures/real_data_prd.pdf}
\hfill
\belowbaseline[-0.23\textheight]{
  \includegraphics[page=9, width=0.35\textwidth]{figures/real_data_prd.pdf}%
}
\hfill

\caption{Measured differential cross sections for block \#0
($p_\mu, \cos\theta_\mu$).
Statistical (shape-only systematic) uncertainties are included in the inner
(outer) error bars. The remainder of the total uncertainty is shown by the gray
band along the $x$-axis.}

\label{fig:XSecResultsblock0}
\end{figure*}

\begin{figure*}
\centering
\includegraphics[page=12, width=0.49\textwidth]{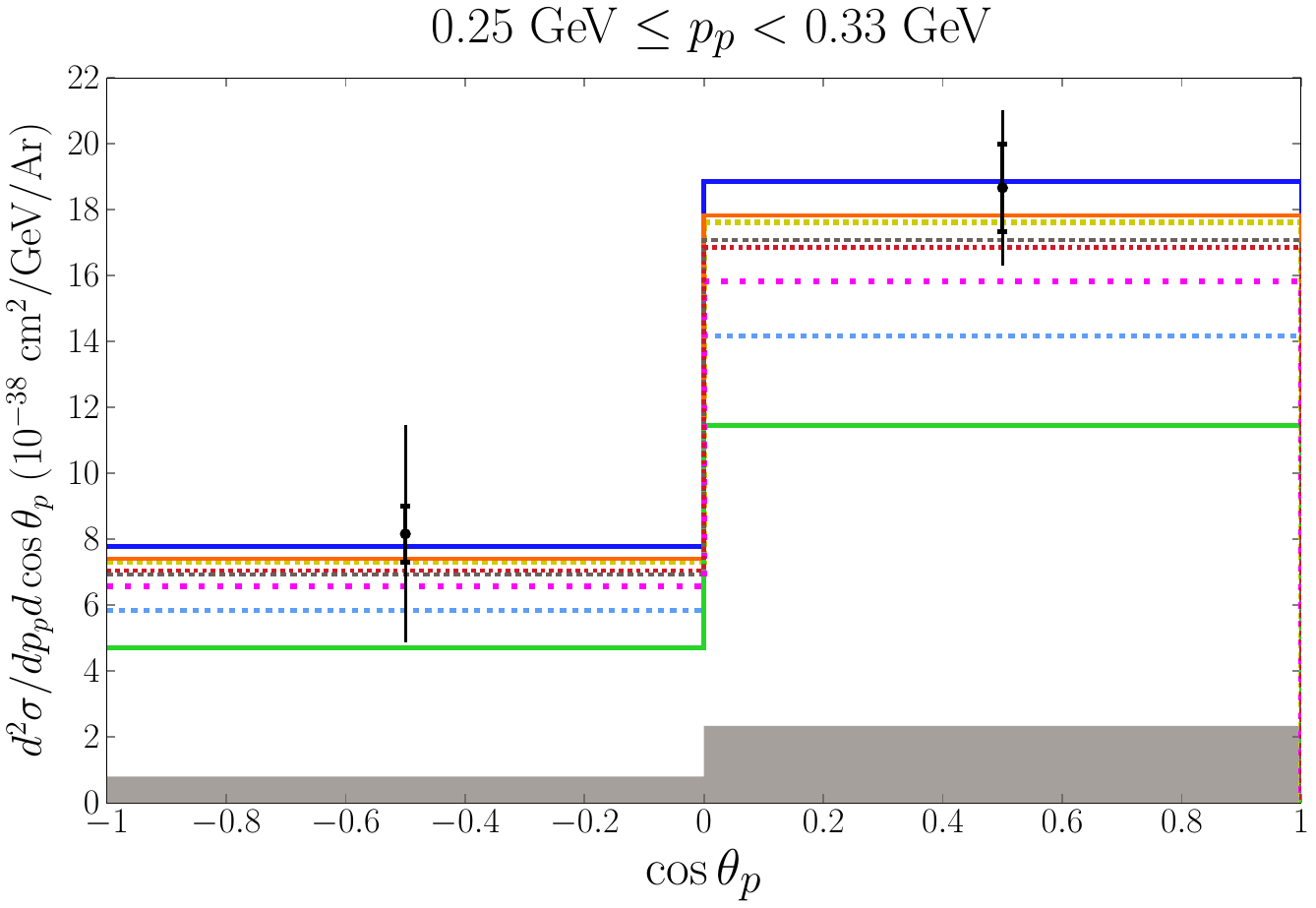}
\hfill
\includegraphics[page=13, width=0.49\textwidth]{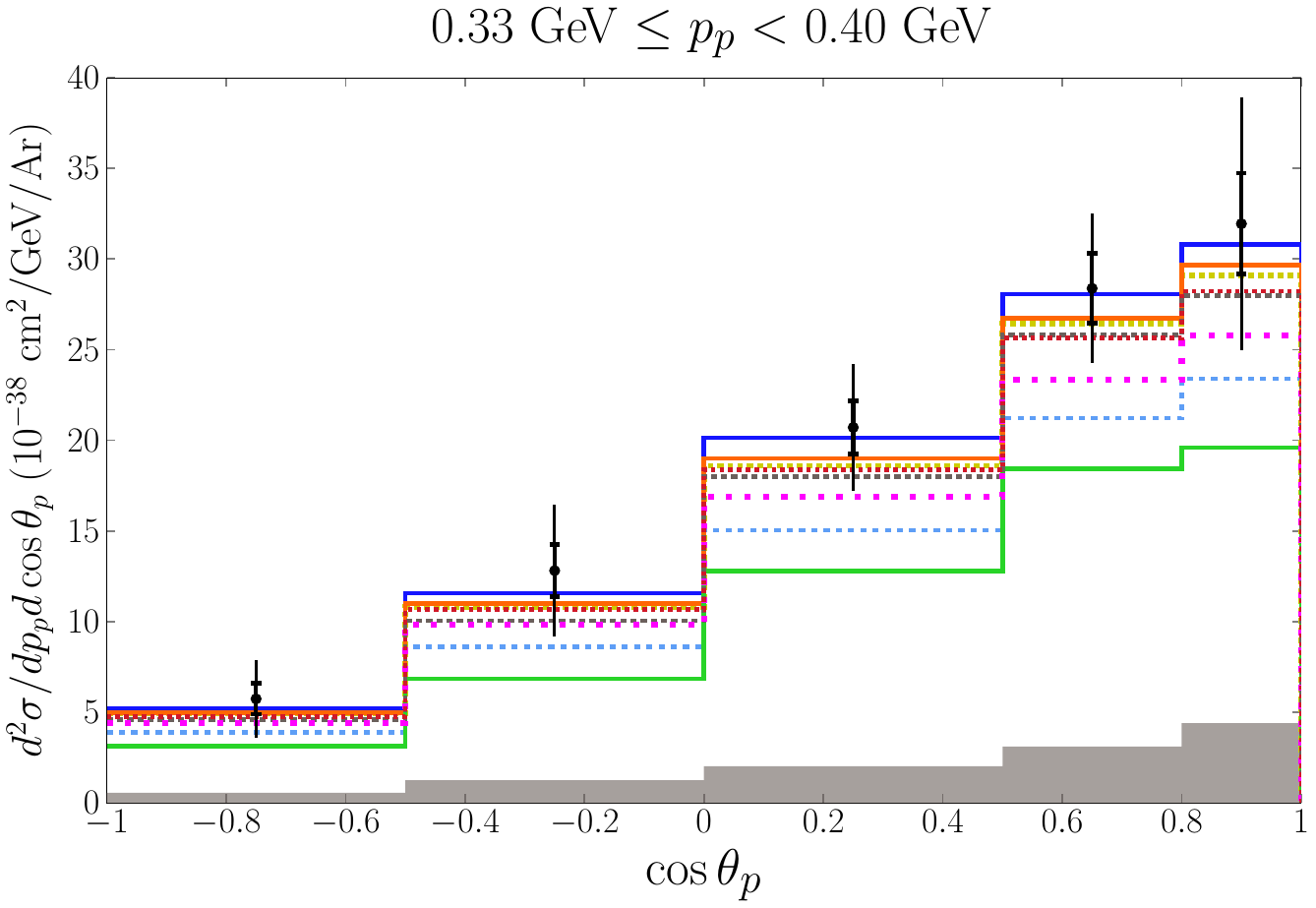}
\vspace{0.2cm}

\includegraphics[page=14, width=0.49\textwidth]{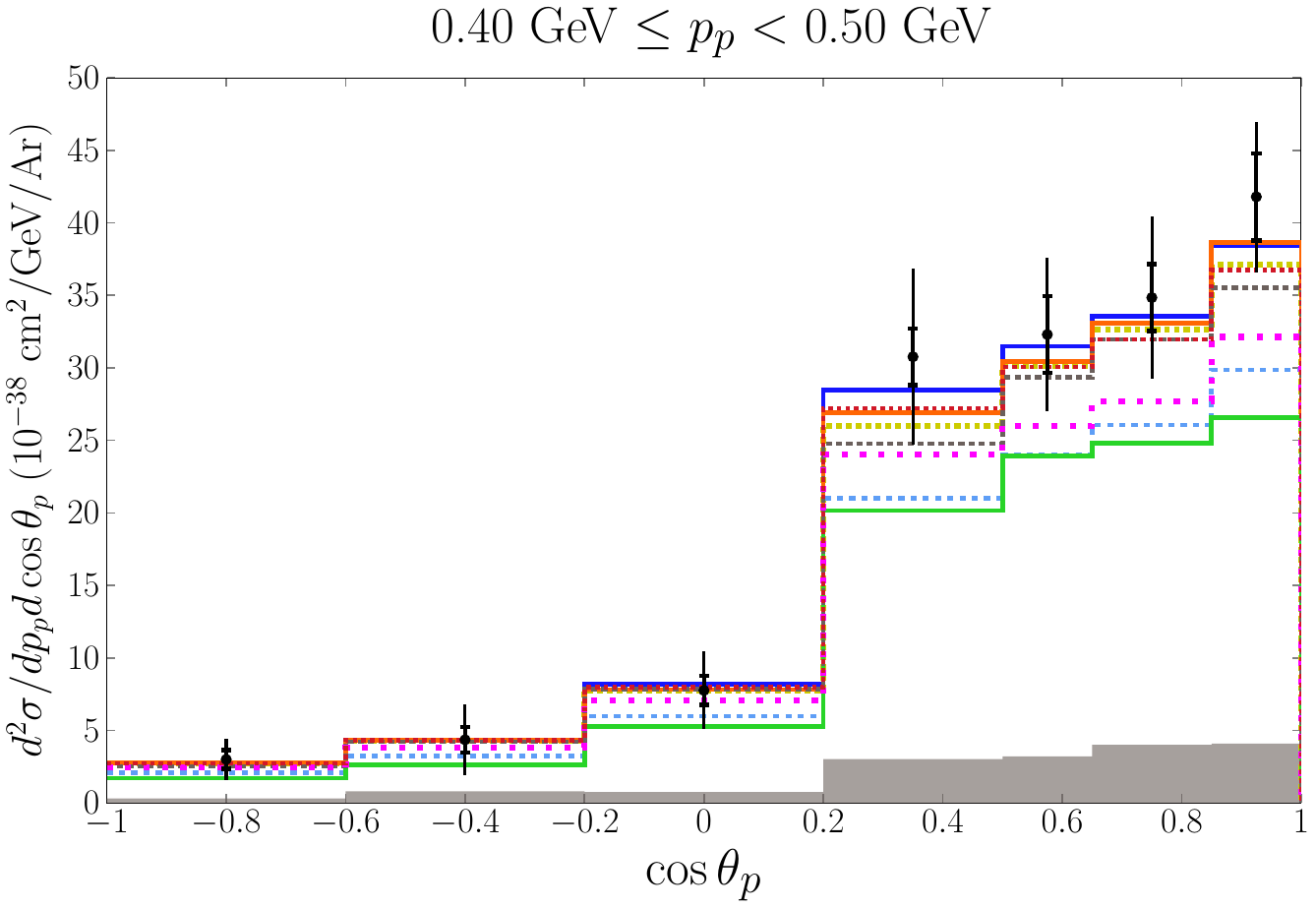}
\hfill
\includegraphics[page=15, width=0.49\textwidth]{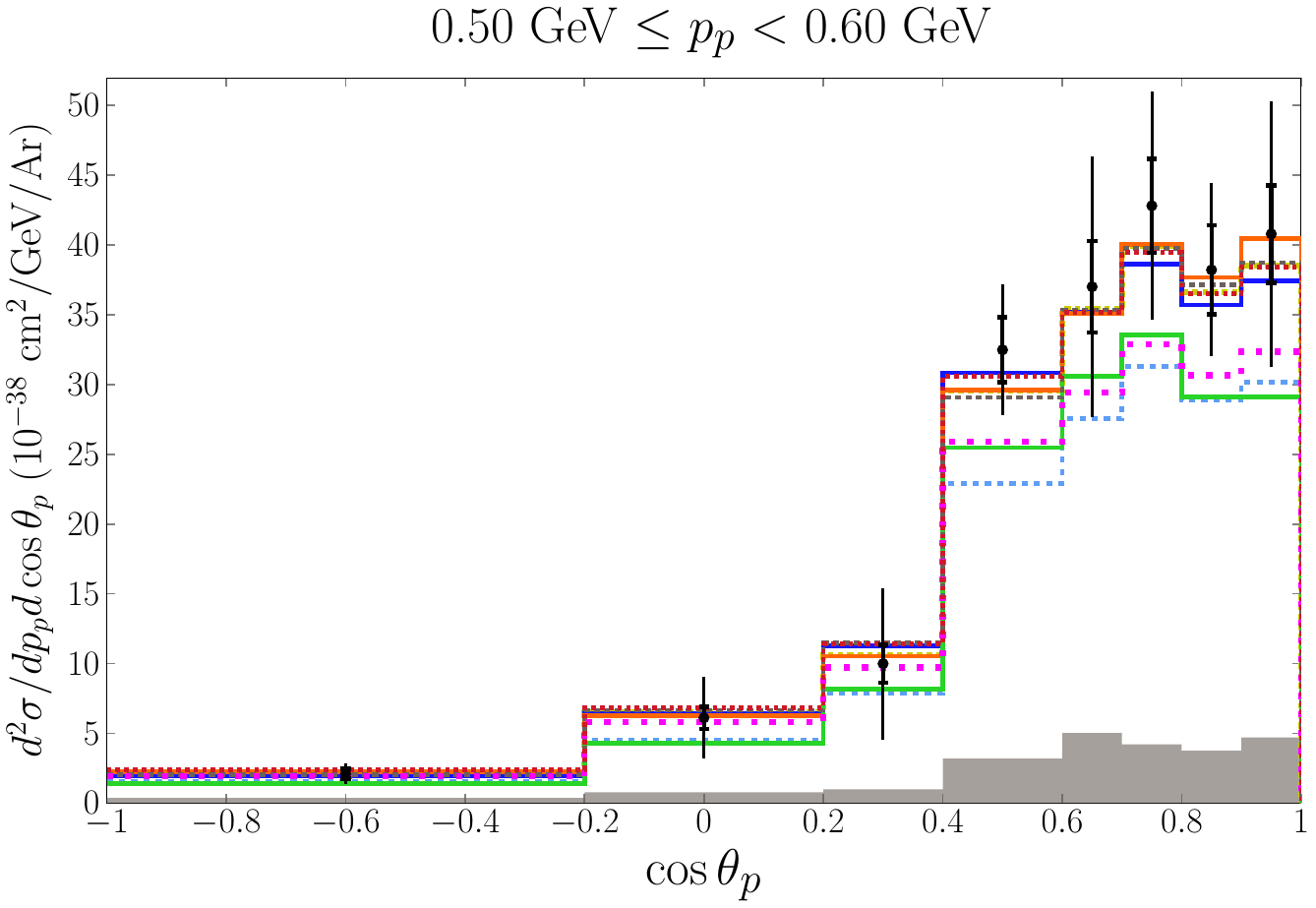}
\vspace{0.2cm}

\includegraphics[page=16, width=0.49\textwidth]{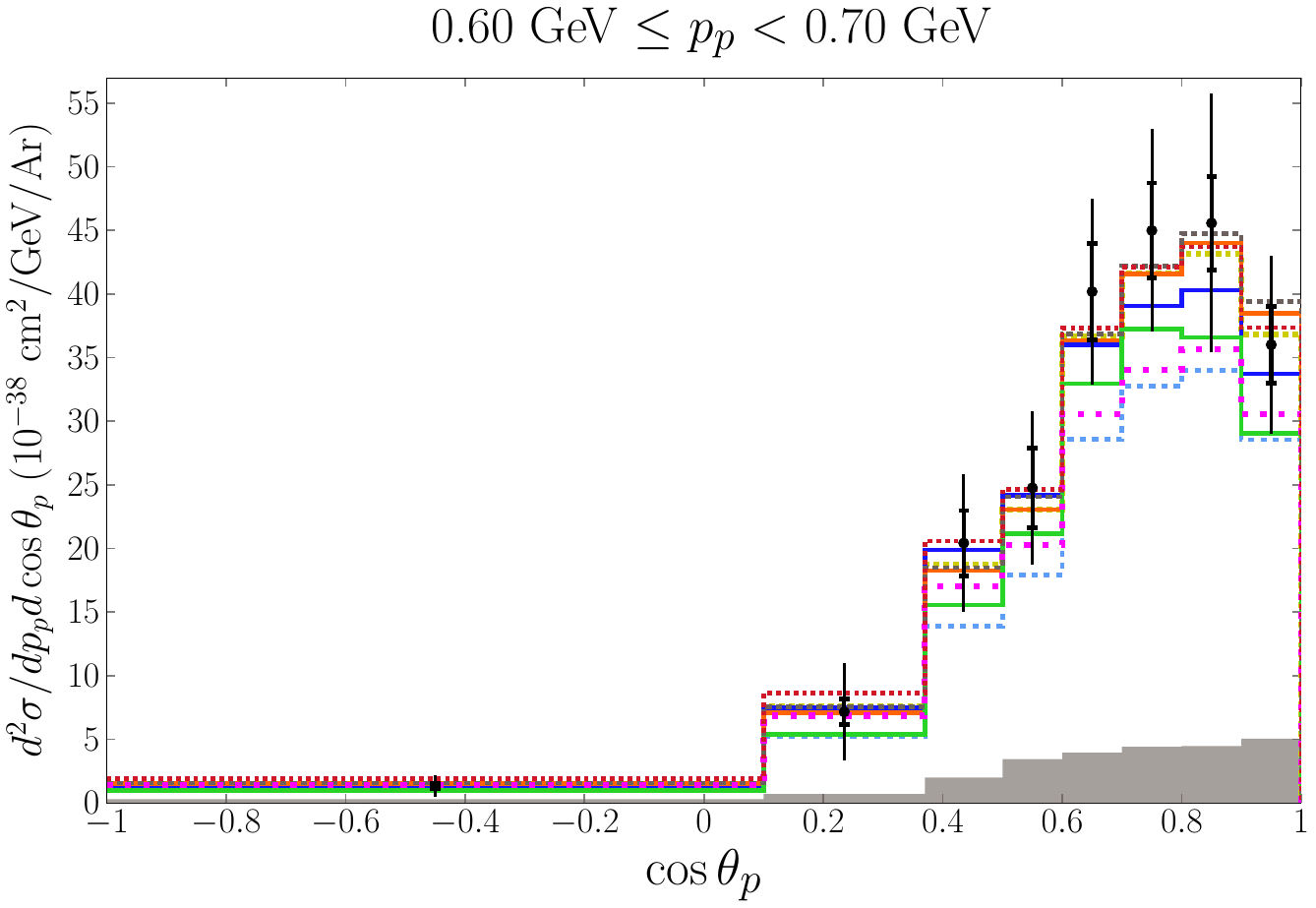}
\hfill
\includegraphics[page=17, width=0.49\textwidth]{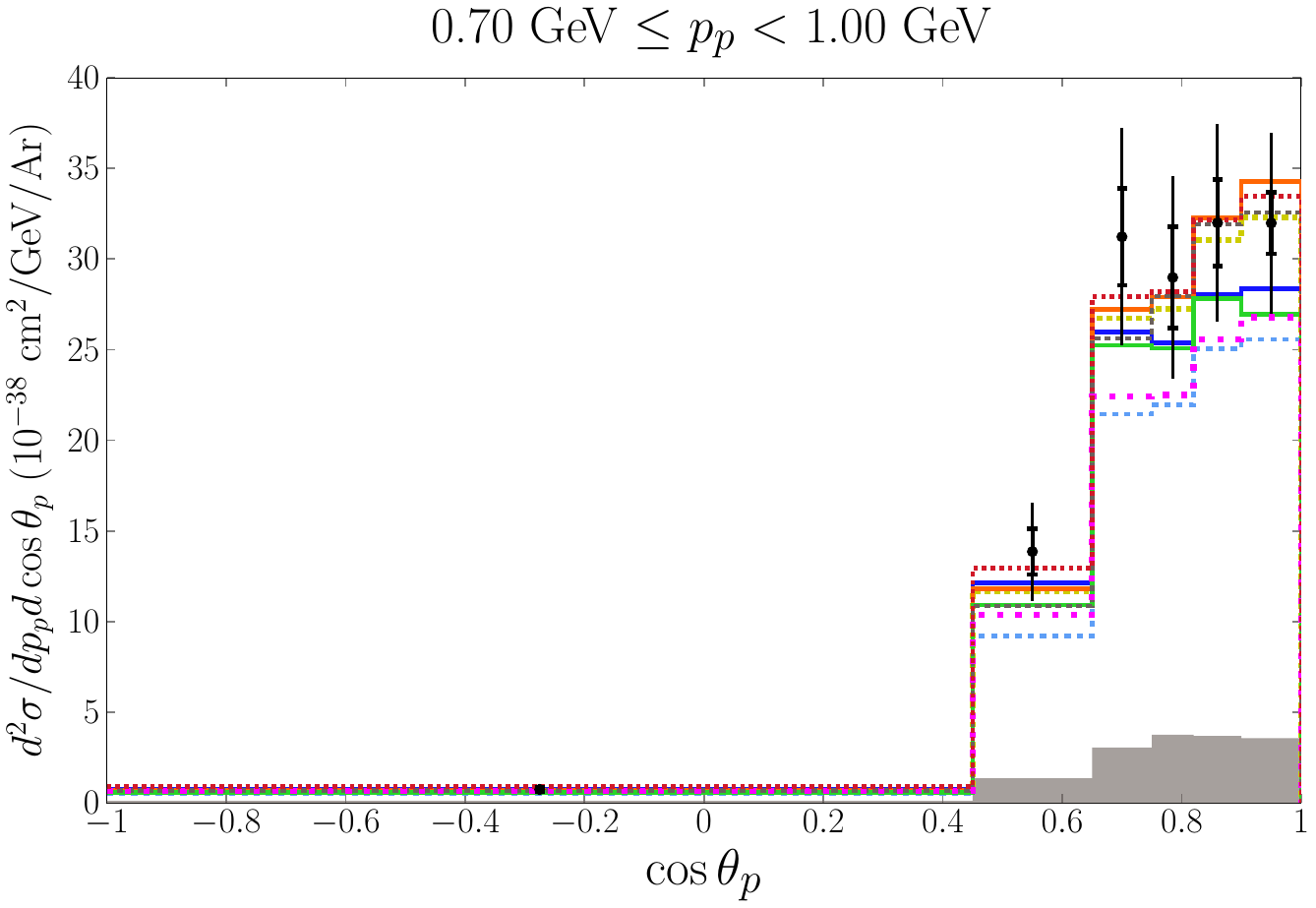}
\vspace{0.2cm}

\includegraphics[page=19, width=0.35\textwidth]{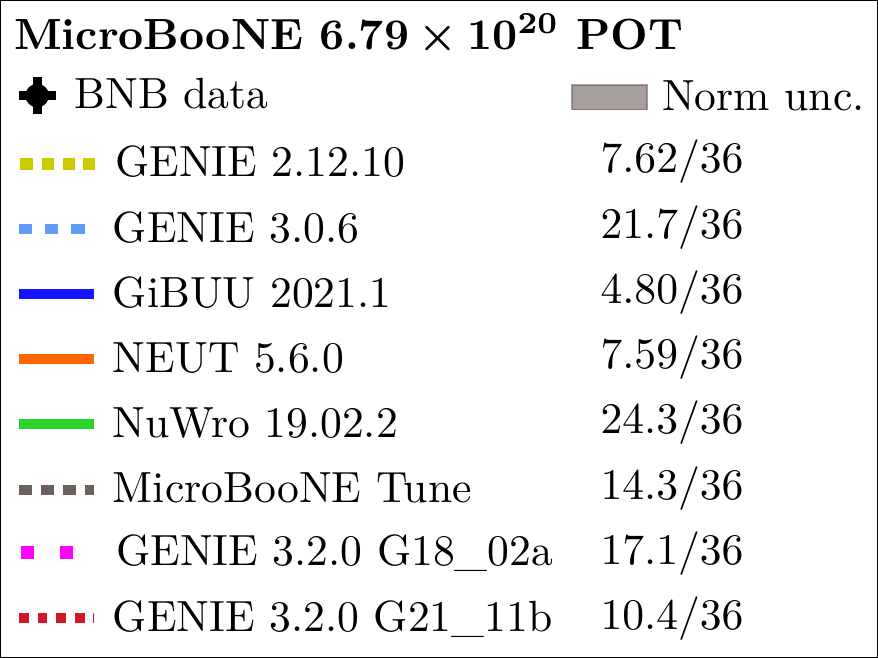}%
\hfill

\caption{Measured differential cross sections for block \#1
($p_p, \cos\theta_p$).
Statistical (shape-only systematic) uncertainties are included in the inner
(outer) error bars. The remainder of the total uncertainty is shown by the gray
band along the $x$-axis.}
\label{fig:XSecResultsblock1}
\end{figure*}

\begin{figure*}
\centering

\includegraphics[page=20, width=0.64\textwidth]{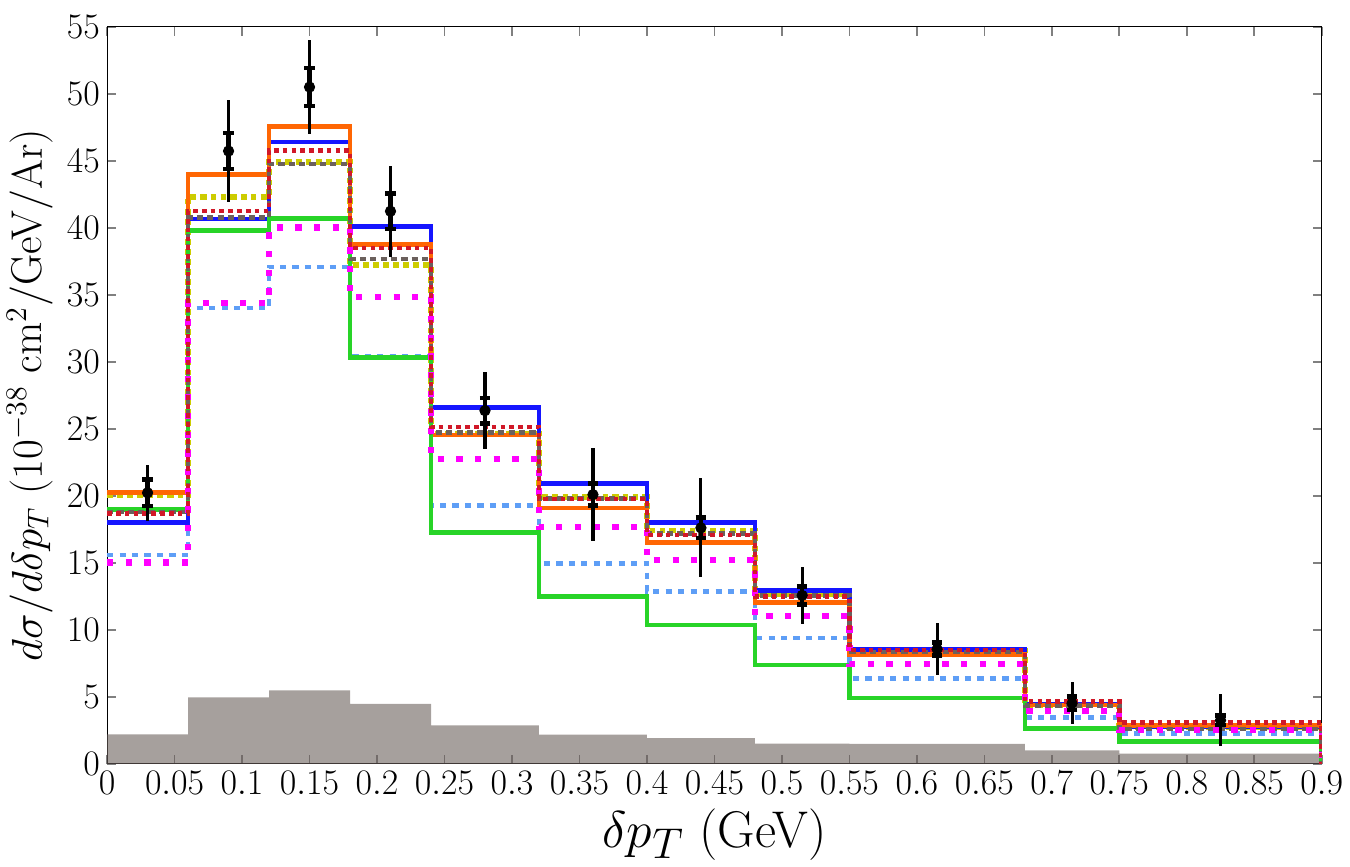}
\hfill
\belowbaseline[-0.23\textheight]{
  \includegraphics[page=21, width=0.35\textwidth]{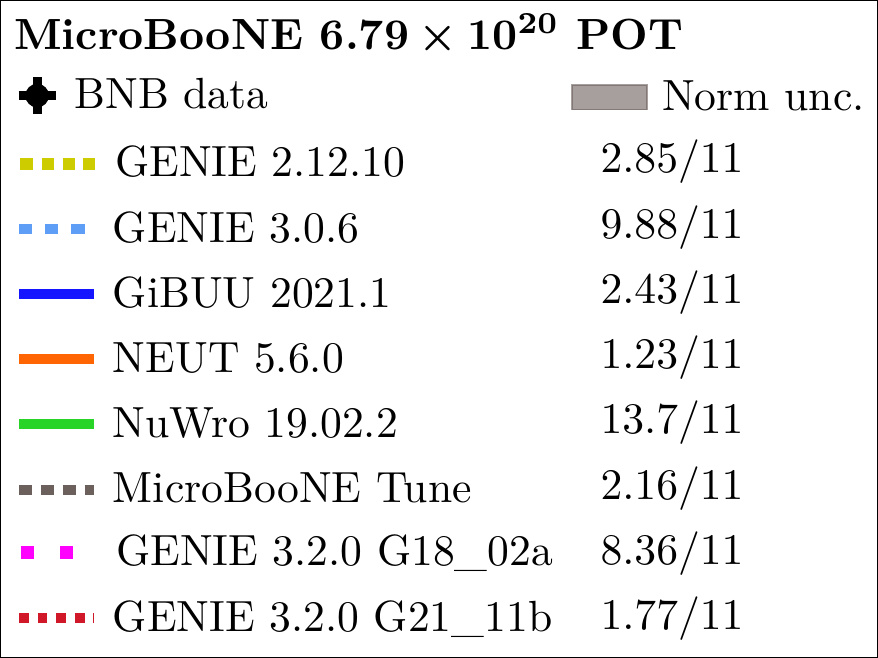}%
}
\hfill

\caption{Measured differential cross sections for block \#2
($\delta p_T$).
Statistical (shape-only systematic) uncertainties are included in the inner
(outer) error bars. The remainder of the total uncertainty is shown by the gray
band along the $x$-axis.}
\label{fig:XSecResultsblock2}
\end{figure*}

\begin{figure*}
\centering
\includegraphics[page=22, width=0.49\textwidth]{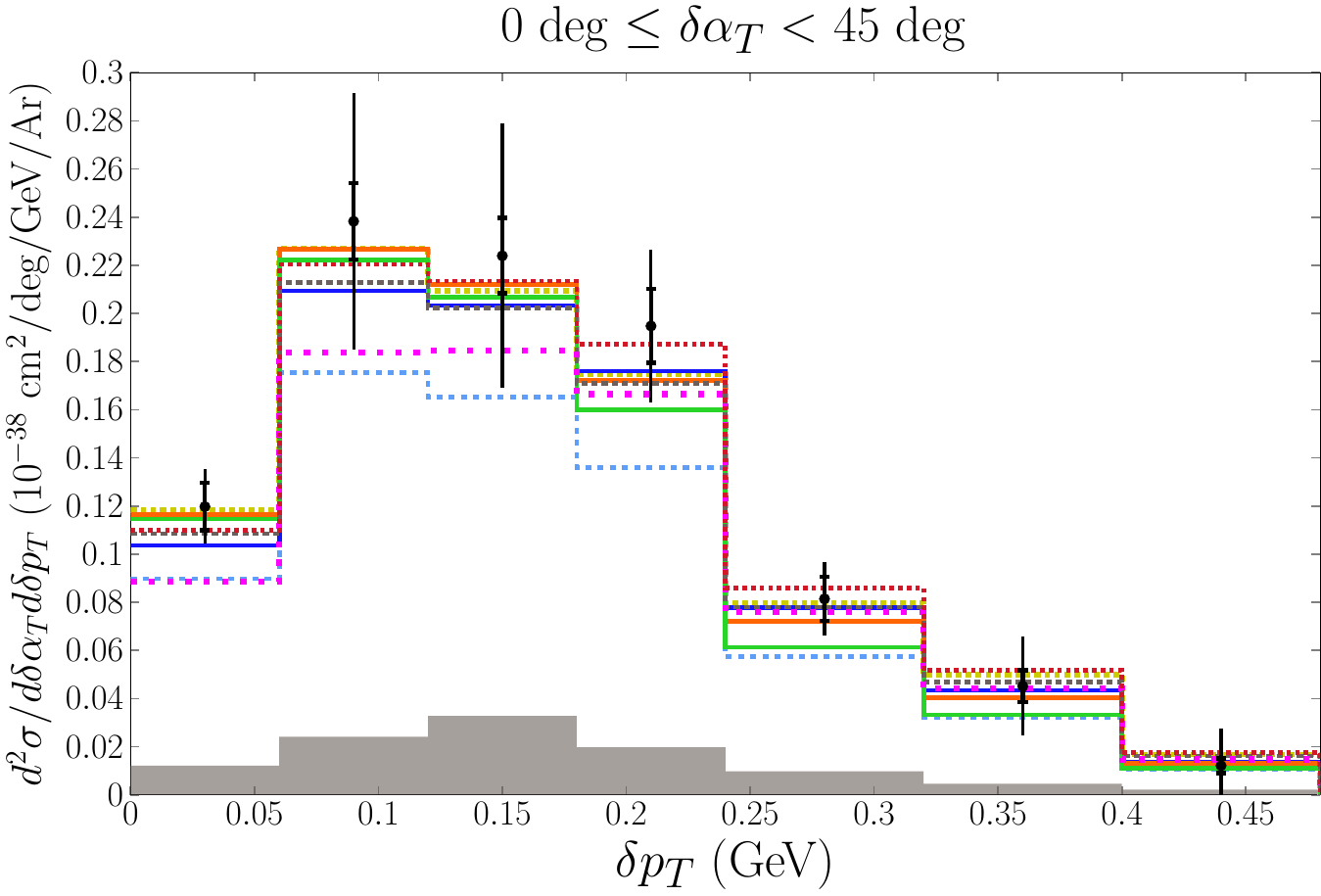}
\hfill
\includegraphics[page=23, width=0.49\textwidth]{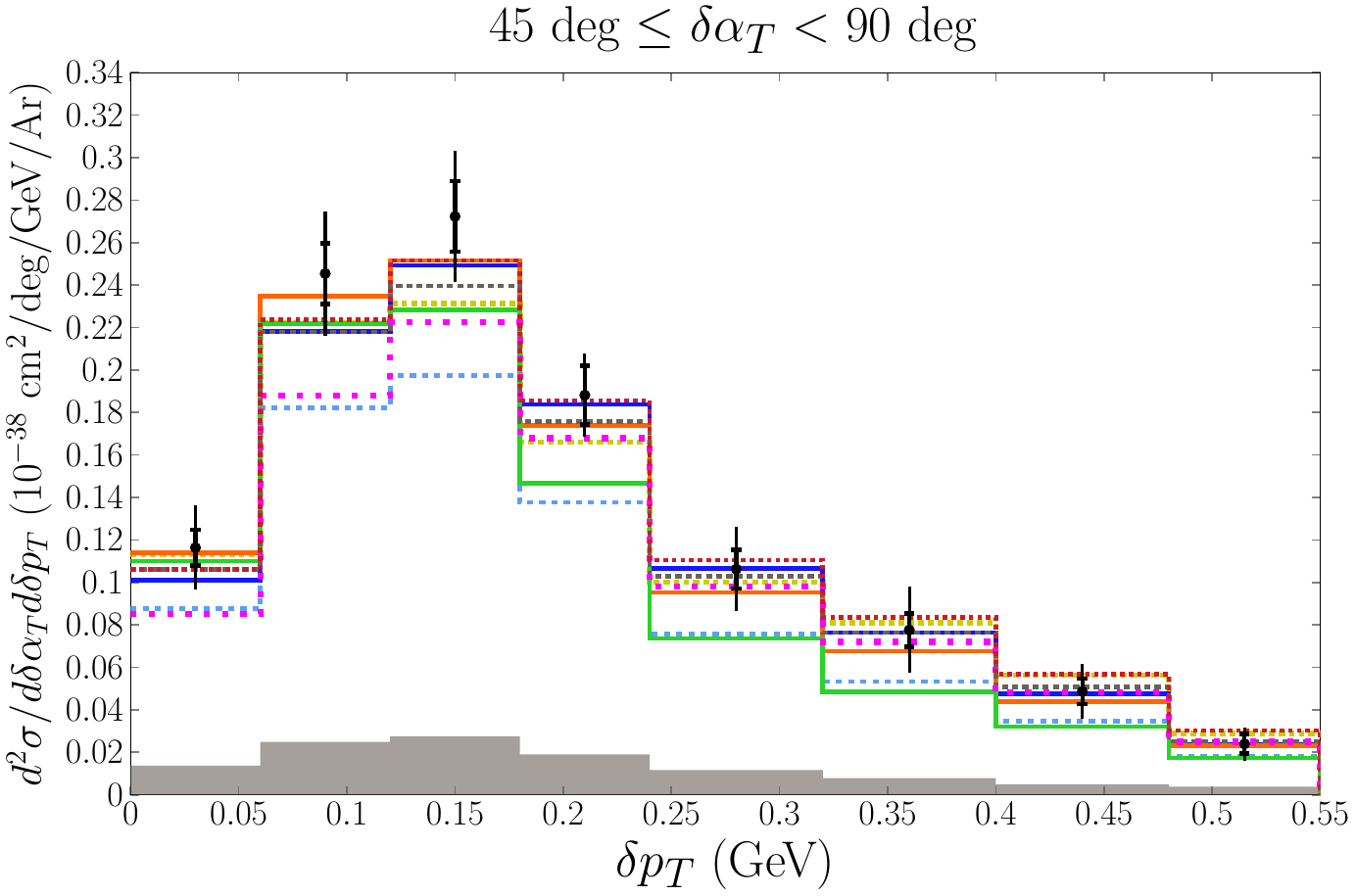}
\vspace{0.2cm}

\includegraphics[page=24, width=0.49\textwidth]{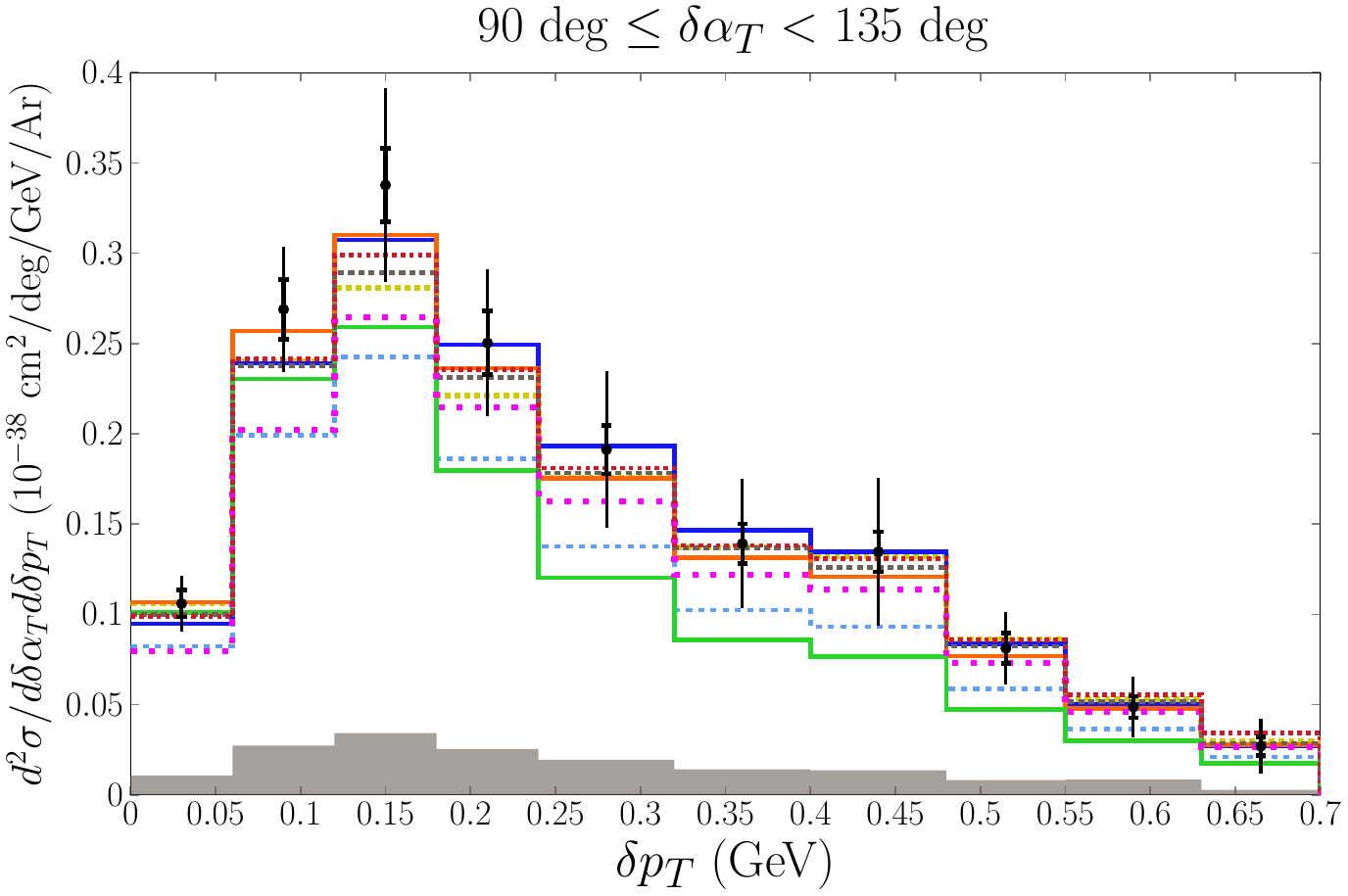}
\hfill
\includegraphics[page=25, width=0.49\textwidth]{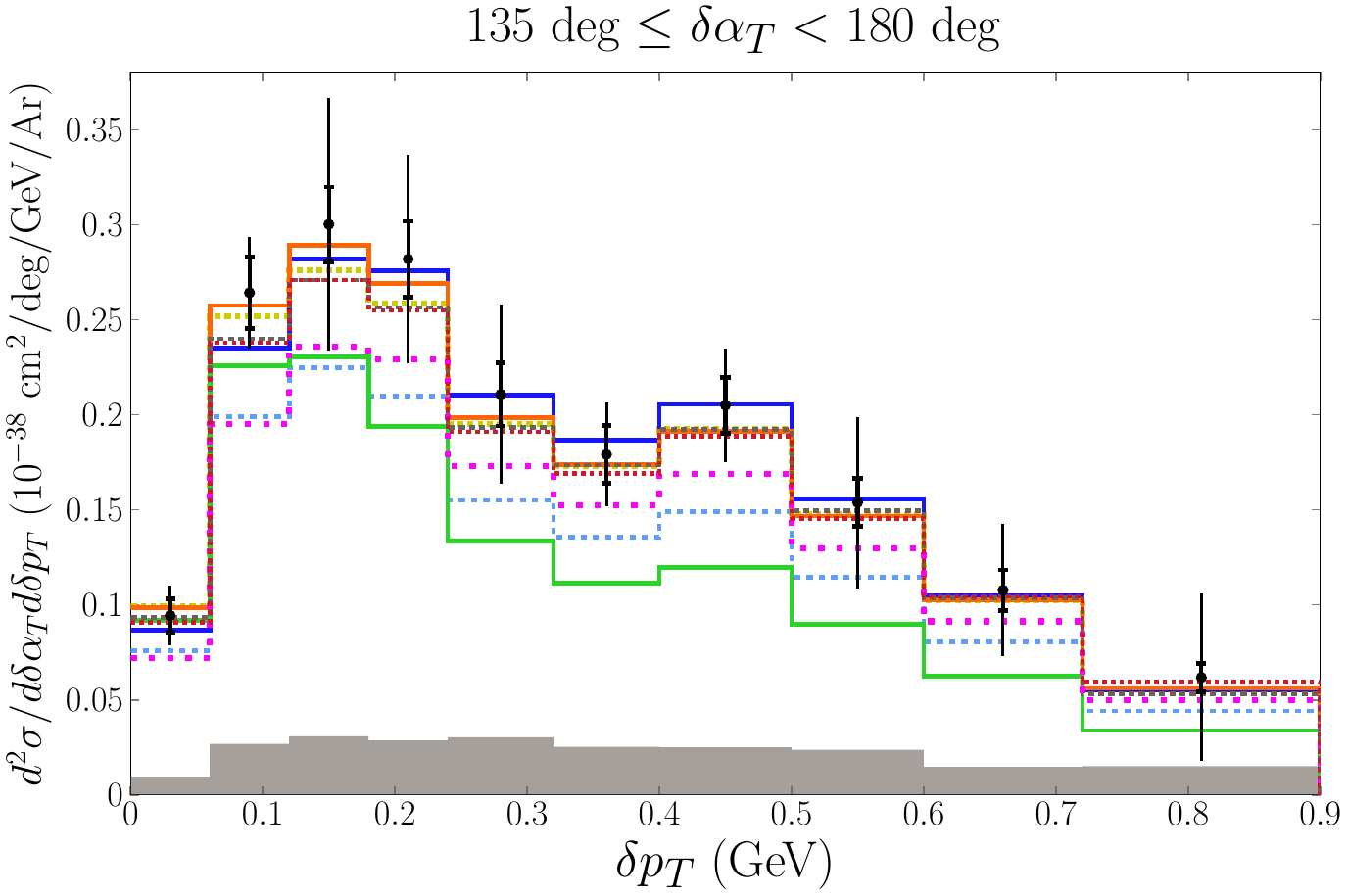}
\vspace{0.2cm}

\includegraphics[page=27, width=0.35\textwidth]{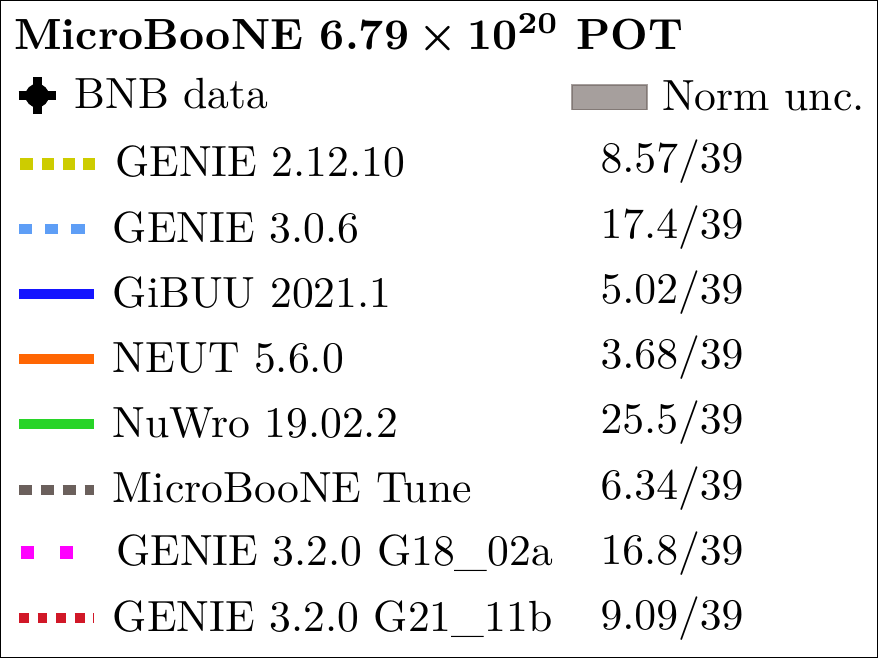}%
\hfill

\caption{Measured differential cross sections for block \#3
$(\delta \alpha_T, \delta p_T)$. The overall
$\chi^2$ value includes contributions from four $\delta p_T$ overflow bins that
are not plotted.
Statistical (shape-only systematic) uncertainties are included in the inner
(outer) error bars. The remainder of the total uncertainty is shown by the gray
band along the $x$-axis.}
\label{fig:XSecResultsblock3}
\end{figure*}

\begin{figure*}
\centering
\includegraphics[page=30, width=0.49\textwidth]{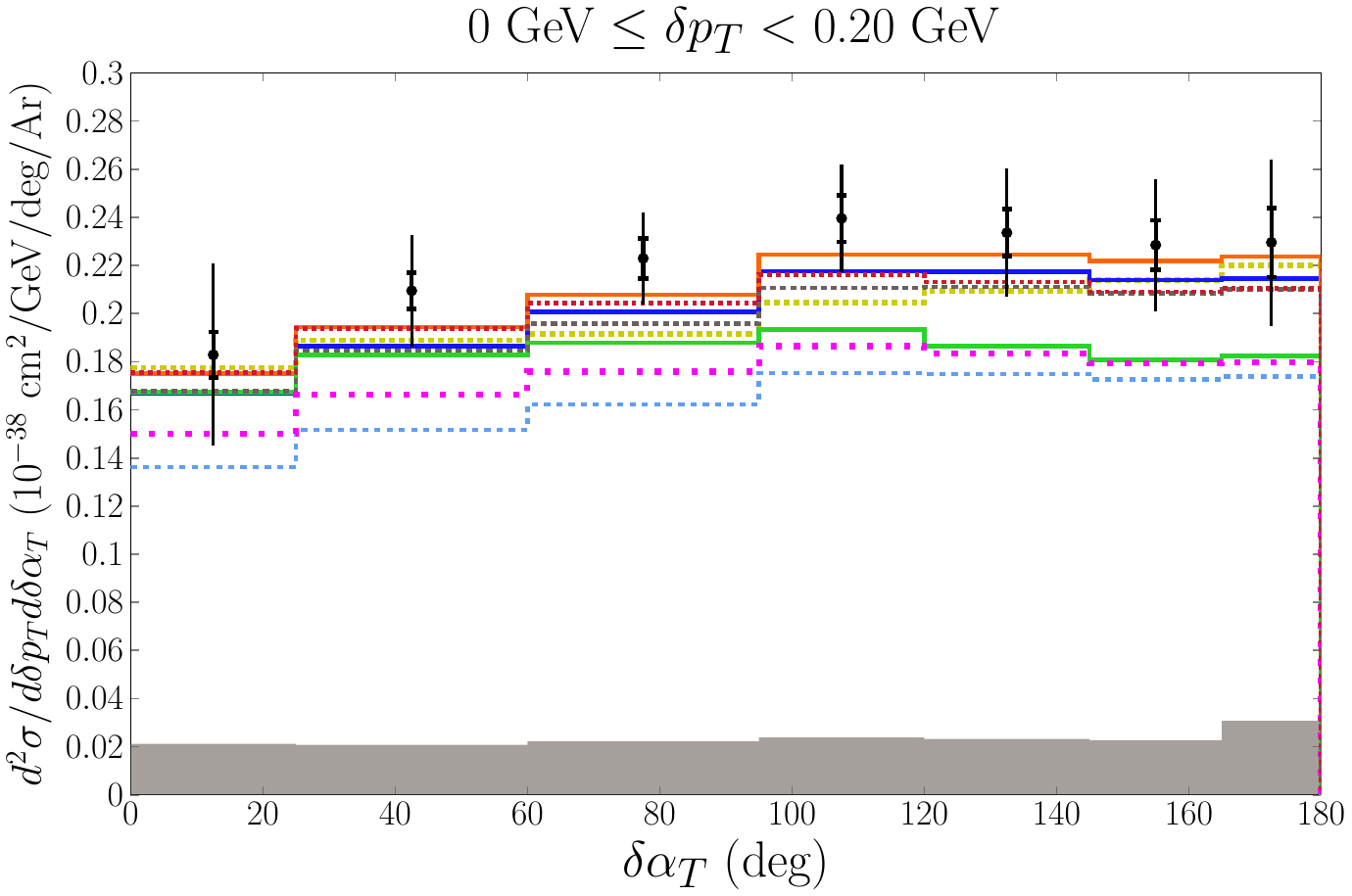}
\hfill
\includegraphics[page=31, width=0.49\textwidth]{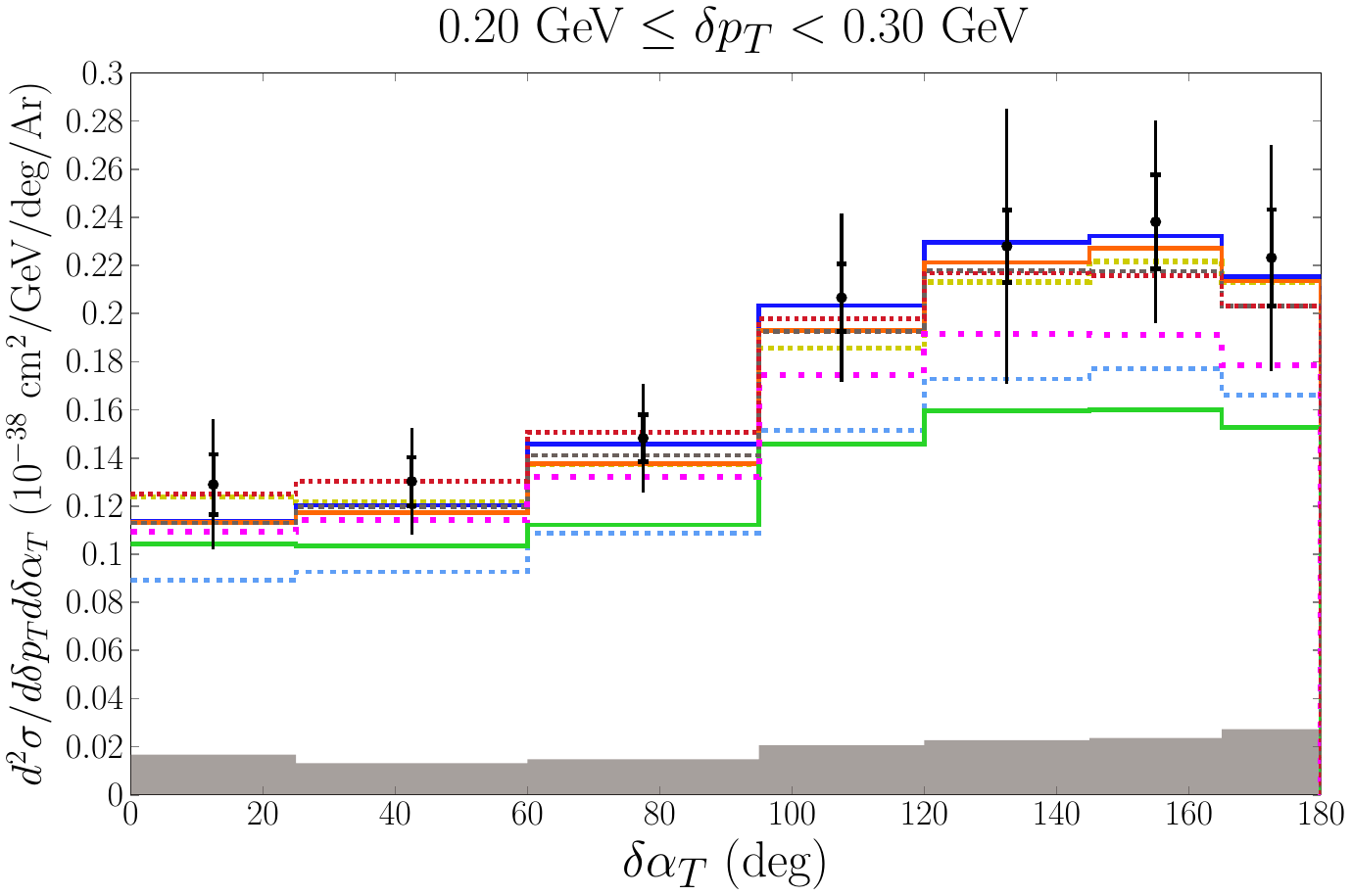}
\vspace{0.2cm}

\includegraphics[page=32, width=0.49\textwidth]{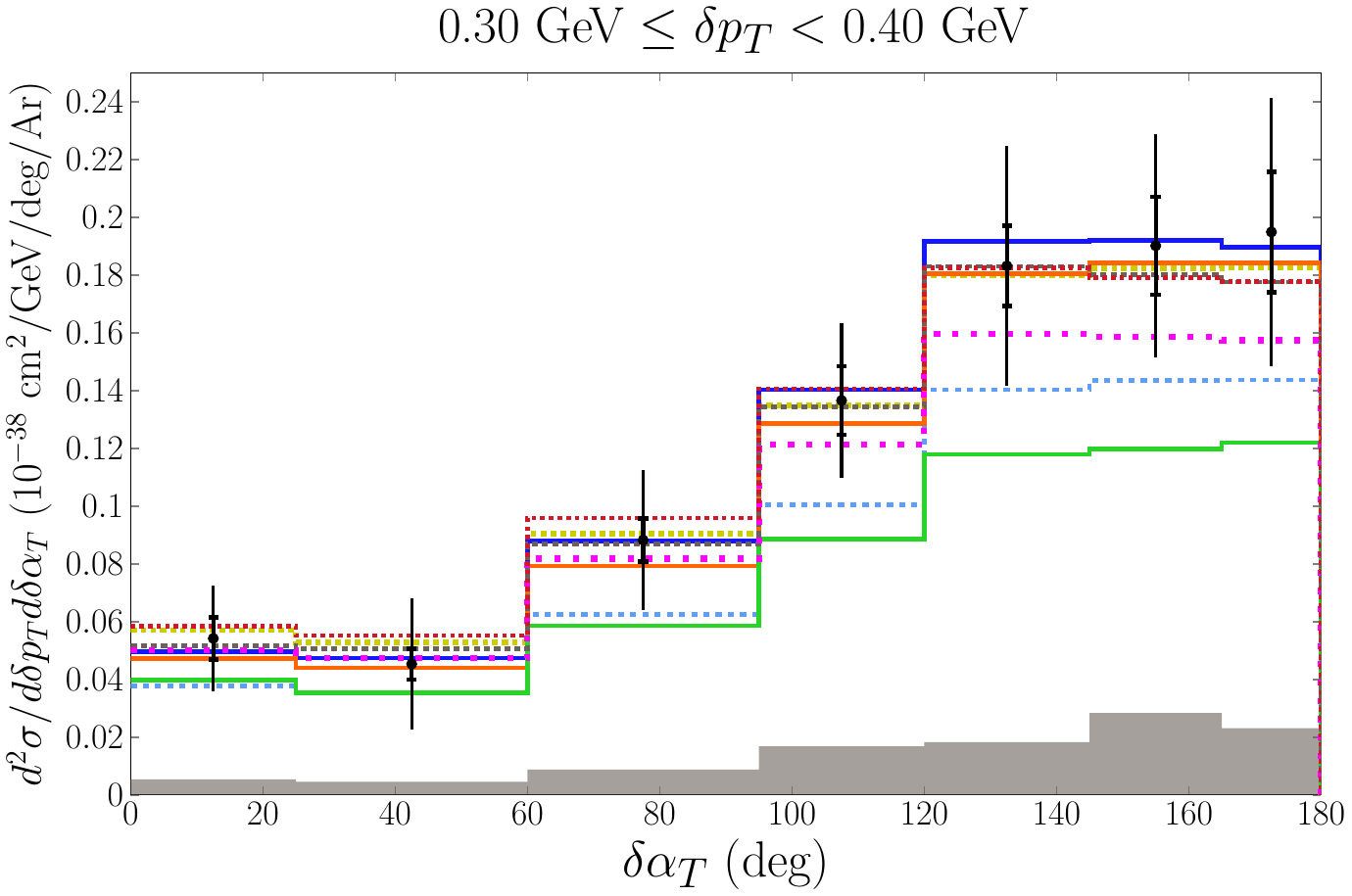}
\hfill
\includegraphics[page=33, width=0.49\textwidth]{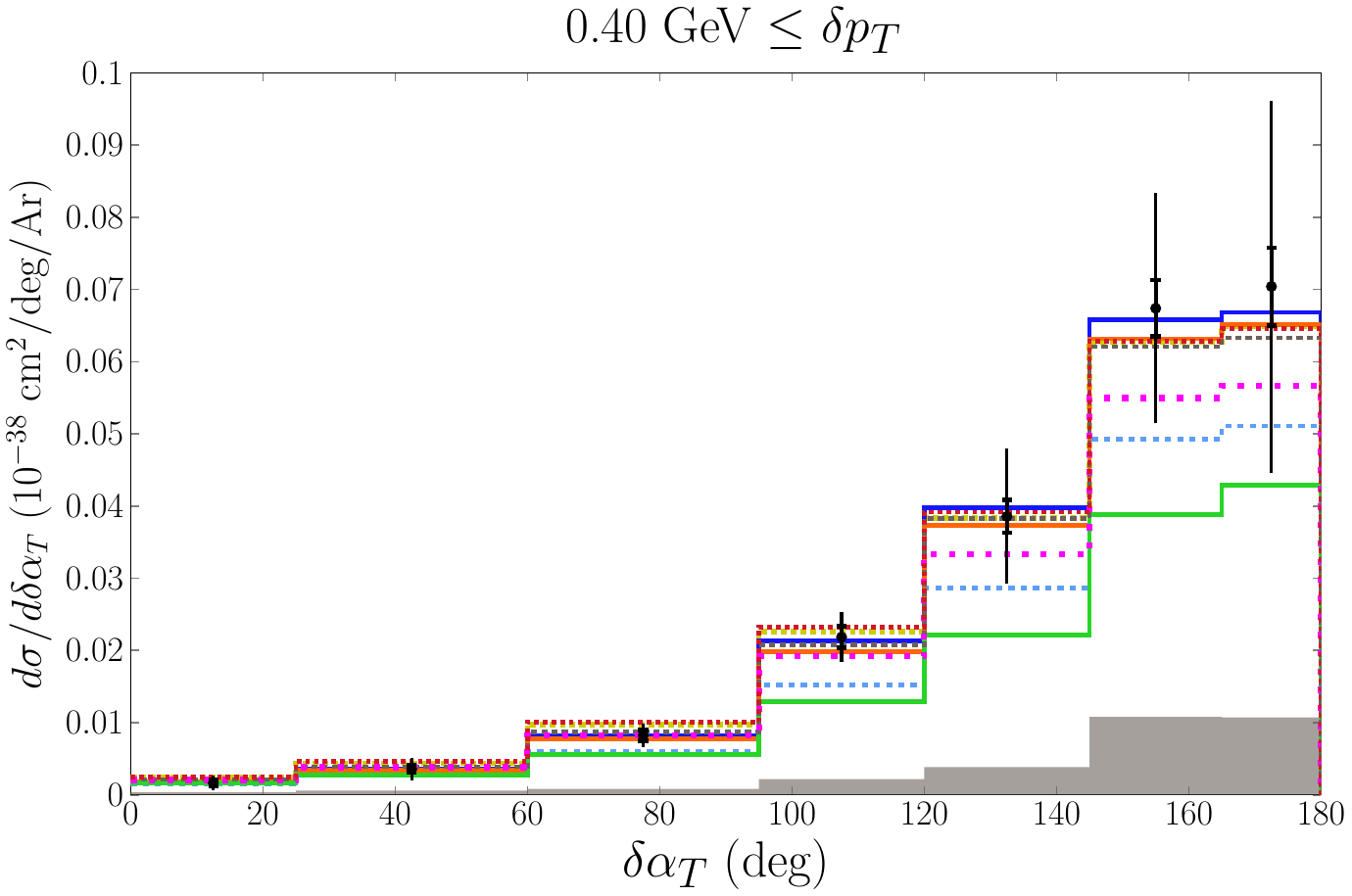}
\vspace{0.2cm}

\includegraphics[page=35, width=0.35\textwidth]{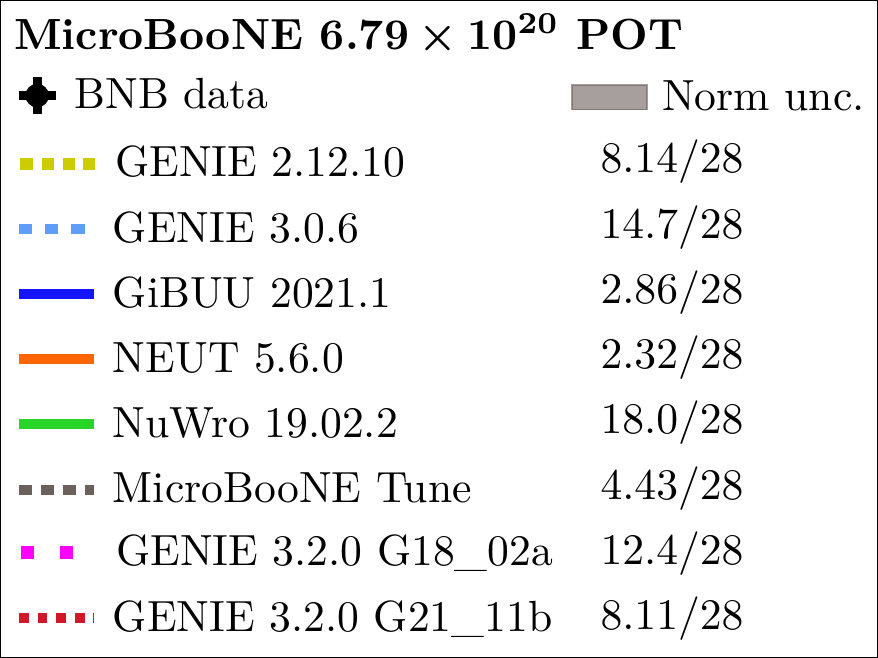}%
\hfill

\caption{Measured differential cross sections for block \#4
$(\delta p_T, \delta \alpha_T)$.
Statistical (shape-only systematic) uncertainties are included in the inner
(outer) error bars. The remainder of the total uncertainty is shown by the gray
band along the $x$-axis.}
\label{fig:XSecResultsblock4}
\end{figure*}

\begin{figure*}
\centering

\includegraphics[page=36, width=0.64\textwidth]{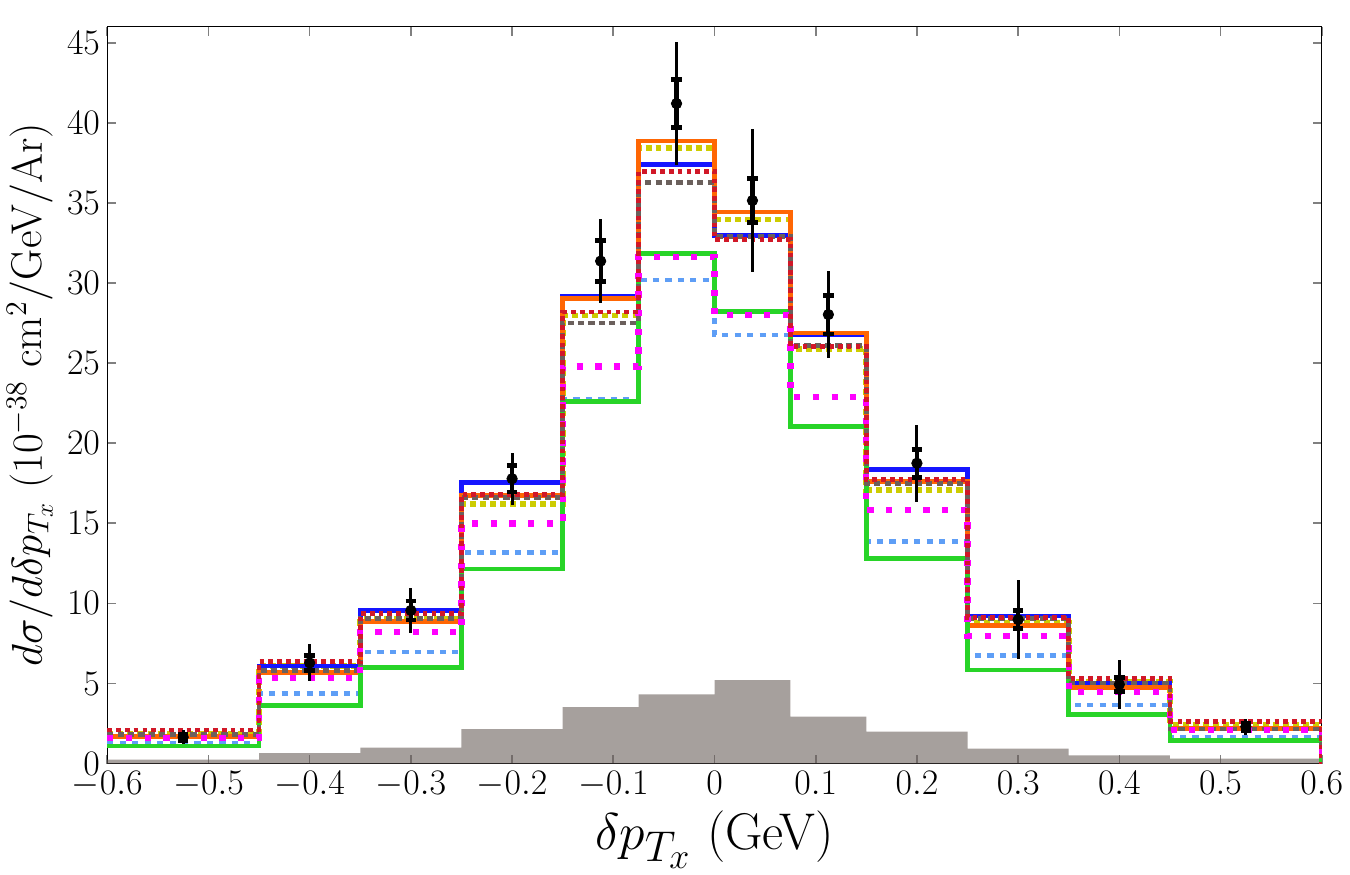}
\hfill
\belowbaseline[-0.23\textheight]{
  \includegraphics[page=37, width=0.35\textwidth]{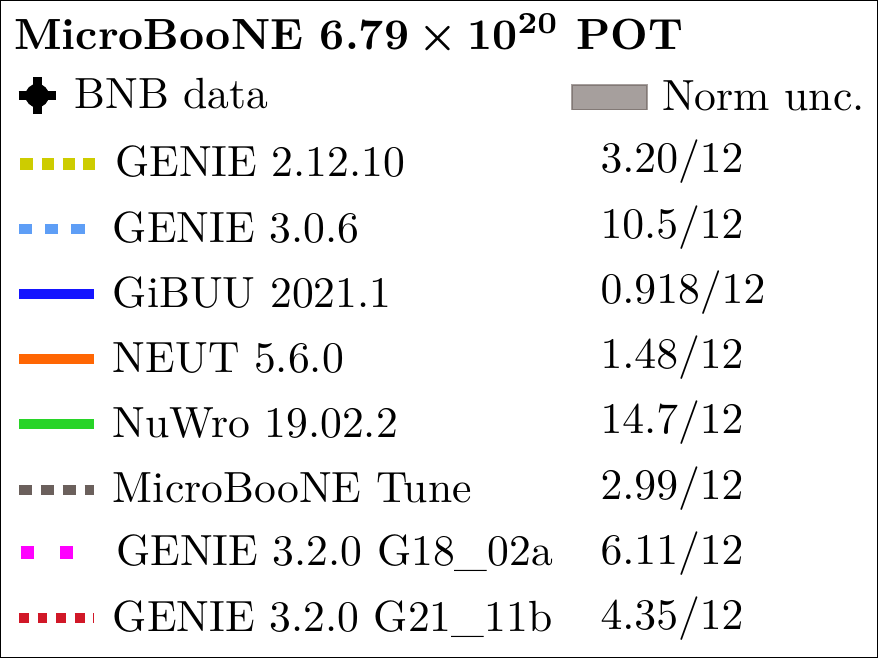}%
}
\hfill

\caption{Measured differential cross sections for block \#5
($\delta p_{T_x}$).
Statistical (shape-only systematic) uncertainties are included in the inner
(outer) error bars. The remainder of the total uncertainty is shown by the gray
band along the $x$-axis.}
\label{fig:XSecResultsblock5}
\end{figure*}

\begin{figure*}
\centering
\includegraphics[page=38, width=0.49\textwidth]{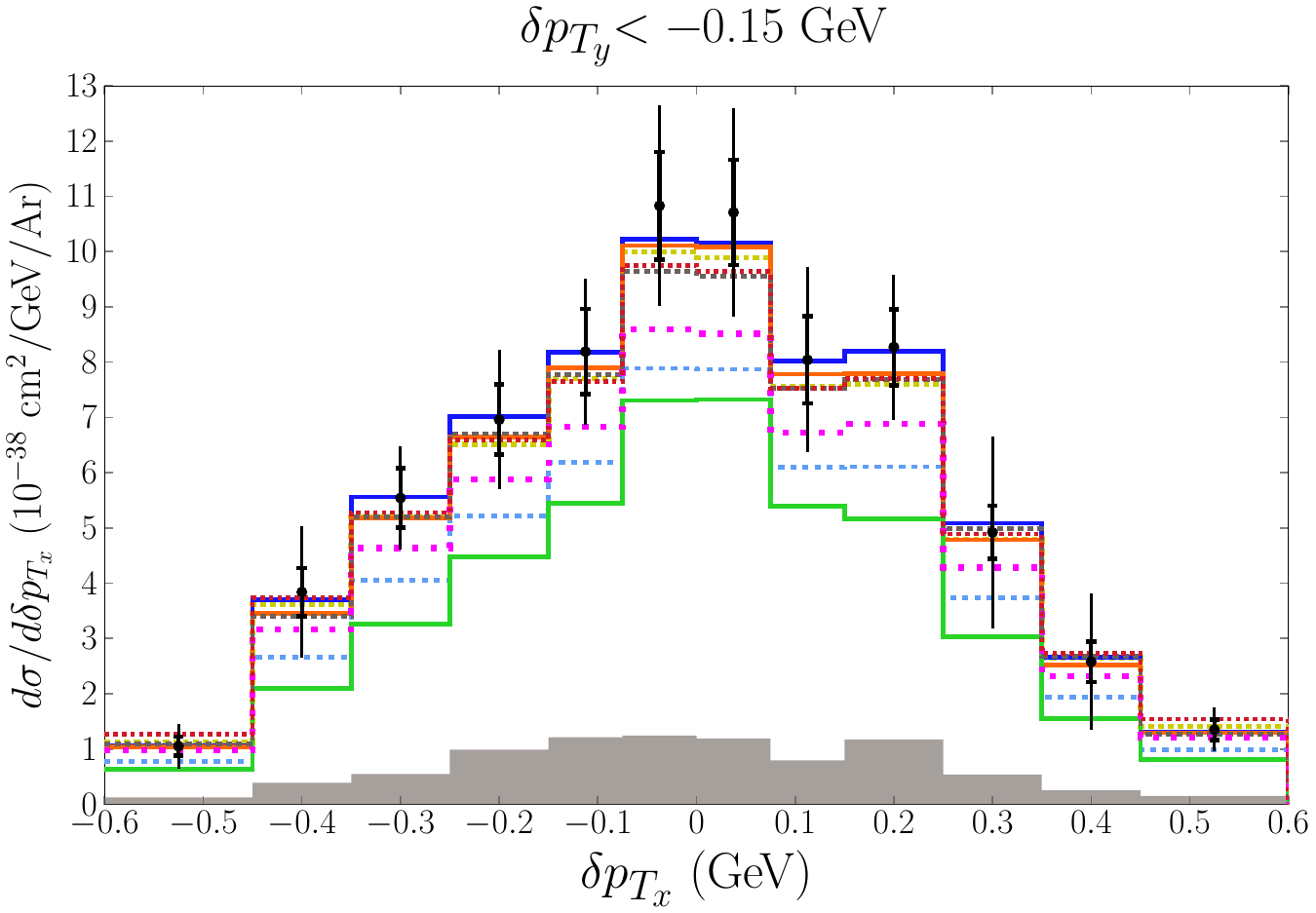}
\hfill
\includegraphics[page=39, width=0.49\textwidth]{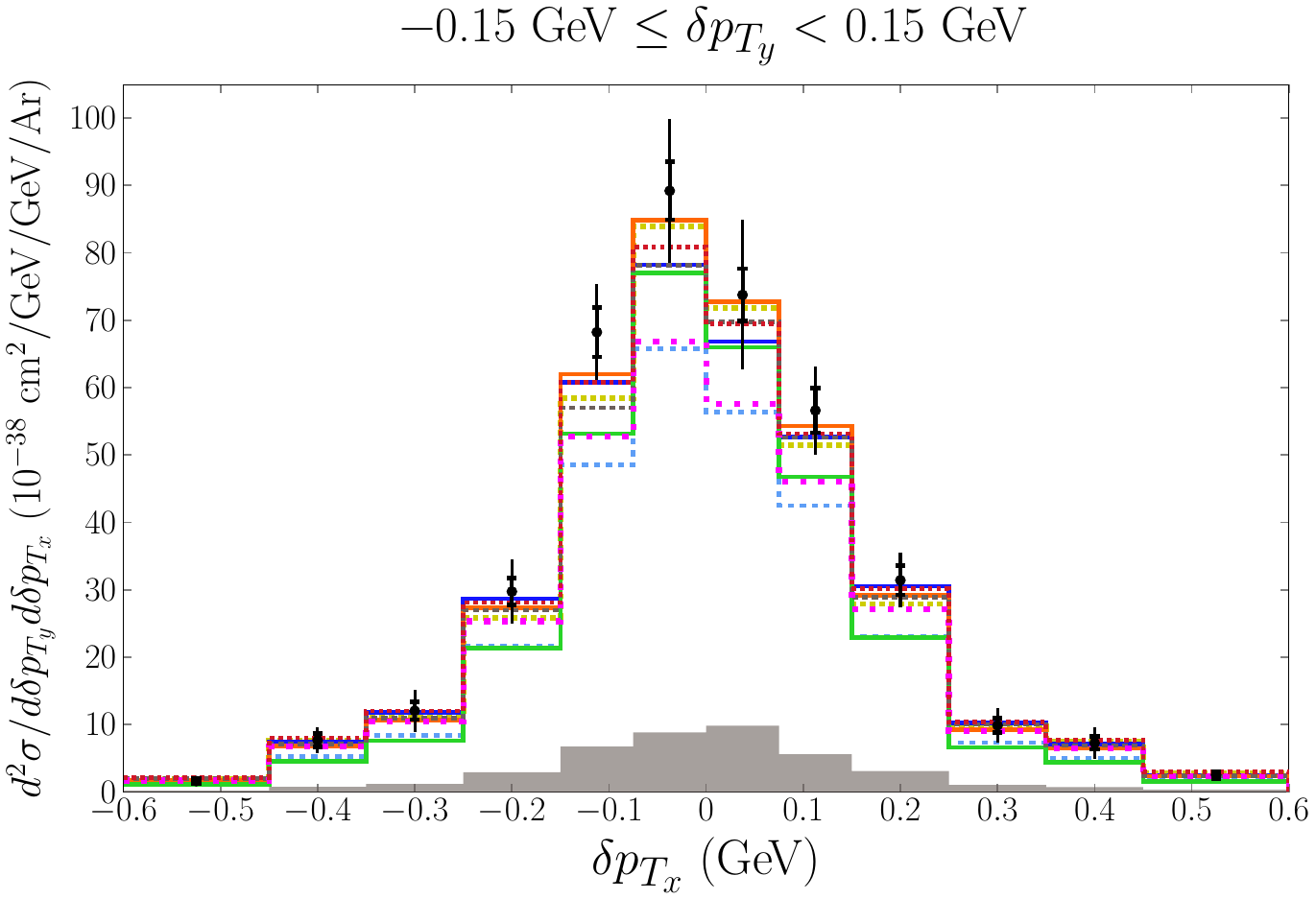}
\vspace{0.2cm}

\includegraphics[page=40, width=0.49\textwidth]{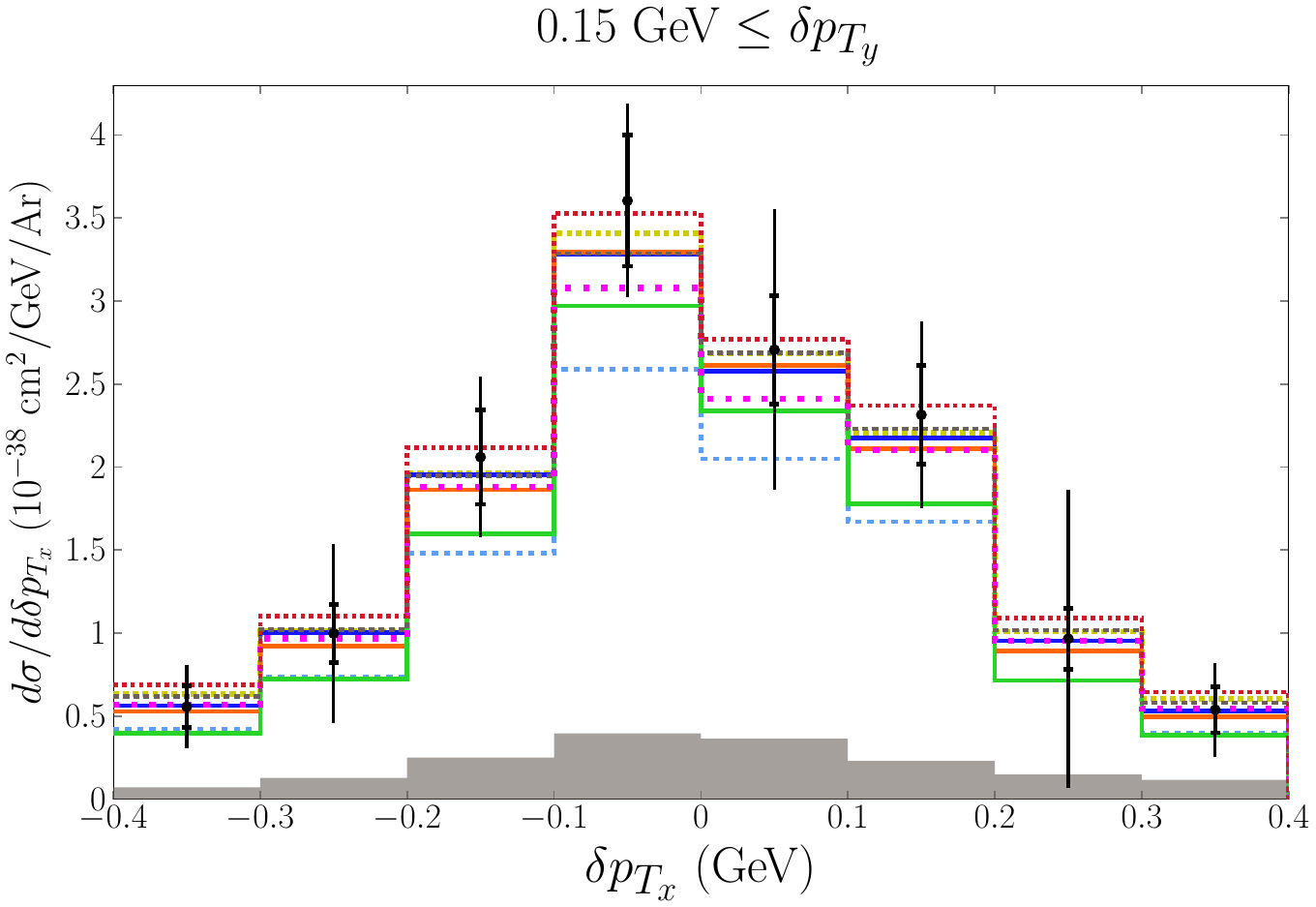}
\hfill
\belowbaseline[-0.23\textheight]{
  \includegraphics[page=42, width=0.35\textwidth]{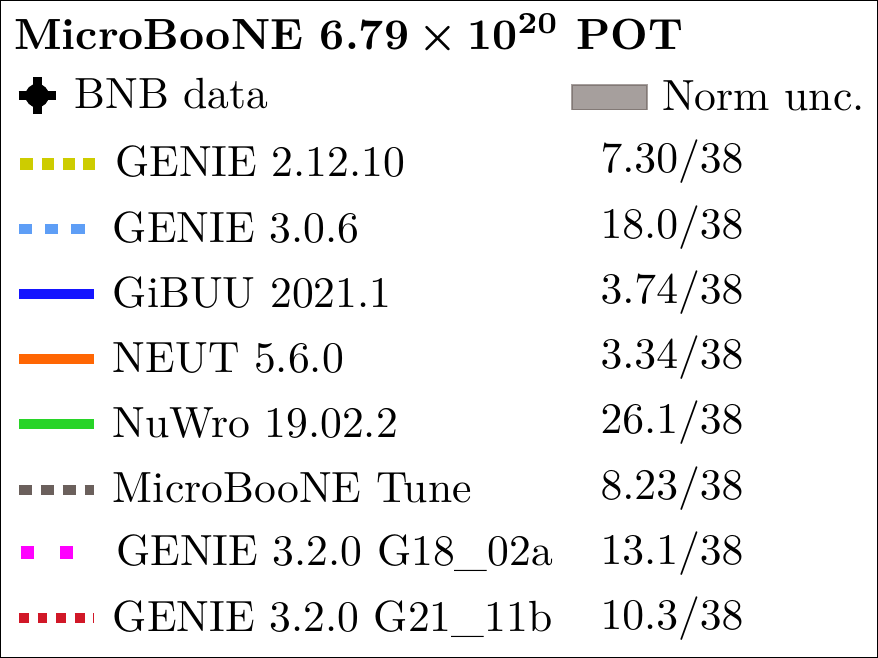}%
}
\hfill

\caption{Measured differential cross sections for block \#6
$(\delta p_{T_y}, \delta p_{T_x})$. The overall
$\chi^2$ value includes contributions from three underflow and three overflow
$\delta p_{T_x}$ bins that are not plotted.
Statistical (shape-only systematic) uncertainties are included in the inner
(outer) error bars. The remainder of the total uncertainty is shown by the gray
band along the $x$-axis.}
\label{fig:XSecResultsblock6}
\end{figure*}

\begin{figure*}
\centering

\includegraphics[page=43, width=0.64\textwidth]{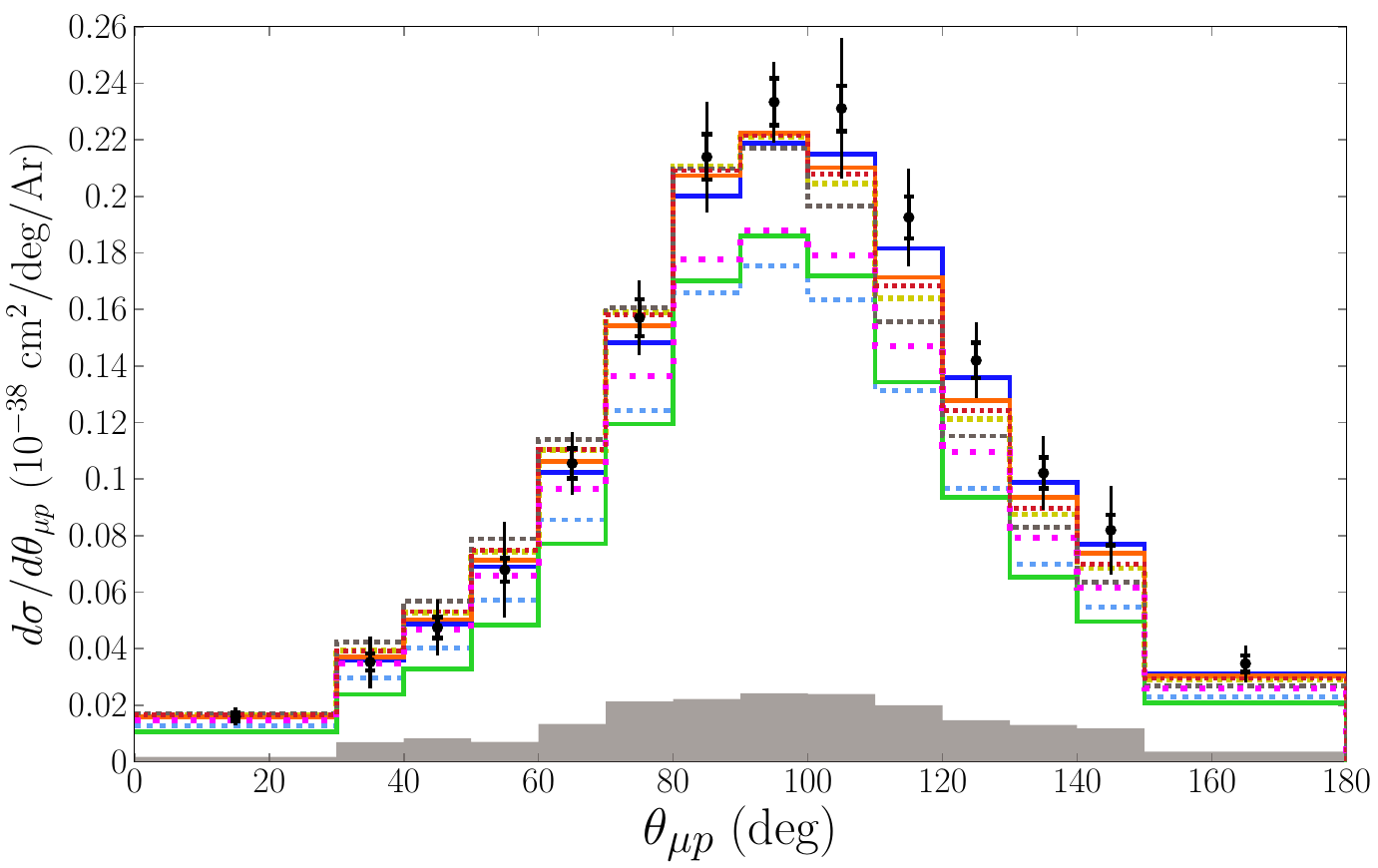}
\hfill
\belowbaseline[-0.23\textheight]{
  \includegraphics[page=44, width=0.35\textwidth]{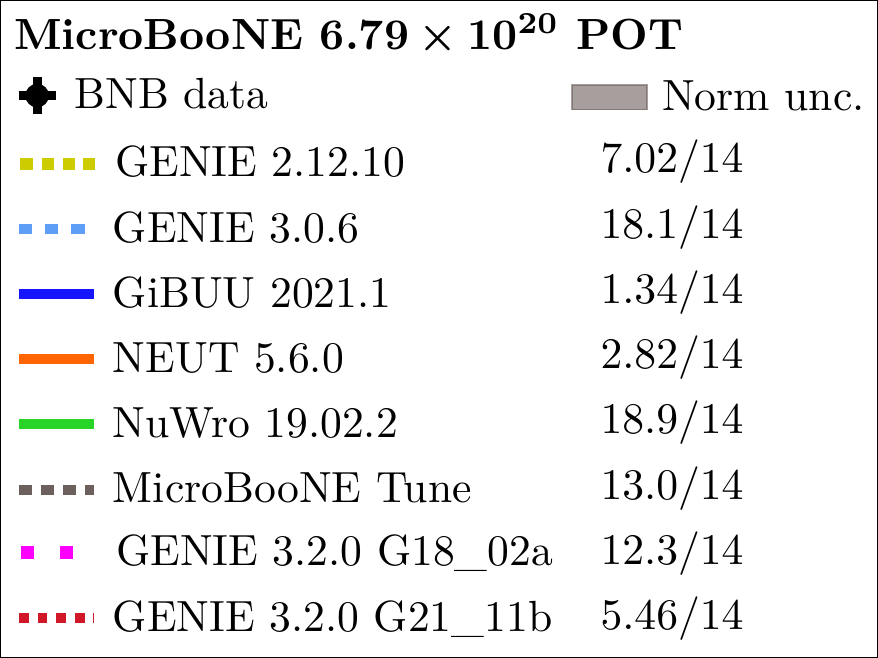}%
}
\hfill

\caption{Measured differential cross sections for block \#7
($\theta_{\mu p}$).
Statistical (shape-only systematic) uncertainties are included in the inner
(outer) error bars. The remainder of the total uncertainty is shown by the gray
band along the $x$-axis.}
\label{fig:XSecResultsblock7}
\end{figure*}

\begin{figure*}
\centering

\includegraphics[page=45, width=0.64\textwidth]{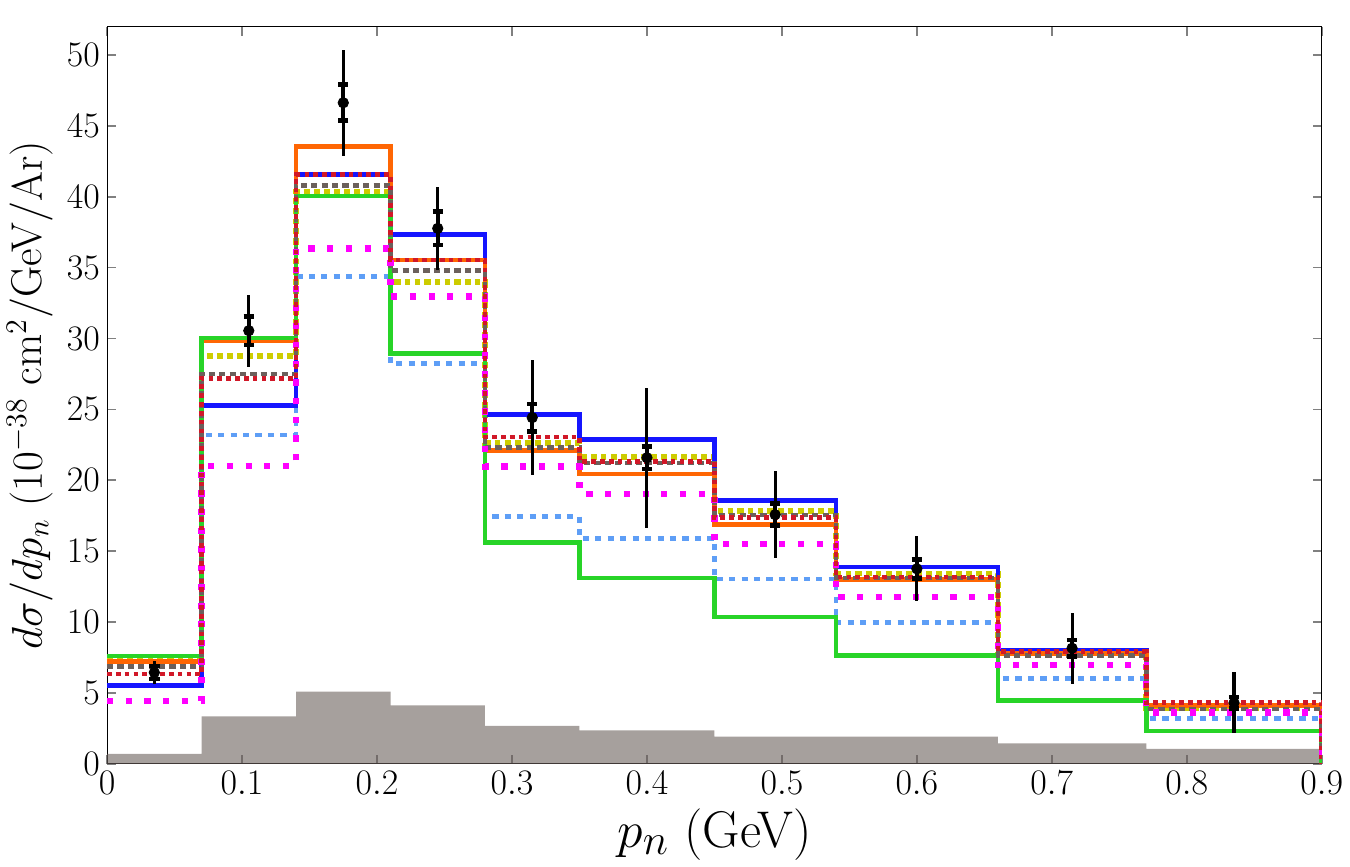}
\hfill
\belowbaseline[-0.23\textheight]{
  \includegraphics[page=46, width=0.35\textwidth]{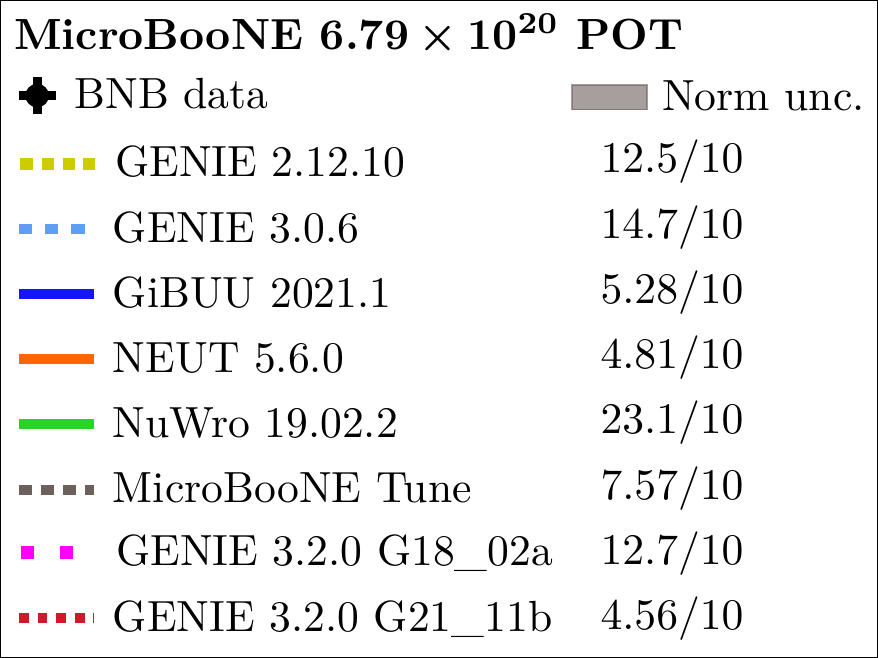}%
}
\hfill

\caption{Measured differential cross sections for block \#8
($p_n$).
Statistical (shape-only systematic) uncertainties are included in the inner
(outer) error bars. The remainder of the total uncertainty is shown by the gray
band along the $x$-axis.}
\label{fig:XSecResultsblock8}
\end{figure*}

\begin{figure*}
\centering
\includegraphics[page=47, width=0.49\textwidth]{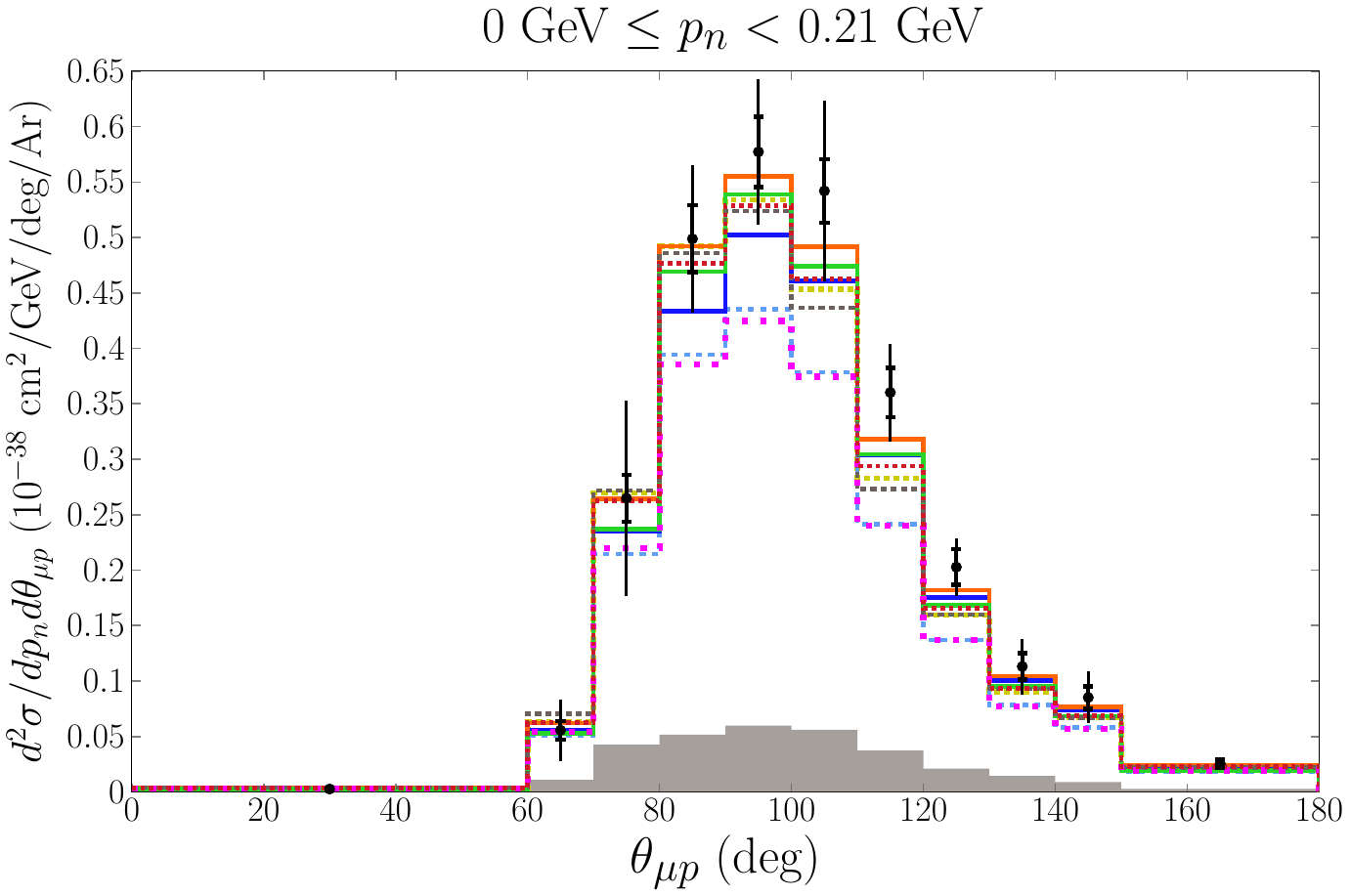}
\hfill
\includegraphics[page=48, width=0.49\textwidth]{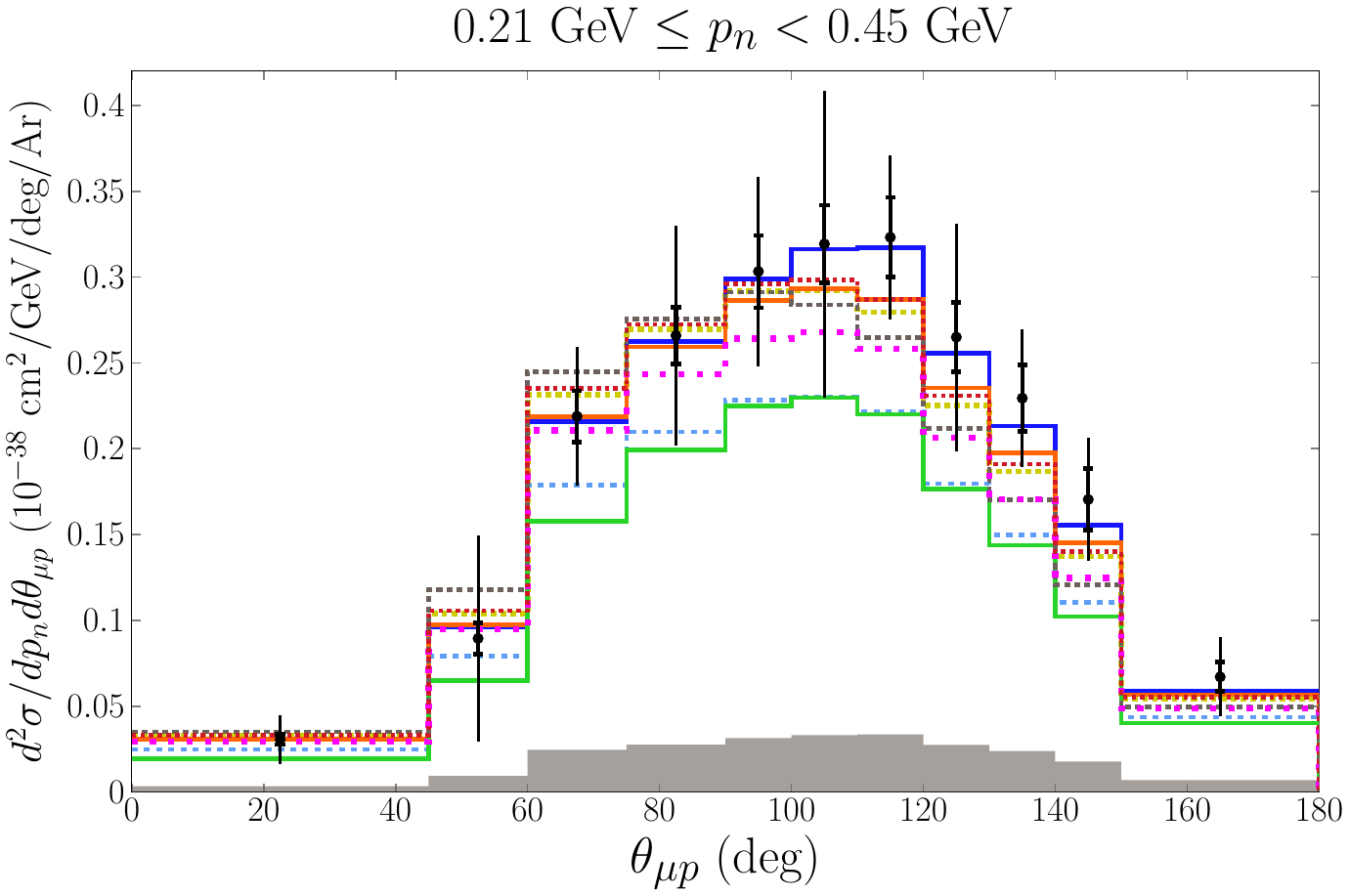}
\vspace{0.2cm}

\includegraphics[page=49, width=0.49\textwidth]{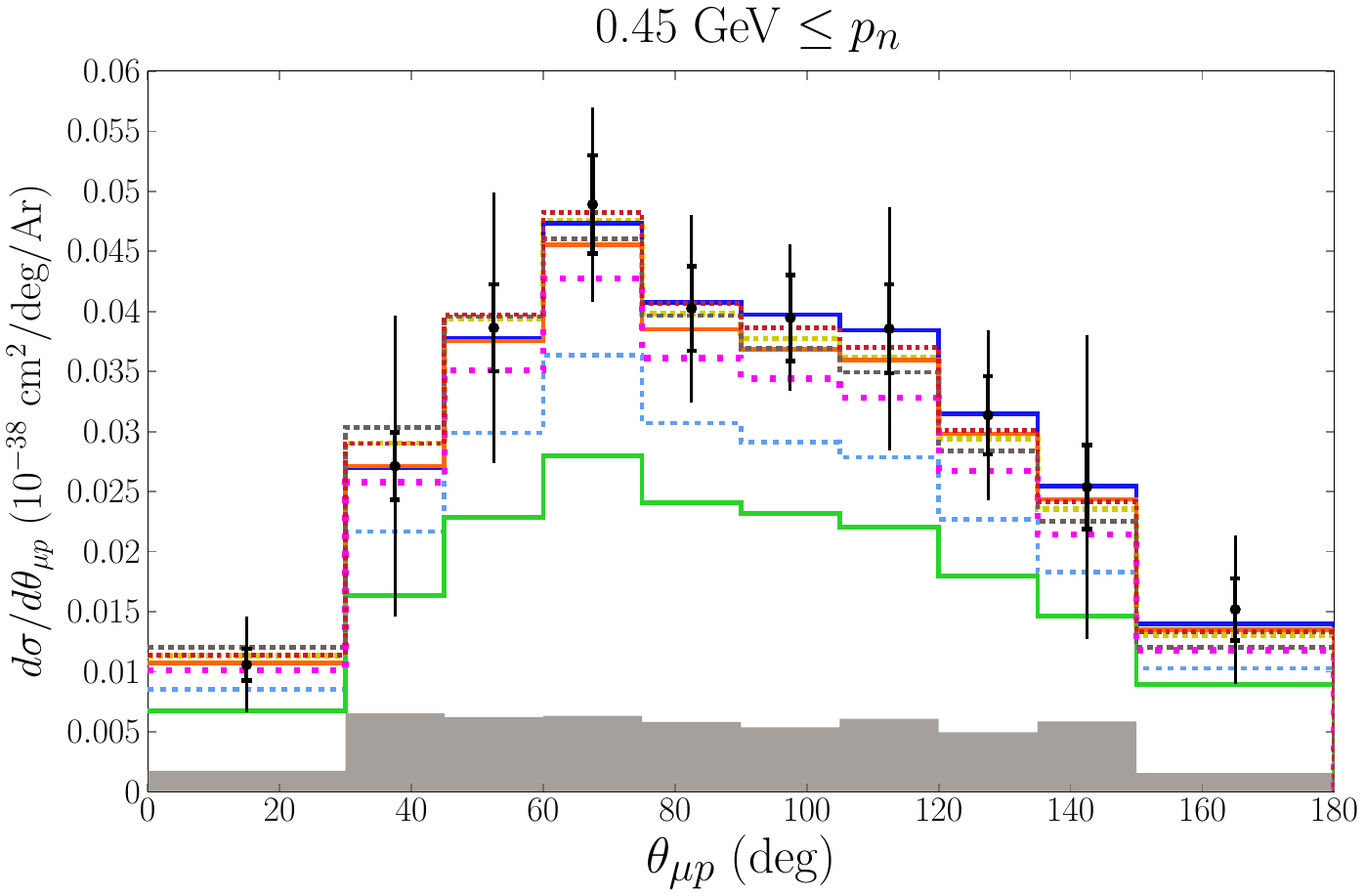}
\hfill
\belowbaseline[-0.23\textheight]{
  \includegraphics[page=51, width=0.35\textwidth]{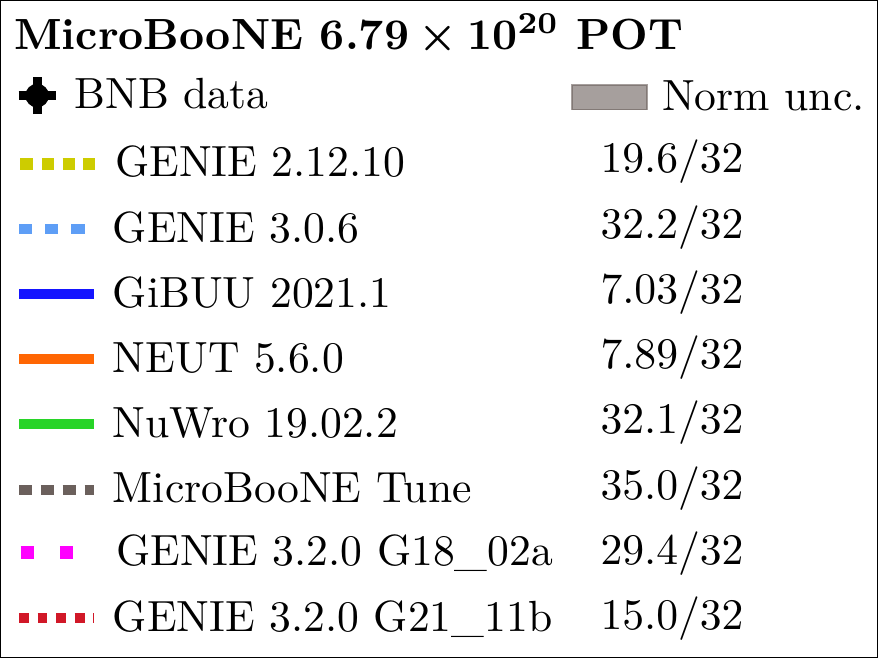}%
}
\hfill

\caption{Measured differential cross sections for block \#9
$(p_n, \theta_{\mu p})$.
Statistical (shape-only systematic) uncertainties are included in the inner
(outer) error bars. The remainder of the total uncertainty is shown by the gray
band along the $x$-axis.}
\label{fig:XSecResultsblock9}
\end{figure*}

\begin{figure*}
\centering

\includegraphics[page=10, width=0.64\textwidth]{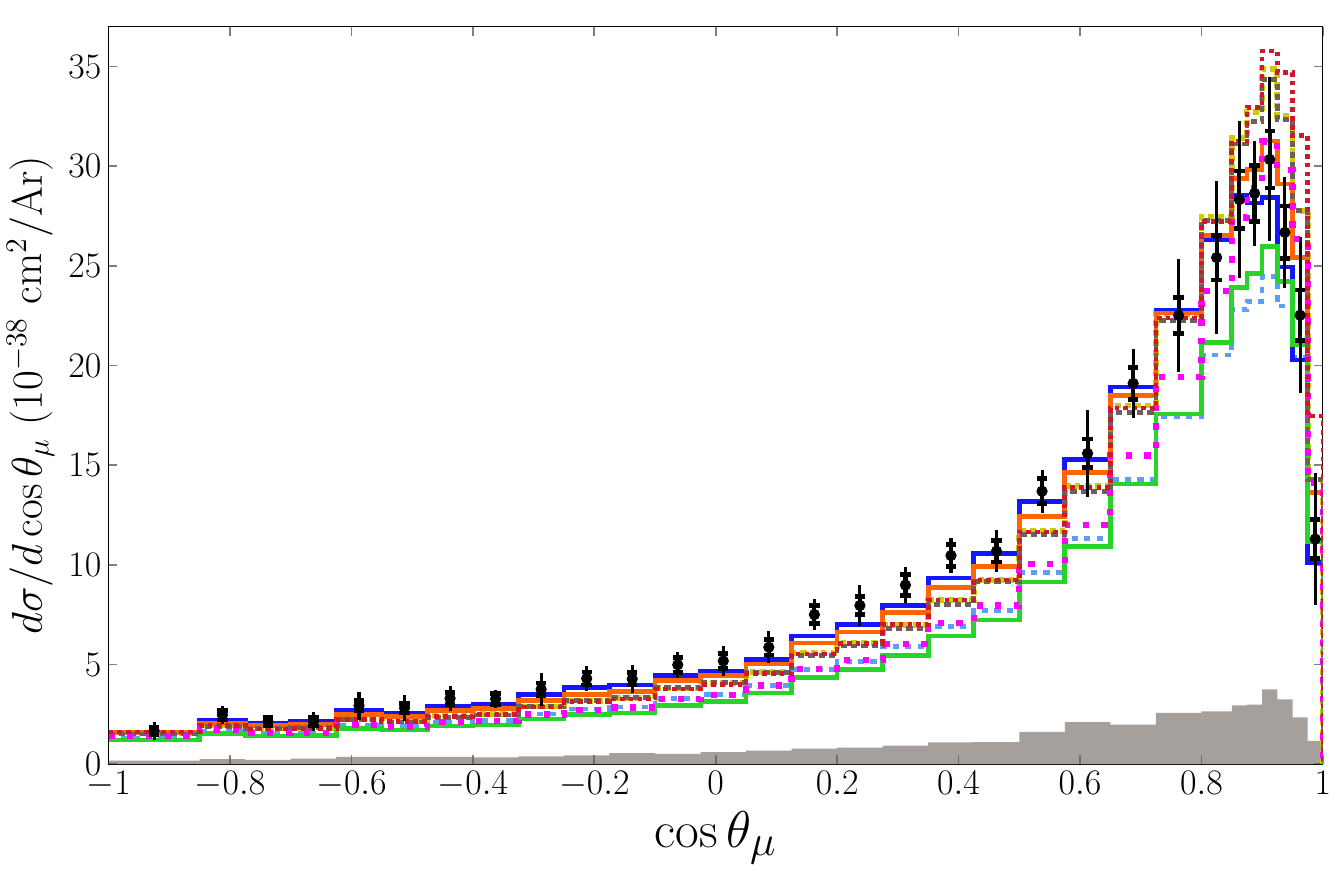}
\hfill
\belowbaseline[-0.23\textheight]{
  \includegraphics[page=11, width=0.35\textwidth]{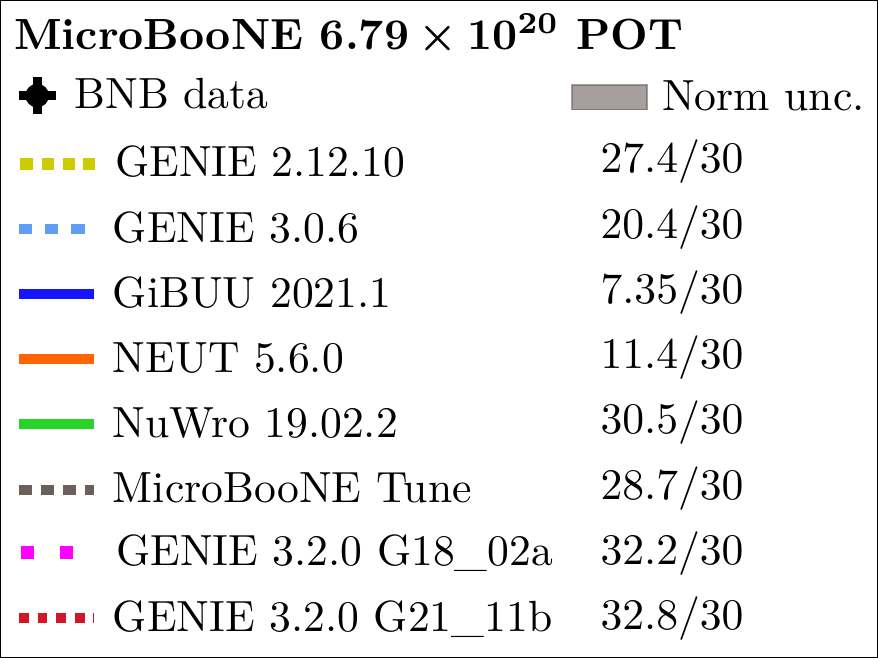}%
}
\hfill

\caption{Measured differential cross sections for block \#10
($\cos\theta_\mu$).
Statistical (shape-only systematic) uncertainties are included in the inner
(outer) error bars. The remainder of the total uncertainty is shown by the gray
band along the $x$-axis.}
\label{fig:XSecResultsblock10}
\end{figure*}

\begin{figure*}
\centering

\includegraphics[page=52, width=0.64\textwidth]{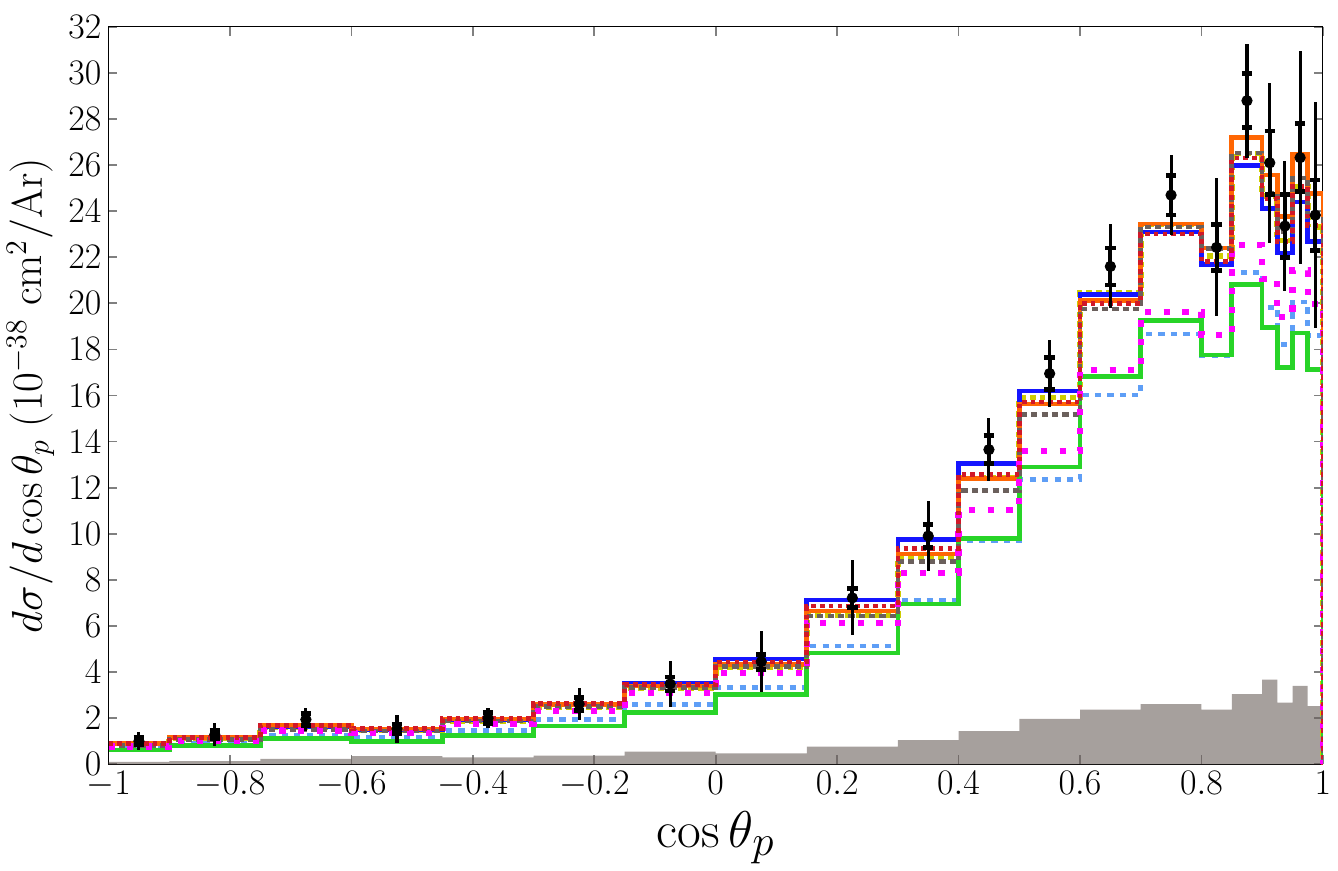}
\hfill
\belowbaseline[-0.23\textheight]{
  \includegraphics[page=53, width=0.35\textwidth]{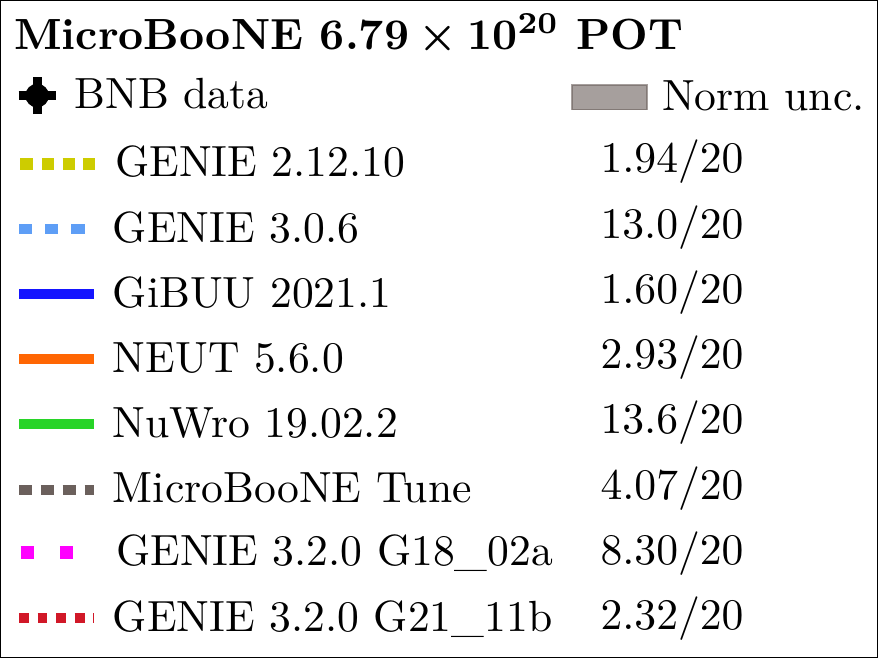}%
}
\hfill

\caption{Measured differential cross sections for block \#11
($\cos\theta_p$).
Statistical (shape-only systematic) uncertainties are included in the inner
(outer) error bars. The remainder of the total uncertainty is shown by the gray
band along the $x$-axis.}
\label{fig:XSecResultsblock11}
\end{figure*}

\begin{figure*}
\centering

\includegraphics[page=54, width=0.64\textwidth]{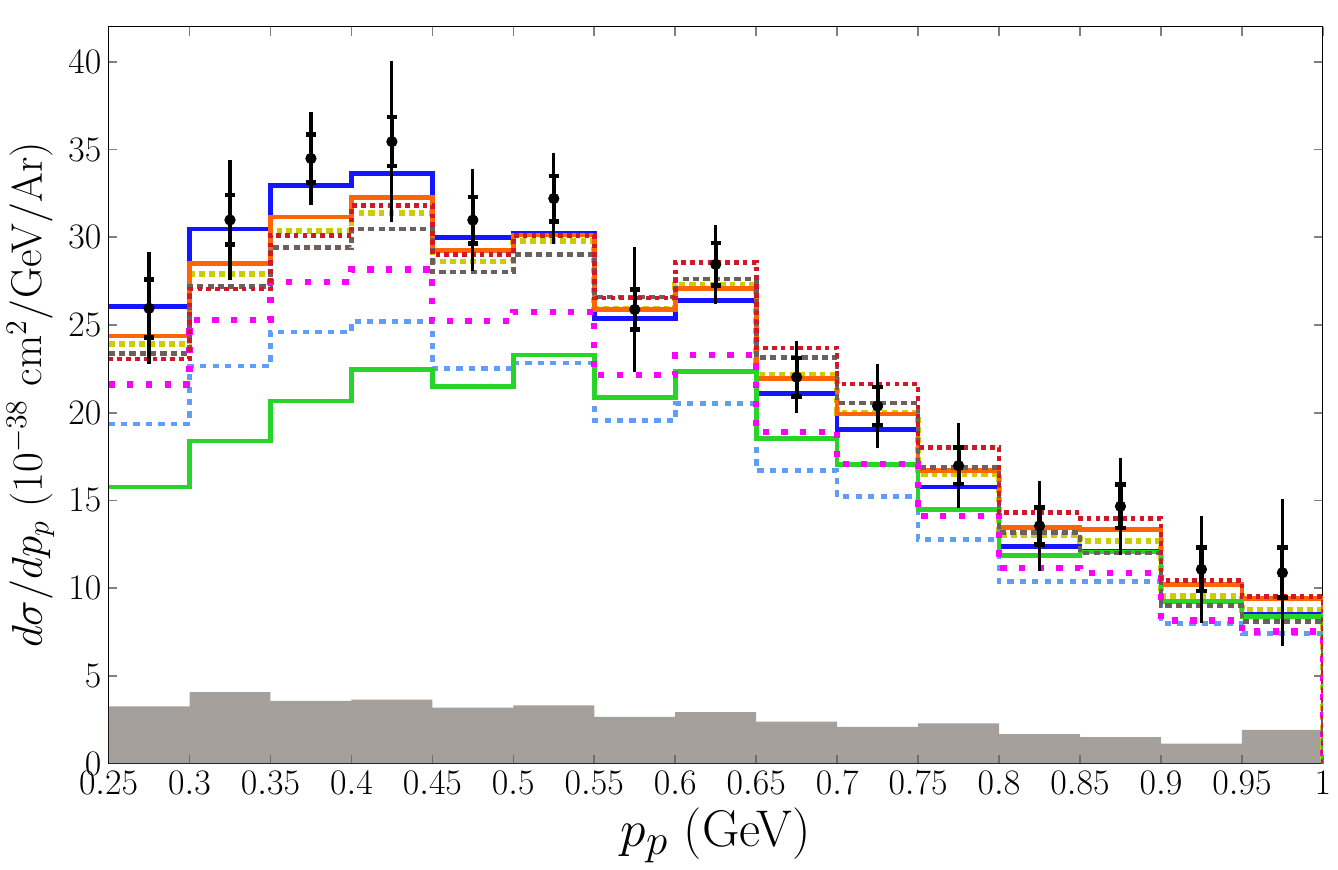}
\hfill
\belowbaseline[-0.23\textheight]{
  \includegraphics[page=55, width=0.35\textwidth]{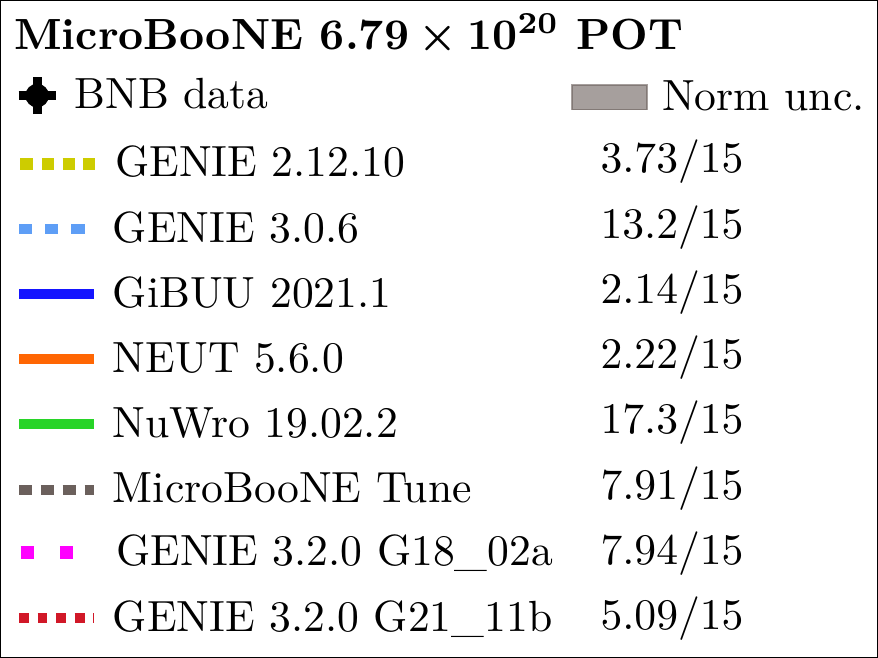}%
}
\hfill

\caption{Measured differential cross sections for block \#12
($p_p$).
Statistical (shape-only systematic) uncertainties are included in the inner
(outer) error bars. The remainder of the total uncertainty is shown by the gray
band along the $x$-axis.}
\label{fig:XSecResultsblock12}
\end{figure*}

\begin{figure*}
\centering

\includegraphics[page=56, width=0.64\textwidth]{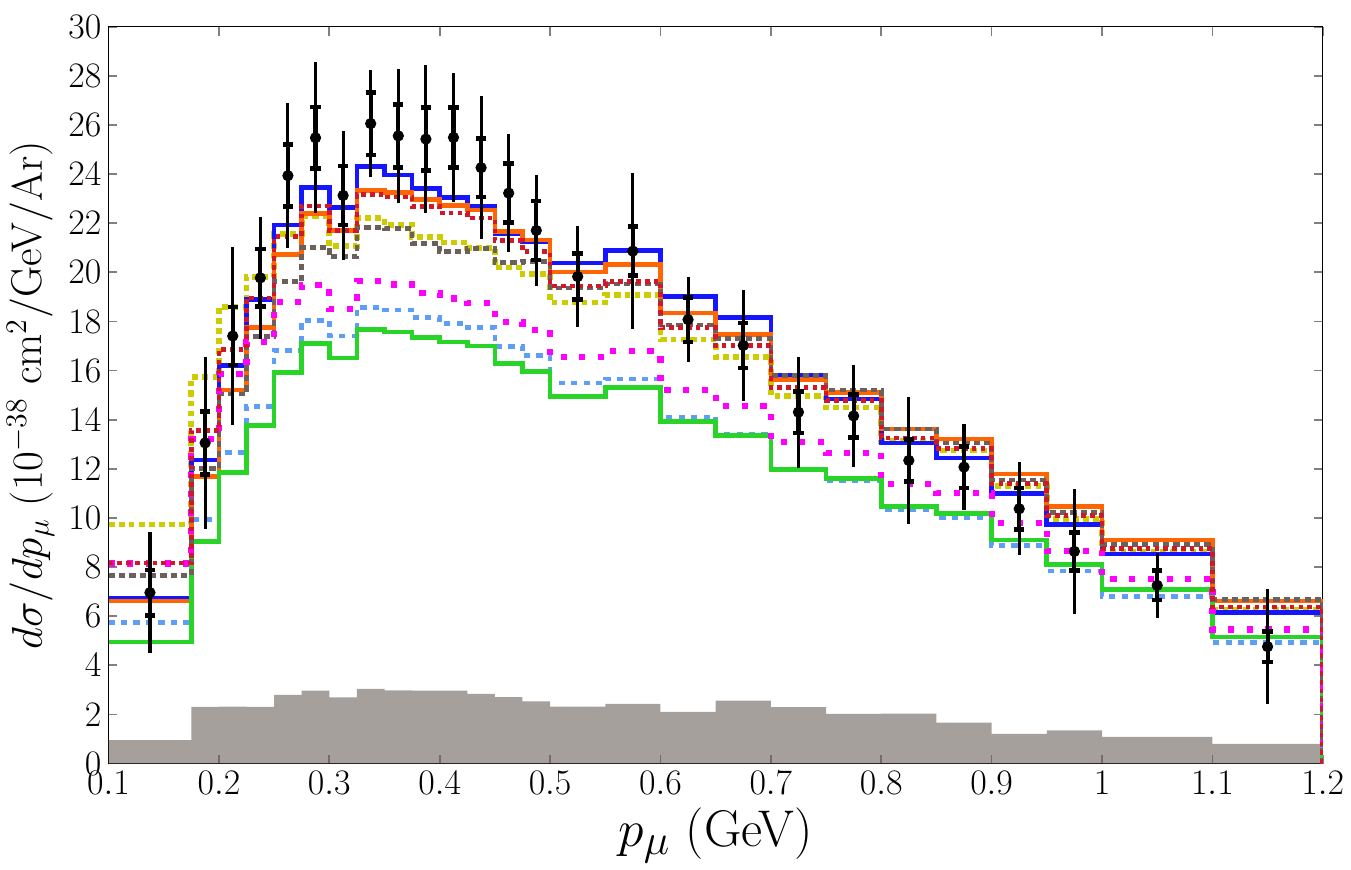}
\hfill
\belowbaseline[-0.23\textheight]{
  \includegraphics[page=57, width=0.35\textwidth]{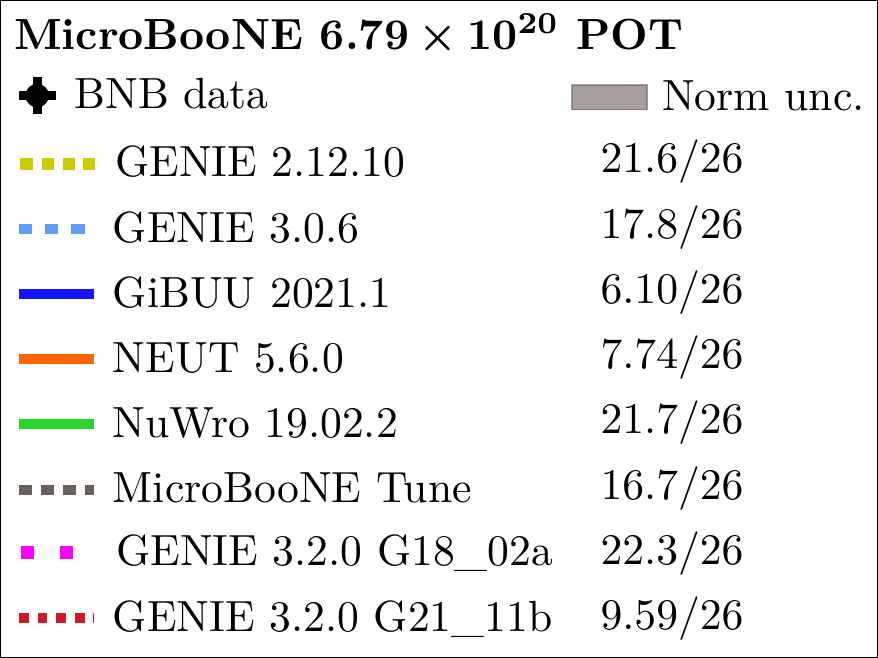}%
}
\hfill

\caption{Measured differential cross sections for block \#13
($p_\mu$).
Statistical (shape-only systematic) uncertainties are included in the inner
(outer) error bars. The remainder of the total uncertainty is shown by the gray
band along the $x$-axis.}
\label{fig:XSecResultsblock13}
\end{figure*}

Figures~\ref{fig:XSecResultsblock0}~to~\ref{fig:XSecResultsblock13} present the
final results of the analysis as flux-averaged differential cross sections. The
measurements are compared to the event generator models described above, each
rendered in the plots as a distinct colored line. The legends accompanying each
figure list the models together with their overall $\chi^2$ scores describing
the level of agreement with data. Each $\chi^2$ score is separated from the
number of bins for which it was calculated by a \texttt{/} character. A separate
figure is included for each block of bins defined in Table~\ref{tab:bin_defs}.

Following recent MicroBooNE publications that report cross sections for an
exclusive single-proton final state~\cite{uBTKIPRL, uBTKIPRD, uBGKIPRD}, the
covariance matrix decomposition described in Sec.~\ref{sec:cov_mat_decomp} is
applied when displaying the measurement uncertainties. For the data points
shown in each plot, the inner error bars represent the statistical uncertainty
only, while the outer error bars also include shape-only systematic
uncertainties. The remaining portion of the total uncertainty, composed of both
normalization and mixed terms, is shown by the gray band along the $x$-axis.

Table~\ref{tab:overall_chi2_scores} presents the $\chi^2$ scores quantifying
goodness-of-fit with the full group of measurements for each of the event
generator models studied. None of the simulation predictions provides a
satisfactory treatment of the entire data set ($\chi^2 \approx 1/$bin). The
best $\chi^2$ value is obtained by the GENIE~3.0.6 calculation, although
poor agreement is seen between this model and the data in some of
the individual blocks of bins, and its prediction is seen to be systematically
low across the entire measured phase space. The modest preference for
GENIE~3.0.6 may thus be attributed to a manifestation of Peelle's Pertinent
Puzzle~\cite{Carlson2009} (in which implausibly low predictions are favored
using a least-squares metric like $\chi^2$) rather than a particularly
high-quality description of the data. The substantially larger $\chi^2$ values
seen for GENIE~2.12.10 and GENIE~3.2.0~G18\_02a, however, do suggest that some
of their common model components (e.g., a global rather than local Fermi gas
description of the nuclear ground state) are disfavored by the present
measurements. As shown in the plots below, GiBUU~2021.1 has the best $\chi^2$
score for many of the individual blocks of bins despite performing more poorly
when confronted with the combined results.

\begin{table}[h]
\caption{Overall $\chi^2$ scores for each of the neutrino interaction
models studied.}
\includegraphics{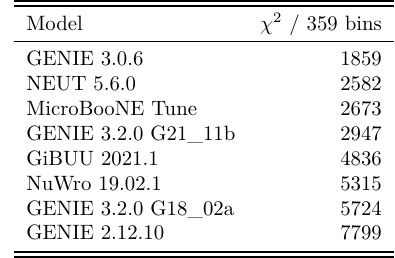}
\label{tab:overall_chi2_scores}
\end{table}

Figure~\ref{fig:XSecResultsblock0} presents the double-differential
cross-section measurement in terms of muon momentum ($p_\mu$) and scattering
angle ($\theta_\mu$) obtained using the kinematic bins from block \#0. The
GiBUU~2021.1 model achieves the best agreement with this distribution,
partially due to its larger cross section relative to most of the other plotted
models in the region of moderate $p_\mu$ (especially between
0.3~and~0.38~GeV$/c$) and intermediate scattering angles. Interestingly, the
data are compatible with an even larger cross section in this region, to the
extent that all studied predictions lie noticeably below the measured central
value in several relevant bins.

Figure~\ref{fig:XSecResultsblock1} presents the double-differential
cross-section measurement in terms of the leading proton momentum ($p_p$) and
scattering angle ($\theta_p$) obtained using the bins from block \#1. At low
$p_p$, the NuWro~19.02.2 calculation markedly undershoots the data,
particularly at forward proton angles. However, this model prediction becomes
more similar to the others with increasing $p_p$, as can also be seen in the
single-differential $p_p$ results shown in Fig.~\ref{fig:XSecResultsblock12}.
Significant model differences are seen in the shape of the forward-angle region
at relatively high $p_p$, but the systematic uncertainty of the present
measurement allows for only limited sensitivity to these details.

Figure~\ref{fig:XSecResultsblock2} shows the single-differential cross section
in the transverse momentum imbalance $\delta p_T$ as obtained from the analysis
bins in block \#2. A clear preference in the $\chi^2$ values is seen for models
which can simultaneously provide a relatively good treatment of the low-$\delta
p_T$ region (dominated by CCQE events without proton FSI) and the high-$\delta
p_T$ tail (dominated by more inelastic events and CCQE with FSI). Notably, two
of the event generator models that show the poorest agreement with this
distribution, NuWro~19.02.2 and GENIE~3.2.0~G18\_02a, have their greatest
tension with the data in opposite kinematic regions corresponding to high- and
low-$\delta p_T$, respectively.

A double-differential result in which the $\delta p_T$ measurement is
sub-divided into four $\delta \alpha_T$ bins (as defined in block \#3) is
plotted in Fig.~\ref{fig:XSecResultsblock3}. Here the NuWro~19.02.2 model shows
good agreement with data in the lowest $\delta \alpha_T$ region, which
gradually worsens with increasing $\delta \alpha_T$. A similar trend is seen in
Fig.~12 of Ref.~\cite{uBTKIPRD}, which reports a similar MicroBooNE measurement
studying a more exclusive single-proton final state. Because the effect of FSI
is to make the $\delta \alpha_T$ distribution more asymmetric and weighted
towards high values~\cite{uBTKIPRD,Lars2022}, these results may indicate an
underestimation of proton FSI in the NuWro~19.02.2 configuration considered
herein. The GENIE~3.2.0~G18\_02a model, on the other hand, shows a roughly
consistent deficit in the lowest $\delta p_T$ bins across all $\delta \alpha_T$
values, which is suggestive of missing CCQE strength. Similar conclusions
follow from Fig.~\ref{fig:XSecResultsblock4}, in which the bins from
block~\#4 are used to report a measurement of double-differential $\delta
\alpha_T$ distributions in four coarse bins of $\delta p_T$.

Figure~\ref{fig:XSecResultsblock5} presents a single-differential cross-section
measurement in $\delta p_{T_x}$, the component of $\delta \mathbf{p}_T$ that is
orthogonal to the transverse projection of the outgoing muon momentum. The
GiBUU~2021.1 and NEUT~5.6.0 models provide particularly good predictions for
this observable, which follows a distribution that is nearly symmetric around
zero. Figure~\ref{fig:XSecResultsblock6} subdivides this measurement into three
coarse bins of $\delta p_{T_y}$, the other component of $\delta \mathbf{p}_T$.
Here the effects of FSI enhance the cross section in the region where $\delta
p_{T_y}$ is negative. The strong deficit seen in the NuWro~19.02.2 prediction
in the lowest $\delta p_{T_y}$ bin is thus symptomatic of the same modeling
deficiencies seen previously at high values of $\delta p_T$ and $\delta
\alpha_T$.

Figure~\ref{fig:XSecResultsblock7} displays the measured single-differential
\ccnp~cross section in bins of $\theta_{\mu p}$, the opening angle between the
outgoing muon and leading proton. Although all of the interaction models studied
in this article agree on the rough shape of this angular distribution, the
position of the peak is sensitive to the relative contributions of QE and 2p2h
events, as seen in the lower-left panel of
Fig.~\ref{fig:DataMCMultipleBlocks1}. The data points provide a peak location
that is shifted slightly to the right of the bulk of the event generator
predictions, with NEUT~5.6.0 and especially GiBUU~2021.1 achieving the best
agreement.

Figure~\ref{fig:XSecResultsblock8} reports the measured distribution of $p_n$, a
three-dimensional analog of $\delta p_T$ which includes the component of missing
momentum $\delta p_L$ longitudinal to the neutrino beam. The addition of this
new direction noticeably worsens the $\chi^2 /$bin for all models studied,
particularly for GENIE~2.12.10 and NEUT~5.6.0, although the latter still
achieves the second-best $\chi^2$ score. When one considers the
single-differential cross section for $p_n$ in isolation, the GiBUU~2021.1,
GENIE~3.2.0~G21\_11b, and MicroBooNE Tune models also describe the data fairly
well ($\chi^2/$bin~$< 1$) despite using distinct theoretical treatments of both
the QE and 2p2h interaction modes.

Figure~\ref{fig:XSecResultsblock9} presents double-differential cross sections
as a function of $\theta_{\mu p}$ in three bins of $p_n$. Here the preference
for the peak location predicted by GiBUU~2021.1 in
Fig.~\ref{fig:XSecResultsblock7} is partially explained by this model's
remarkably good description of the data in the bin of moderate $p_n$
(\SI{0.21}{\GeV\per\clight} to \SI{0.45}{\GeV\per\clight}). As shown in
Fig.~\ref{fig:DataMCblock9}, this is also the kinematic region in which
differences in the relative contributions of QE and 2p2h events have the
greatest impact on the shape of the $\theta_{\mu p}$ distribution. Although it
obtains one of the larger $\chi^2$ values for this group of bins (block \#9),
the NuWro~19.02.2 model provides nearly the best agreement ($\chi^2 =
3.85/11$~bins) in the QE-dominated region of $p_n <$
\SI{0.21}{\GeV\per\clight}. Only the NEUT~5.6.0 model achieves slightly better
performance ($\chi^2 = 3.32/11$~bins) for this region of phase space.

Figure~\ref{fig:XSecResultsblock10} shows the measured single-differential
cross section as a function of the muon scattering angle $\theta_\mu$. The
lowest $\chi^2$ values obtained by GiBUU~2021.1 and NEUT~5.6.0 can be
attributed to two features of these model predictions. First, the data favor
significantly more cross-section strength in the $\cos\theta_\mu \lesssim 0.4$
region than is predicted by the majority of the studied event generators. Here
the GiBUU~2021.1 and NEUT~5.6.0 predictions are noticeably higher than the
others. A similar tendency for models to underpredict data in this angular
region has also been seen in recent MicroBooNE cross-section results for
quasielastic-like interactions~\cite{uBTKIPRD} and inclusive charged-current
$\nu_\mu$ scattering with at least one proton in the final
state~\cite{uBCC0pNpPRD}. Second, both GiBUU~2021.1 and NEUT~5.6.0 provide a
good description within uncertainties of the forward-angle region
($\cos\theta_\mu \gtrsim 0.8$), although GiBUU~2021.1 lies substantially closer
to the measured data point in the most-forward bin.

Finally, Figs.~\ref{fig:XSecResultsblock11}, \ref{fig:XSecResultsblock12}, and
\ref{fig:XSecResultsblock13} present measured single-differential cross
sections in terms of the leading proton scattering angle ($\theta_p$), leading
proton momentum ($p_p$), and muon momentum ($p_\mu$), respectively. For all
three distributions, GiBUU~2021.1 provides the best quantitative agreement with
the data, driven in part by its prediction of a higher \ccnp~cross section at
moderate proton scattering angles ($\cos\theta_p \in [0,0.6]$), low proton
momenta ($p_p < \SI{0.5}{\GeV\per\clight}$) and moderate muon momenta ($p_\mu
\in [0.25, 0.5]~\si{\GeV\per\clight}$).

For the single-differential $p_\mu$ and $p_p$ results, similar trends can be
seen in MicroBooNE data for inclusive CC $\nu_\mu$ events with one or more
final-state protons~\cite{uBCC0pNpPRD}. For both the \ccnp\ cross sections
shown here and the inclusive results, data indicate a higher cross section near
the peak of the $p_\mu$ distribution than is predicted by a variety of neutrino
event generators, and models with a larger prediction at low $p_p$ are also
preferred.

\section{Summary and Conclusions}
\label{sec:summary_and_conclusions}

This article presents a detailed study of charged-current muon-neutrino
interactions with argon leading to mesonless final states containing one or
more protons. Flux-averaged cross sections for this \ccnp~process were measured
using the Fermilab Booster Neutrino Beam and MicroBooNE detector. The results
are reported as a function of ten observables related to the three-momenta of
the outgoing muon and leading proton. A larger data set and significant
improvements to MicroBooNE's software tools since the experiment's first
\ccnp~cross-section analysis was reported~\cite{MCC8ccnpPaper} allow the
present work to achieve higher precision and explore new kinematic
distributions, including the first double-differential measurements in the
\ccnp~channel for an argon target.

A covariance matrix describing correlated uncertainties between all
\TotalBinCount~kinematic bins is provided in the supplemental materials,
allowing goodness-of-fit to be quantitatively assessed for theoretical
predictions describing the entire data set. This data release strategy
represents an improvement over the typical procedure employed to date in the
neutrino scattering literature, in which correlations between distinct
kinematic distributions are not disclosed.

The predictions of several standard neutrino event generators are presented and
compared to the MicroBooNE data. The GiBUU~2021.1 model provides a relatively
good description of all of the individual differential cross sections studied,
but its poor agreement with the combined set of measurements (quantified with
the $\chi^2$ score given in Table~\ref{tab:overall_chi2_scores}) suggests that
the correlations between kinematic distributions are less well-modeled. Overall
agreement is likewise unsatisfactory for the other neutrino interaction models
considered herein, although those with the least sophisticated treatments of
nuclear effects (GENIE~2.12.10 and GENIE~3.2.0~G18\_02a) obtain substantially
higher $\chi^2$ values than the others. The GENIE~3.0.6 prediction achieves the
best overall goodness-of-fit of the models studied, but it is seen to
systematically underpredict the data across a large fraction of the measured
phase space.

Some of the kinematic regions for which the present measurements give the
greatest model discrimination power include moderate $p_\mu$, low to moderate
$\cos\theta_\mu$, and low $p_p$. The data show a clear preference for the
greater cross-section strength assigned to these regions by GiBUU~2021.1 and
NEUT~5.6.0 compared with the other event generators studied. These two models
also describe the distribution of the muon-proton opening angle $\theta_{\mu p}$
noticeably better than their counterparts. Especially when combined with recent
MicroBooNE measurements examining more exclusive~\cite{uBTKIPRL,uBTKIPRD,uB2p}
and inclusive~\cite{CCinclPRL,MicroBooNE:2023foc} final-state topologies in
charged-current $\nu_\mu$-argon scattering, the present data set provides a
highly detailed benchmark for the ongoing effort to improve event generators to
the precision needed for the future accelerator-based neutrino oscillation
program.

\section{Acknowledgments}

This document was prepared by the MicroBooNE collaboration using the resources
of the Fermi National Accelerator Laboratory (Fermilab), a U.S. Department of
Energy, Office of Science, HEP User Facility. Fermilab is managed by Fermi
Research Alliance, LLC (FRA), acting under Contract No. DE-AC02-07CH11359.
MicroBooNE is supported by the following: the U.S. Department of Energy, Office
of Science, Offices of High Energy Physics and Nuclear Physics; the U.S.
National Science Foundation; the Swiss National Science Foundation; the Science
and Technology Facilities Council (STFC), part of the United Kingdom Research
and Innovation; the Royal Society (United Kingdom); the UK Research and
Innovation (UKRI) Future Leaders Fellowship; and the NSF AI Institute for
Artificial Intelligence and Fundamental Interactions. Additional support for
the laser calibration system and cosmic ray tagger was provided by the Albert
Einstein Center for Fundamental Physics, Bern, Switzerland. We also acknowledge
the contributions of technical and scientific staff to the design,
construction, and operation of the MicroBooNE detector as well as the
contributions of past collaborators to the development of MicroBooNE analyses,
without whom this work would not have been possible. For the purpose of open
access, the authors have applied a Creative Commons Attribution (CC BY) public
copyright license to any Author Accepted Manuscript version arising from this
submission.

\appendix

\section{Bin definitions}
\label{sec:bin_defs}

Table~\ref{tab:bin_defs} presents the full set of \TotalBinCount\ kinematic
bins used to report the \ccnp\ cross section measurements. For an observable
$x$, the bin with lower limit $x^\mathrm{low}$ and upper limit
$x^\mathrm{high}$ will include events with $x^\mathrm{low} \leq x <
x^\mathrm{high}$. The blocks of related bins (indicated in the table) are
defined so that a selected event will belong to a unique bin within each block.

\newcommand{\doubletoprule}%
  {\toprule\\[-1.3em]\toprule}
\newcommand{\doublebottomrule}[1]%
  {\toprule\\[-1.3em]\toprule\\[#1]}

\afterpage{\onecolumngrid
\renewcommand{\doublerulesep}{0pt}
\begin{longtable}{@{\extracolsep{\fill}}SSSSS}
\caption{Bin definitions used in the analysis.\label{tab:bin_defs}}
\\
\multicolumn{5}{c}{} \\
\multicolumn{5}{c}{Block 0: $(p_\mu, \cos\theta_\mu)$} \\[0.3mm]
\doubletoprule
{bin number}
& {$p_\mu^\mathrm{low}$ (\si{\GeV\per\clight})}
& {$p_\mu^\mathrm{high}$ (\si{\GeV\per\clight})}
& {$\cos\theta_\mu^\mathrm{low}$}
& {$\cos\theta_\mu^\mathrm{high}$}
\\*
\midrule

0 & 0.1 & 0.24 & -1 & -0.55 \\*
1 &  & & -0.55 & -0.25 \\*
2 &  & & -0.25 &  0 \\*
3 &  & & 0 & 0.25 \\*
4 &  & & 0.25 & 0.45 \\*
5 &  & & 0.45 & 0.7 \\*
6 &  & & 0.7 & 1 \\[3mm]

7 & 0.24 & 0.3 & -1 & -0.55 \\*
8 &  & & -0.55 & -0.25 \\*
9 &  & & -0.25 & 0 \\*
10 &  & & 0 & 0.25 \\*
11 &  & & 0.25 & 0.45 \\*
12 &  & & 0.45 & 0.7 \\*
13 &  & & 0.7 & 1 \\[3mm]

14 & 0.3 & 0.38 & -1 & -0.4 \\*
15 &  & & -0.4 & -0.1 \\*
16 &  & & -0.1 &  0.1 \\*
17 &  & &  0.1 &  0.35 \\*
18 &  & &  0.35 &  0.5 \\*
19 &  & &  0.5  &  0.7 \\*
20 &  & &  0.7  &  0.85 \\*
21 &  & &  0.85 &  1 \\[3mm]

22 & 0.38 & 0.48 & -1 & 0 \\*
23 &  & &  0   & 0.5 \\*
24 &  & &  0.5 & 0.65 \\*
25 &  & &  0.65 & 0.8 \\*
26 &  & &  0.8  & 0.92 \\*
27 &  & &  0.92 & 1 \\[3mm]

28 & 0.48 & 0.7 & -1 & 0.2 \\*
29 &  & & 0.2 & 0.5 \\*
30 &  & & 0.5 & 0.65 \\*
31 &  & & 0.65 & 0.8 \\*
32 &  & & 0.8 & 0.875 \\*
33 &  & & 0.875 & 0.95 \\*
34 &  & & 0.95 & 1 \\[3mm]

35 & 0.7 & 0.85 & -1 & 0.65 \\*
36 &  & & 0.65 & 0.8 \\*
37 &  & & 0.8 & 0.875 \\*
38 &  & & 0.875 & 0.95 \\*
39 &  & & 0.95 & 1 \\[3mm]

40 & 0.85 & 1.2 & -1 & 0.85 \\*
41 &  & & 0.85 & 0.9 \\*
42 &  & & 0.9 & 0.95 \\*
43 &  & & 0.95 & 1 \\
\doublebottomrule{0.5em}
\multicolumn{5}{c}{Block 1: $(p_{p}, \cos\theta_{p})$} \\[0.3mm]
\doubletoprule
{bin number}
& {$p_p^\mathrm{low}$ (\si{\GeV\per\clight})}
& {$p_p^\mathrm{high}$ (\si{\GeV\per\clight})}
& {$\cos\theta_p^\mathrm{low}$}
& {$\cos\theta_p^\mathrm{high}$}
\\*
\midrule

44 & 0.25 & 0.325 & -1 & 0 \\*
45 &  & & 0 & 1 \\[3mm]

46 & 0.325 & 0.4 & -1 & -0.5 \\*
47 &  & & -0.5 & 0 \\*
48 &  & & 0 & 0.5 \\*
49 &  & & 0.5 & 0.8 \\*
50 &  & & 0.8 & 1 \\[3mm]

51 & 0.4 & 0.5 & -1 & -0.6 \\*
52 &  & & -0.6 & -0.2 \\*
53 &  & & -0.2 &  0.2 \\*
54 &  & &  0.2 &  0.5 \\*
55 &  & &  0.5 &  0.65 \\*
56 &  & &  0.65 & 0.85 \\*
57 &  & &  0.85 & 1 \\[3mm]

58 & 0.5 & 0.6 & -1 & -0.2 \\*
59 &  & & -0.2 & 0.2 \\*
60 &  & &  0.2 & 0.4 \\*
61 &  & &  0.4 & 0.6 \\*
62 &  & &  0.6 & 0.7 \\*
63 &  & &  0.7 & 0.8 \\*
64 &  & &  0.8 & 0.9 \\*
65 &  & &  0.9 & 1 \\[3mm]

66 & 0.6 & 0.7 & -1 & 0.1 \\*
67 &  & & 0.1 & 0.37 \\*
68 &  & & 0.37 & 0.5 \\*
69 &  & & 0.5 & 0.6 \\*
70 &  & & 0.6 & 0.7 \\*
71 &  & & 0.7 & 0.8 \\*
72 &  & & 0.8 & 0.9 \\*
73 &  & & 0.9 & 1 \\[3mm]

74 & 0.7 & 1 & -1 & 0.45 \\*
75 &  & & 0.45 & 0.65 \\*
76 &  & & 0.65 & 0.75 \\*
77 &  & & 0.75 & 0.82 \\*
78 &  & & 0.82 & 0.9 \\*
79 &  & & 0.9 & 1 \\
\doublebottomrule{0.5em}
\multicolumn{5}{c}{Block 2: $\delta p_T$} \\[0.3mm]
\doubletoprule
{bin number}
& {$\delta p_T^\mathrm{low}$ (\si{\GeV\per\clight})}
& {$\delta p_T^\mathrm{high}$ (\si{\GeV\per\clight})}
\\*
\midrule
80 & 0 & 0.06 \\*
81 & 0.06 & 0.12 \\*
82 & 0.12 & 0.18 \\*
83 & 0.18 & 0.24 \\*
84 & 0.24 & 0.32 \\*
85 & 0.32 & 0.40 \\*
86 & 0.40 & 0.48 \\*
87 & 0.48 & 0.55 \\*
88 & 0.55 & 0.68 \\*
89 & 0.68 & 0.75 \\*
90 & 0.75 & 0.90 \\*
91 & 0.90 & {$\infty$} \\
\doublebottomrule{0.5em}
\multicolumn{5}{c}{Block 3: $(\delta \alpha_T, \delta p_T)$} \\[0.3mm]
\doubletoprule
{bin number}
& {$\delta \alpha_T^\mathrm{low}$ ($^\circ$)}
& {$\delta \alpha_T^\mathrm{high}$ ($^\circ$)}
& {$\delta p_T^\mathrm{low}$ (\si{\GeV\per\clight})}
& {$\delta p_T^\mathrm{high}$ (\si{\GeV\per\clight})}
\\*
\midrule

92 & 0 & 45 & 0 & 0.06 \\*
93 &  & & 0.06 & 0.12 \\*
94 &  & & 0.12 & 0.18 \\*
95 &  & & 0.18 & 0.24 \\*
96 &  & & 0.24 & 0.32 \\*
97 &  & & 0.32 & 0.40 \\*
98 &  & & 0.40 & 0.48 \\*
99 &  & & 0.48 & {$\infty$} \\[3mm]

100 & 45 & 90 & 0 & 0.06 \\*
101 &  & & 0.06 & 0.12 \\*
102 &  & & 0.12 & 0.18 \\*
103 &  & & 0.18 & 0.24 \\*
104 &  & & 0.24 & 0.32 \\*
105 &  & & 0.32 & 0.40 \\*
106 &  & & 0.40 & 0.48 \\*
107 &  & & 0.48 & 0.55 \\*
108 &  & & 0.55 & {$\infty$} \\[3mm]

109 & 90 & 135 & 0 & 0.06 \\*
110 &  & & 0.06 & 0.12 \\*
111 &  & & 0.12 & 0.18 \\*
112 &  & & 0.18 & 0.24 \\*
113 &  & & 0.24 & 0.32 \\*
114 &  & & 0.32 & 0.40 \\*
115 &  & & 0.40 & 0.48 \\*
116 &  & & 0.48 & 0.55 \\*
117 &  & & 0.55 & 0.63 \\*
118 &  & & 0.63 & 0.70 \\*
119 &  & & 0.70 & {$\infty$} \\[3mm]

120 & 135 & 180 & 0 & 0.06 \\*
121 &  & & 0.06 & 0.12 \\*
122 &  & & 0.12 & 0.18 \\*
123 &  & & 0.18 & 0.24 \\*
124 &  & & 0.24 & 0.32 \\*
125 &  & & 0.32 & 0.40 \\*
126 &  & & 0.40 & 0.50 \\*
127 &  & & 0.50 & 0.60 \\*
128 &  & & 0.60 & 0.72 \\*
129 &  & & 0.72 & 0.90 \\*
130 &  & & 0.90 & {$\infty$} \\
\doublebottomrule{0.5em}
\multicolumn{5}{c}{Block 4: $(\delta p_T, \delta \alpha_T)$} \\[0.3mm]
\doubletoprule
{bin number}
& {$\delta p_T^\mathrm{low}$ (\si{\GeV\per\clight})}
& {$\delta p_T^\mathrm{high}$ (\si{\GeV\per\clight})}
& {$\delta \alpha_T^\mathrm{low}$ ($^\circ$)}
& {$\delta \alpha_T^\mathrm{high}$ ($^\circ$)}
\\*
\midrule

131 & 0 & 0.2 & 0 & 25 \\*
132 &  & & 25 & 60 \\*
133 &  & & 60 & 95 \\*
134 &  & & 95 & 120 \\*
135 &  & & 120 & 145 \\*
136 &  & & 145 & 165 \\*
137 &  & & 165 & 180 \\[3mm]

138 & 0.2 & 0.3 & 0 & 25 \\*
139 &  & & 25 & 60 \\*
140 &  & & 60 & 95 \\*
141 &  & & 95 & 120 \\*
142 &  & & 120 & 145 \\*
143 &  & & 145 & 165 \\*
144 &  & & 165 & 180 \\[3mm]

145 & 0.3 & 0.4 & 0 & 25 \\*
146 &  & & 25 & 60 \\*
147 &  & & 60 & 95 \\*
148 &  & & 95 & 120 \\*
149 &  & & 120 & 145 \\*
150 &  & & 145 & 165 \\*
151 &  & & 165 & 180 \\[3mm]

152 & 0.4 & {$\infty$} & 0 & 25 \\*
153 &  & & 25 & 60 \\*
154 &  & & 60 & 95 \\*
155 &  & & 95 & 120 \\*
156 &  & & 120 & 145 \\*
157 &  & & 145 & 165 \\*
158 &  & & 165 & 180 \\
\doublebottomrule{0.5em}
\multicolumn{5}{c}{Block 5: $\delta p_{T_x}$} \\[0.3mm]
\doubletoprule
{bin number}
& {$\delta p_{T_x}^\mathrm{low}$ (\si{\GeV\per\clight})}
& {$\delta p_{T_x}^\mathrm{high}$ (\si{\GeV\per\clight})}
\\*
\midrule

159 & {$-\infty$} & -0.60 \\*
160 & -0.60 & -0.45 \\*
161 & -0.45 & -0.35 \\*
162 & -0.35 & -0.25 \\*
163 & -0.25 & -0.15 \\*
164 & -0.15 & -0.075 \\*
165 & -0.075 & 0 \\*
166 & 0 & 0.075 \\*
167 & 0.075 & 0.15 \\*
168 & 0.15 & 0.25 \\*
169 & 0.25 & 0.35 \\*
170 & 0.35 & 0.45 \\*
171 & 0.45 & 0.6 \\*
172 & 0.60 & {$\infty$} \\
\doublebottomrule{0.9em}
\multicolumn{5}{c}{Block 6: $(\delta p_{T_y}, \delta p_{T_x})$} \\[0.3mm]
\doubletoprule
{bin number}
& {$\delta p_{T_y}^\mathrm{low}$ (\si{\GeV\per\clight})}
& {$\delta p_{T_y}^\mathrm{high}$ (\si{\GeV\per\clight})}
& {$\delta p_{T_x}^\mathrm{low}$ (\si{\GeV\per\clight})}
& {$\delta p_{T_x}^\mathrm{high}$ (\si{\GeV\per\clight})}
\\*
\midrule

173 & {$-\infty$} & -0.15 & {$-\infty$} & -0.6 \\*
174 &  & & -0.6 & -0.45 \\*
175 &  & & -0.45 & -0.35 \\*
176 &  & & -0.35 & -0.25 \\*
177 &  & & -0.25 & -0.15 \\*
178 &  & & -0.15 & -0.075 \\*
179 &  & & -0.075 & 0 \\*
180 &  & &  0 & 0.075 \\*
181 &  & &  0.075 & 0.15 \\*
182 &  & &  0.15 & 0.25 \\*
183 &  & &  0.25 & 0.35 \\*
184 &  & &  0.35 & 0.45 \\*
185 &  & &  0.45 & 0.60 \\*
186 &  & &  0.60 & {$\infty$} \\[3mm]

187 & -0.15 & 0.15 & {$-\infty$} & -0.6 \\*
188 &  & & -0.6 & -0.45 \\*
189 &  & & -0.45 & -0.35 \\*
190 &  & & -0.35 & -0.25 \\*
191 &  & & -0.25 & -0.15 \\*
192 &  & & -0.15 & -0.075 \\*
193 &  & & -0.075 & 0 \\*
194 &  & &  0 & 0.075 \\*
195 &  & &  0.075 & 0.15 \\*
196 &  & &  0.15 & 0.25 \\*
197 &  & &  0.25 & 0.35 \\*
198 &  & &  0.35 & 0.45 \\*
199 &  & &  0.45 & 0.60 \\*
200 &  & &  0.60 & {$\infty$} \\[3mm]

201 & 0.15 & {$\infty$} & {$-\infty$} & -0.4 \\*
202 &  & & -0.4 & -0.3 \\*
203 &  & & -0.3 & -0.2 \\*
204 &  & & -0.2 & -0.1 \\*
205 &  & & -0.1 & 0 \\*
206 &  & & 0 & 0.1 \\*
207 &  & & 0.1 & 0.2 \\*
208 &  & & 0.2 & 0.3 \\*
209 &  & & 0.3 & 0.4 \\*
210 &  & & 0.4 & {$\infty$} \\
\doublebottomrule{0.5em}
\multicolumn{5}{c}{Block 7: $\theta_{\mu p}$} \\[0.3mm]
\doubletoprule
{bin number}
& {$\theta_{\mu p}^\mathrm{low}$ ($^\circ$)}
& {$\theta_{\mu p}^\mathrm{high}$ ($^\circ$)}
\\*
\midrule

211 & 0 & 30 \\*
212 & 30 & 40 \\*
213 & 40 & 50 \\*
214 & 50 & 60 \\*
215 & 60 & 70 \\*
216 & 70 & 80 \\*
217 & 80 & 90 \\*
218 & 90 & 100 \\*
219 & 100 & 110 \\*
220 & 110 & 120 \\*
221 & 120 & 130 \\*
222 & 130 & 140 \\*
223 & 140 & 150 \\*
224 & 150 & 180 \\
\doublebottomrule{0.5em}
\multicolumn{5}{c}{Block 8: $p_n$} \\[0.3mm]
\doubletoprule
{bin number}
& {$p_n^\mathrm{low}$ (\si{\GeV\per\clight})}
& {$p_n^\mathrm{high}$ (\si{\GeV\per\clight})}
\\*
\midrule

225 & 0 & 0.07 \\*
226 & 0.07 & 0.14 \\*
227 & 0.14 & 0.21 \\*
228 & 0.21 & 0.28 \\*
229 & 0.28 & 0.35 \\*
230 & 0.35 & 0.45 \\*
231 & 0.45 & 0.54 \\*
232 & 0.54 & 0.66 \\*
233 & 0.66 & 0.77 \\*
234 & 0.77 & 0.9 \\*
235 & 0.9 & {$\infty$} \\
\doublebottomrule{0.5em}
\multicolumn{5}{c}{Block 9: $(p_n, \theta_{\mu p})$} \\[0.3mm]
\doubletoprule
{bin number}
& {$p_n^\mathrm{low}$ (\si{\GeV\per\clight})}
& {$p_n^\mathrm{high}$ (\si{\GeV\per\clight})}
& {$\theta_{\mu p}^\mathrm{low}$ ($^\circ$)}
& {$\theta_{\mu p}^\mathrm{high}$ ($^\circ$)}
\\*
\midrule

236 & 0 & 0.21 & 0 & 60 \\*
237 &  & & 60 & 70 \\*
238 &  & & 70 & 80 \\*
239 &  & & 80 & 90 \\*
240 &  & & 90 & 100 \\*
241 &  & & 100 & 110 \\*
242 &  & & 110 & 120 \\*
243 &  & & 120 & 130 \\*
244 &  & & 130 & 140 \\*
245 &  & & 140 & 150 \\*
246 &  & & 150 & 180 \\[3mm]

247 & 0.21 & 0.45 & 0 & 45 \\*
248 &  & & 45 & 60 \\*
249 &  & & 60 & 75 \\*
250 &  & & 75 & 90 \\*
251 &  & & 90 & 100 \\*
252 &  & & 100 & 110 \\*
253 &  & & 110 & 120 \\*
254 &  & & 120 & 130 \\*
255 &  & & 130 & 140 \\*
256 &  & & 140 & 150 \\*
257 &  & & 150 & 180 \\[3mm]

258 & 0.45 & {$\infty$} & 0 & 30 \\*
259 &  & & 30 & 45 \\*
260 &  & & 45 & 60 \\*
261 &  & & 60 & 75 \\*
262 &  & & 75 & 90 \\*
263 &  & & 90 & 105 \\*
264 &  & & 105 & 120 \\*
265 &  & & 120 & 135 \\*
266 &  & & 135 & 150 \\*
267 &  & & 150 & 180 \\
\doublebottomrule{0.5em}
\multicolumn{5}{c}{Block 10: $\cos\theta_\mu$} \\[0.3mm]
\doubletoprule
{bin number}
& {$\cos\theta_\mu^\mathrm{low}$}
& {$\cos\theta_\mu^\mathrm{high}$}
\\*
\midrule

268 & -1 &  -0.85 \\*
269 & -0.85 &  -0.775 \\*
270 & -0.775 &  -0.7 \\*
271 & -0.7 &  -0.625 \\*
272 & -0.625 &  -0.55 \\*
273 & -0.55 &  -0.475 \\*
274 & -0.475 &  -0.4 \\*
275 & -0.4 &  -0.325 \\*
276 & -0.325 &  -0.25 \\*
277 & -0.25 &  -0.175 \\*
278 & -0.175 &  -0.1 \\*
279 & -0.1 &  -0.025 \\*
280 & -0.025 &  0.05 \\*
281 & 0.05 &  0.125 \\*
282 & 0.125 &  0.2 \\*
283 & 0.2 &  0.275 \\*
284 & 0.275 &  0.35 \\*
285 & 0.35 &  0.425 \\*
286 & 0.425 &  0.5 \\*
287 & 0.5 &  0.575 \\*
288 & 0.575 &  0.65 \\*
289 & 0.65 &  0.725 \\*
290 & 0.725 &  0.8 \\*
291 & 0.8 &  0.85 \\*
292 & 0.85 &  0.875 \\*
293 & 0.875 &  0.9 \\*
294 & 0.9 &  0.925 \\*
295 & 0.925 &  0.95 \\*
296 & 0.95 &  0.975 \\*
297 & 0.975 &  1 \\
\doublebottomrule{0.5em}
\multicolumn{5}{c}{Block 11: $\cos\theta_p$} \\[0.3mm]
\doubletoprule
{bin number}
& {$\cos\theta_p^\mathrm{low}$}
& {$\cos\theta_p^\mathrm{high}$}
\\*
\midrule

298 & -1 & -0.9 \\*
299 & -0.9 & -0.75 \\*
300 & -0.75 & -0.6 \\*
301 & -0.6 & -0.45 \\*
302 & -0.45 & -0.3 \\*
303 & -0.3 & -0.15 \\*
304 & -0.15 & 0 \\*
305 & 0 & 0.15 \\*
306 & 0.15 & 0.3 \\*
307 & 0.3 & 0.4 \\*
308 & 0.4 & 0.5 \\*
309 & 0.5 & 0.6 \\*
310 & 0.6 & 0.7 \\*
311 & 0.7 & 0.8 \\*
312 & 0.8 & 0.85 \\*
313 & 0.85 & 0.9 \\*
314 & 0.9 & 0.925 \\*
315 & 0.925 & 0.95 \\*
316 & 0.95 & 0.975 \\*
317 & 0.975 & 1 \\*
\doublebottomrule{0.5em}
\multicolumn{5}{c}{Block 12: $p_p$} \\[0.3mm]
\doubletoprule
{bin number}
& {$p_p^\mathrm{low}$ (\si{\GeV\per\clight})}
& {$p_p^\mathrm{high}$ (\si{\GeV\per\clight})}
\\*
\midrule

318 & 0.25 & 0.3\\*
319 & 0.3 & 0.35\\*
320 & 0.35 & 0.4\\*
321 & 0.4 & 0.45\\*
322 & 0.45 & 0.5\\*
323 & 0.5 & 0.55\\*
324 & 0.55 & 0.6\\*
325 & 0.6 & 0.65\\*
326 & 0.65 & 0.7\\*
327 & 0.7 & 0.75\\*
328 & 0.75 & 0.8\\*
329 & 0.8 & 0.85\\*
330 & 0.85 & 0.9\\*
331 & 0.9 & 0.95\\*
332 & 0.95 & 1 \\*
\doublebottomrule{0.5em}
\multicolumn{5}{c}{Block 13: $p_\mu$} \\[0.3mm]
\doubletoprule
{bin number}
& {$p_\mu^\mathrm{low}$ (\si{\GeV\per\clight})}
& {$p_\mu^\mathrm{high}$ (\si{\GeV\per\clight})}
\\*
\midrule

333 & 0.1 & 0.175 \\*
334 & 0.175 & 0.2 \\*
335 & 0.2 & 0.225 \\*
336 & 0.225 & 0.25 \\*
337 & 0.25 & 0.275 \\*
338 & 0.275 & 0.3 \\*
339 & 0.3 & 0.325 \\*
340 & 0.325 & 0.35 \\*
341 & 0.35 & 0.375 \\*
342 & 0.375 & 0.4 \\*
343 & 0.4 & 0.425 \\*
344 & 0.425 & 0.45 \\*
345 & 0.45 & 0.475 \\*
346 & 0.475 & 0.5 \\*
347 & 0.5 & 0.55 \\*
348 & 0.55 & 0.6 \\*
349 & 0.6 & 0.65 \\*
350 & 0.65 & 0.7 \\*
351 & 0.7 & 0.75 \\*
352 & 0.75 & 0.8 \\*
353 & 0.8 & 0.85 \\*
354 & 0.85 & 0.9 \\*
355 & 0.9 & 0.95 \\*
356 & 0.95 & 1 \\*
357 & 1 & 1.1 \\*
358 & 1.1 & 1.2 \\
\doublebottomrule{0.5em}
\end{longtable}
\twocolumngrid
}
\clearpage

\bibliography{main.bib}

\end{document}


\centering

{\large\textbf{Measurement of double-differential cross sections for mesonless
charged-current muon neutrino interactions on argon with final-state protons
using the MicroBooNE detector}}

\justify
\section{Basic data release}
\label{sec:basic_data_release}

The compressed tar archive file \texttt{basic\_data\_release.tar.bz2} contains
the information required to allow new model predictions to be compared to the
data reported in this paper. On Unix-like operating systems, the file contents
can be extracted by running the command
\begin{center}
\texttt{tar xvfj basic\_data\_release.tar.bz2}
\end{center}
in a terminal. All files discussed in the remainder of this section will be
made available by this procedure. A separate, more detailed set of supplemental
files related to this analysis is described in
Sec.~\ref{sec:extended_data_release}.

Table~\ref{tab:FATXsecDataTable} reports the flux-averaged \ccnp\ total cross
section $\langle \sigma \rangle_\trueBinIdx$ measured in each of the analysis
bins $\trueBinIdx$ defined in Table~III from the main text. In the notation of
Sec.~VI~F from the main text, the flux-averaged total cross sections and their
covariances are obtained from the unfolded event counts
$\MeasuredSignalEvtCount_\trueBinIdx$ via the relations
\begin{align}
\langle \sigma \rangle_\trueBinIdx
&= \frac{ \MeasuredSignalEvtCount_\trueBinIdx }
{ \IntegratedFlux \, \NumTargets }
&
\mathrm{Cov}\big(
\langle \sigma \rangle_\trueBinIdx,
\langle \sigma \rangle_\secondTrueBinIdx \big)
&= \frac{ \mathrm{Cov}(\MeasuredSignalEvtCount_\trueBinIdx,
\MeasuredSignalEvtCount_\secondTrueBinIdx) }
{ \IntegratedFlux^2 \, \NumTargets^2 } \,.
\end{align}
The $\langle \sigma \rangle_\trueBinIdx$ values given here (in units of
\SI{e-38}{\centi\meter\squared} per argon nucleus) can thus be converted to the
differential cross sections shown in the main text by dividing by the relevant
bin width(s) $\PhaseSpaceVecWidths_\trueBinIdx$. For two-dimensional bins,
$\PhaseSpaceVecWidths_\trueBinIdx$ is just the product of the widths along each
axis, e.g., bin 0 from this analysis has $\PhaseSpaceVecWidths_0 =
\SI{0.063}{\GeV\per\clight}$. The choice to tabulate total rather than
differential cross sections, as explained and recommended in
Ref.~\cite[Sec.~III$\,$D]{GardinerXSecExtract}, greatly simplifies handling of
units when covariances are reported between measurements of different
observables. Statistical and total uncertainties on the measurements
(calculated by taking the square root of the diagonal elements of the relevant
covariance matrix) are also reported in the last two columns of
Table~\ref{tab:FATXsecDataTable}.

\pgfplotstableread{data_files/all_unf_slice_047_hist.txt}{\myDataTable}

\pgfplotstabletypeset[
  row predicate/.code={%
    \pgfplotstablegetelem{#1}{x0}\of{\myDataTable}%
    \pgfmathparse{#1}%
    \ifnum\pgfmathresult<359\relax%
    \else\pgfplotstableuserowfalse\fi%
  },
  every head row/.style={ output empty row },
  columns={x0, UnfData, UnfData_DataStats_error, UnfData_total_error},
  columns/x0/.style={column type = {c@{\hskip 1em}}},
  columns/UnfData/.style={string type,
    column type = {S[round-mode=places, round-precision=3]@{\hskip 1em}}},
  columns/UnfData_DataStats_error/.style={
    string type, column type = {S[round-mode=places, round-precision=3]@{\hskip 1em}}},
  columns/UnfData_total_error/.style={string type,
    column type = {S[round-mode=places, round-precision=3]}},
  begin table=\begin{longtable},
  every first row/.append style={before row={%
    \caption{Measured flux-averaged \ccnp\ total cross sections}
    \label{tab:FATXsecDataTable}\\
    \toprule
    \multirow{2}{*}{bin number} & {total cross section}
    & {stat. unc.} & {total unc.} \\
    & {($10^{-38}$~\si{\centi\meter\squared}$/$Ar)}
    & {($10^{-38}$~\si{\centi\meter\squared}$/$Ar)}
    & {($10^{-38}$~\si{\centi\meter\squared}$/$Ar)} \\
    \midrule
    \endfirsthead
    \multicolumn{4}{c}%
    {Table \thetable\ continued from previous page} \\
    \toprule
    \multirow{2}{*}{bin number} & {total cross section}
    & {stat. unc.} & {total unc.} \\
    & {($10^{-38}$~\si{\centi\meter\squared}$/$Ar)}
    & {($10^{-38}$~\si{\centi\meter\squared}$/$Ar)}
    & {($10^{-38}$~\si{\centi\meter\squared}$/$Ar)} \\
    \midrule
    \endhead
    \bottomrule \\
    \endfoot
    \bottomrule \\
    \endlastfoot
    }
  },
  end table=\end{longtable},
]{\myDataTable}

A machine-readable version of Table~\ref{tab:FATXsecDataTable} is provided as
the text file \texttt{xsec\_summary\_table.txt}. The first line of this file
contains the string \texttt{numXbins} followed by the number of bins
(\TotalBinCount) used to present the final measurement. The following lines
match the contents of Table~\ref{tab:FATXsecDataTable} (without the column
headings), but the numerical values are reported to the full precision adopted
in the \cpp~code used to execute the analysis.

\begin{figure}[h]
\centering
\includegraphics[width=0.49\textwidth]{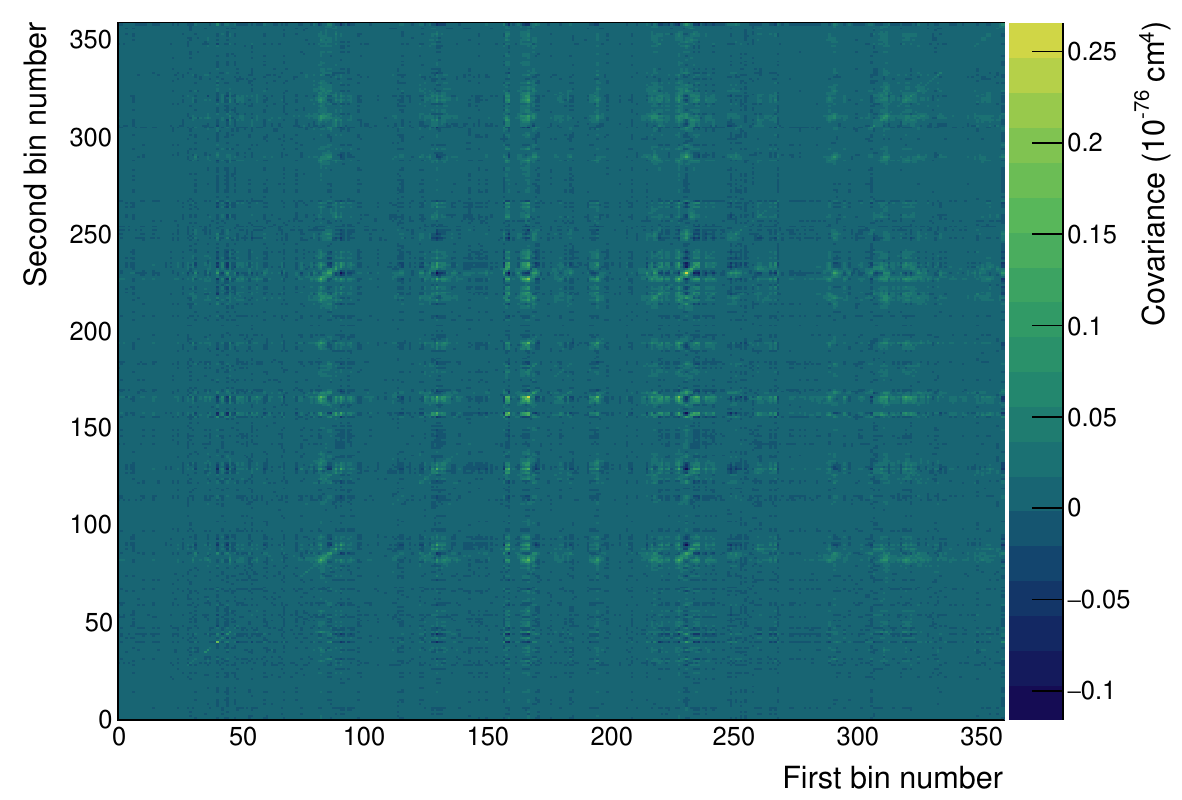}
\hfill
\includegraphics[width=0.49\textwidth]{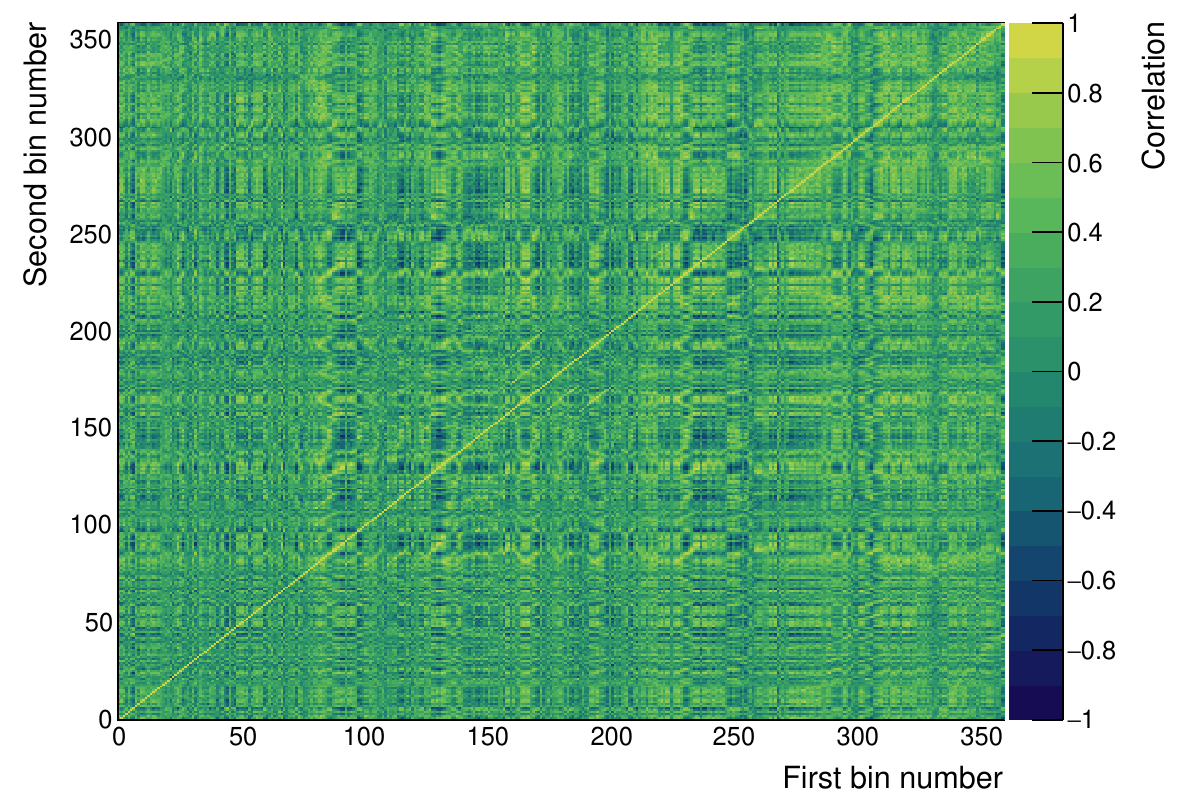}
\caption{The total covariance matrix (left) and correlation matrix (right) for
the flux-averaged total cross sections measured in the analysis.}
\label{fig:cov_and_corr_matrices}
\end{figure}

The $\TotalBinCount \times \TotalBinCount$ covariance matrix describing the
total uncertainty on the measured $\langle \sigma \rangle_\trueBinIdx$ values
is too large to conveniently tabulate in this document. However,
the left-hand panel of Fig.~\ref{fig:cov_and_corr_matrices} presents a plot
of the total covariance matrix elements. The right-hand panel gives the
corresponding correlation matrix, whose elements are computed according to
the formula
\begin{equation}
\mathrm{Corr}\big(
\langle \sigma \rangle_\trueBinIdx,
\langle \sigma \rangle_\secondTrueBinIdx \big)
=
\frac{ \mathrm{Cov}\big(
\langle \sigma \rangle_\trueBinIdx,
\langle \sigma \rangle_\secondTrueBinIdx \big) }
{
\mathrm{Cov}\big(
\langle \sigma \rangle_\trueBinIdx,
\langle \sigma \rangle_\trueBinIdx \big)
\cdot
\mathrm{Cov}\big(
\langle \sigma \rangle_\secondTrueBinIdx,
\langle \sigma \rangle_\secondTrueBinIdx \big)
} \,.
\end{equation}
A machine-readable table of the elements of the total covariance matrix is
provided in the \texttt{mat\_table\_cov\_total.txt} file. The first two lines
of the file contain the strings \texttt{numXbins} and \texttt{numYbins},
respectively, each followed by the total number of bins reported
(\TotalBinCount). The third line provides column labels (\texttt{xbin},
\texttt{ybin}, and \texttt{z}) for the numerical data that follow. The
remaining lines of the file list a row ($\trueBinIdx$) and column
($\secondTrueBinIdx$) index followed by the corresponding total covariance
matrix element $\mathrm{Cov}( \langle \sigma \rangle_\trueBinIdx, \langle
\sigma \rangle_\secondTrueBinIdx)$ in units of
\SI{e-76}{\centi\meter\tothe{4}}.

In addition to the total covariance matrix, several of its most important
components are also provided in separate text files. These include
\texttt{mat\_table\_cov\_detVar\_total.txt},
\texttt{mat\_table\_cov\_flux.txt}, and
\texttt{mat\_table\_cov\_xsec\_total.txt}, which respectively tabulate the
covariances on the measured cross sections due to systematic uncertainties
related to the detector response, neutrino flux prediction, and neutrino
interaction model. In many bins, the portion of the interaction model
covariance estimated by using NuWro as an alternative event generator (see
Sec.~VA of main text) is the leading source of systematic uncertainty. For
reference, this contribution to the total covariance matrix is tabulated
individually in the text file \texttt{mat\_table\_cov\_NuWroGenie.txt}. It is
already included in the overall interaction model covariance matrix reported in
\texttt{mat\_table\_cov\_xsec\_total.txt}. All of these partial covariance
matrix files use the same units (\SI{e-76}{\centi\meter\tothe{4}}) as the total
covariance matrix.

\begin{figure}[h]
\centering
\includegraphics[width=0.49\textwidth]{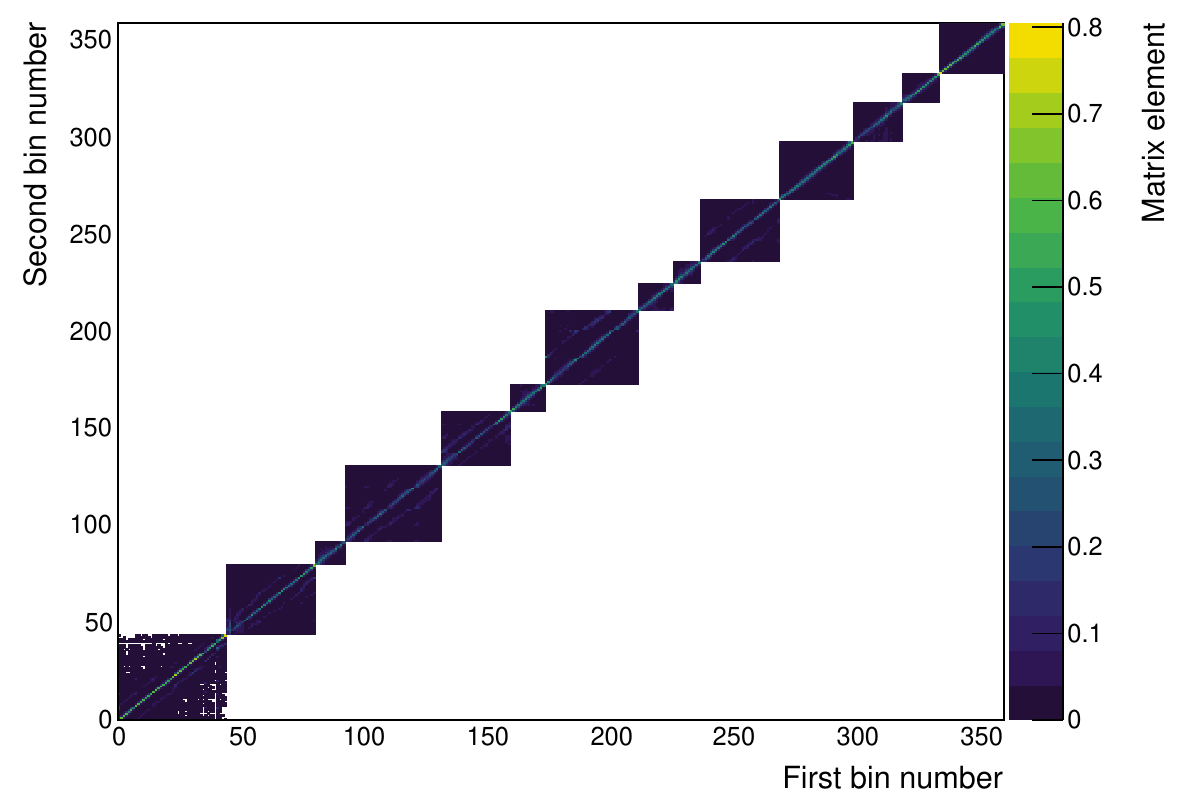}
\caption{The additional smearing matrix $\AddSmearMatrix$.}
\label{fig:add_smear_matrix}
\end{figure}

As discussed in Sec.~VI~A of the main text, a theoretical prediction of the
vector of flux-averaged total cross sections $\langle \sigma
\rangle_\trueBinIdx$ should be multiplied by the additional smearing matrix
$\AddSmearMatrix$ before being directly compared to the data in
Table~\ref{tab:FATXsecDataTable}. Figure~\ref{fig:add_smear_matrix} displays
the block diagonal structure of $\AddSmearMatrix$ using a color scheme in which
matrix elements that are identically zero are rendered as white squares. The
file \texttt{mat\_table\_add\_smear.txt} provides a machine-readable list of
the additional smearing matrix elements using the same format as the total
covariance matrix file (\texttt{mat\_table\_cov\_total.txt}). However, note
that the numerical values in the final column are now dimensionless.

To illustrate how these text files may be manipulated to compare a theoretical
prediction to the full data set, an example \cpp\ program is provided in the
file \texttt{calc\_chi2.C}. The program relies on the \texttt{TMatrixD} class
defined by ROOT and must either be executed using the ROOT \cpp\ interpreter or
be compiled against the ROOT shared libraries. The \texttt{calc\_chi2.C}
program loads tables of the measured $\langle \sigma \rangle_\trueBinIdx$
values, their covariances, and the elements of $\AddSmearMatrix$. It also loads
a table of theoretical $\langle \sigma \rangle_\trueBinIdx$ values (stored in
the text file \texttt{vec\_table\_uBTune.txt}) predicted by the MicroBooNE Tune
interaction model described in the main text. The overall $\chi^2$
goodness-of-fit metric obtained for the MicroBooNE Tune model in the main text
is then reproduced and printed to the terminal.

A similar program is provided in the file \texttt{calc\_chi2.py}
for users who prefer to work with the Python programming language. In this
case, the program depends on the NumPy package but not on ROOT. Within the
numerical precision printed to the terminal, it is expected to produce
the same output as the example \cpp\ program.

By replacing the numerical $\langle \sigma \rangle_\trueBinIdx$ values in the
file \texttt{vec\_table\_uBTune.txt} and rerunning either \texttt{calc\_chi2.C}
or \texttt{calc\_chi2.py}, one may immediately compute an overall $\chi^2$
score for a comparison of an alternate cross-section calculation to the
measured data points. The replacement $\langle \sigma \rangle_\trueBinIdx$
values need not have the same numerical precision as the original ones, and
they may be represented using either scientific or fixed-point notation.

The expected $\nu_\mu$ flux generated by the Fermilab Booster Neutrino Beam at
the location of MicroBooNE is tabulated in the file
\texttt{microboone\_numu\_flux.txt}, which has been duplicated from the
supplemental materials included with Ref.~\cite{uBCCincl2D}. Theoretical
predictions of the total cross sections $\langle \sigma \rangle_\trueBinIdx$
should be averaged over the neutrino energy distribution given therein.

\section{Migration matrices}
\label{sec:mig_mat}

The \textit{migration matrix} $\MigrationMatrix_{\recoBinIdx\trueBinIdx}$
defined in Eq.~12 of the main text quantifies smearing effects in the
reconstruction of physics observables measured in the present analysis. Plots
of the migration matrices for each block of bins used to report the final
results are given in the figures below.

\begin{figure}[h]
\centering
\includegraphics[width=\MigMatWidth]{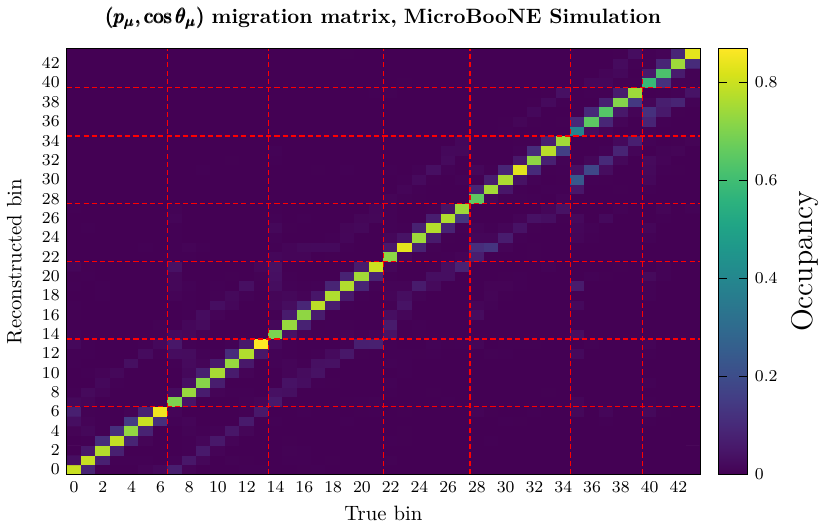}
\caption{Migration matrix for block \#0 of bins used for the
double-differential measurement of ($p_\mu, \cos\theta_\mu$). Dashed red lines
indicate momentum bin boundaries.}
\label{fig:migration_block0}
\end{figure}

\begin{figure}
\centering
\includegraphics[width=\MigMatWidth]{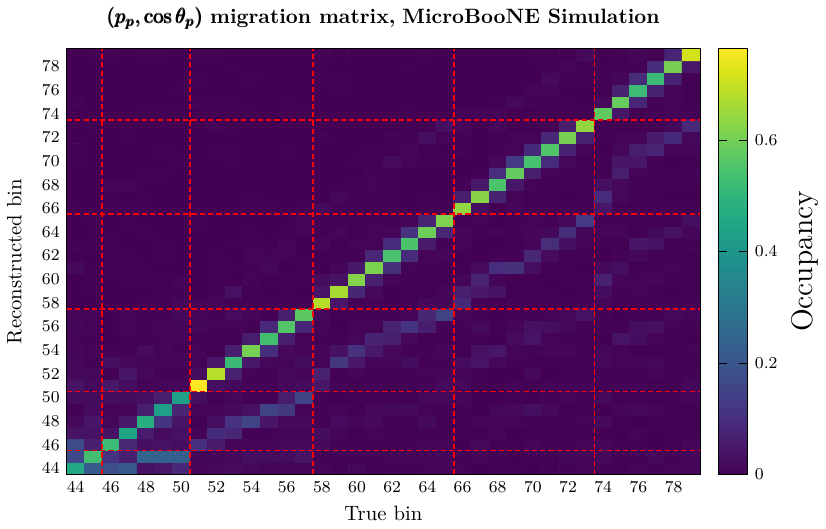}
\caption{Migration matrix for block \#1 of bins used for the
double-differential measurement of ($p_p, \cos\theta_p$). Dashed red lines
indicate momentum bin boundaries.}
\label{fig:migration_block1}
\end{figure}

\begin{figure}
\centering
\includegraphics[width=\MigMatWidth]{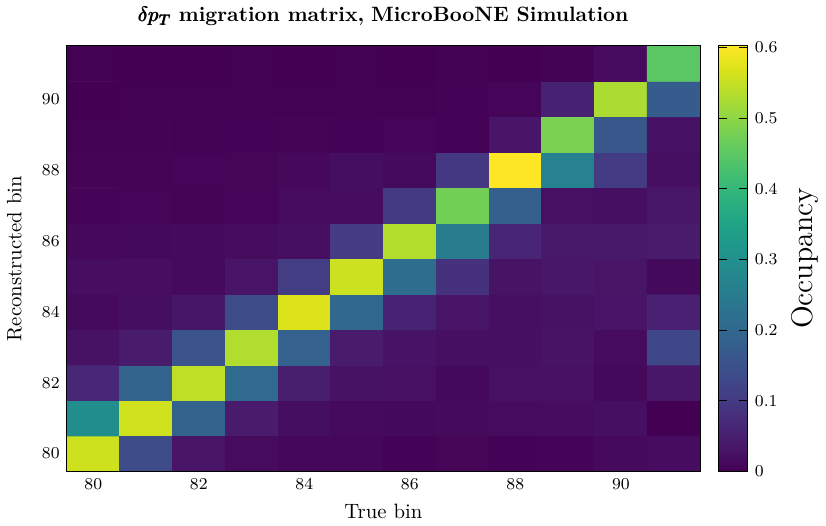}
\caption{Migration matrix for block \#2 of bins used for the
single-differential measurement of $\delta p_T$.}
\label{fig:migration_block2}
\end{figure}

\begin{figure}
\centering
\includegraphics[width=\MigMatWidth]{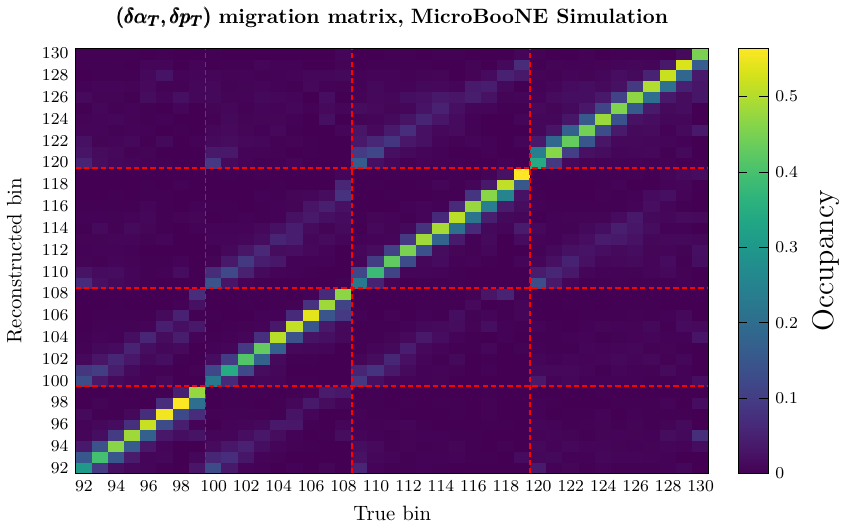}
\caption{Migration matrix for block \#3 of bins used for the
double-differential measurement of $(\delta \alpha_T, \delta p_T)$.
Dashed red lines indicate $\delta \alpha_T$ bin boundaries.}
\label{fig:migration_block3}
\end{figure}

\begin{figure}
\centering
\includegraphics[width=\MigMatWidth]{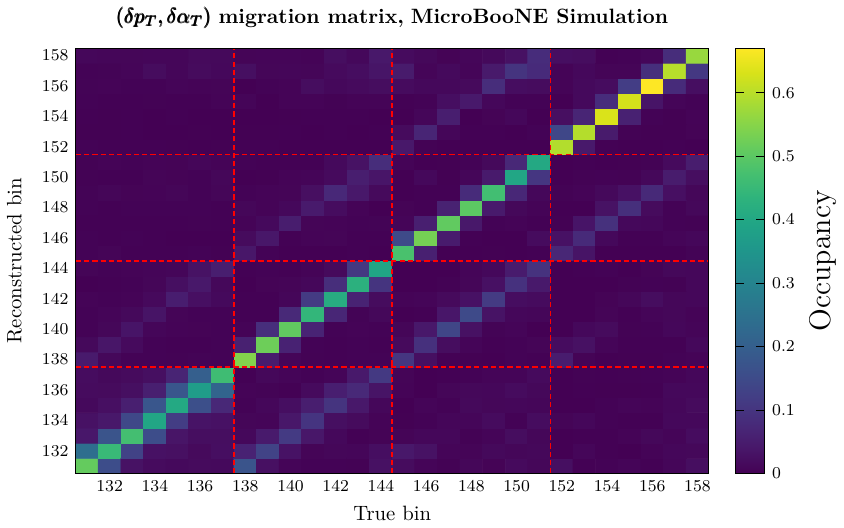}
\caption{Migration matrix for block \#4 of bins used for the
double-differential measurement of $(\delta p_T, \delta \alpha_T)$.
Dashed red lines indicate $\delta p_T$ bin boundaries.}
\label{fig:migration_block4}
\end{figure}

\begin{figure}
\centering
\includegraphics[width=\MigMatWidth]{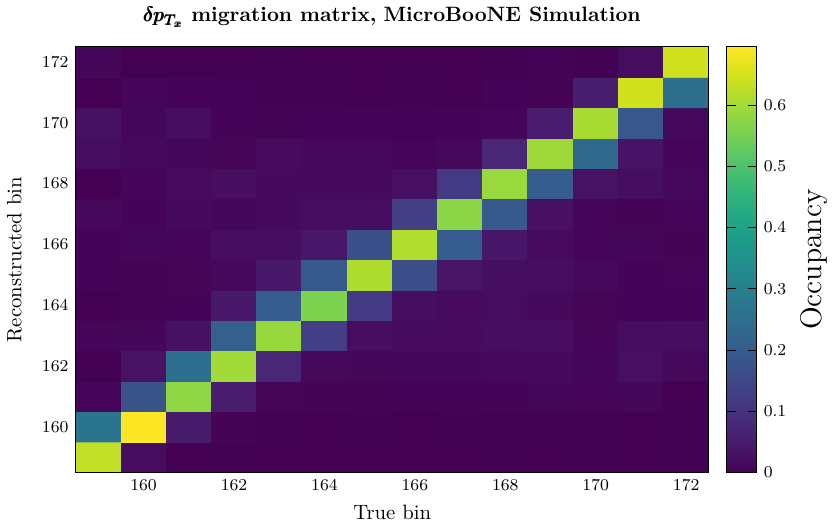}
\caption{Migration matrix for block \#5 of bins used for the
single-differential measurement of $\delta p_{T_x}$.}
\label{fig:migration_block5}
\end{figure}

\begin{figure}
\centering
\includegraphics[width=\MigMatWidth]{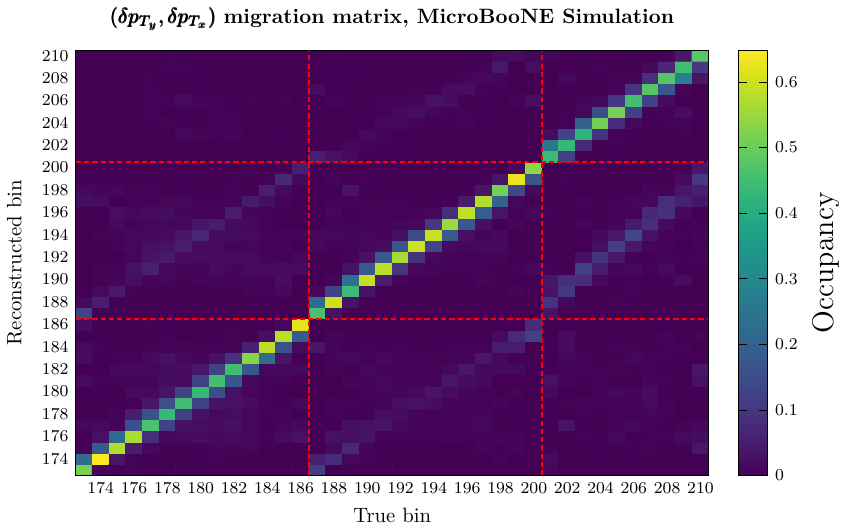}
\caption{Migration matrix for block \#6 of bins used for the
double-differential measurement of $(\delta p_{T_y}, \delta p_{T_x})$.
Dashed red lines indicate $\delta p_{T_y}$ bin boundaries.}
\label{fig:migration_block6}
\end{figure}

\begin{figure}
\centering
\includegraphics[width=\MigMatWidth]{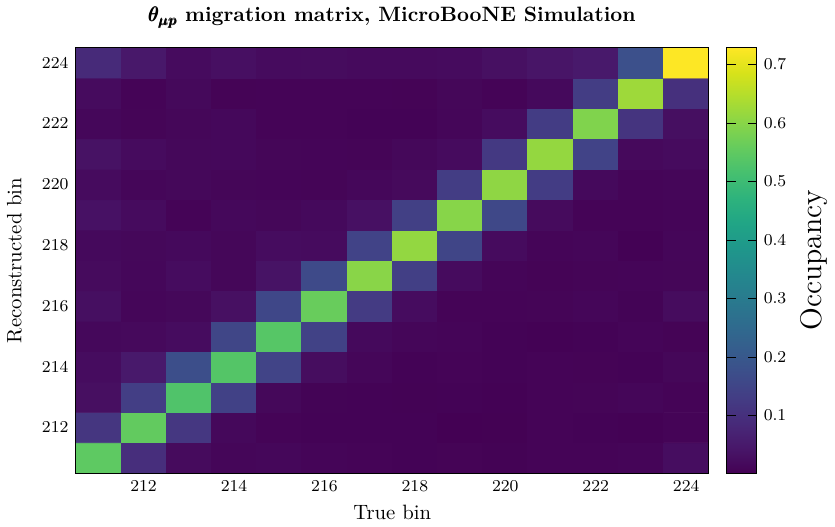}
\caption{Migration matrix for block \#7 of bins used for the
single-differential measurement of $\theta_{\mu p}$.}
\label{fig:migration_block7}
\end{figure}

\begin{figure}
\centering
\includegraphics[width=\MigMatWidth]{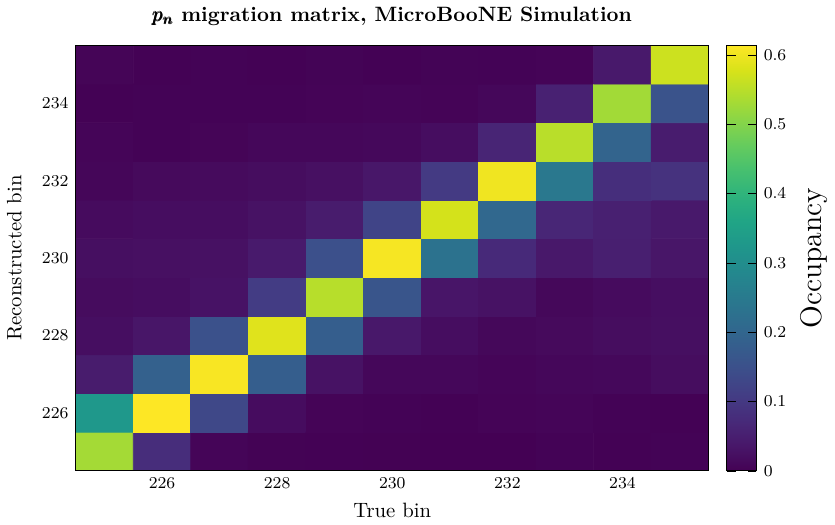}
\caption{Migration matrix for block \#8 of bins used for the
single-differential measurement of $p_n$.}
\label{fig:migration_block8}
\end{figure}

\begin{figure}
\centering
\includegraphics[width=\MigMatWidth]{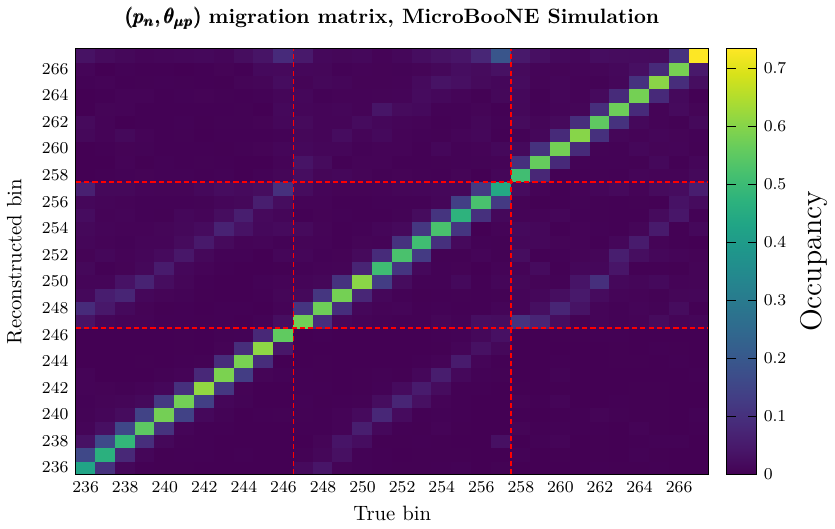}
\caption{Migration matrix for block \#9 of bins used for the
double-differential measurement of $(p_n, \theta_{\mu p})$.
Dashed red lines indicate $p_n$ bin boundaries.}
\label{fig:migration_block9}
\end{figure}

\begin{figure}
\centering
\includegraphics[width=\MigMatWidth]{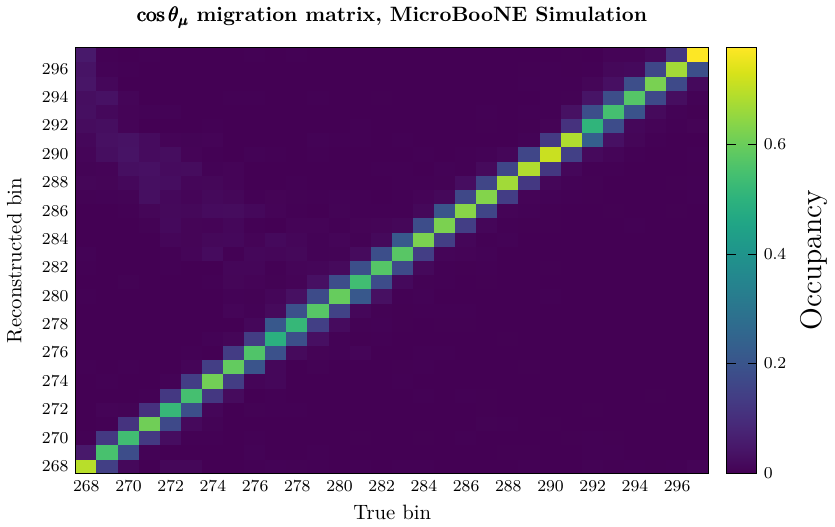}
\caption{Migration matrix for block \#10 of bins used for the
single-differential measurement of $\cos\theta_\mu$.}
\label{fig:migration_block10}
\end{figure}

\begin{figure}
\centering
\includegraphics[width=\MigMatWidth]{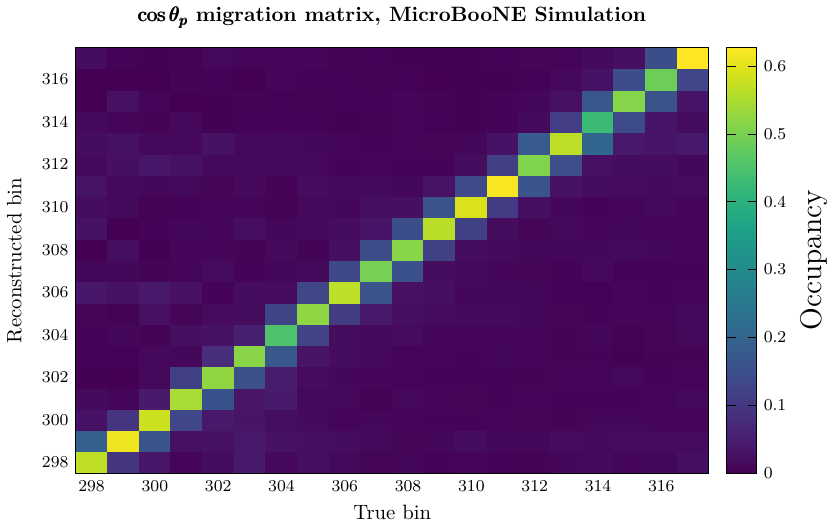}
\caption{Migration matrix for block \#11 of bins used for the
single-differential measurement of $\cos\theta_p$.}
\label{fig:migration_block11}
\end{figure}

\begin{figure}
\centering
\includegraphics[width=\MigMatWidth]{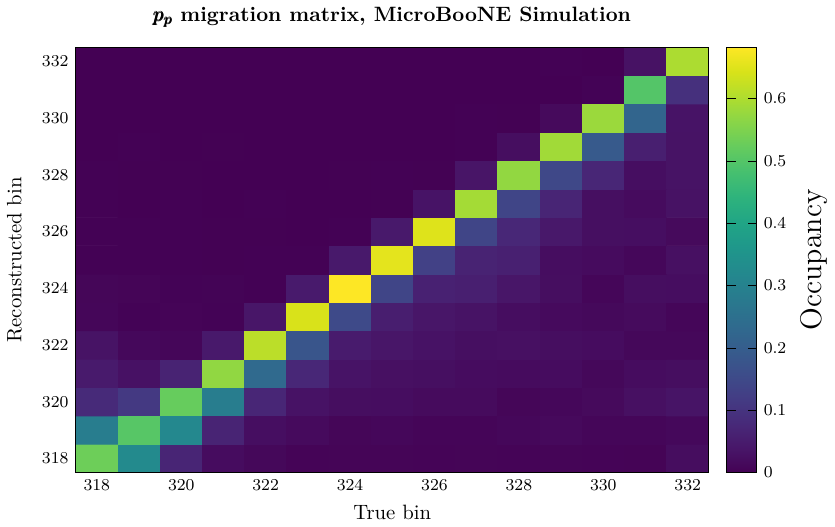}
\caption{Migration matrix for block \#12 of bins used for the
single-differential measurement of $p_p$.}
\label{fig:migration_block12}
\end{figure}

\begin{figure}
\centering
\includegraphics[width=\MigMatWidth]{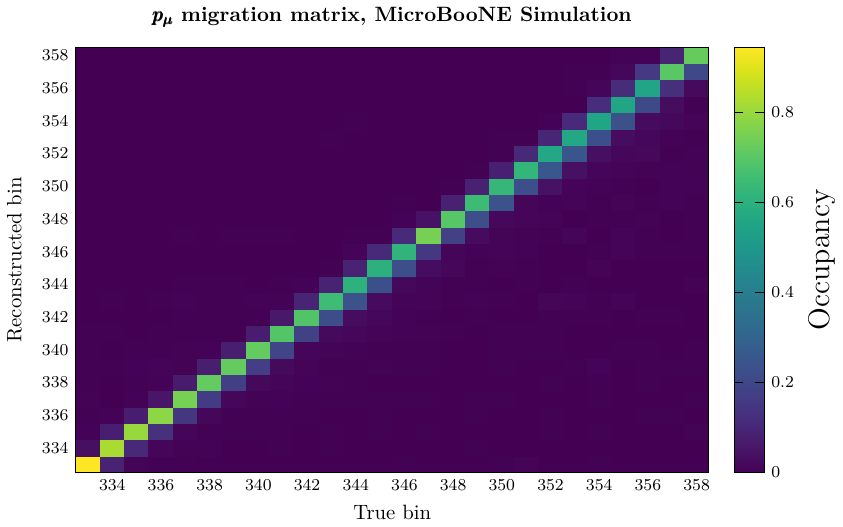}
\caption{Migration matrix for block \#13 of bins used for the
single-differential measurement of $p_\mu$.}
\label{fig:migration_block13}
\end{figure}

\clearpage
\section{Sideband study results}
\label{sec:sideband_reco_plots}

Measured reconstructed event distributions obtained using the combined sideband
selection defined in Sec.~V~F of the main text are shown below. The plots
use the same format as those in Sec.~V~E of the main text.

\begin{figure*}[h]
\centering
\includegraphics[page=16, width=0.49\textwidth]{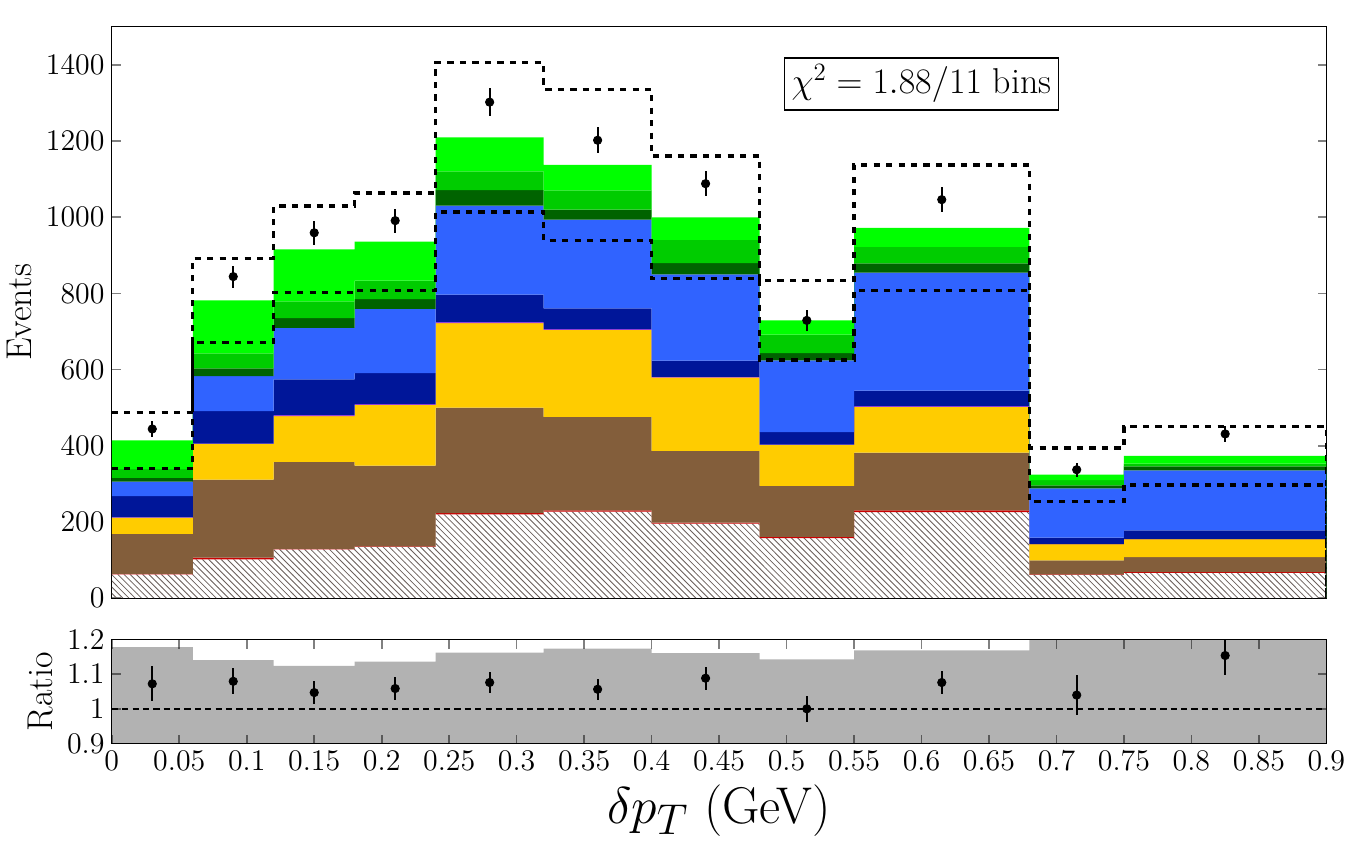}
\hfill
\includegraphics[page=28, width=0.49\textwidth]{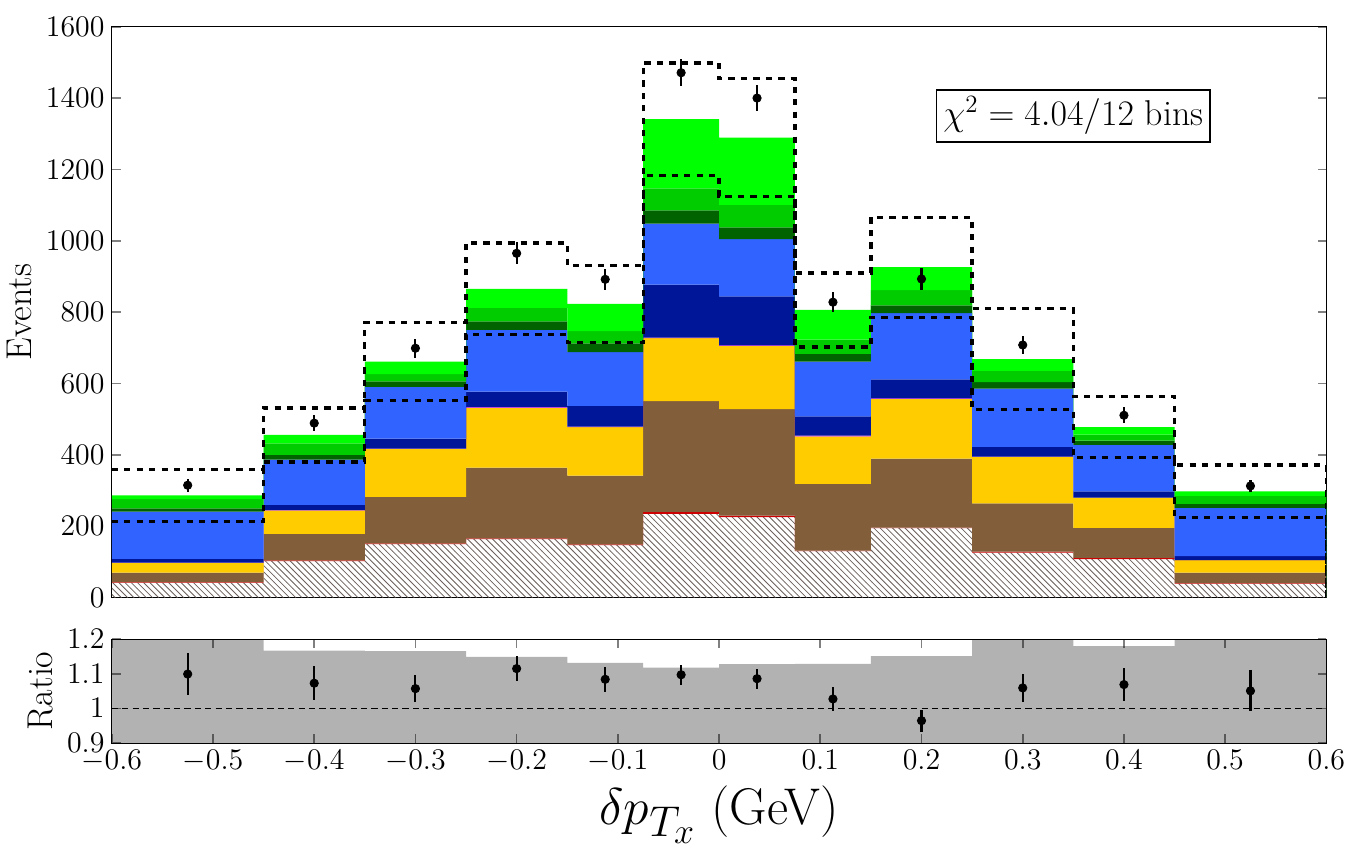}
\vspace{0.2cm}

\includegraphics[page=33, width=0.49\textwidth]{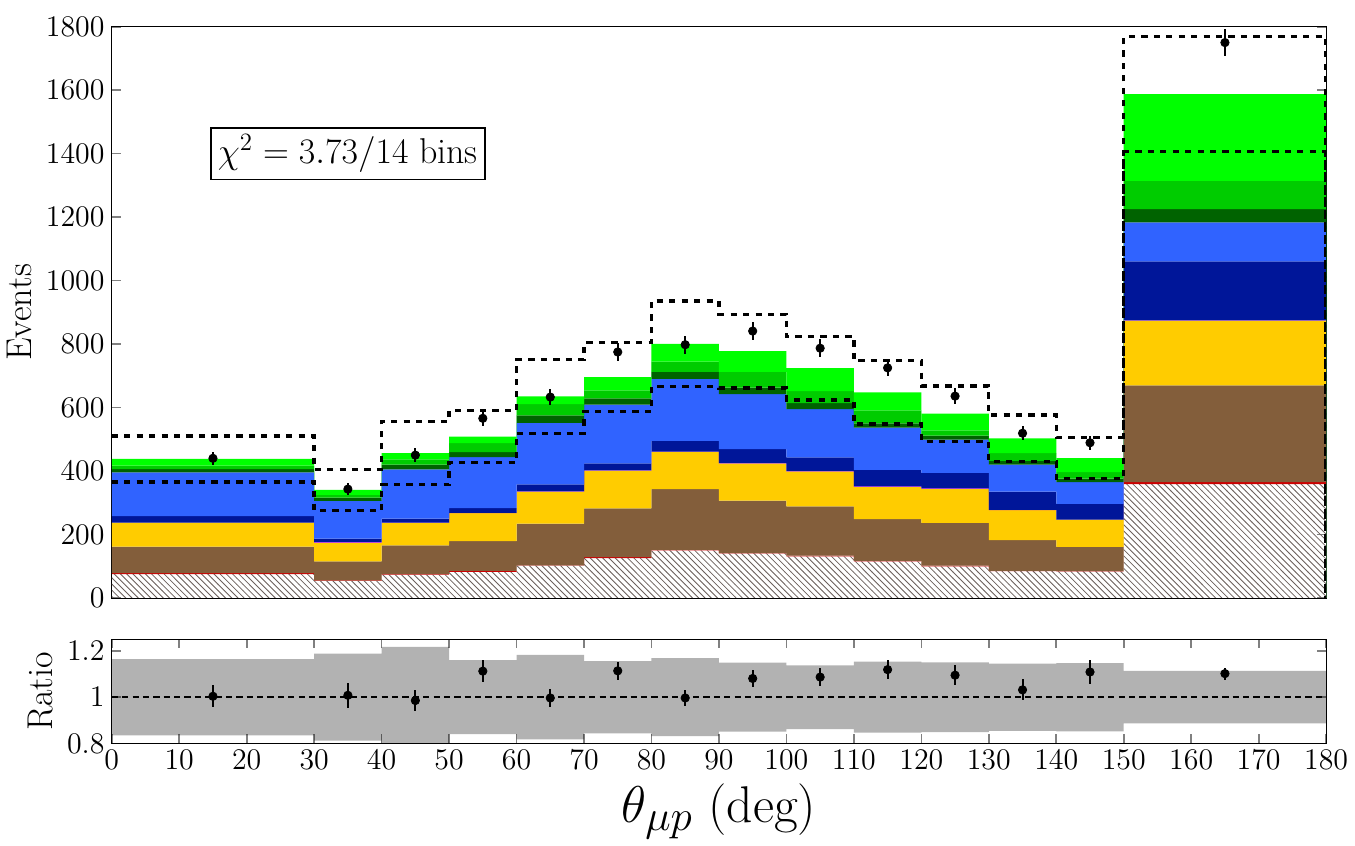}
\hfill
\includegraphics[page=34, width=0.49\textwidth]{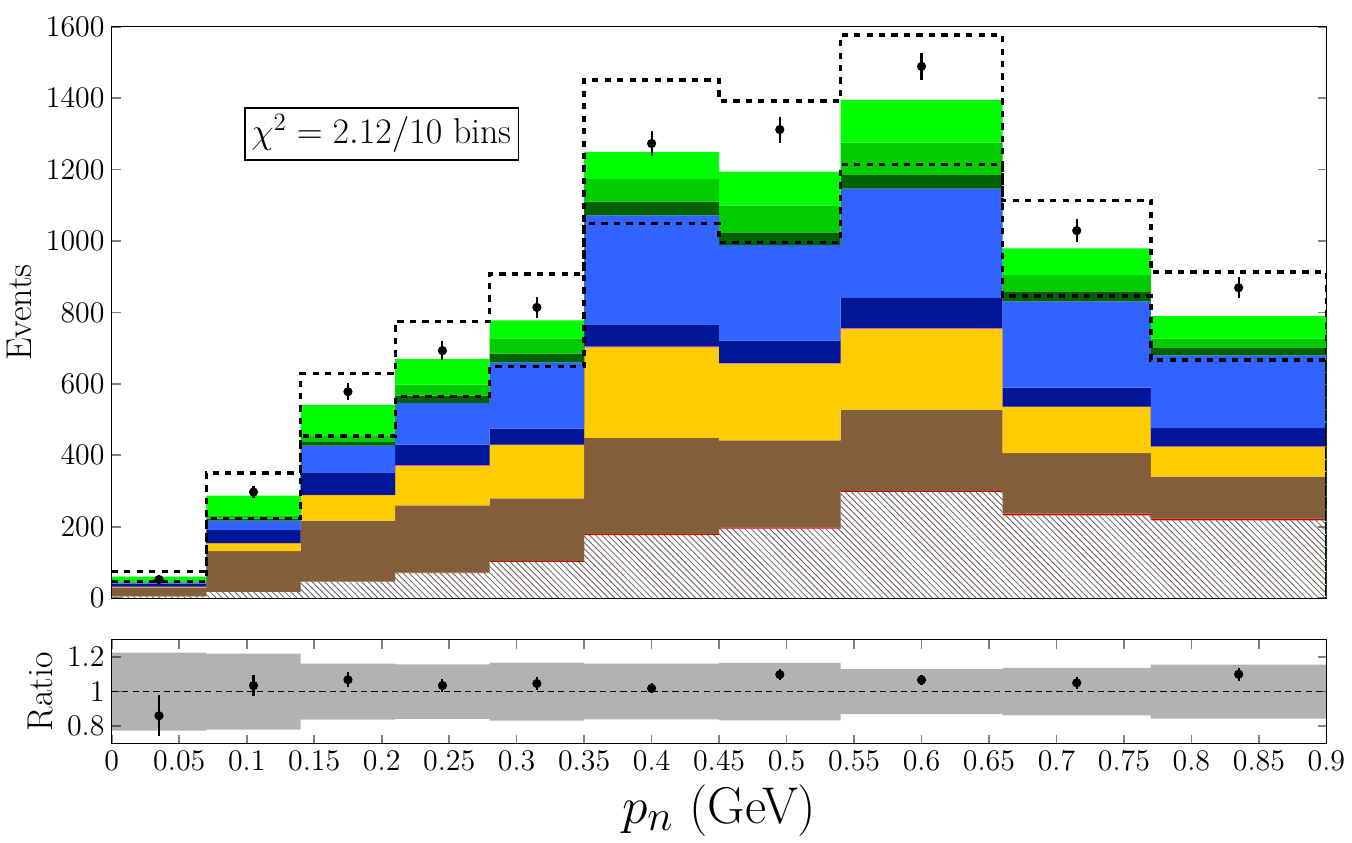}
\vspace{0.2cm}

\includegraphics[page=17, width=0.35\textwidth]{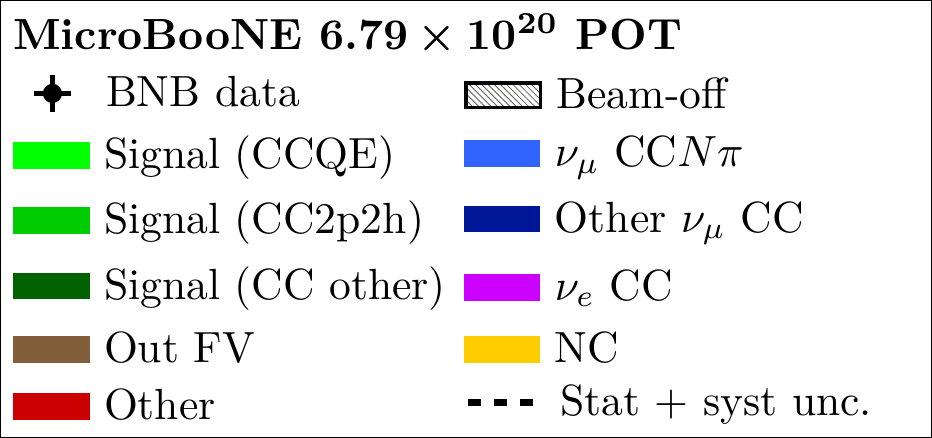}%
\hfill

\caption{Reconstructed event distributions for block~\#2
($\delta p_T$, upper left), block~\#5 ($\delta p_{T_x}$,
upper right), block~\#7 ($\theta_{\mu p}$, lower left),
and block~\#8 ($p_n$, lower right).}
\label{fig:DataSBMCMultipleBlocks1}
\end{figure*}

\begin{figure*}
\centering
\includegraphics[page=1, width=0.49\textwidth]{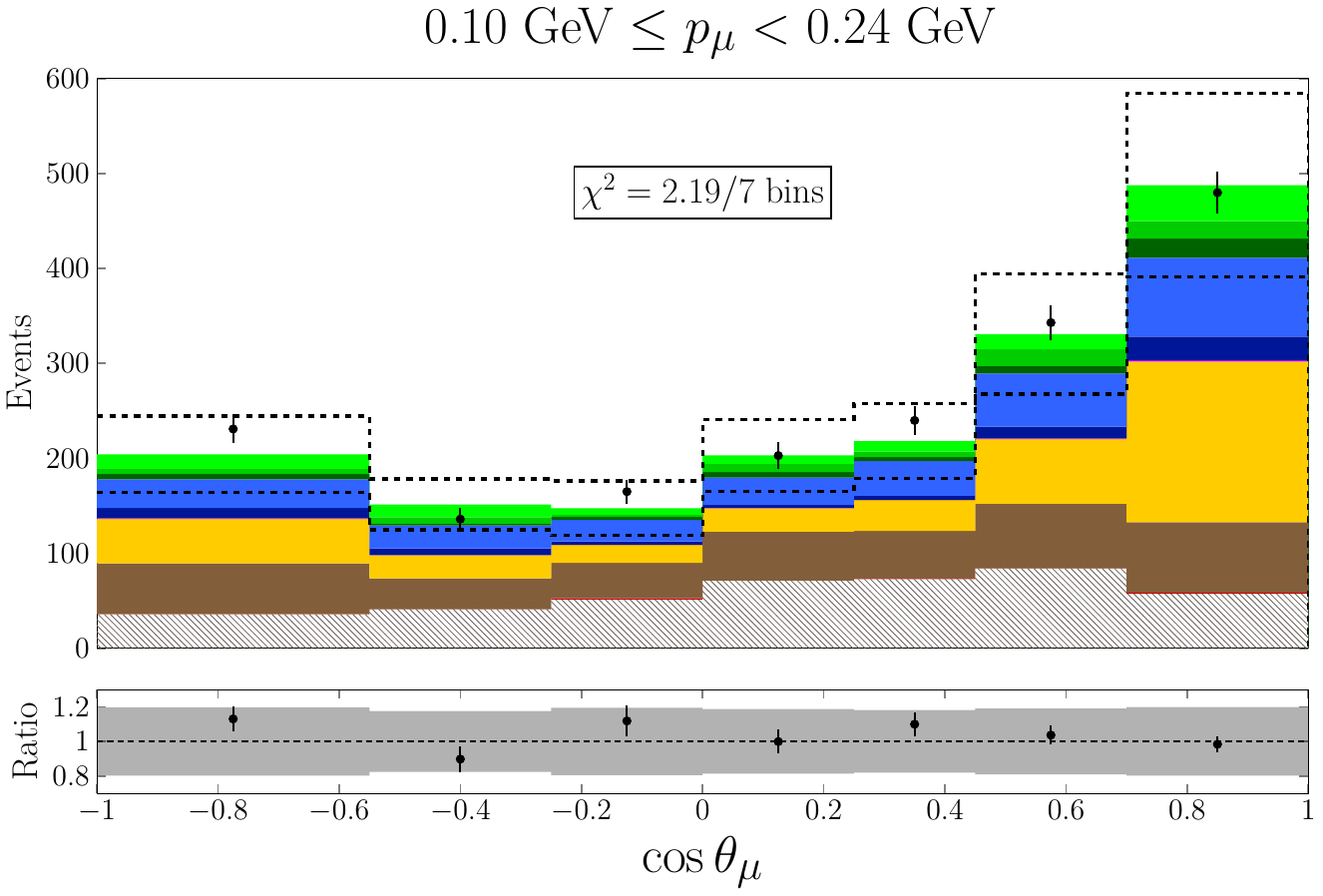}
\hfill
\includegraphics[page=2, width=0.49\textwidth]{figures/reco_results_sidebands_prd.pdf}
\vspace{0.1cm}

\includegraphics[page=3, width=0.49\textwidth]{figures/reco_results_sidebands_prd.pdf}
\hfill
\includegraphics[page=4, width=0.49\textwidth]{figures/reco_results_sidebands_prd.pdf}
\vspace{0.1cm}

\includegraphics[page=5, width=0.49\textwidth]{figures/reco_results_sidebands_prd.pdf}
\hfill
\includegraphics[page=6, width=0.49\textwidth]{figures/reco_results_sidebands_prd.pdf}
\vspace{0.1cm}

\includegraphics[page=7, width=0.49\textwidth]{figures/reco_results_sidebands_prd.pdf}
\hfill
\belowbaseline[-0.23\textheight]{
  \includegraphics[page=8, width=0.35\textwidth]{figures/reco_results_sidebands_prd.pdf}%
}
\hfill

\caption{Reconstructed event distributions for block \#0
($p_\mu, \cos\theta_\mu$).}
\label{fig:DataSBMCblock0}
\end{figure*}

\begin{figure*}
\centering
\includegraphics[page=9, width=0.49\textwidth]{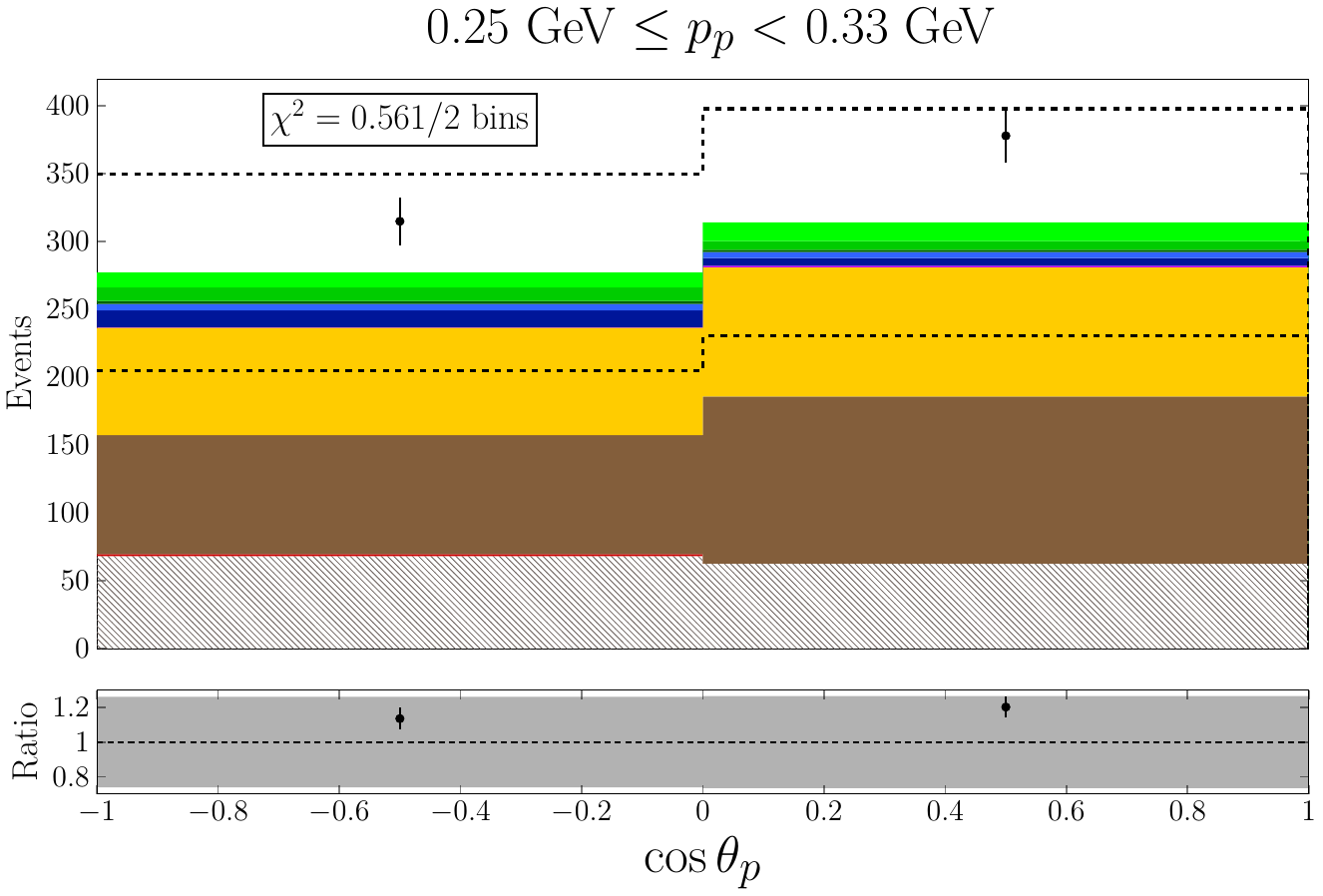}
\hfill
\includegraphics[page=10, width=0.49\textwidth]{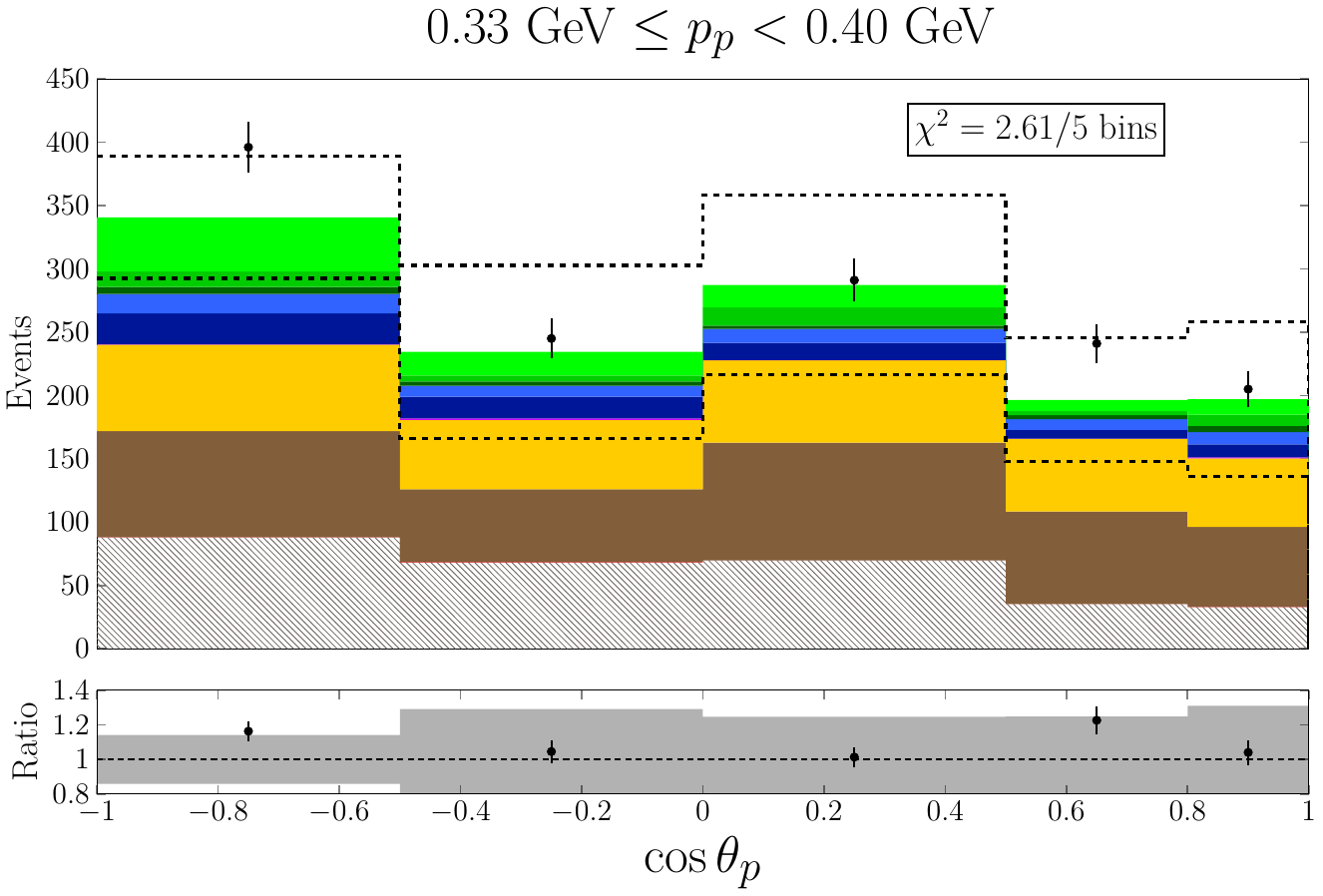}
\vspace{0.2cm}

\includegraphics[page=11, width=0.49\textwidth]{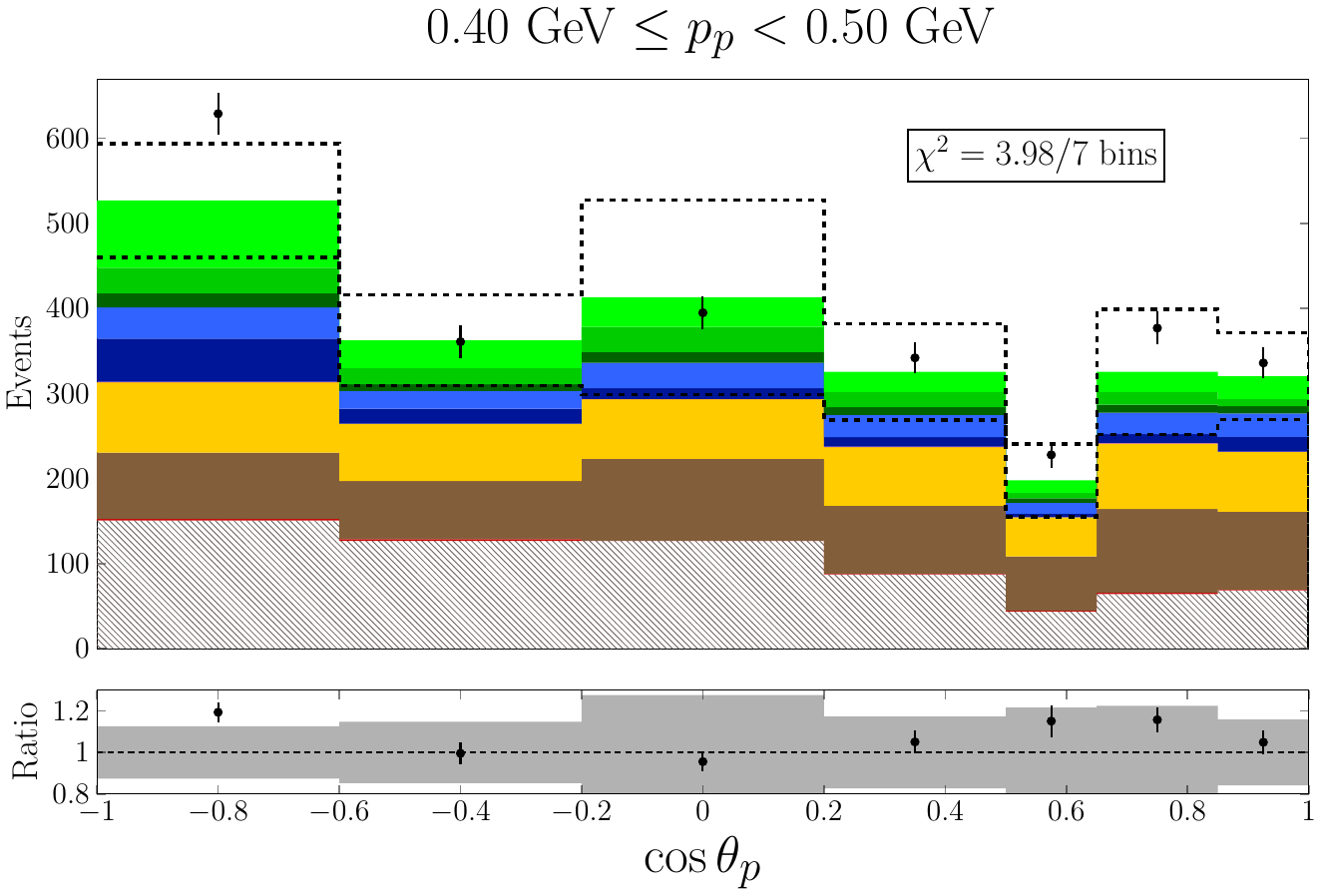}
\hfill
\includegraphics[page=12, width=0.49\textwidth]{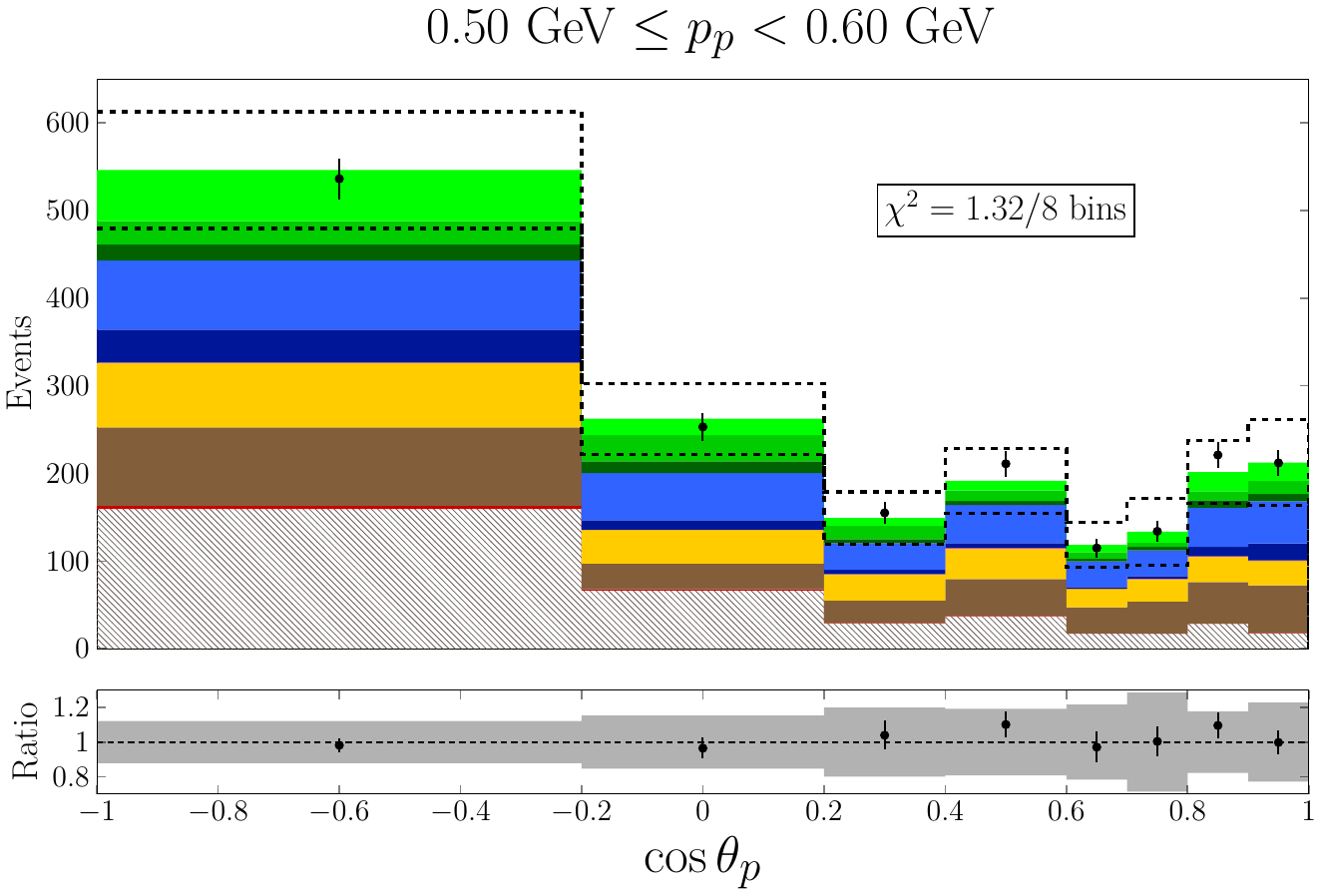}
\vspace{0.2cm}

\includegraphics[page=13, width=0.49\textwidth]{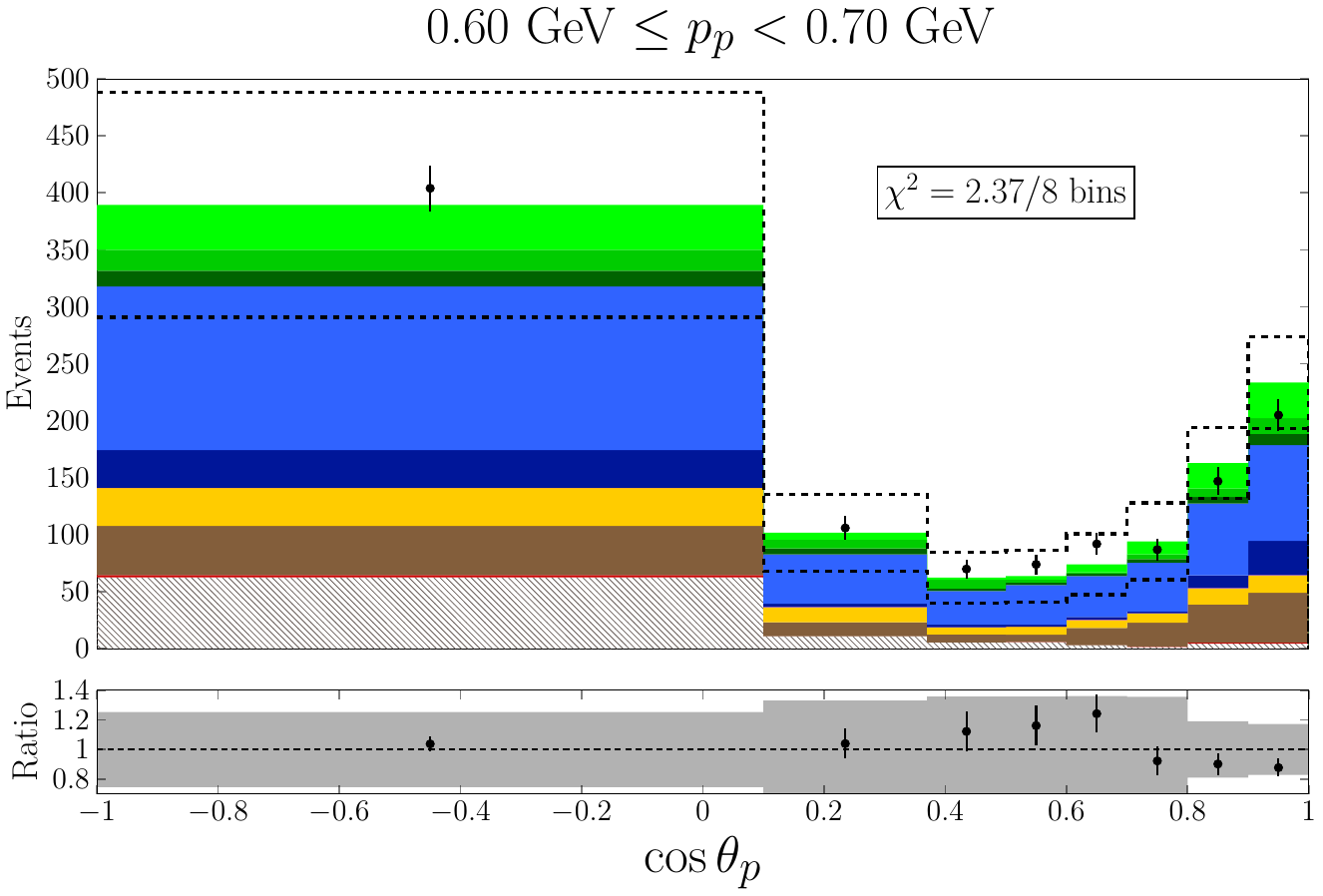}
\hfill
\includegraphics[page=14, width=0.49\textwidth]{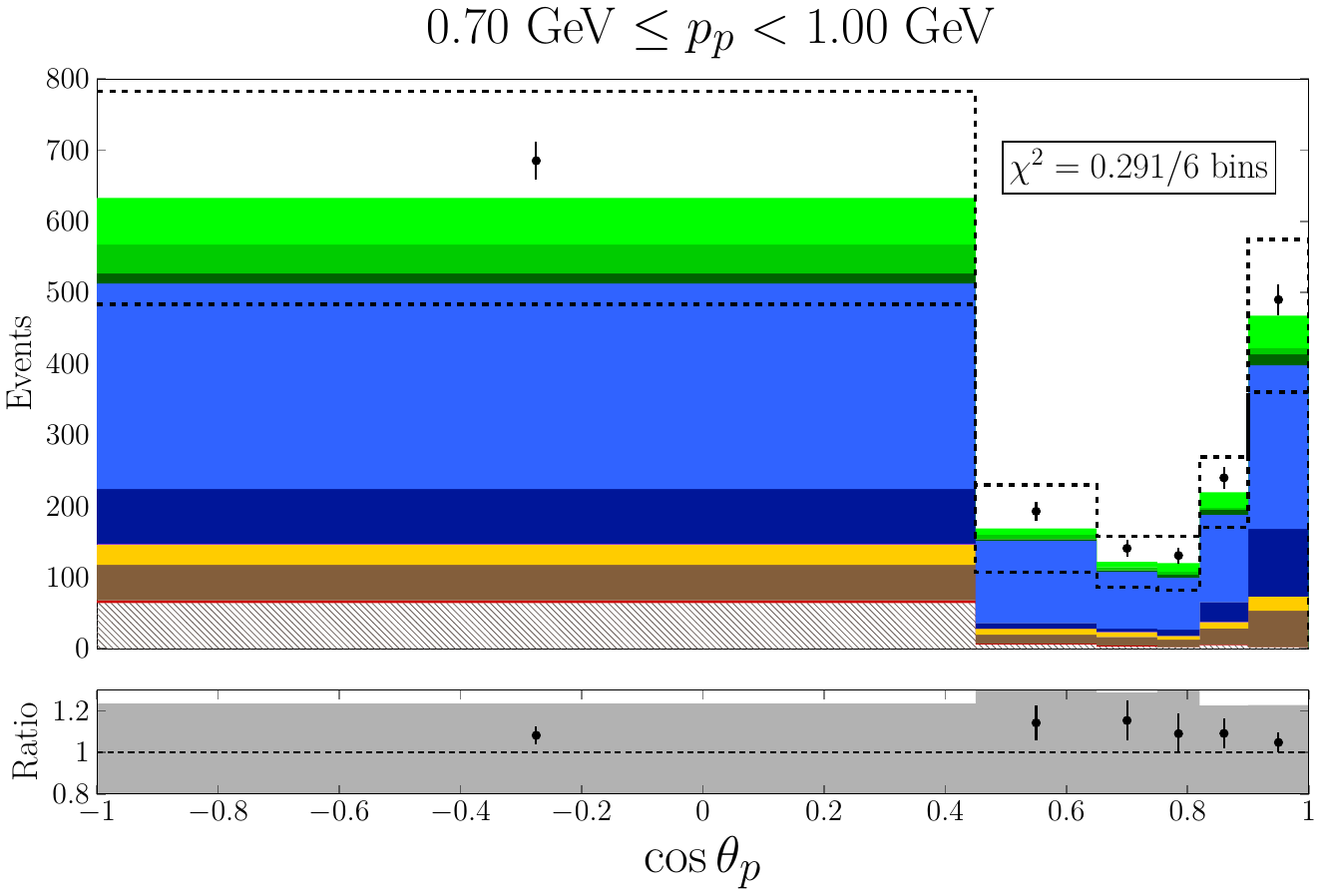}
\vspace{0.2cm}

\includegraphics[page=15, width=0.35\textwidth]{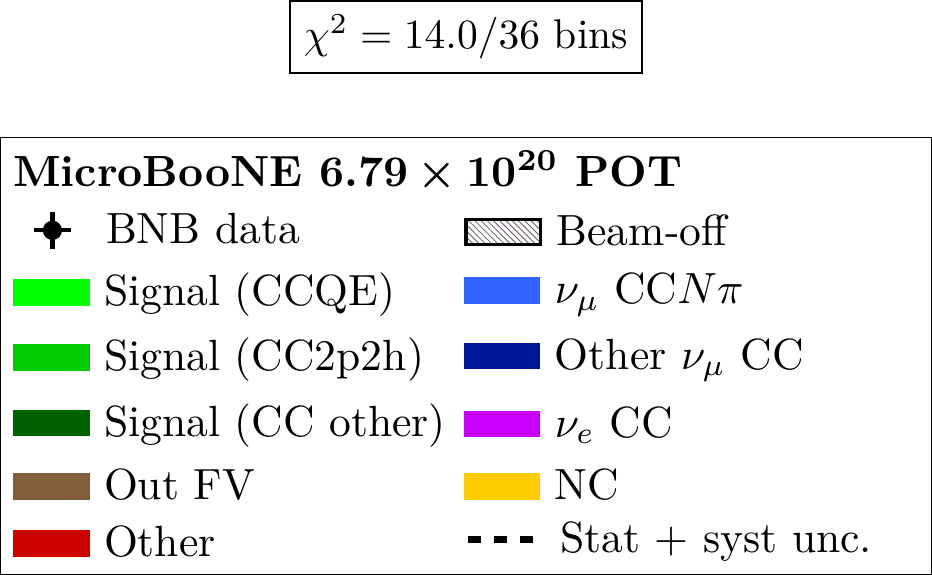}%
\hfill

\caption{Reconstructed event distributions for block \#1
($p_p, \cos\theta_p$).}
\label{fig:DataSBMCblock1}
\end{figure*}

\begin{figure*}
\centering
\includegraphics[page=18, width=0.49\textwidth]{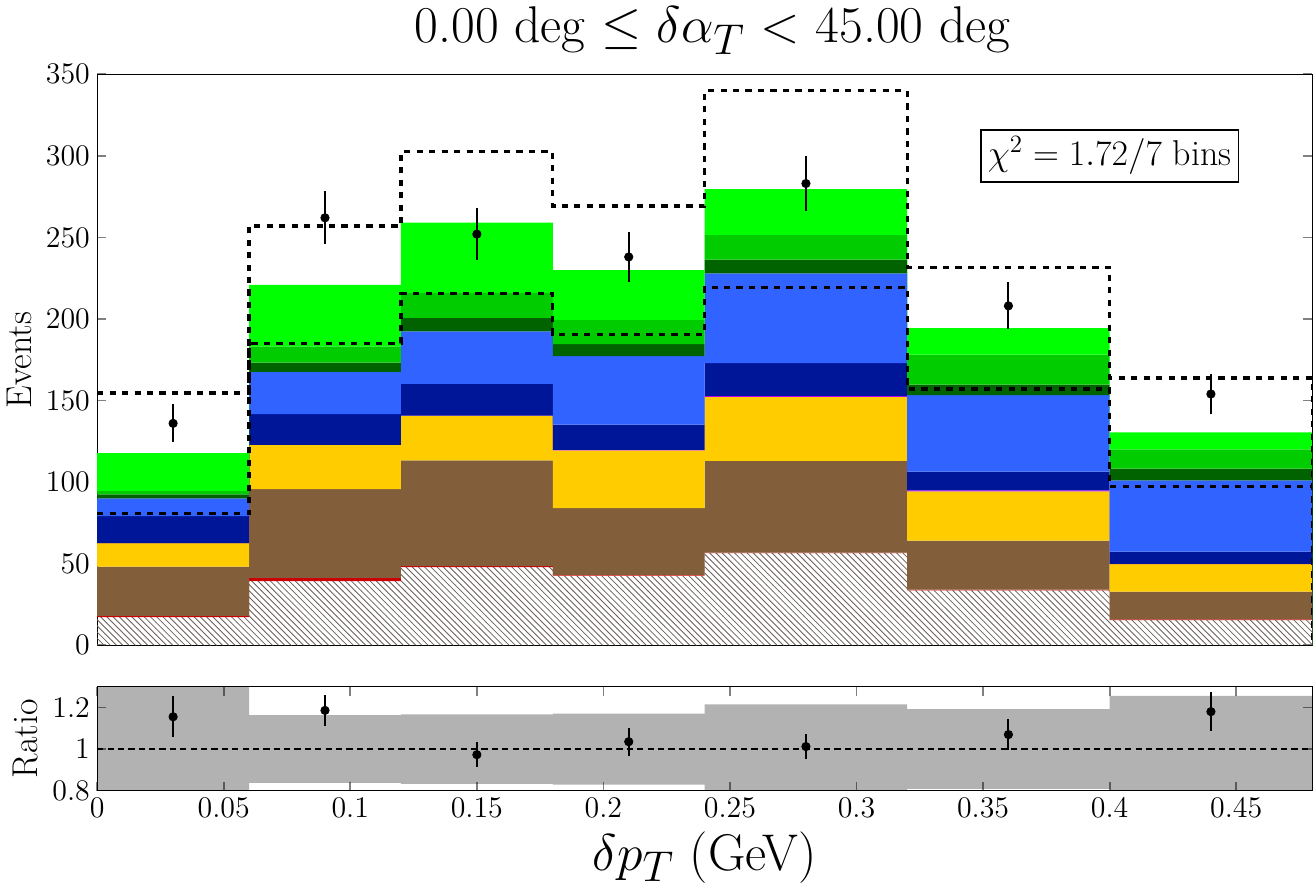}
\hfill
\includegraphics[page=19, width=0.49\textwidth]{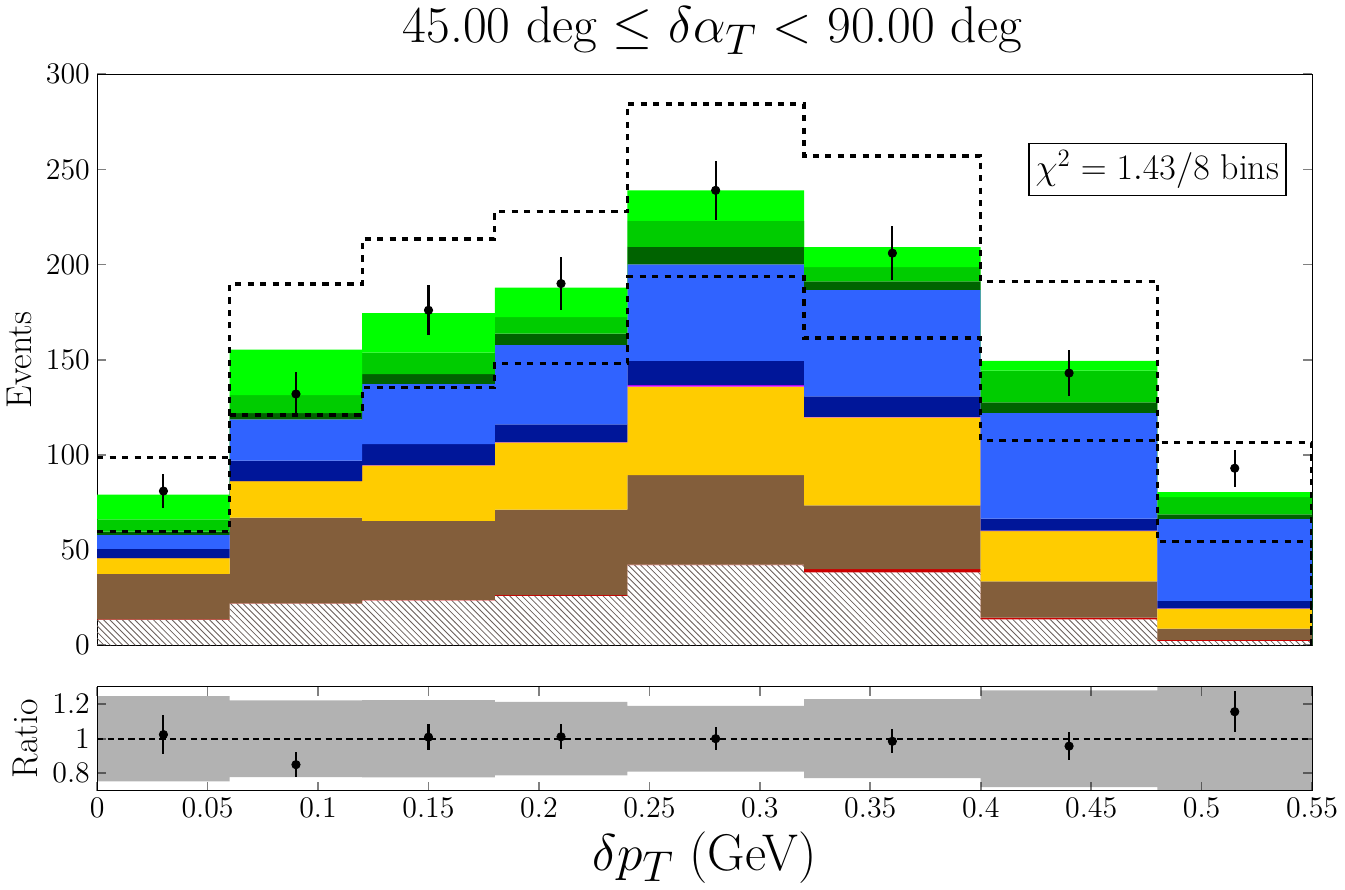}
\vspace{0.2cm}

\includegraphics[page=20, width=0.49\textwidth]{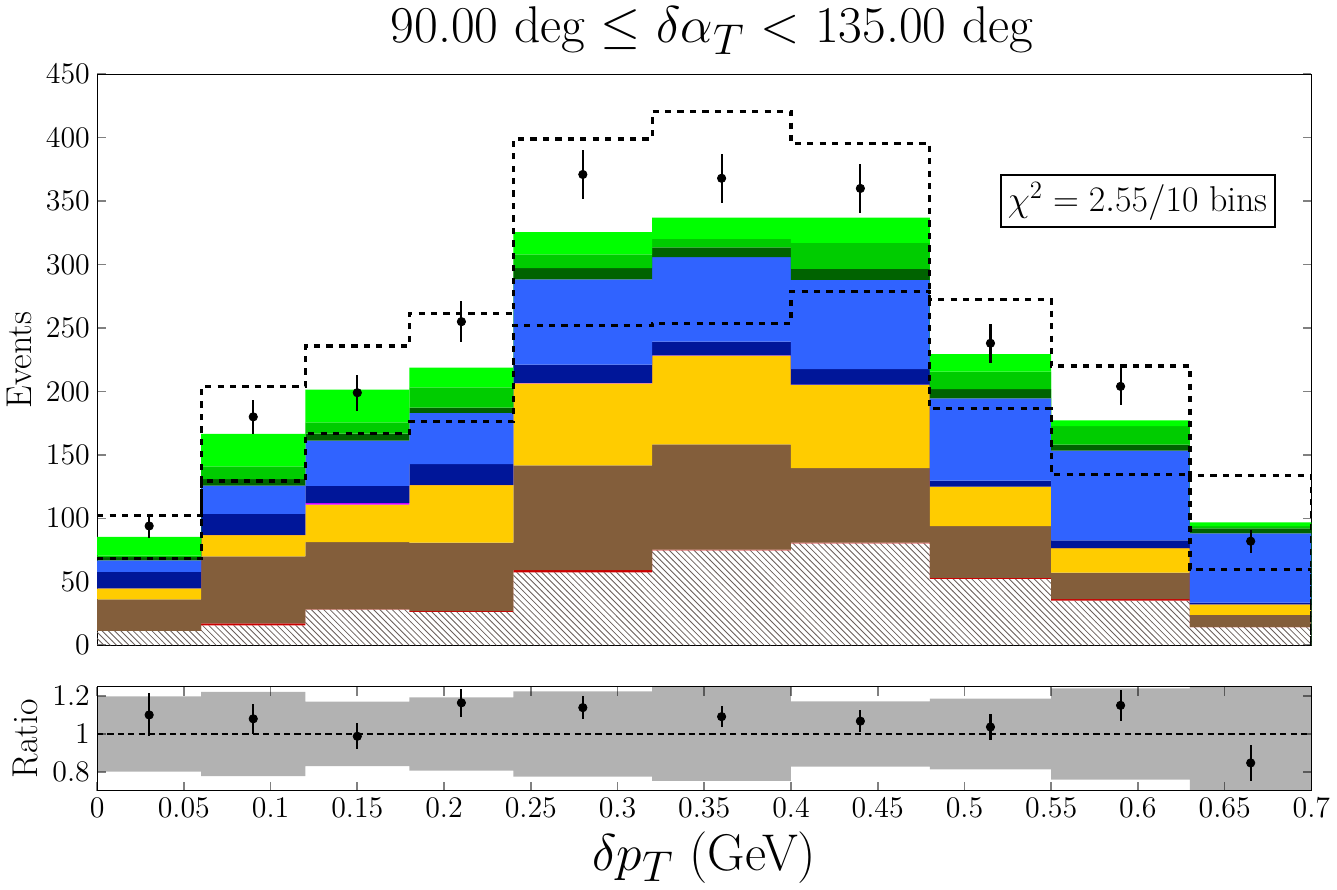}
\hfill
\includegraphics[page=21, width=0.49\textwidth]{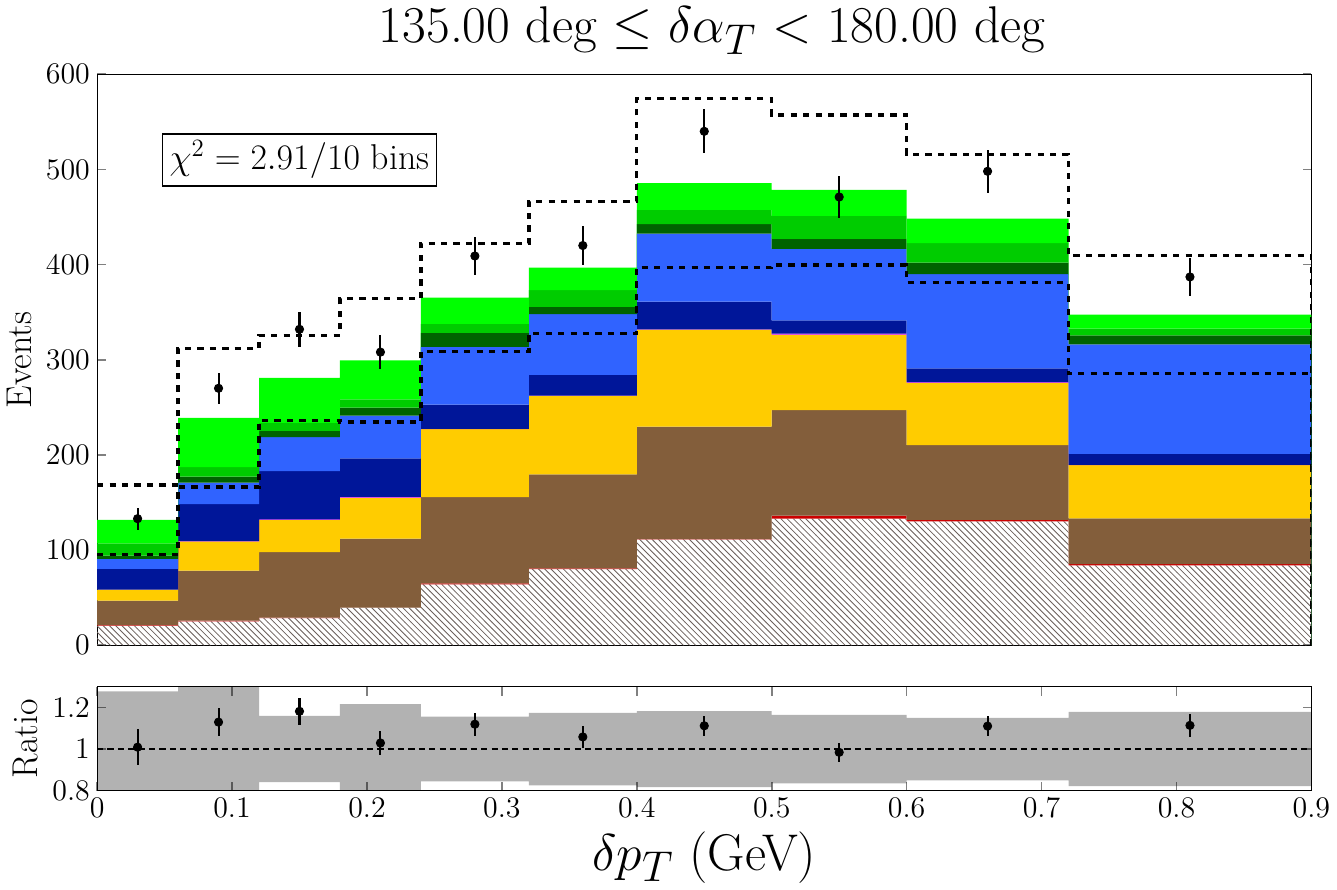}
\vspace{0.2cm}

\includegraphics[page=22, width=0.35\textwidth]{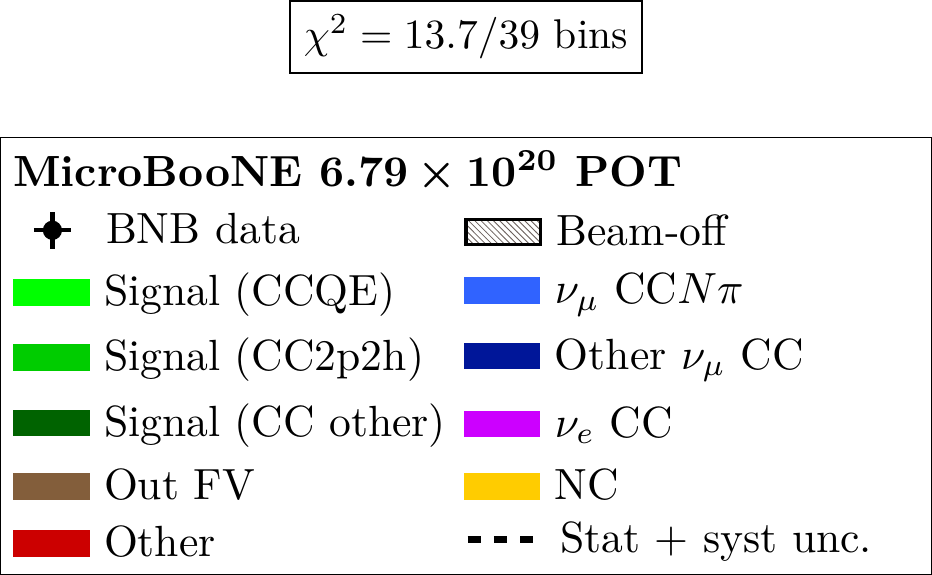}%
\hfill

\caption{Reconstructed event distributions for block \#3
$(\delta \alpha_T, \delta p_T)$. The overall $\chi^2$
value includes contributions from four $\delta p_T$ overflow bins that are not
plotted.}
\label{fig:DataSBMCblock3}
\end{figure*}

\begin{figure*}
\centering
\includegraphics[page=23, width=0.49\textwidth]{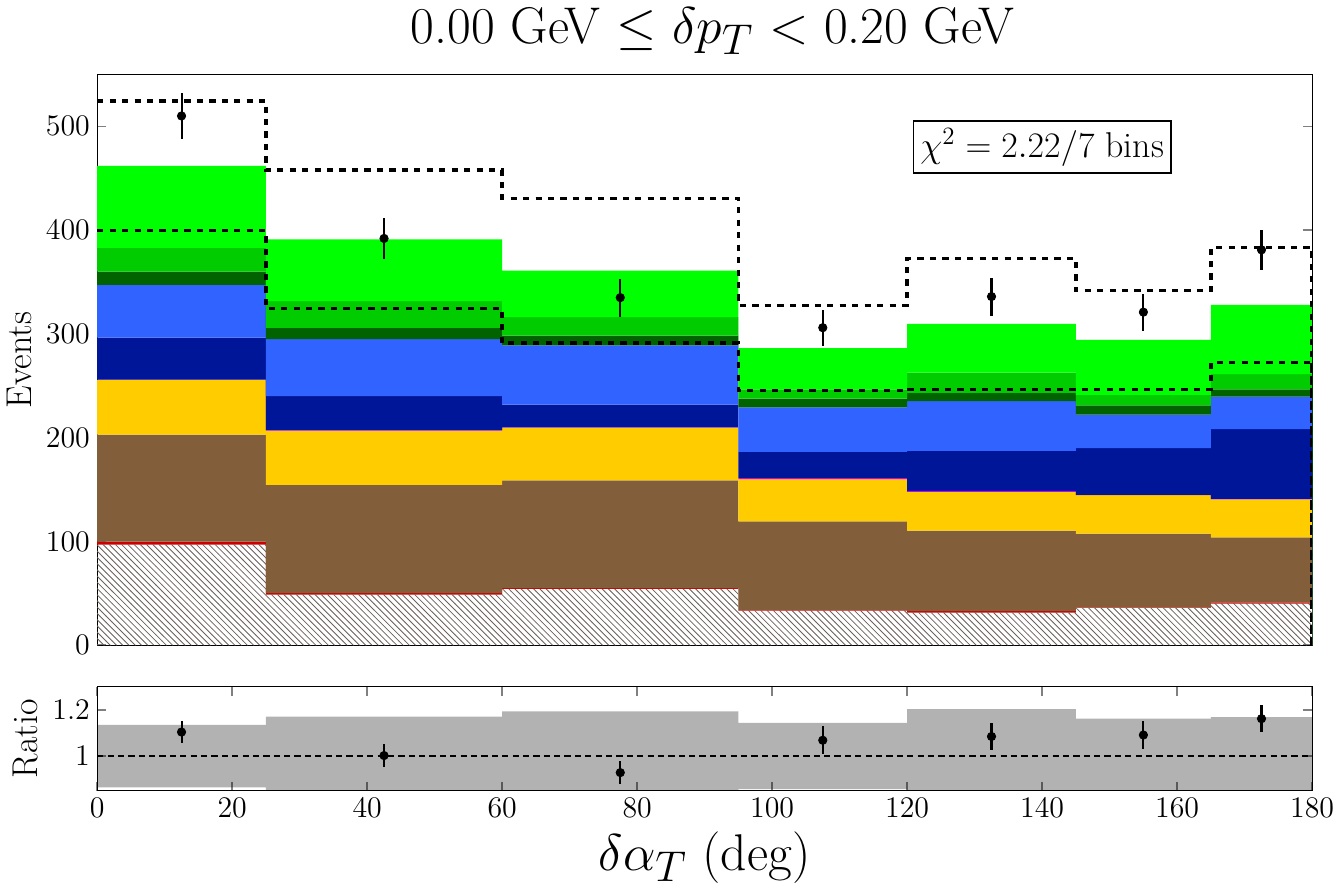}
\hfill
\includegraphics[page=24, width=0.49\textwidth]{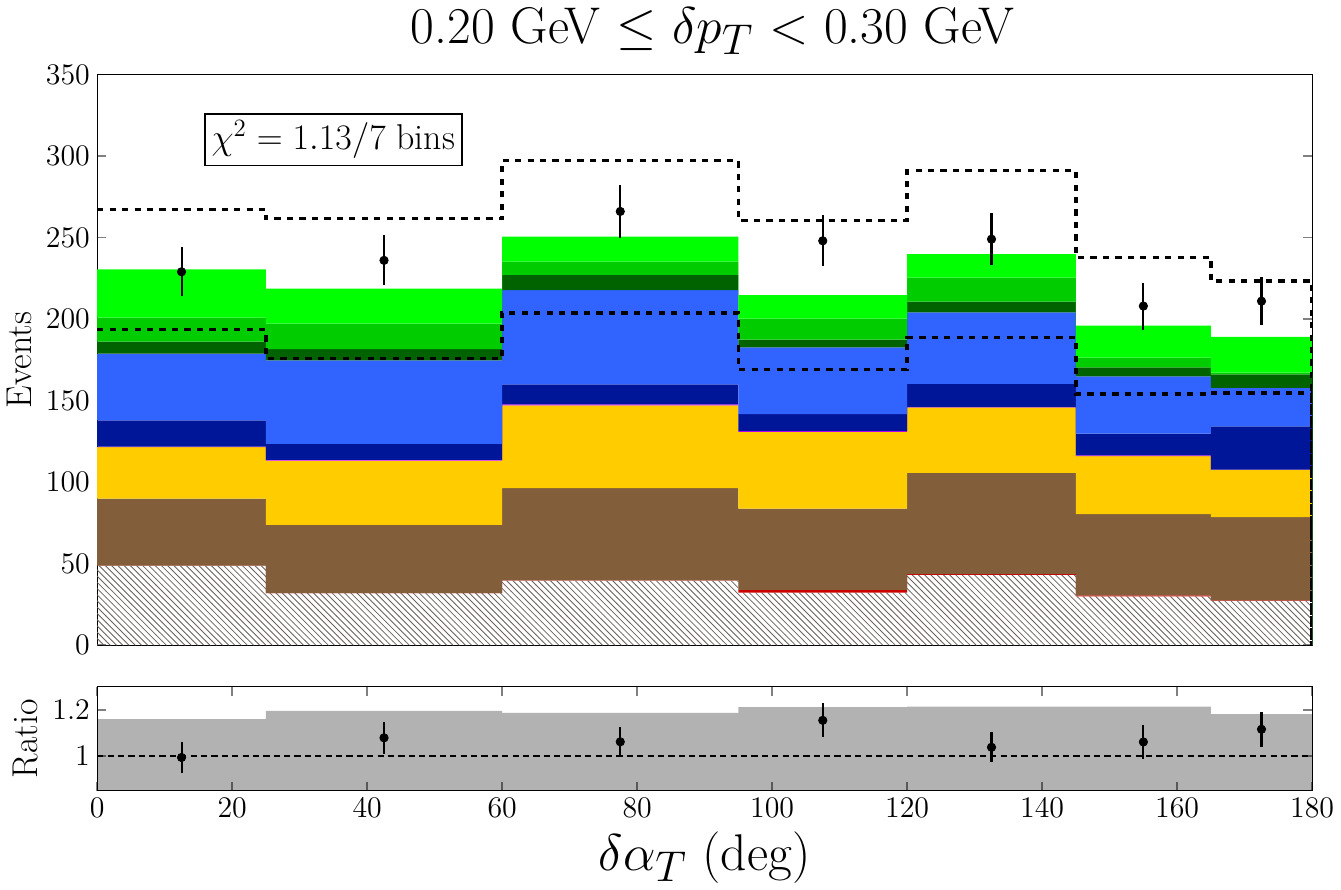}
\vspace{0.2cm}

\includegraphics[page=25, width=0.49\textwidth]{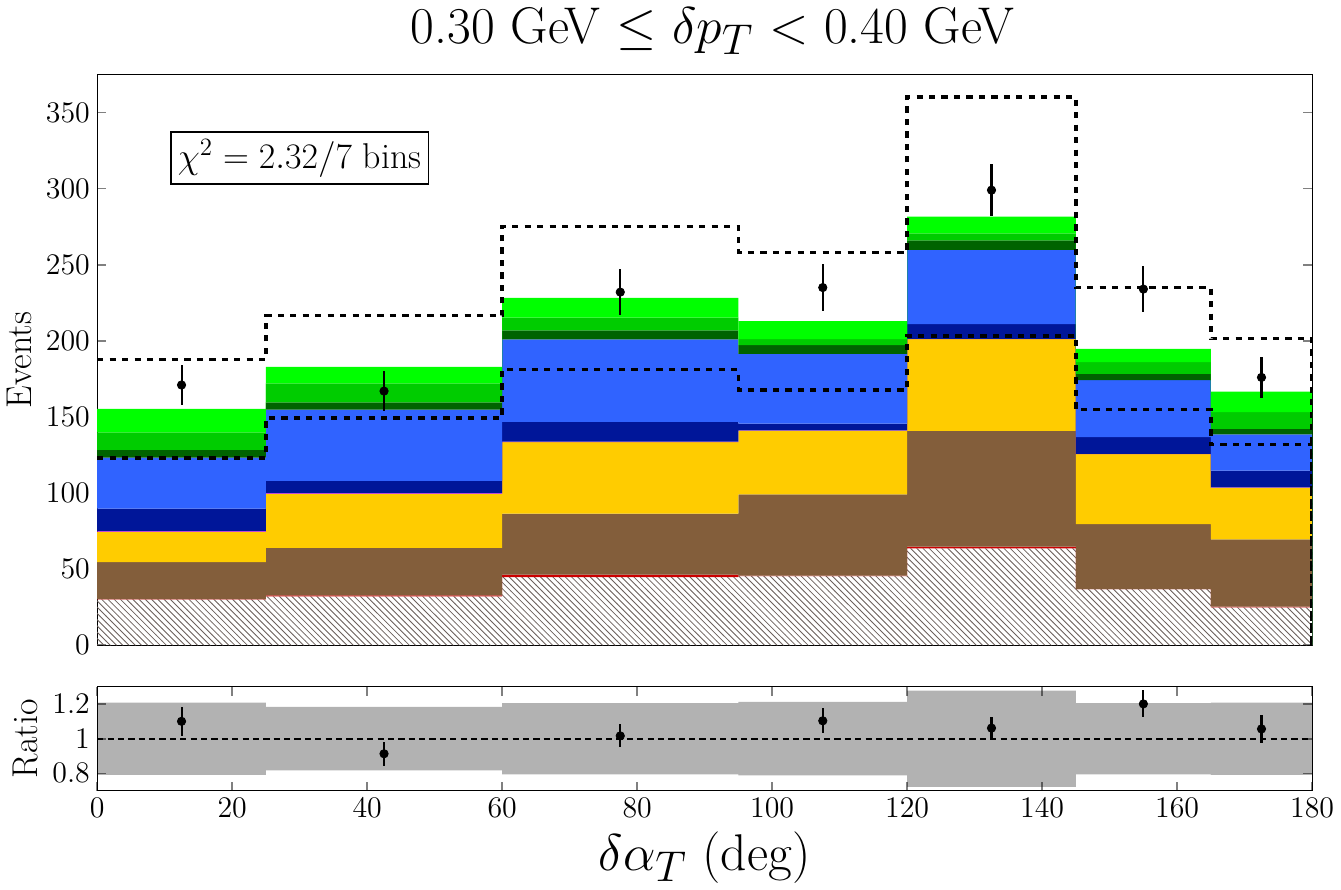}
\hfill
\includegraphics[page=26, width=0.49\textwidth]{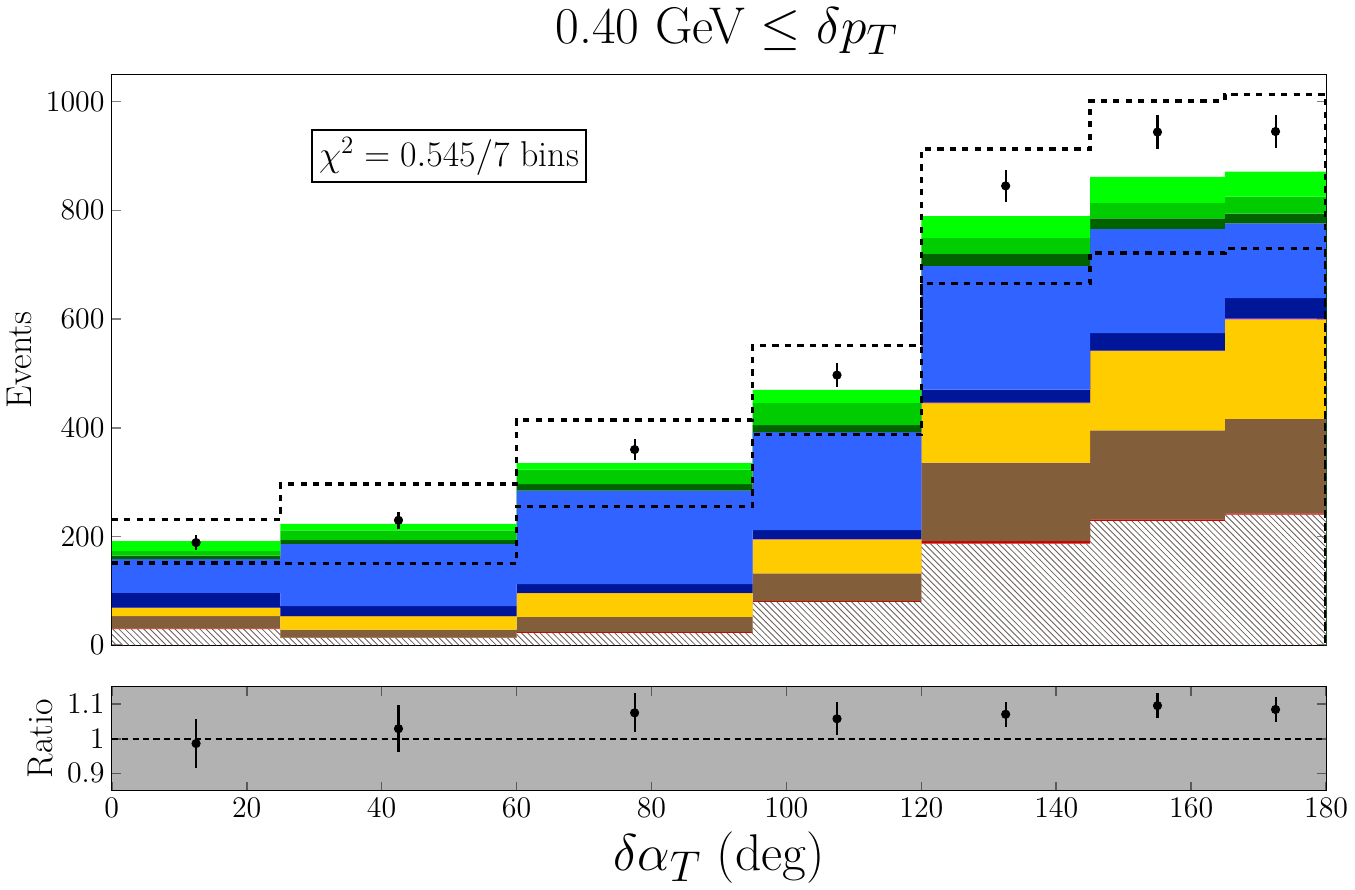}
\vspace{0.2cm}

\includegraphics[page=27, width=0.35\textwidth]{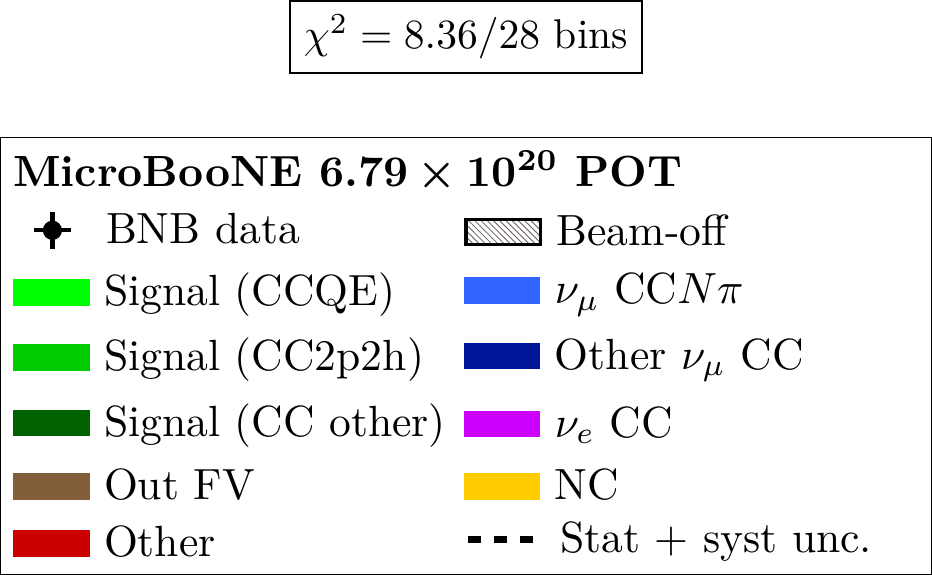}%
\hfill

\caption{Reconstructed event distributions for block \#4
$(\delta p_T, \delta \alpha_T)$.}
\label{fig:DataSBMCblock4}
\end{figure*}

\begin{figure*}
\centering
\includegraphics[page=29, width=0.49\textwidth]{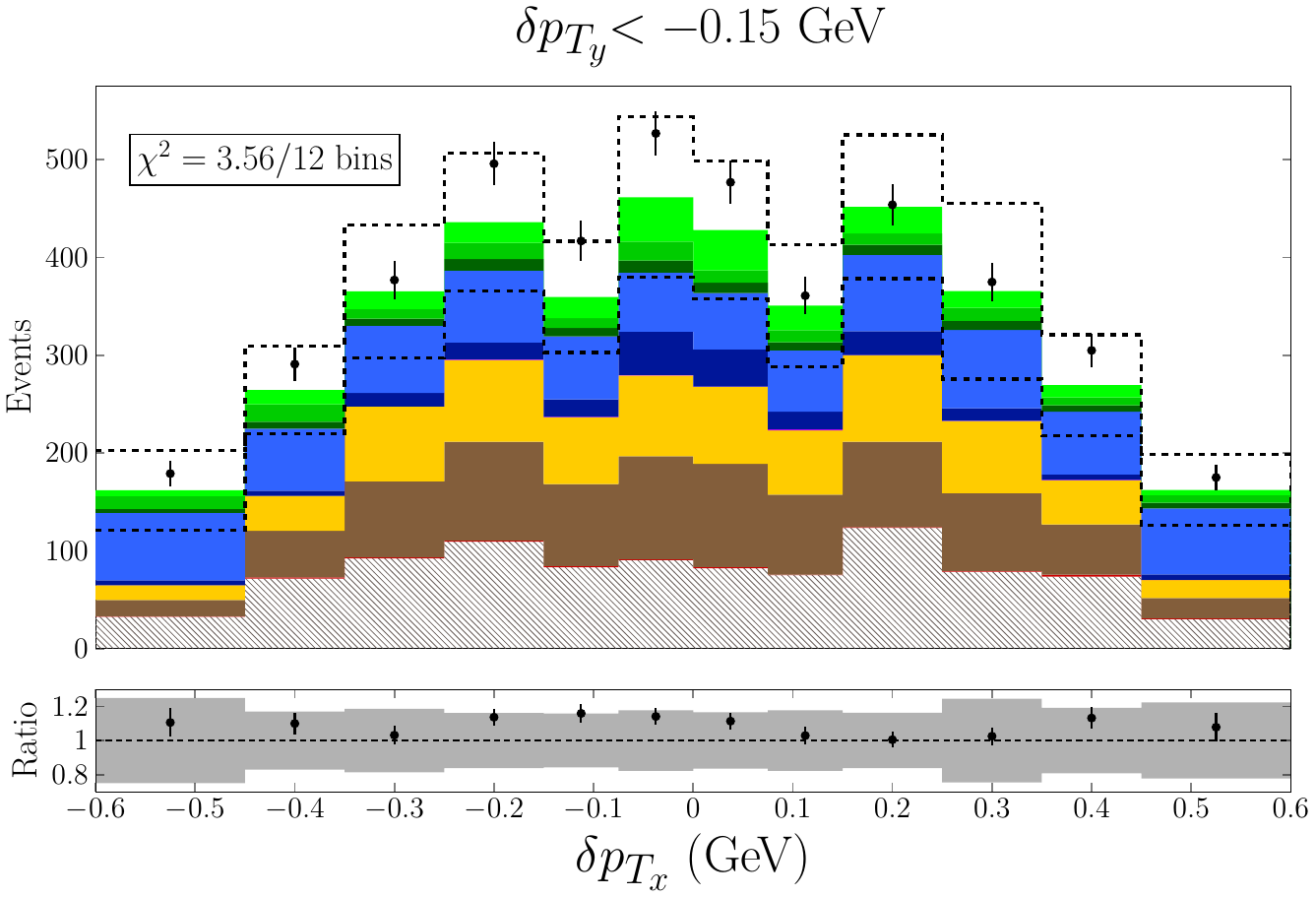}
\hfill
\includegraphics[page=30, width=0.49\textwidth]{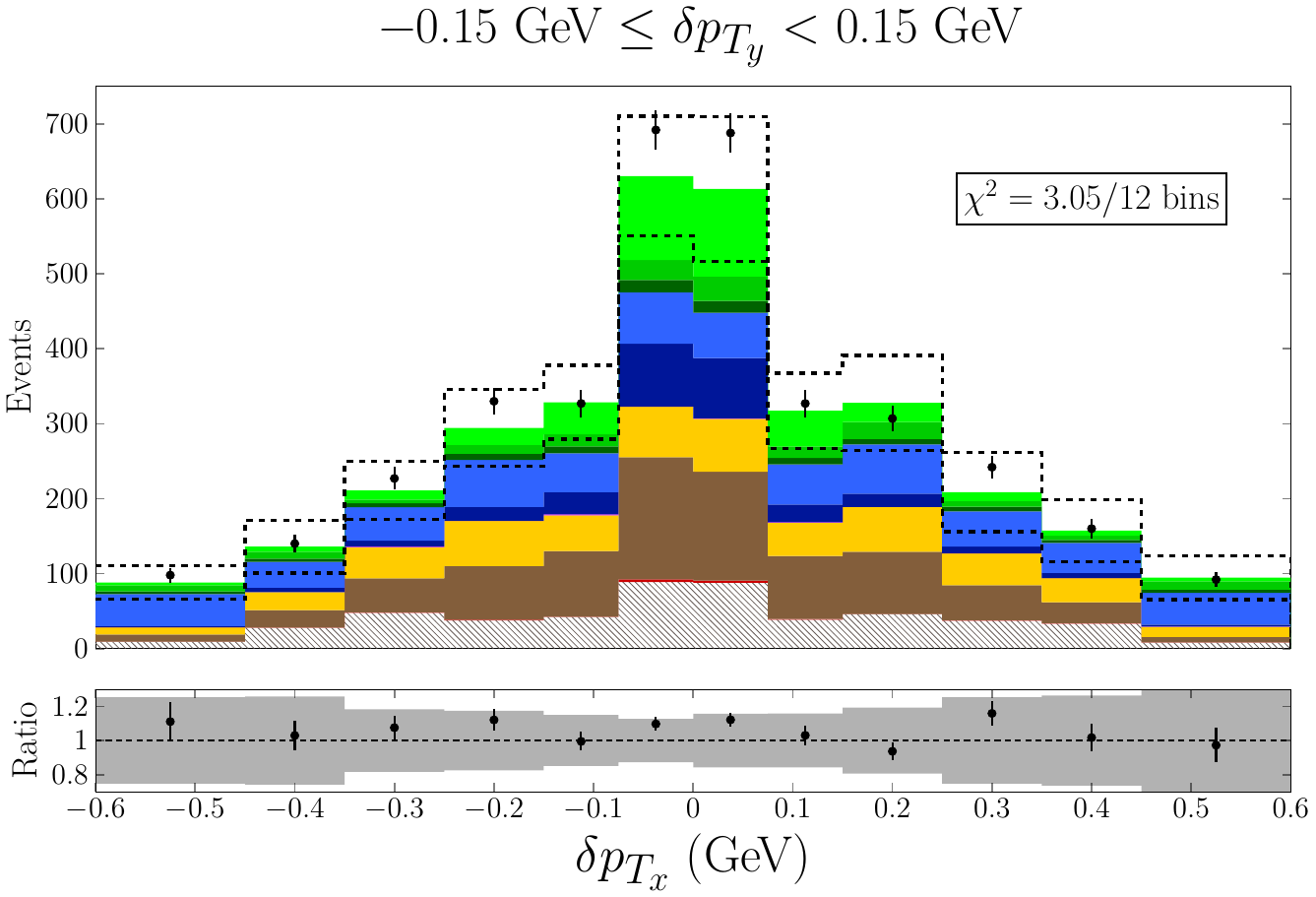}
\vspace{0.2cm}

\includegraphics[page=31, width=0.49\textwidth]{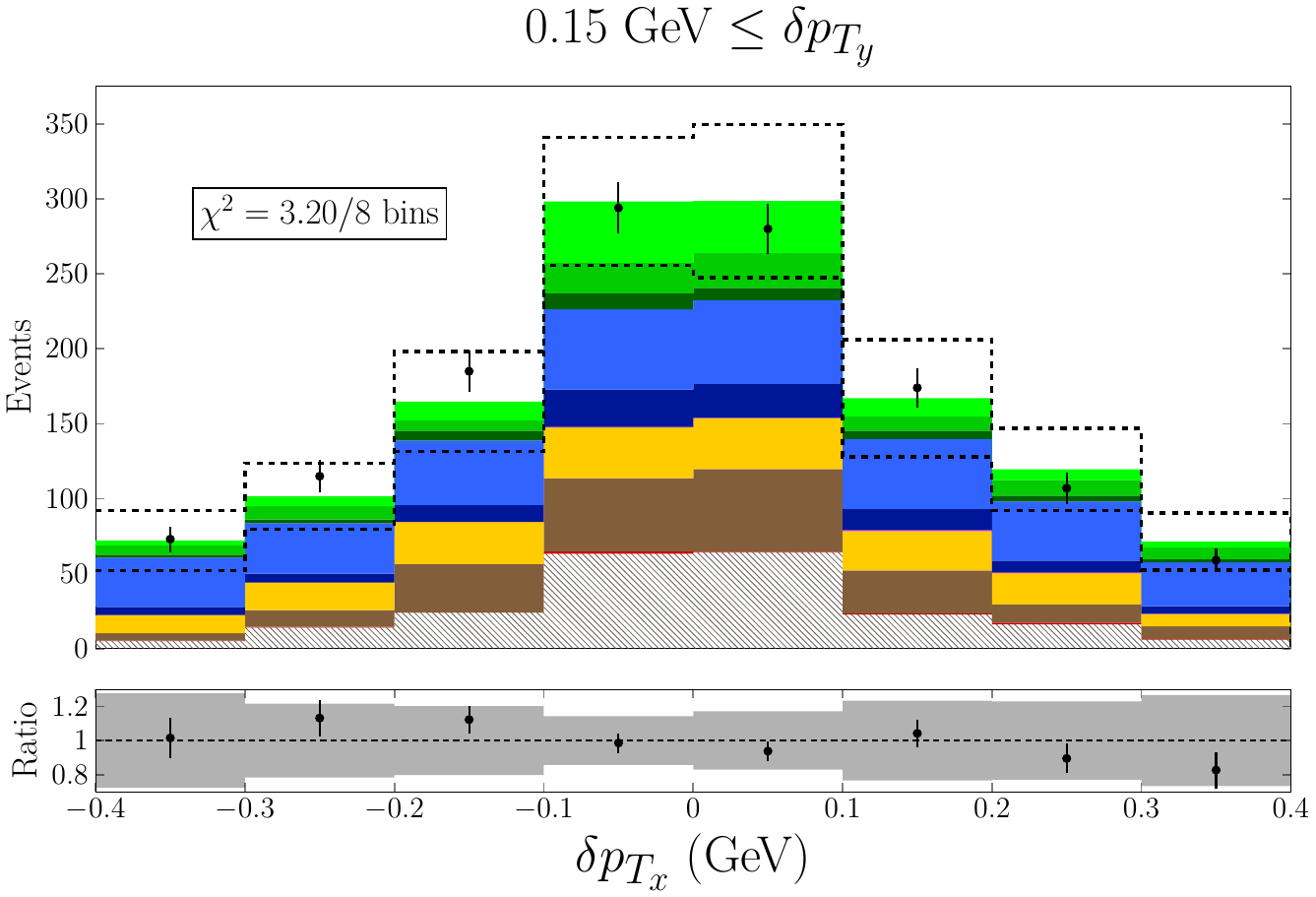}
\hfill
\belowbaseline[-0.23\textheight]{
  \includegraphics[page=32, width=0.35\textwidth]{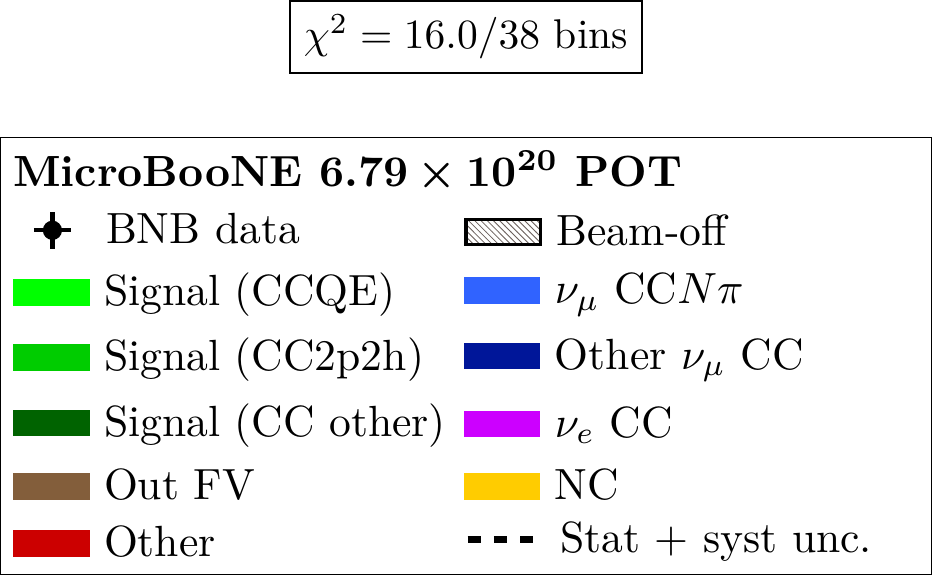}%
}
\hfill

\caption{Reconstructed event distributions for block \#6
$(\delta p_{T_y}, \delta p_{T_x})$. The overall $\chi^2$
value includes contributions from three underflow and three overflow $\delta
p_{T_x}$ bins that are not plotted.}
\label{fig:DataSBMCblock6}
\end{figure*}

\begin{figure*}
\centering
\includegraphics[page=35, width=0.49\textwidth]{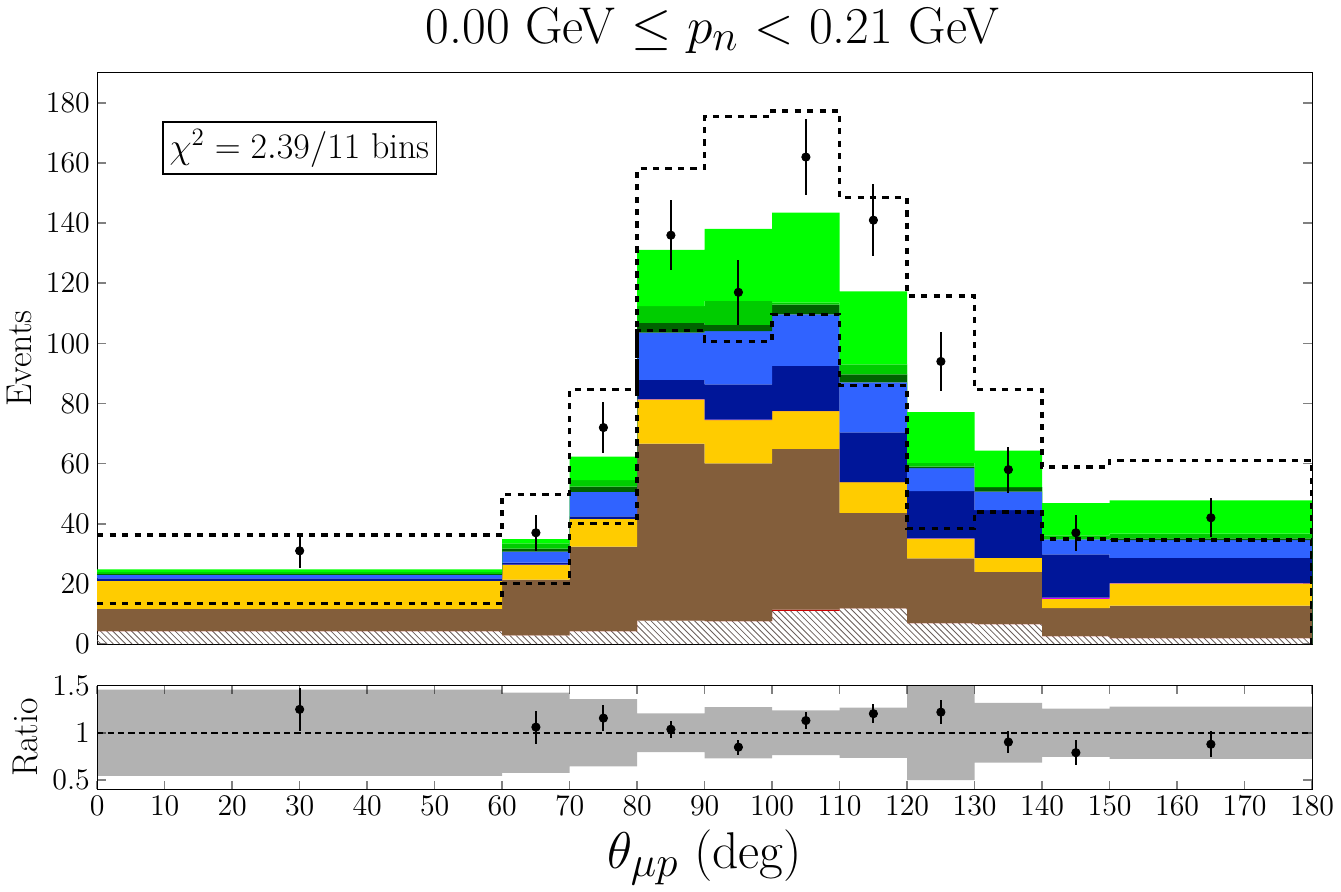}
\hfill
\includegraphics[page=36, width=0.49\textwidth]{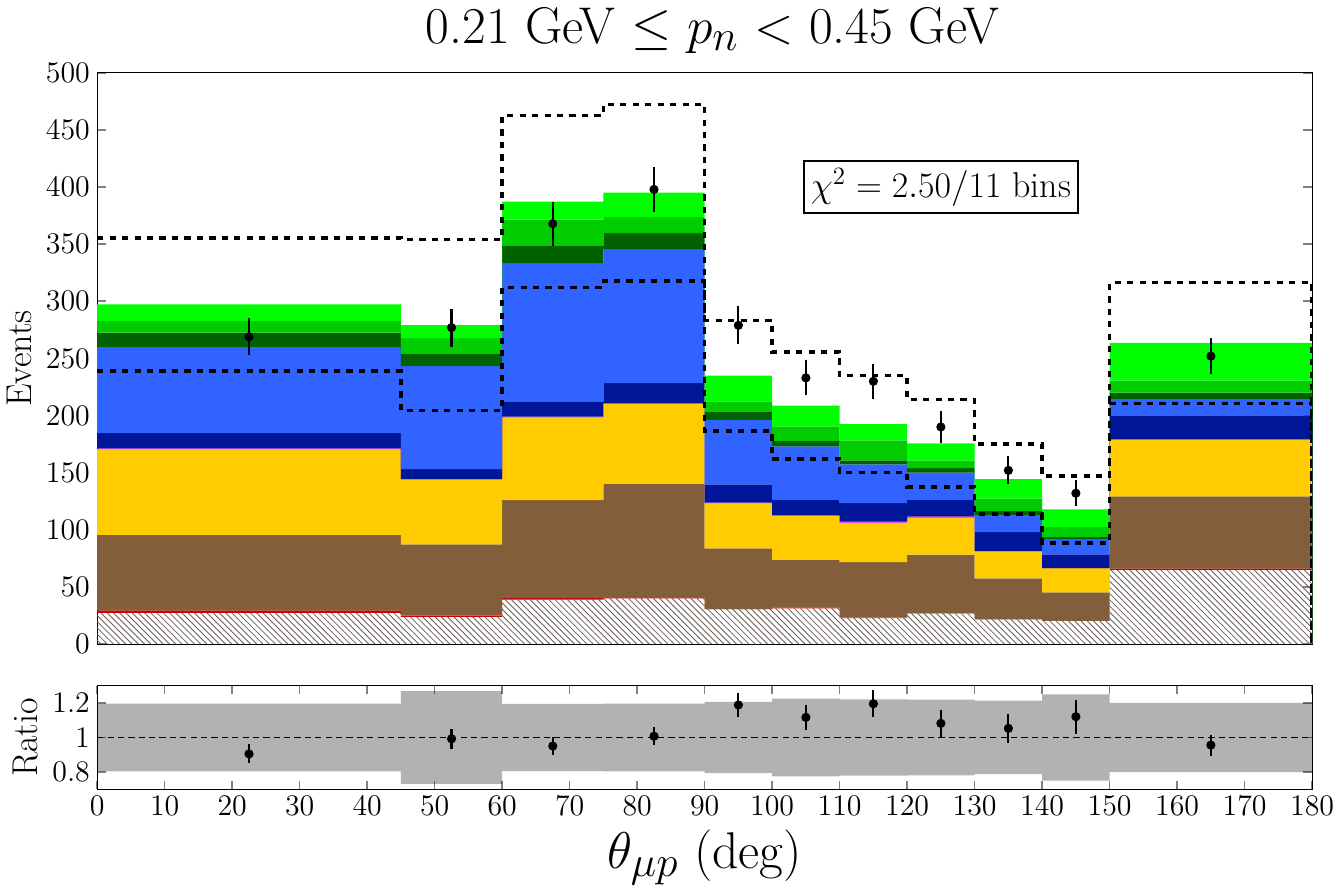}
\vspace{0.2cm}

\includegraphics[page=37, width=0.49\textwidth]{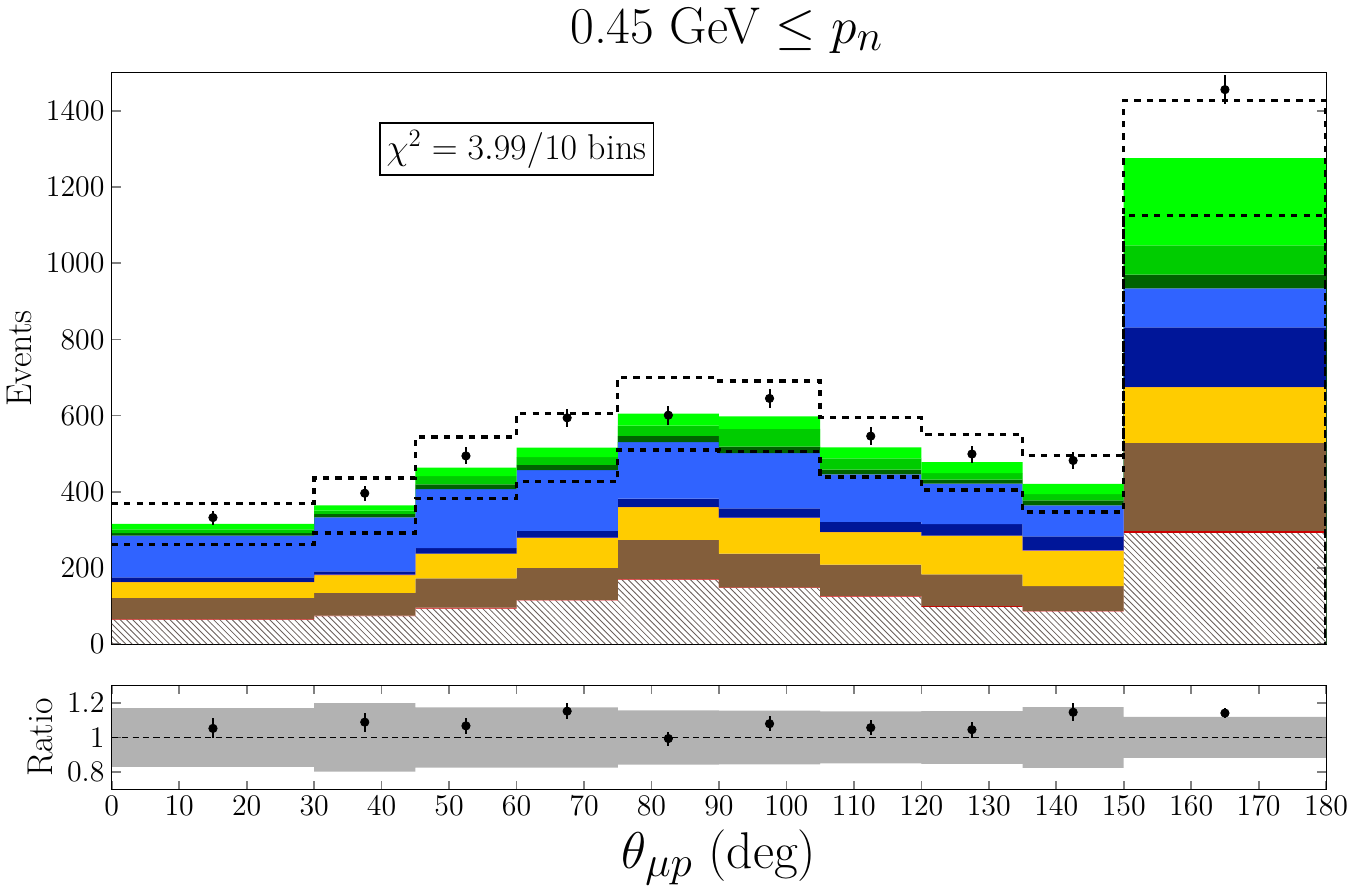}
\hfill
\belowbaseline[-0.23\textheight]{
  \includegraphics[page=38, width=0.35\textwidth]{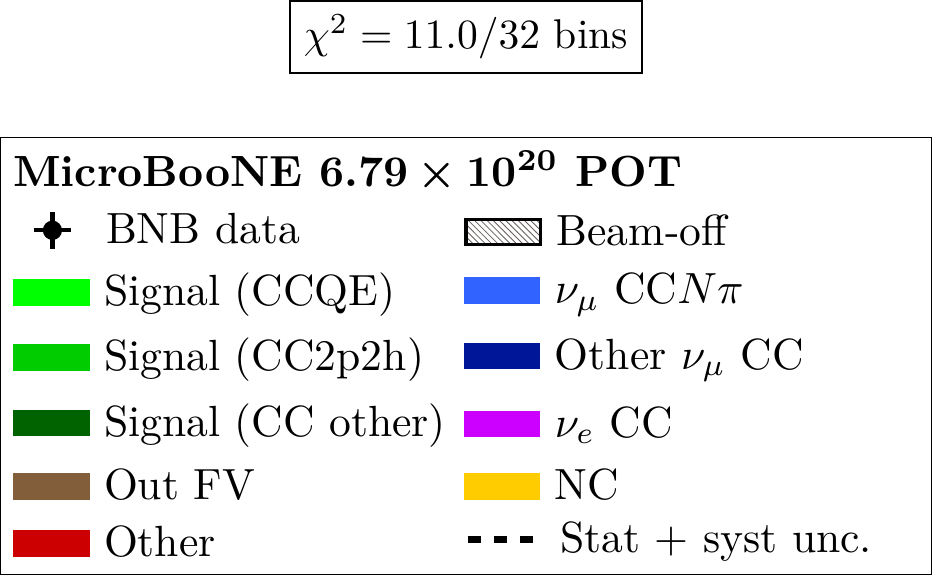}%
}
\hfill

\caption{Reconstructed event distributions for block \#9
$(p_n, \theta_{\mu p})$.}
\label{fig:DataSBMCblock9}
\end{figure*}

\begin{figure*}
\centering
\includegraphics[page=39, width=0.49\textwidth]{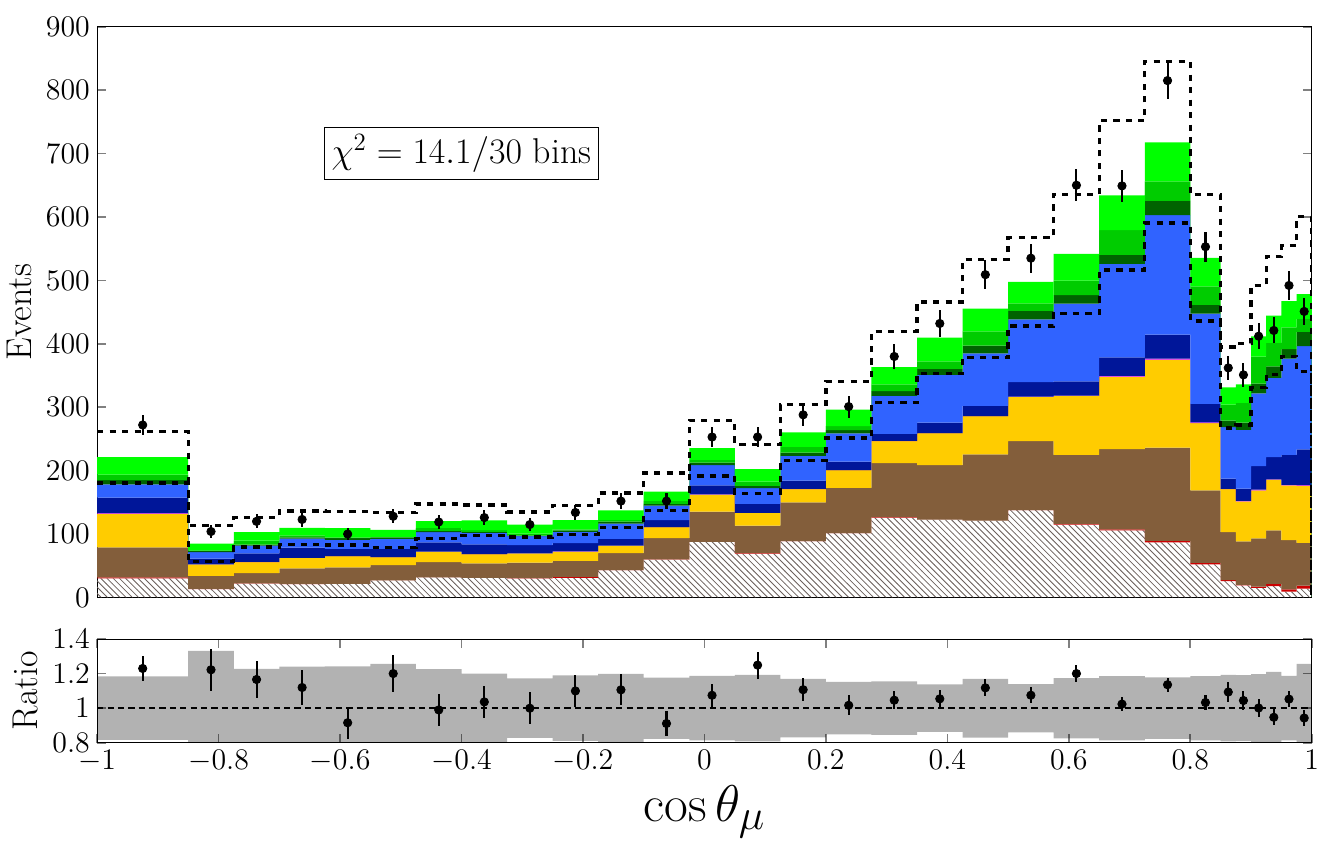}
\hfill
\includegraphics[page=40, width=0.49\textwidth]{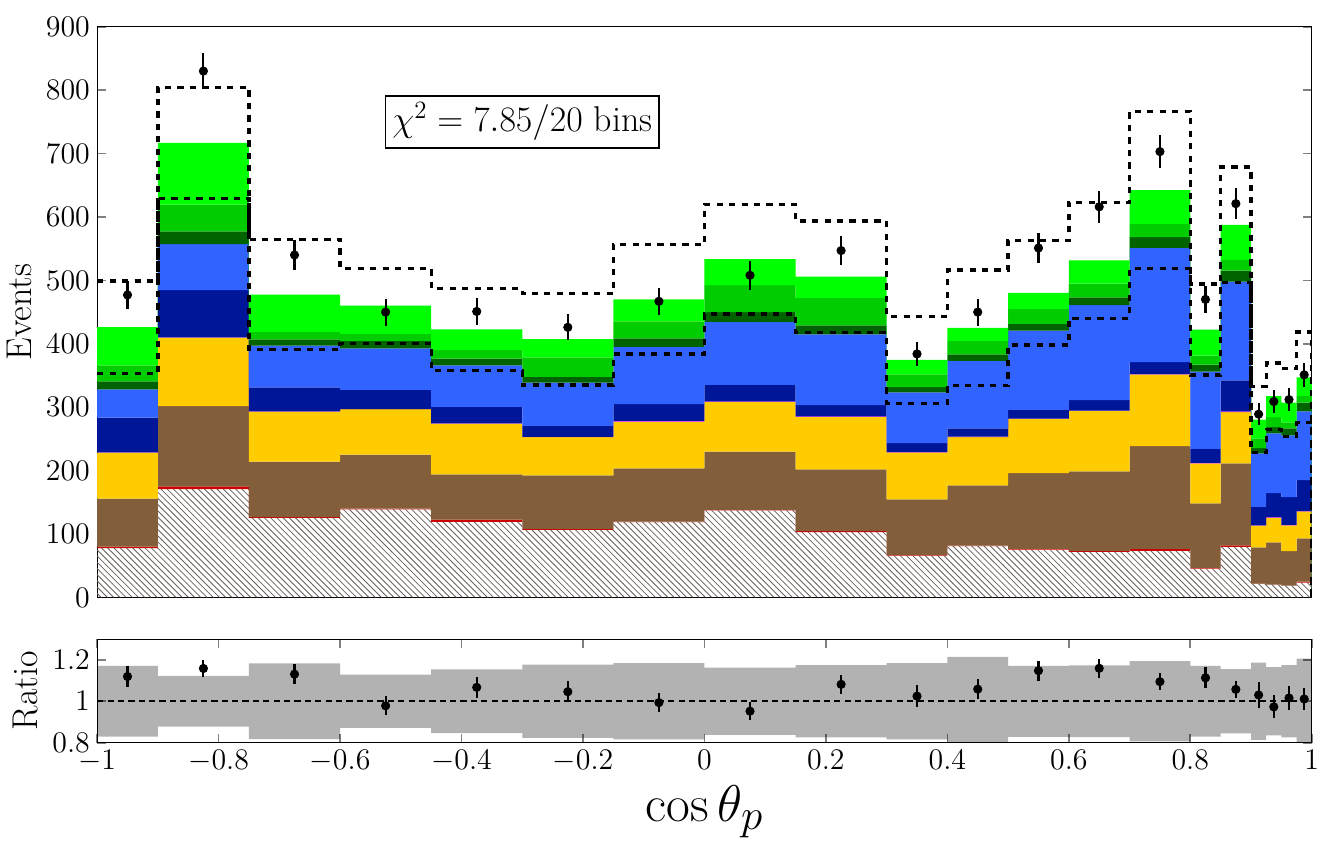}
\vspace{0.2cm}

\includegraphics[page=41, width=0.49\textwidth]{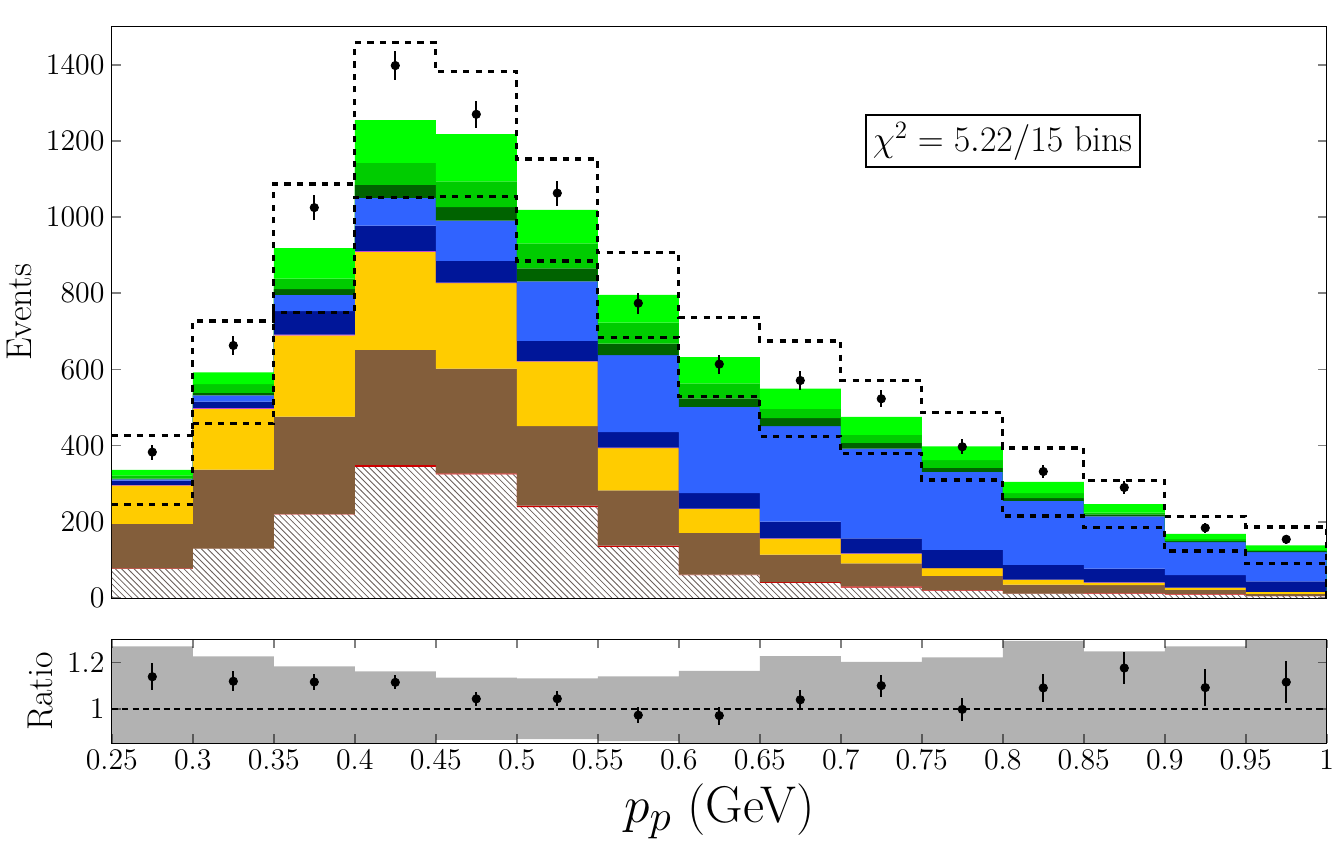}
\hfill
\includegraphics[page=42, width=0.49\textwidth]{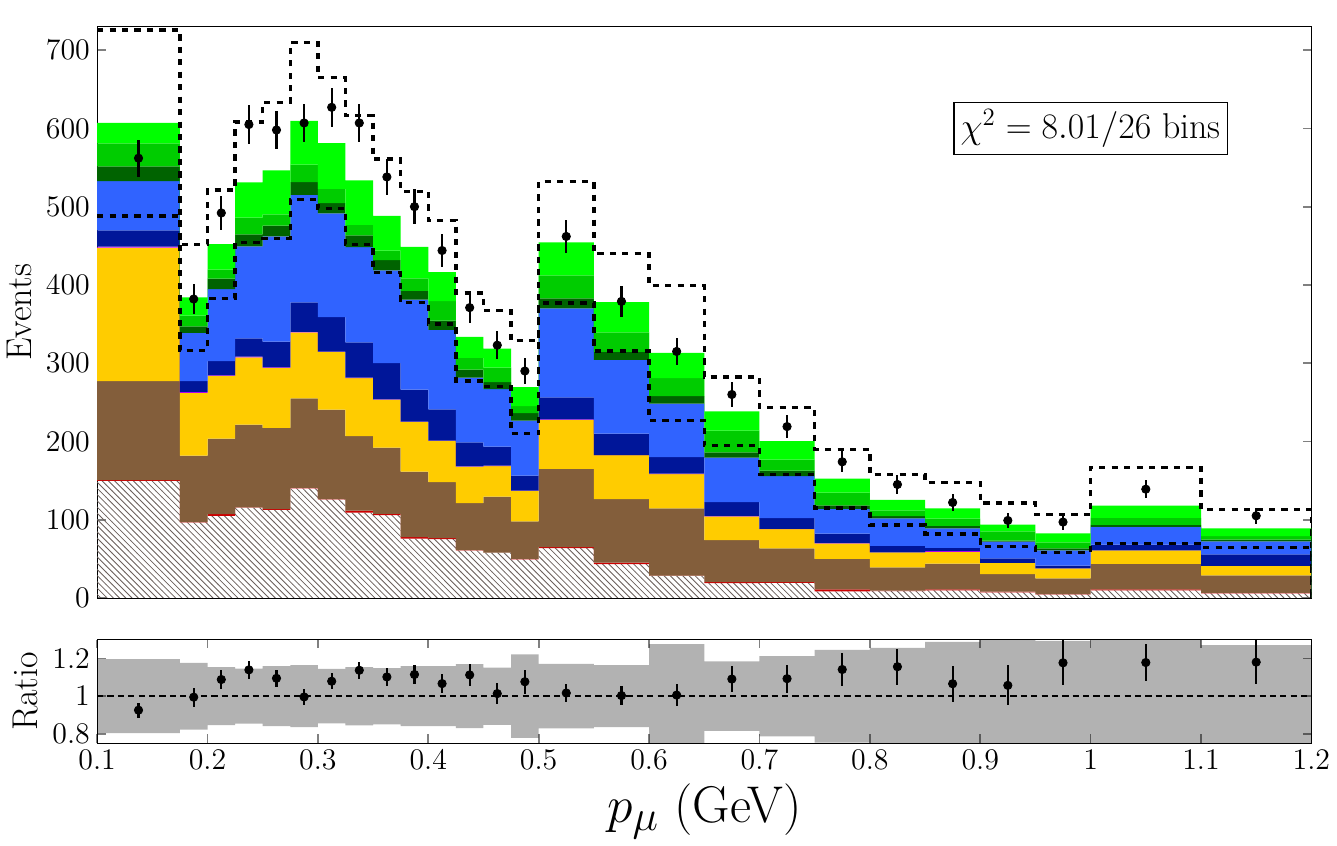}
\vspace{0.2cm}

\includegraphics[page=17, width=0.35\textwidth]{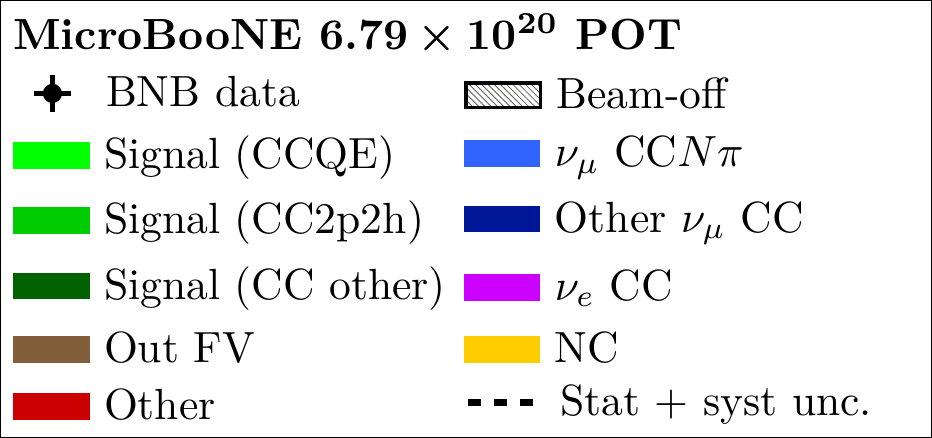}%
\hfill

\caption{Reconstructed event distributions for block~\#10
($\cos\theta_\mu$, upper left),
block~\#11 ($\cos\theta_p$, upper right),
block~\#12 ($p_p$, lower left), and block~\#13 ($p_\mu$, lower right).}
\label{fig:DataSBMCMultipleBlocks2}
\end{figure*}

\clearpage
\section{Response matrix}
\label{sec:resp_mat}

\begin{figure}[h]
\centering
\includegraphics[width=0.49\textwidth]{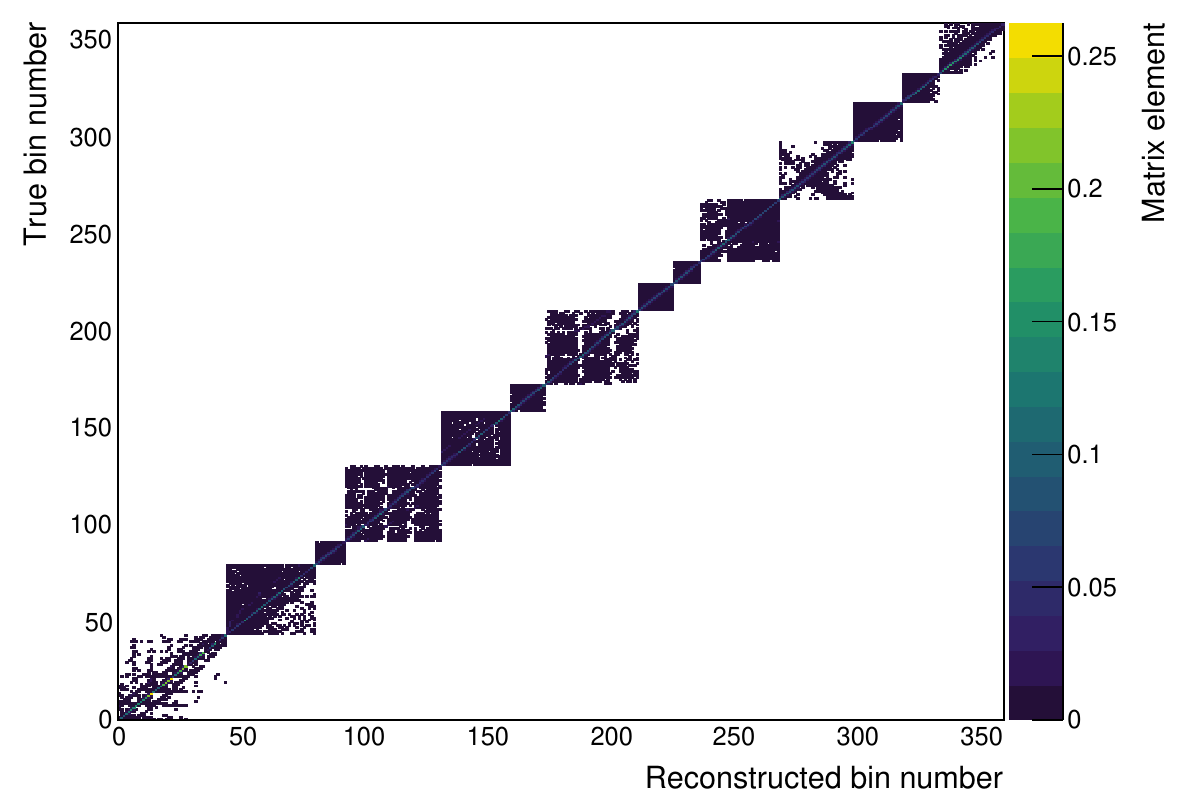}
\caption{The detector response matrix $\ResponseMatrix$.}
\label{fig:response_matrix}
\end{figure}

The \textit{response matrix} $\ResponseMatrix$ defined
in Eq.~11 of the main text is employed in the cross-section unfolding procedure
to estimate both efficiency and bin migration corrections. The elements of the
matrix are tabulated in the text file
\texttt{mat\_table\_detector\_response.txt} using the reconstructed (true) bin
index $\recoBinIdx$ ($\trueBinIdx$) along the $x$ ($y$) axis. The file format
is identical to the one used to report the matrices in
Sec.~\ref{sec:basic_data_release}. Note that, like the additional smearing
matrix $\AddSmearMatrix$, the elements of $\ResponseMatrix$ are dimensionless.
A plot of the response matrix including all blocks for the full measurement is
given in Fig.~\ref{fig:response_matrix}.

\section{Data Comparisons to MicroBooNE Tune with Model Uncertainty}
\label{sec:tune_unc}

\begin{figure*}
\centering
\includegraphics[page=1, width=0.425\textwidth]{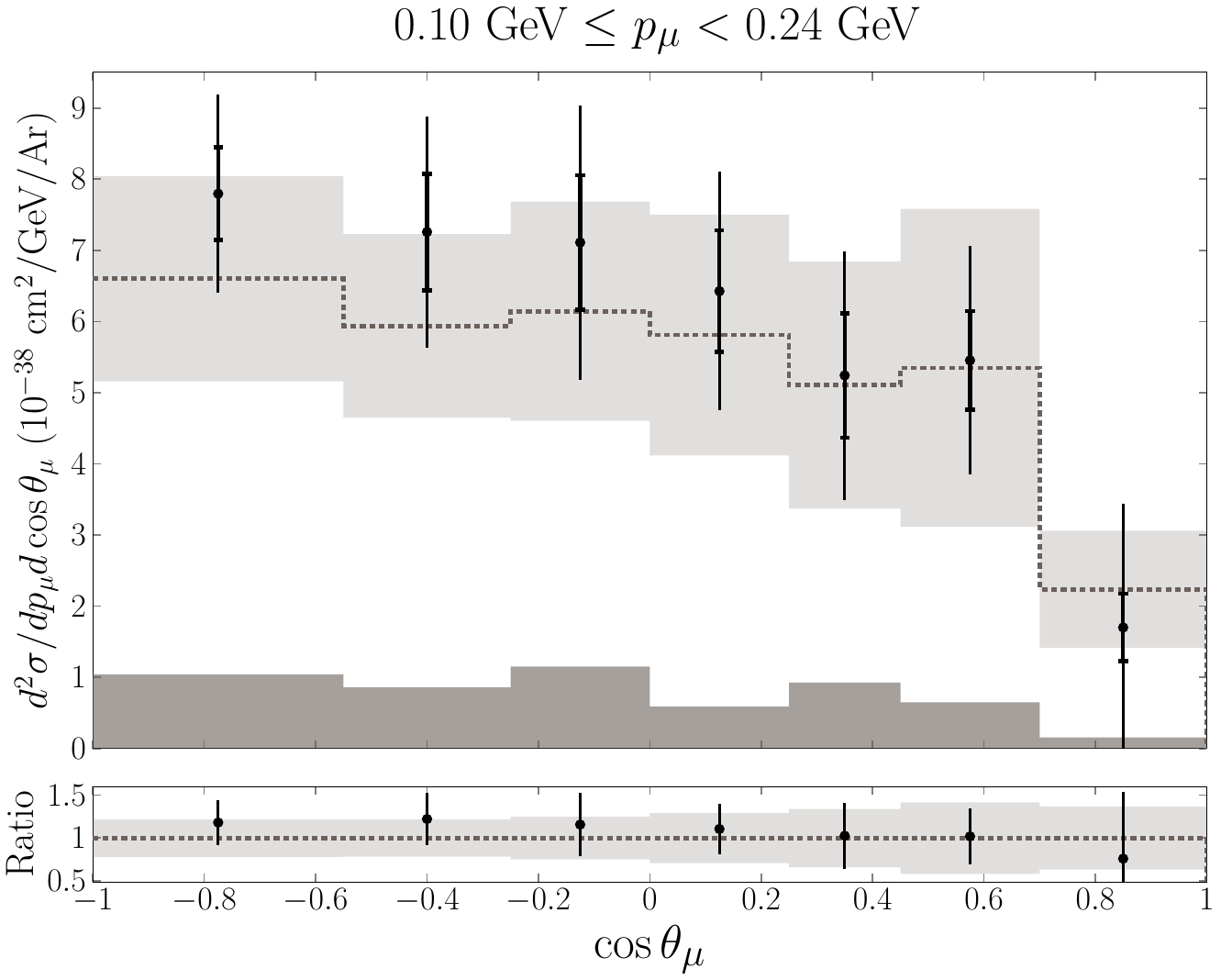}
\hfill
\includegraphics[page=2, width=0.425\textwidth]{figures/real_data_prd_unc.pdf}
\vspace{0.1cm}

\includegraphics[page=3, width=0.425\textwidth]{figures/real_data_prd_unc.pdf}
\hfill
\includegraphics[page=4, width=0.425\textwidth]{figures/real_data_prd_unc.pdf}
\vspace{0.1cm}

\includegraphics[page=5, width=0.425\textwidth]{figures/real_data_prd_unc.pdf}
\hfill
\includegraphics[page=6, width=0.425\textwidth]{figures/real_data_prd_unc.pdf}
\vspace{0.1cm}

\includegraphics[page=7, width=0.425\textwidth]{figures/real_data_prd_unc.pdf}
\hfill
\belowbaseline[-0.23\textheight]{
  \includegraphics[page=9, width=0.35\textwidth]{figures/real_data_prd_unc.pdf}%
}
\hfill

\caption{Measured differential cross sections for block \#0
($p_\mu, \cos\theta_\mu$).
Statistical (shape-only systematic) uncertainties are included in the inner
(outer) error bars. The dark (light) gray band shows the
remainder of the measurement uncertainty (theoretical uncertainty on the
prediction). Error bars in the lower panels show the full measurement
uncertainty.}

\label{fig:ResultsTuneUncblock0}
\end{figure*}

\begin{figure*}
\centering
\includegraphics[page=12, width=0.49\textwidth]{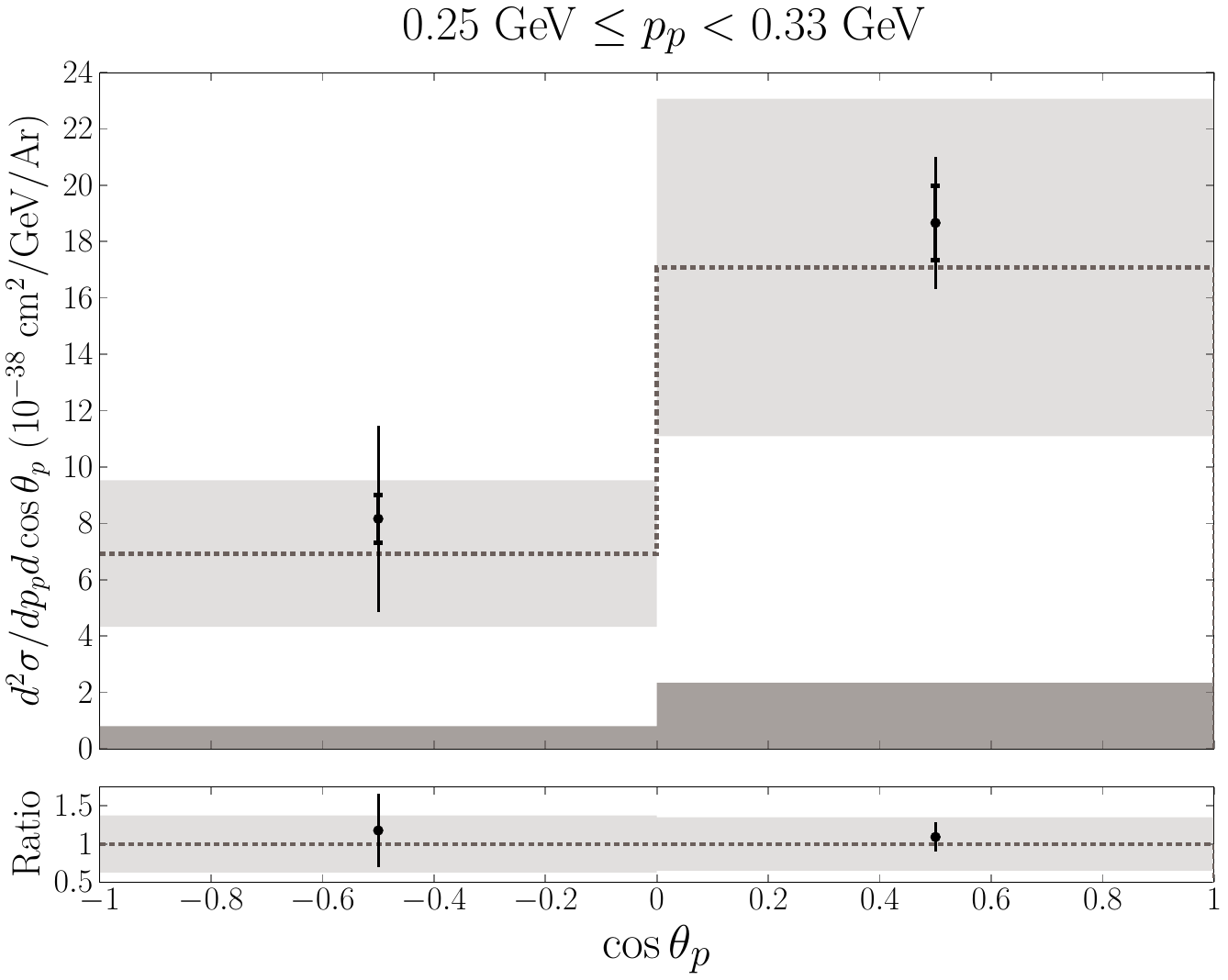}
\hfill
\includegraphics[page=13, width=0.49\textwidth]{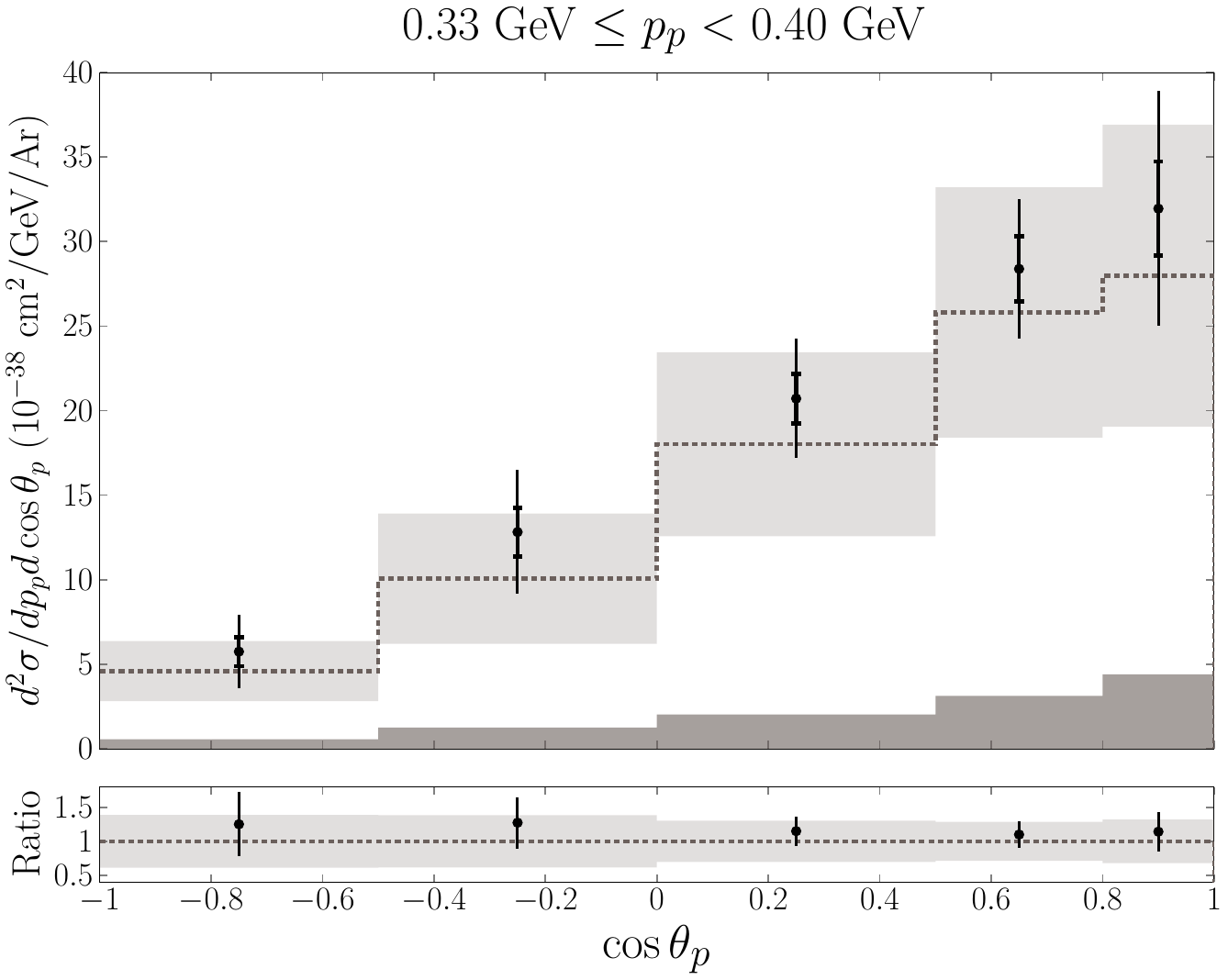}
\vspace{0.2cm}

\includegraphics[page=14, width=0.49\textwidth]{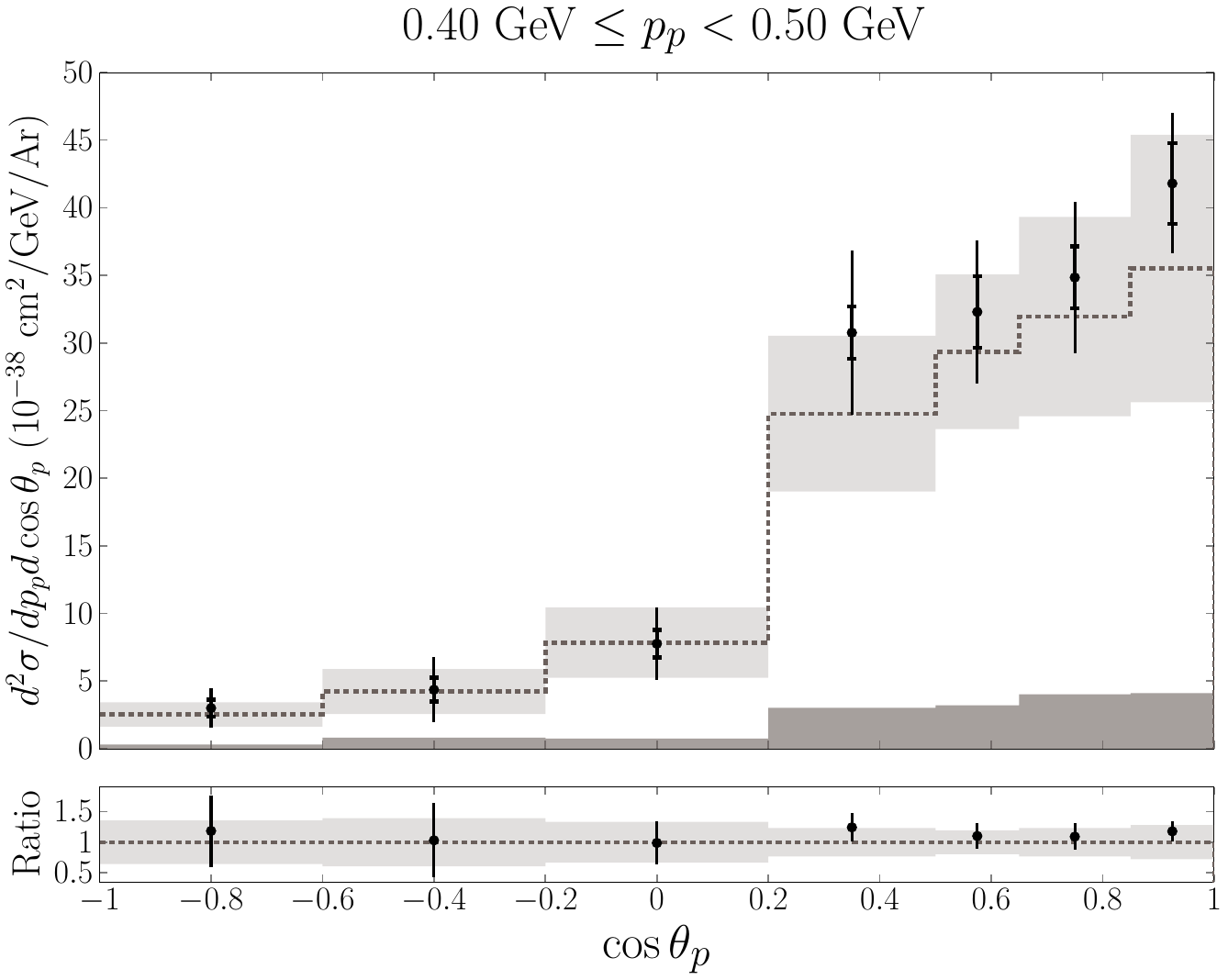}
\hfill
\includegraphics[page=15, width=0.49\textwidth]{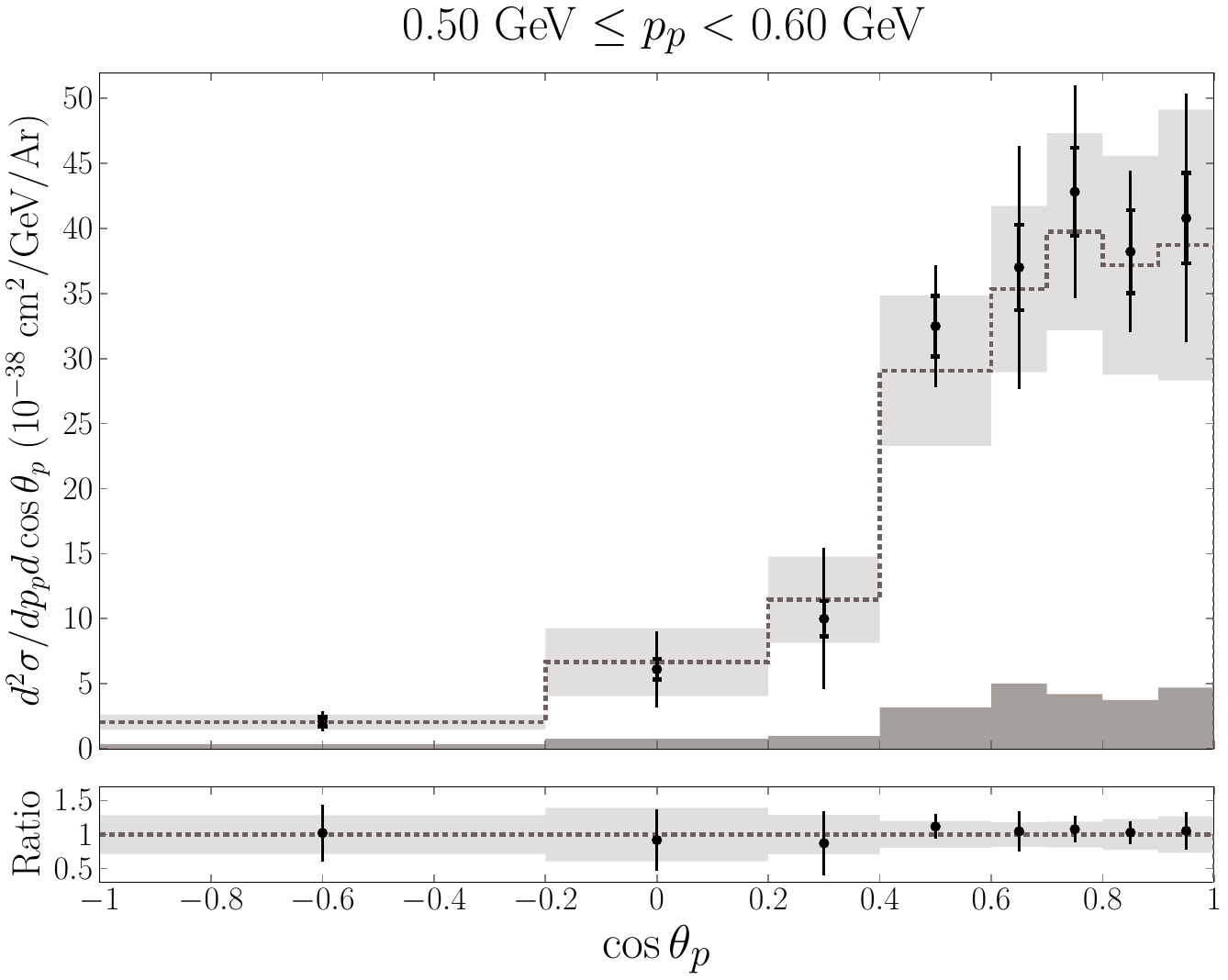}
\vspace{0.2cm}

\includegraphics[page=16, width=0.49\textwidth]{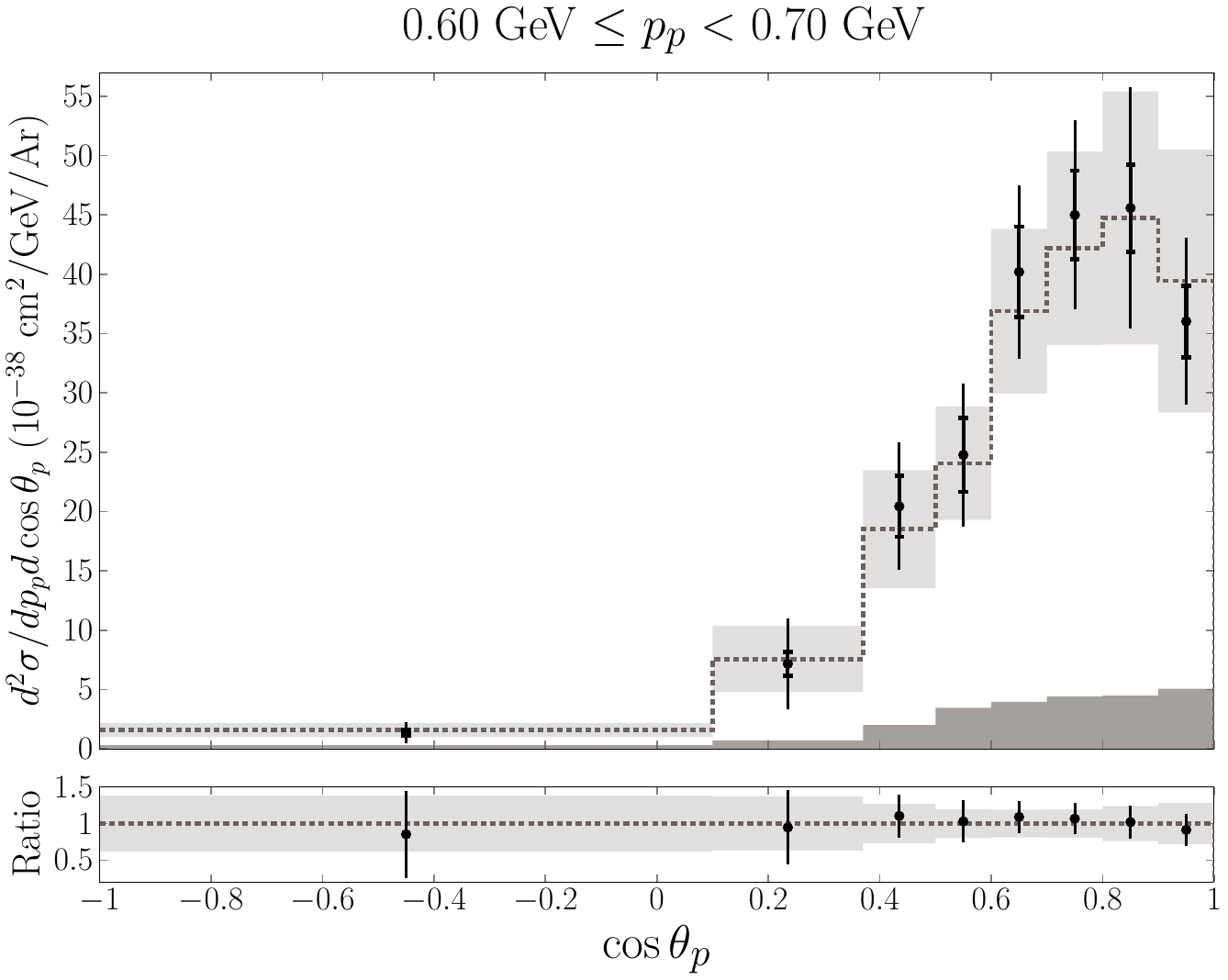}
\hfill
\includegraphics[page=17, width=0.49\textwidth]{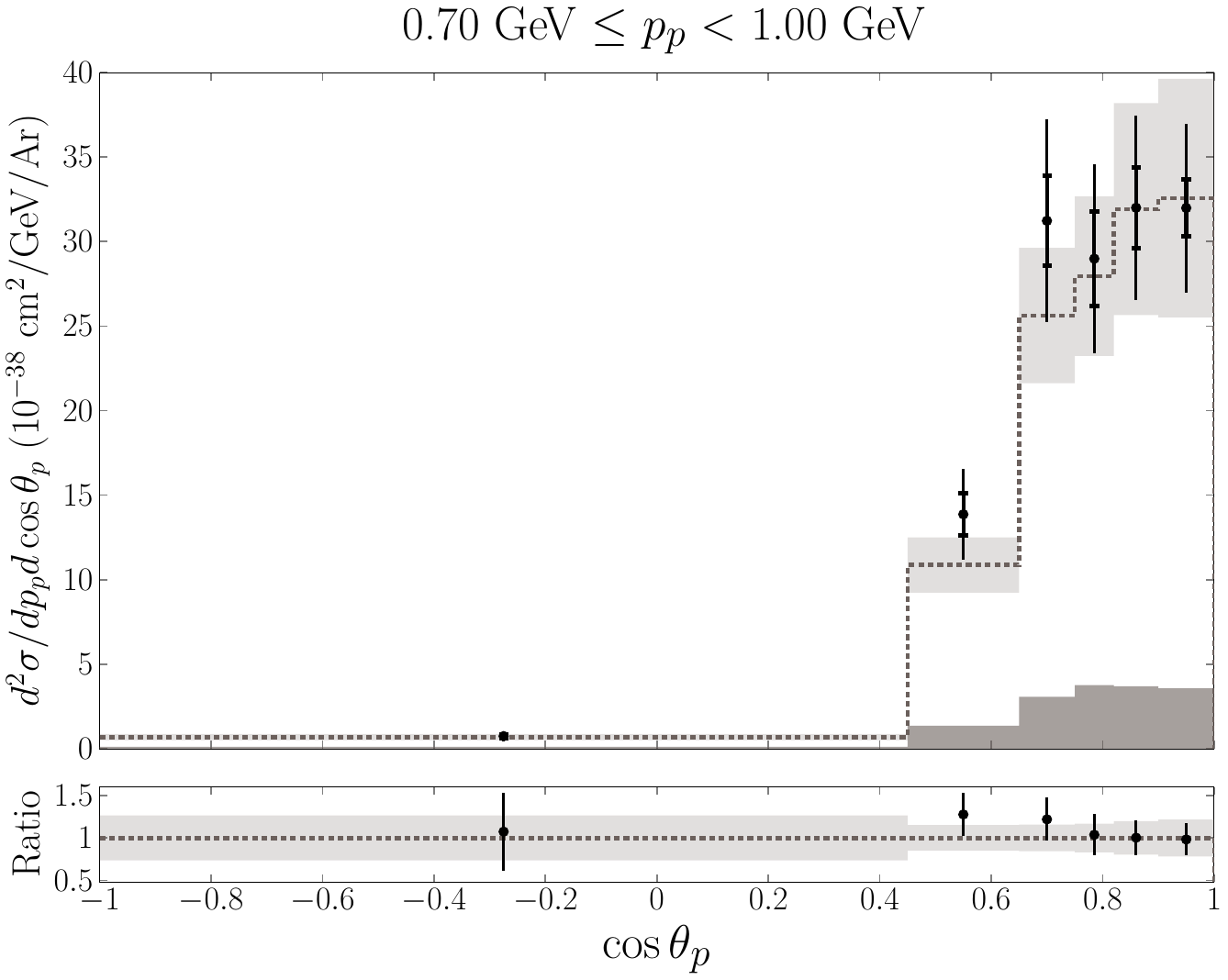}
\vspace{0.2cm}

\includegraphics[page=19, width=0.35\textwidth]{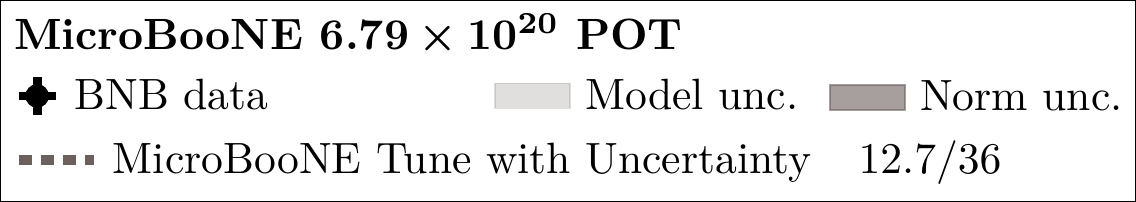}%
\hfill

\caption{Measured differential cross sections for block \#1
($p_p, \cos\theta_p$).
Statistical (shape-only systematic) uncertainties are included in the inner
(outer) error bars. The dark (light) gray band shows the
remainder of the measurement uncertainty (theoretical uncertainty on the
prediction). Error bars in the lower panels show the full measurement
uncertainty.}
\label{fig:ResultsTuneUncblock1}
\end{figure*}

\begin{figure*}
\centering

\includegraphics[page=20, width=0.64\textwidth]{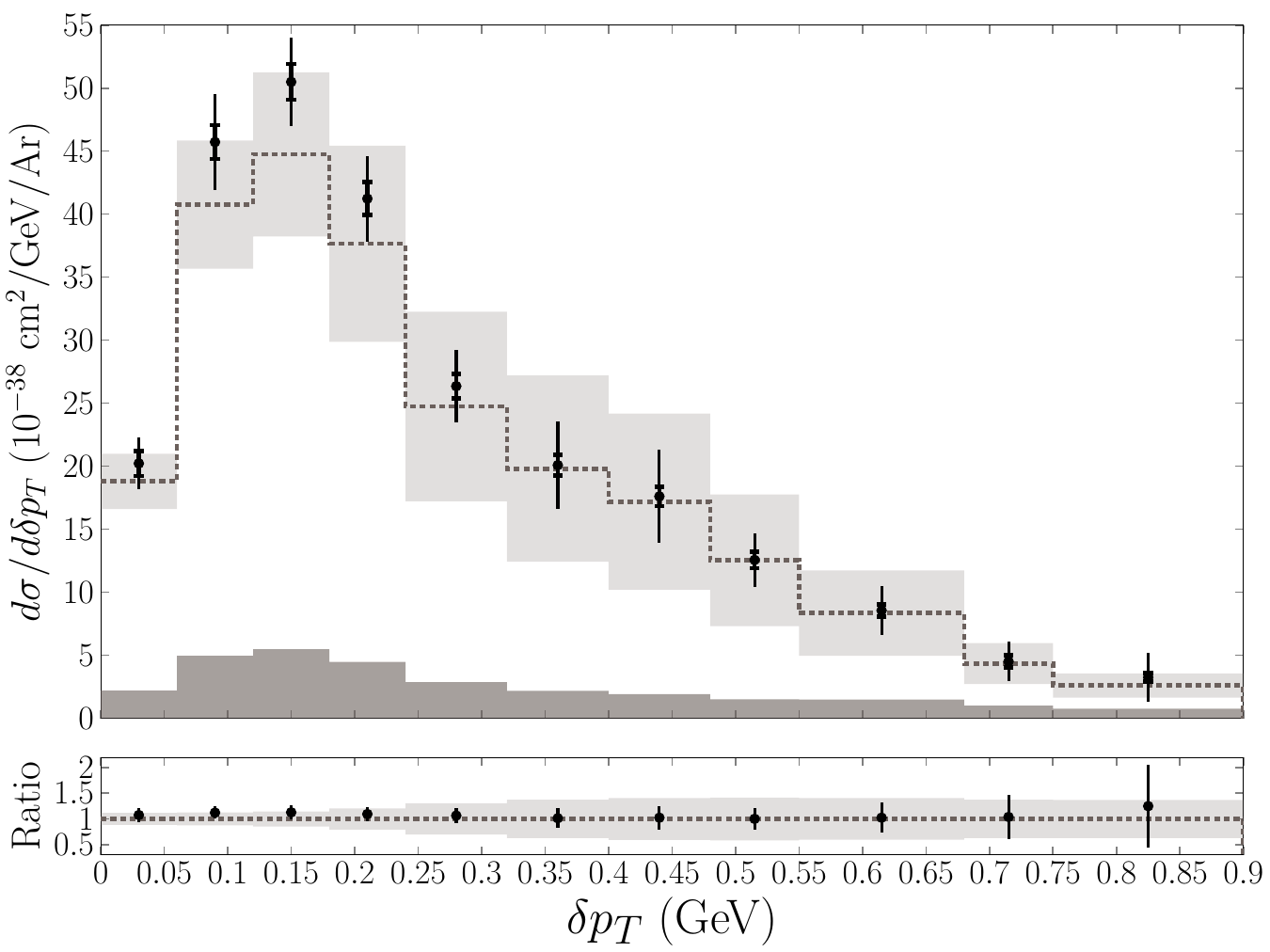}
\hfill
\belowbaseline[-0.23\textheight]{
  \includegraphics[page=21, width=0.35\textwidth]{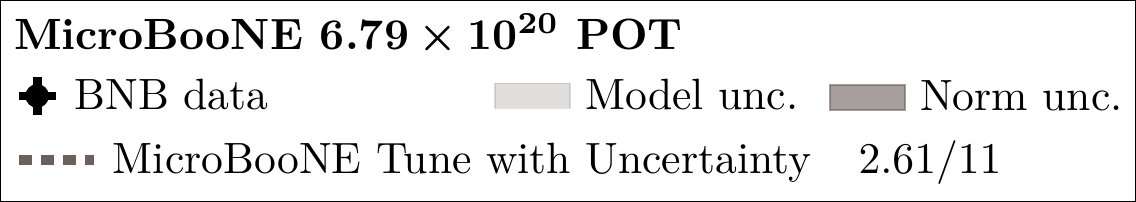}%
}
\hfill

\caption{Measured differential cross sections for block \#2
($\delta p_T$).
Statistical (shape-only systematic) uncertainties are included in the inner
(outer) error bars.
The dark (light) gray band shows the
remainder of the measurement uncertainty (theoretical uncertainty on the
prediction). Error bars in the lower panels show the full measurement
uncertainty.}
\label{fig:ResultsTuneUncblock2}
\end{figure*}

\begin{figure*}
\centering
\includegraphics[page=22, width=0.49\textwidth]{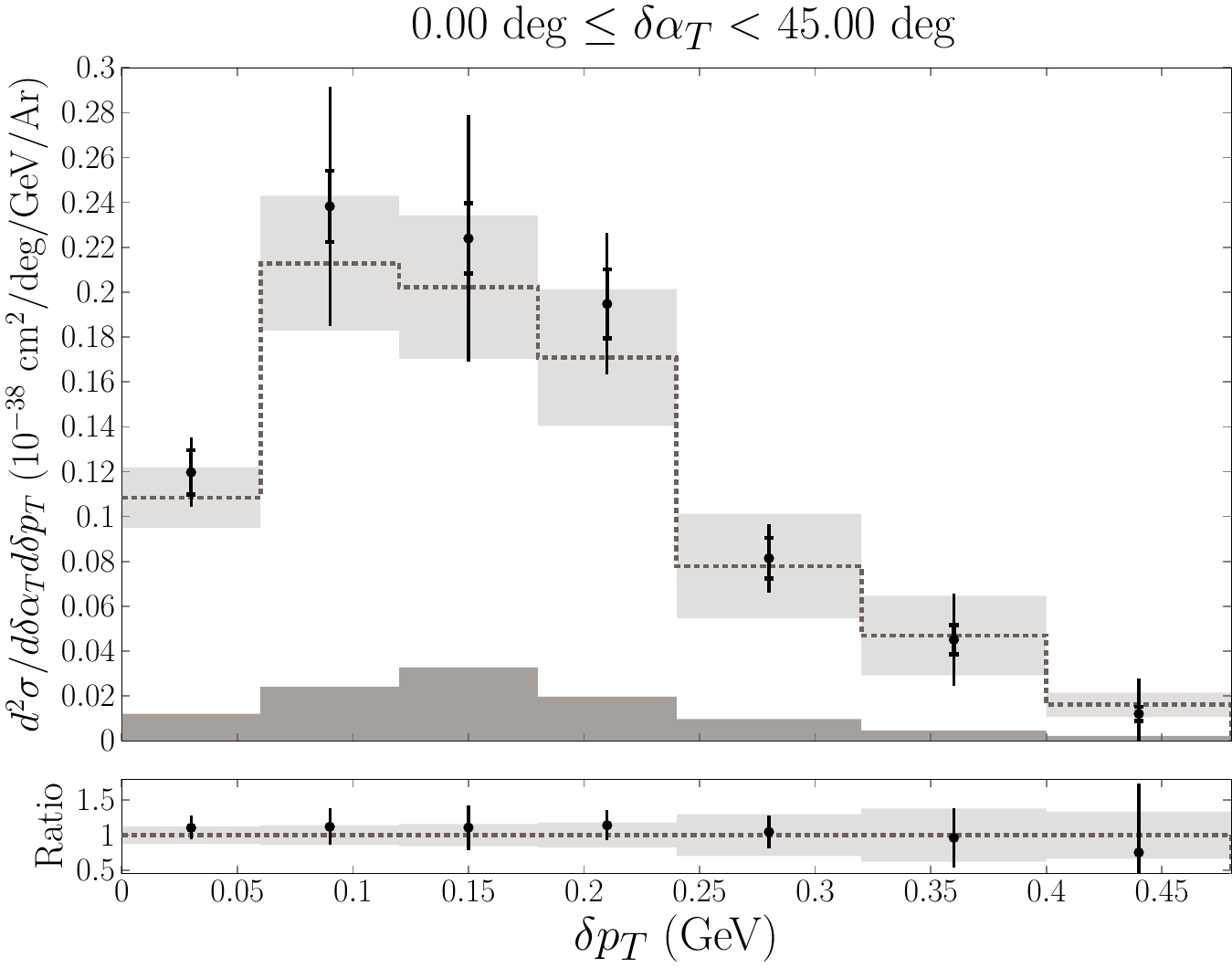}
\hfill
\includegraphics[page=23, width=0.49\textwidth]{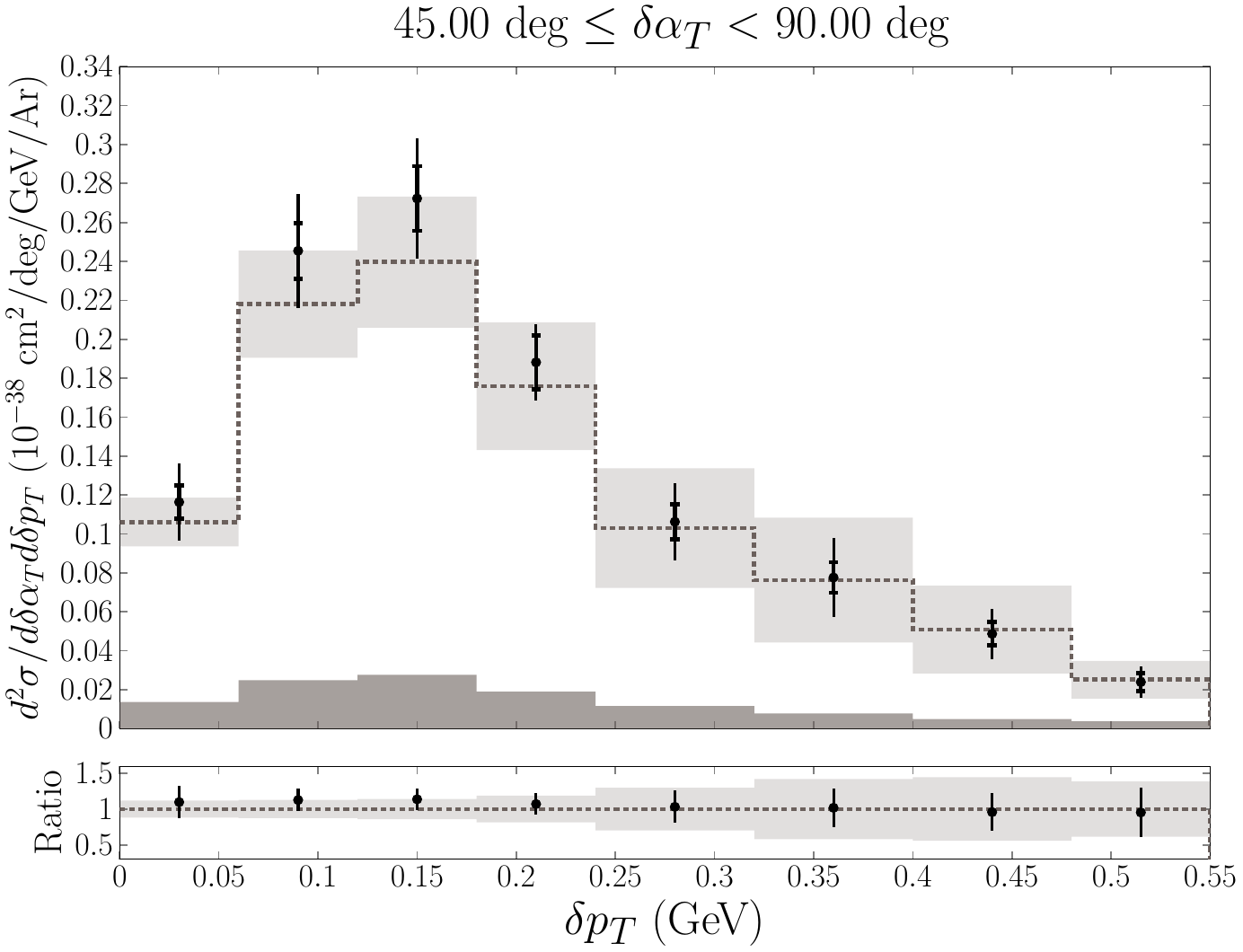}
\vspace{0.2cm}

\includegraphics[page=24, width=0.49\textwidth]{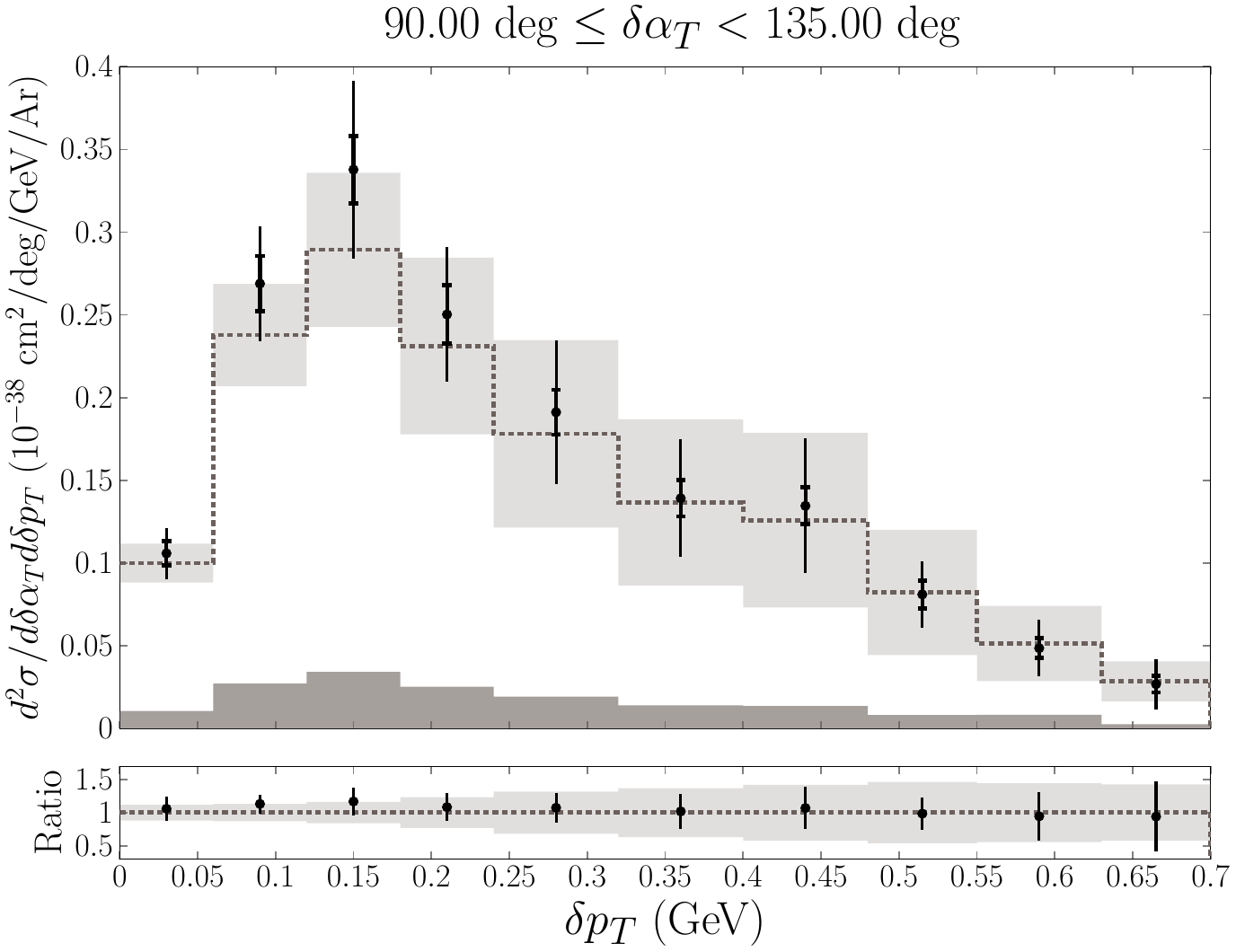}
\hfill
\includegraphics[page=25, width=0.49\textwidth]{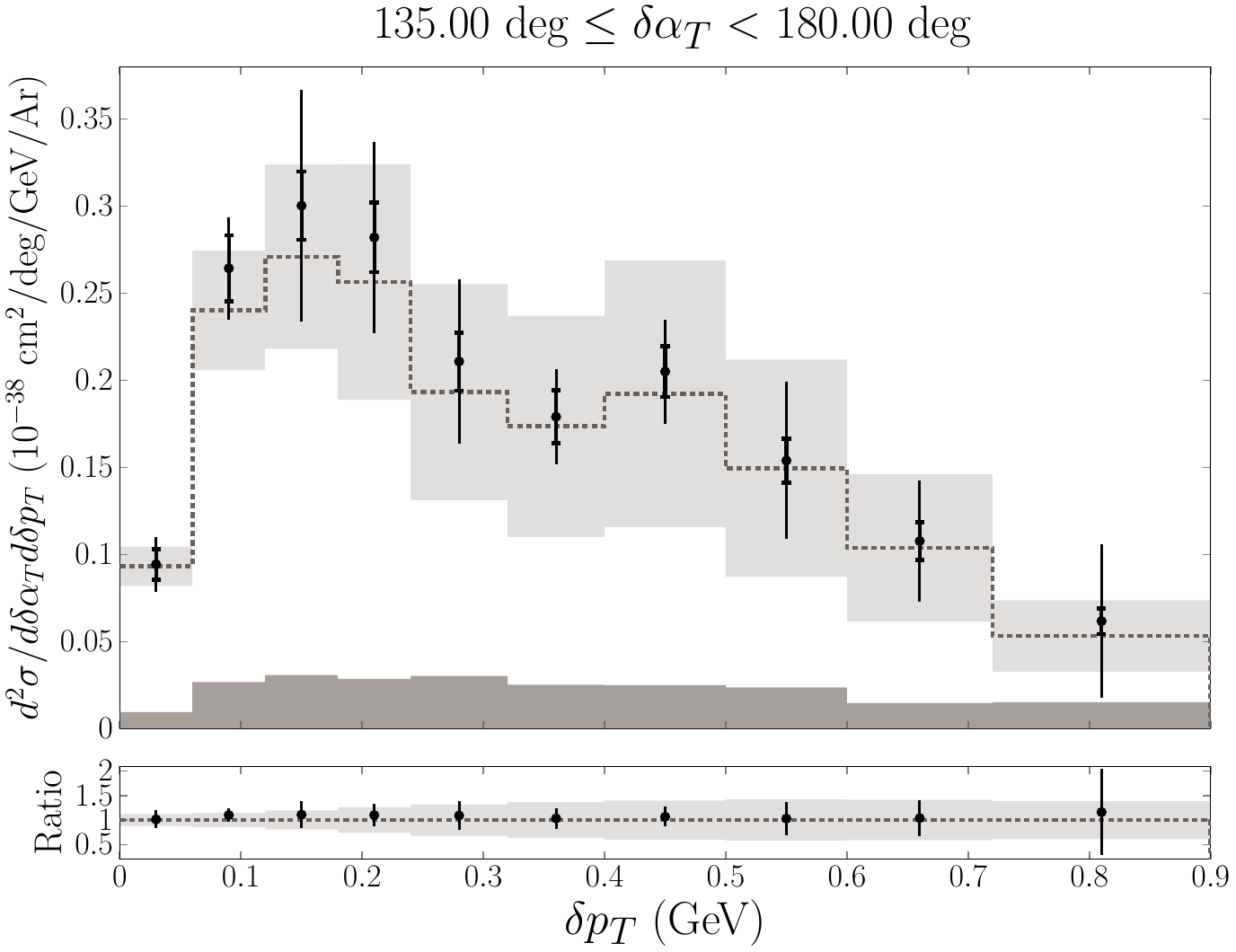}
\vspace{0.2cm}

\includegraphics[page=27, width=0.35\textwidth]{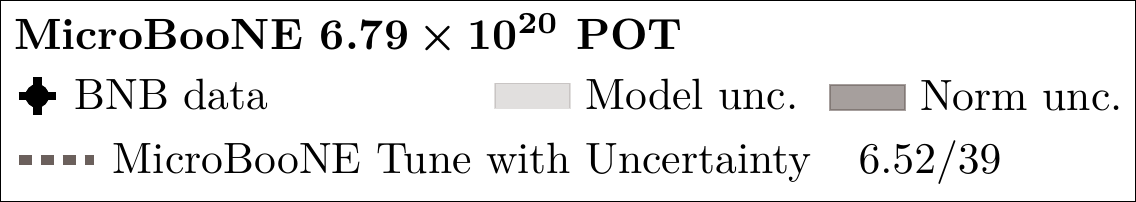}%
\hfill

\caption{Measured differential cross sections for block \#3
$(\delta \alpha_T, \delta p_T)$. The overall
$\chi^2$ value includes contributions from four $\delta p_T$ overflow bins that
are not plotted.
Statistical (shape-only systematic) uncertainties are included in the inner
(outer) error bars.
The dark (light) gray band shows the
remainder of the measurement uncertainty (theoretical uncertainty on the
prediction). Error bars in the lower panels show the full measurement
uncertainty.}
\label{fig:ResultsTuneUncblock3}
\end{figure*}

\begin{figure*}
\centering
\includegraphics[page=30, width=0.49\textwidth]{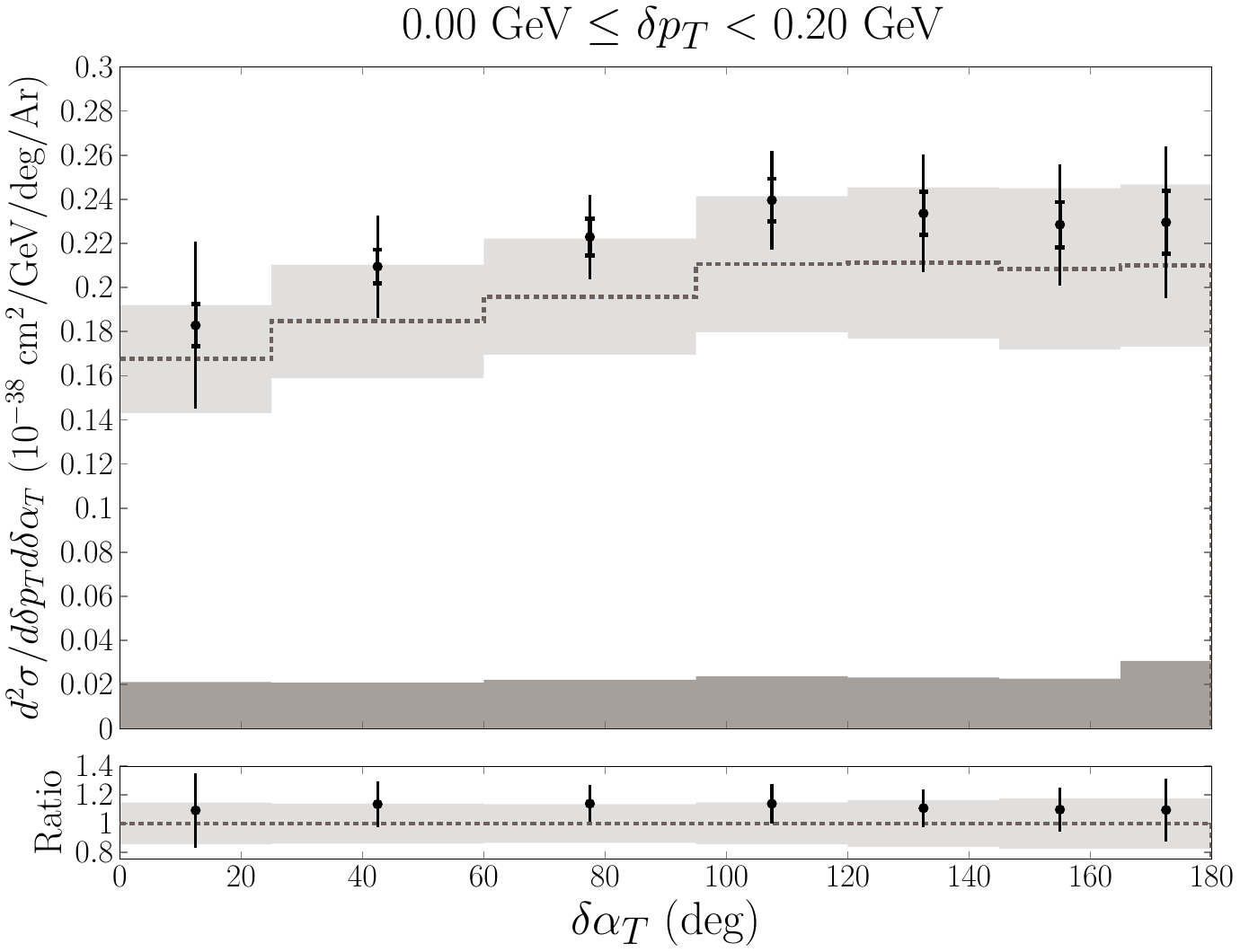}
\hfill
\includegraphics[page=31, width=0.49\textwidth]{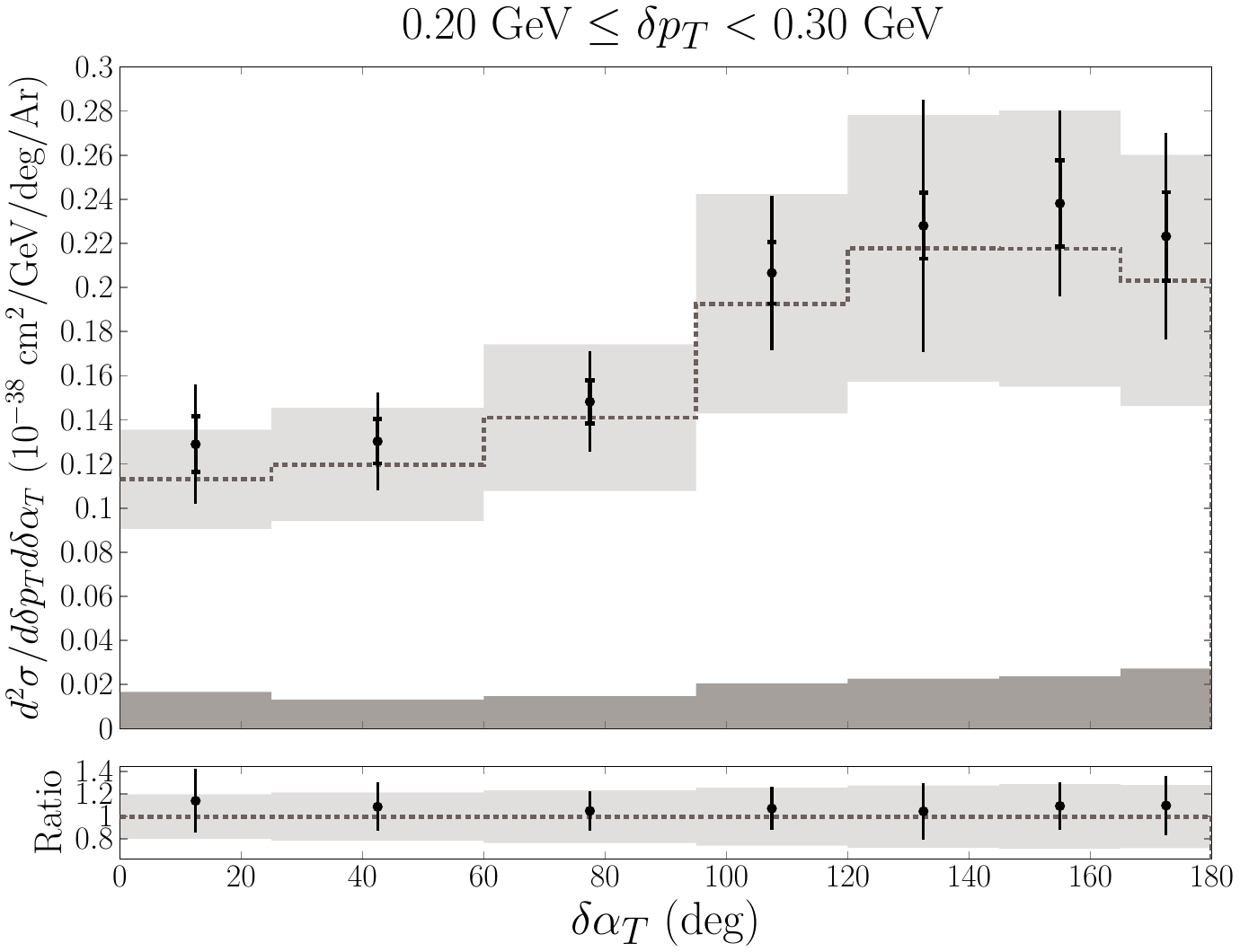}
\vspace{0.2cm}

\includegraphics[page=32, width=0.49\textwidth]{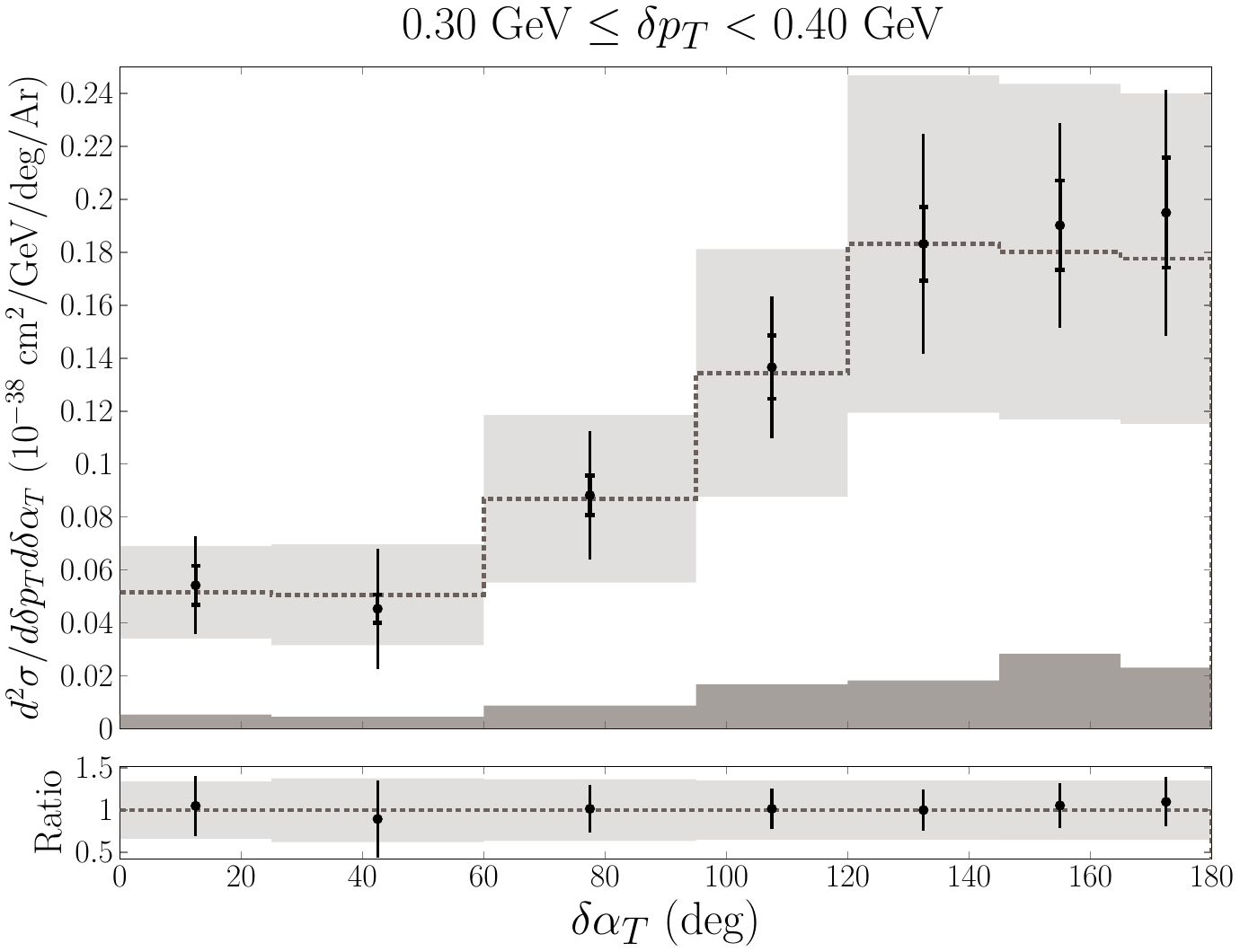}
\hfill
\includegraphics[page=33, width=0.49\textwidth]{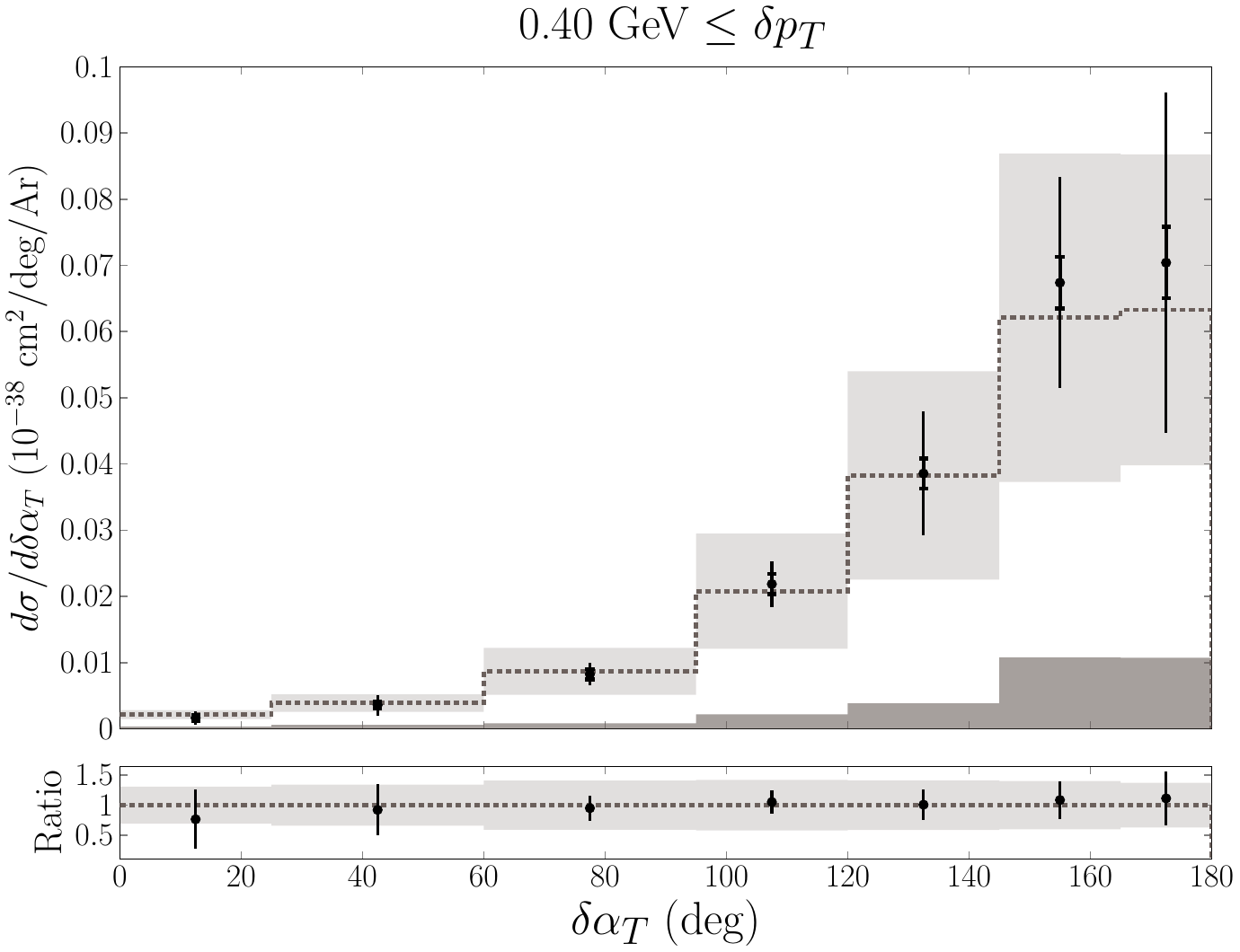}
\vspace{0.2cm}

\includegraphics[page=35, width=0.35\textwidth]{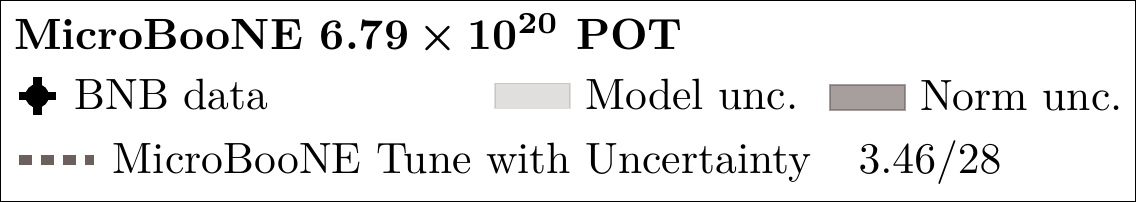}%
\hfill

\caption{Measured differential cross sections for block \#4
$(\delta p_T, \delta \alpha_T)$.
Statistical (shape-only systematic) uncertainties are included in the inner
(outer) error bars.
The dark (light) gray band shows the
remainder of the measurement uncertainty (theoretical uncertainty on the
prediction). Error bars in the lower panels show the full measurement
uncertainty.}
\label{fig:ResultsTuneUncblock4}
\end{figure*}

\begin{figure*}
\centering

\includegraphics[page=36, width=0.64\textwidth]{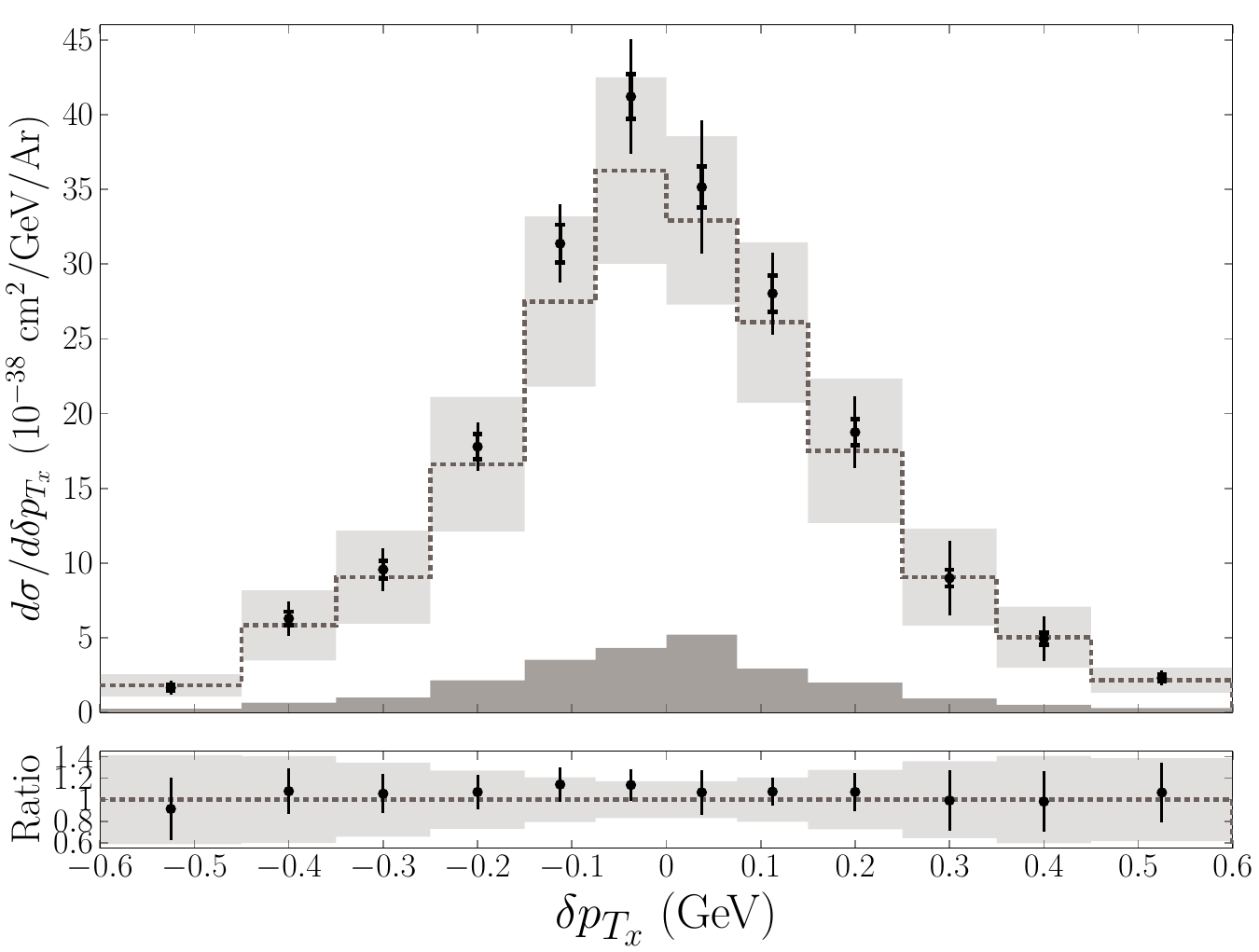}
\hfill
\belowbaseline[-0.23\textheight]{
  \includegraphics[page=37, width=0.35\textwidth]{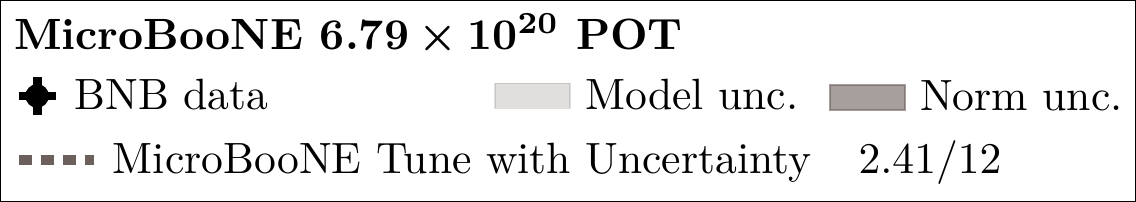}%
}
\hfill

\caption{Measured differential cross sections for block \#5
($\delta p_{T_x}$).
Statistical (shape-only systematic) uncertainties are included in the inner
(outer) error bars.
The dark (light) gray band shows the
remainder of the measurement uncertainty (theoretical uncertainty on the
prediction). Error bars in the lower panels show the full measurement
uncertainty.}
\label{fig:ResultsTuneUncblock5}
\end{figure*}

\begin{figure*}
\centering
\includegraphics[page=38, width=0.49\textwidth]{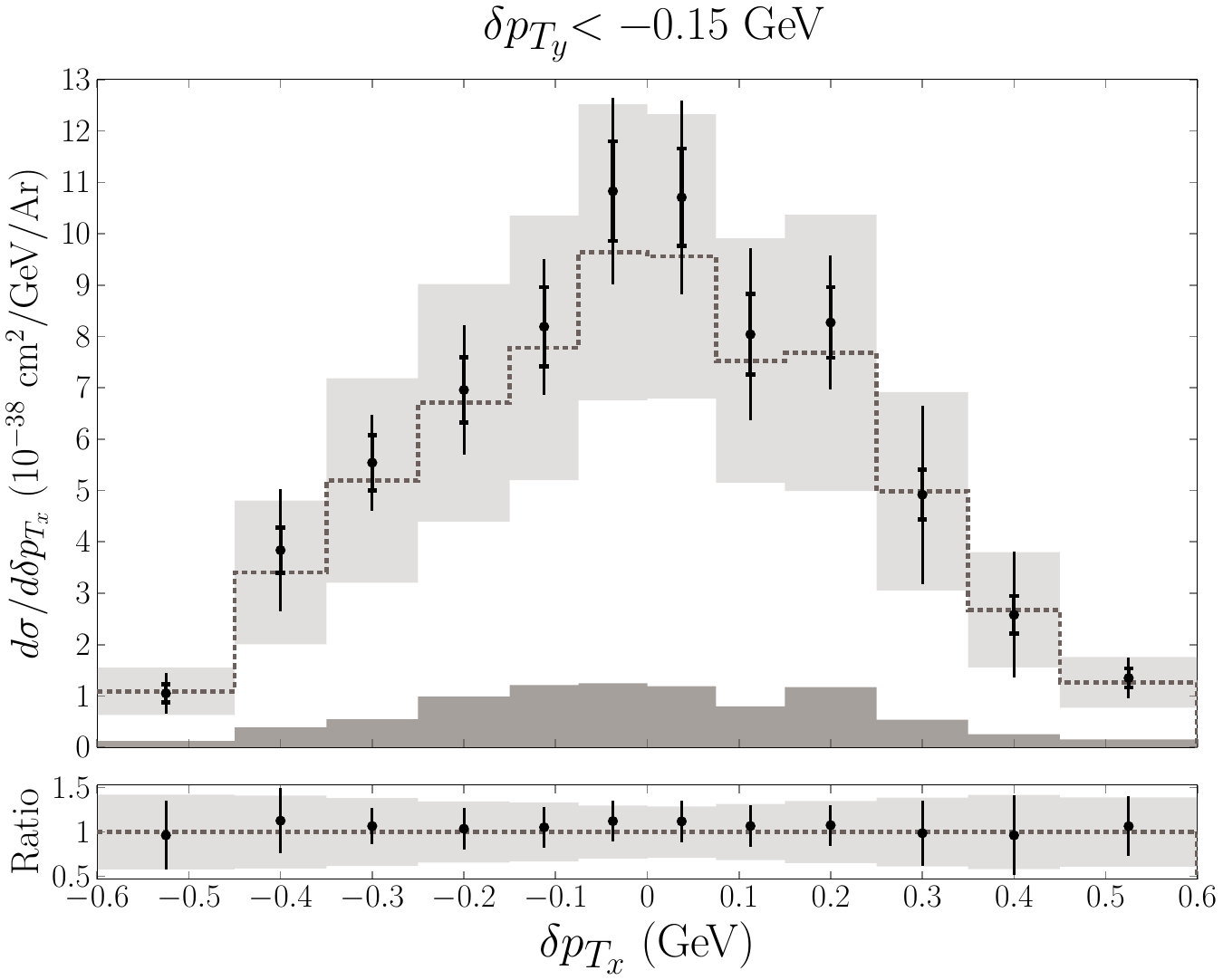}
\hfill
\includegraphics[page=39, width=0.49\textwidth]{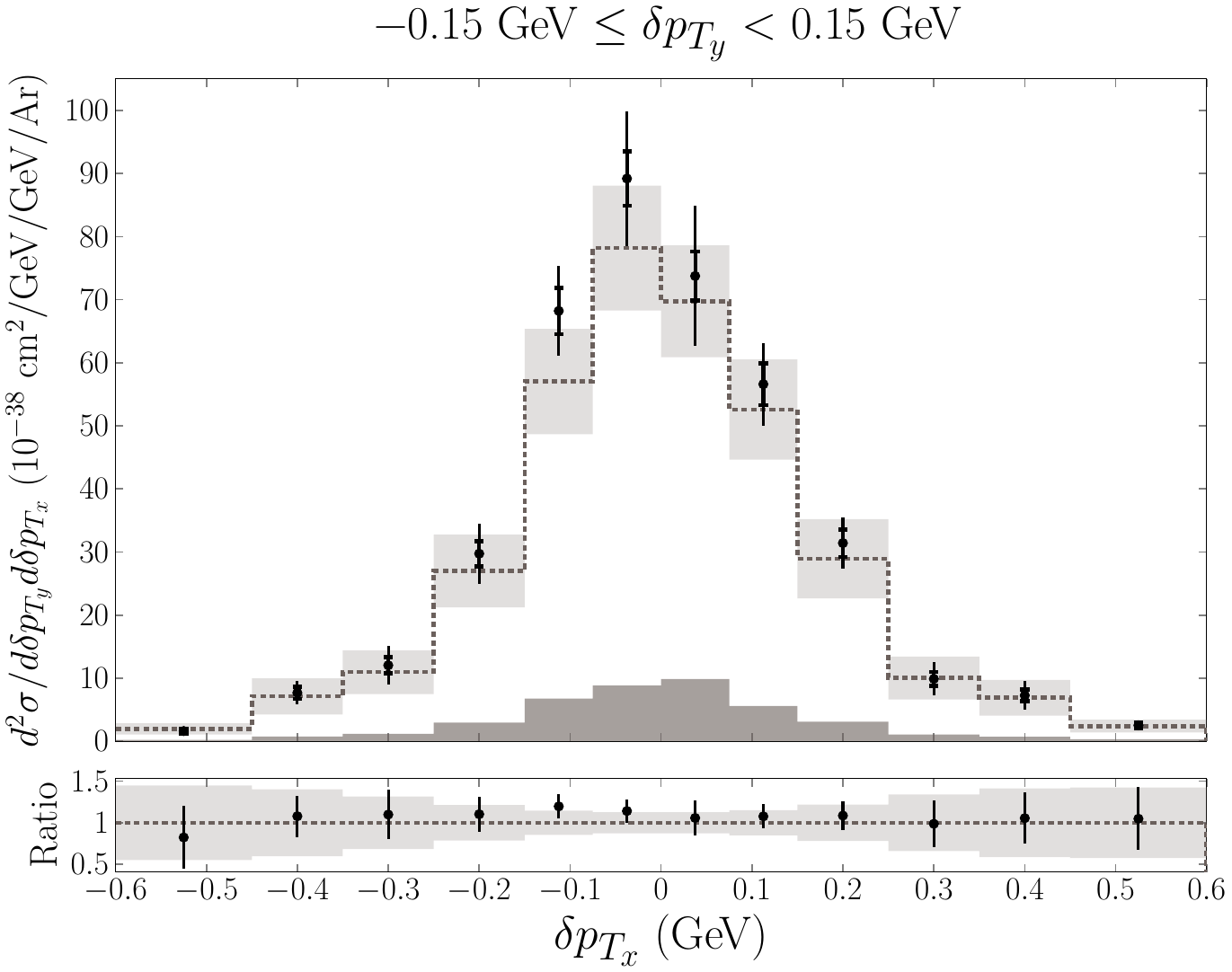}
\vspace{0.2cm}

\includegraphics[page=40, width=0.49\textwidth]{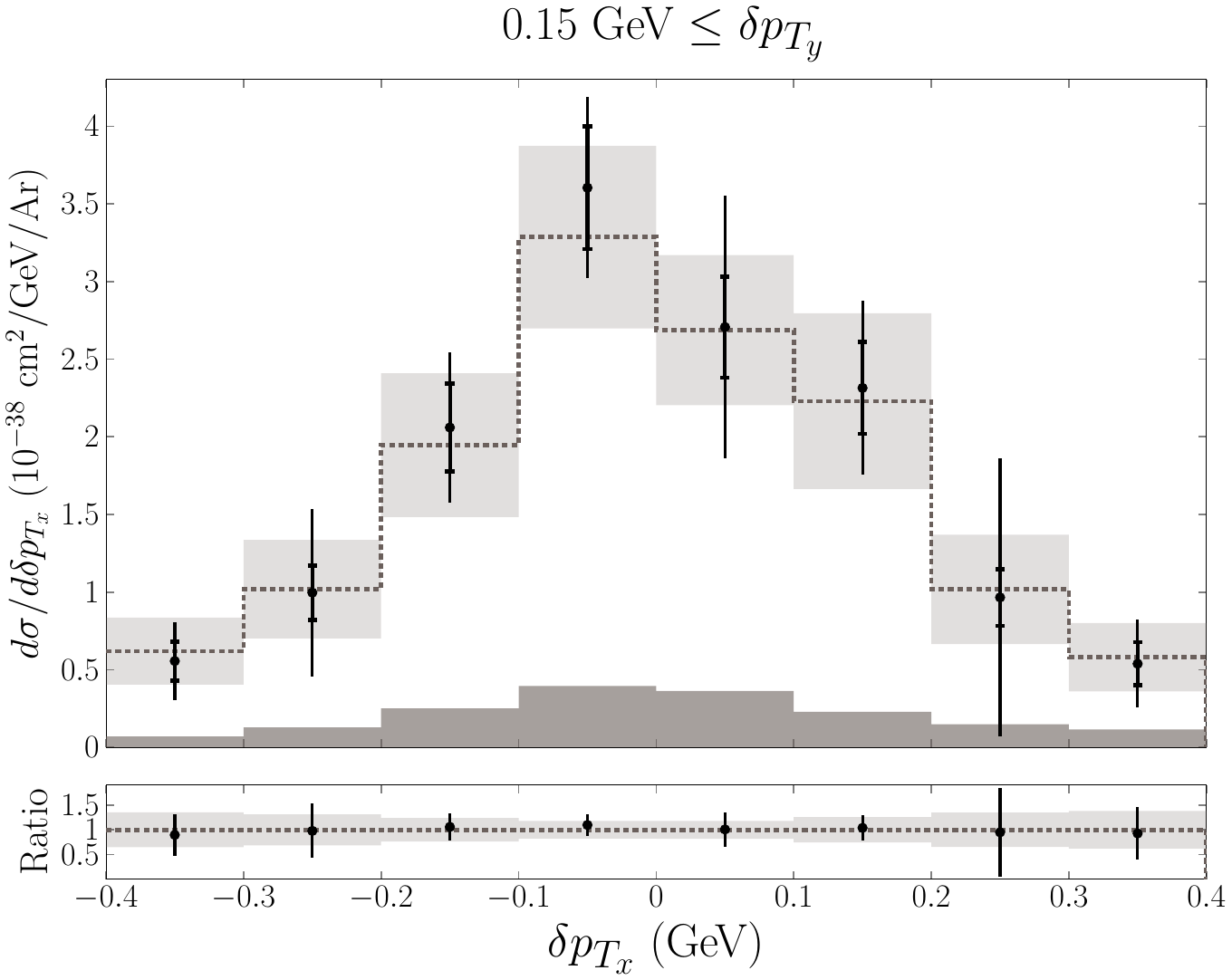}
\hfill
\belowbaseline[-0.23\textheight]{
  \includegraphics[page=42, width=0.35\textwidth]{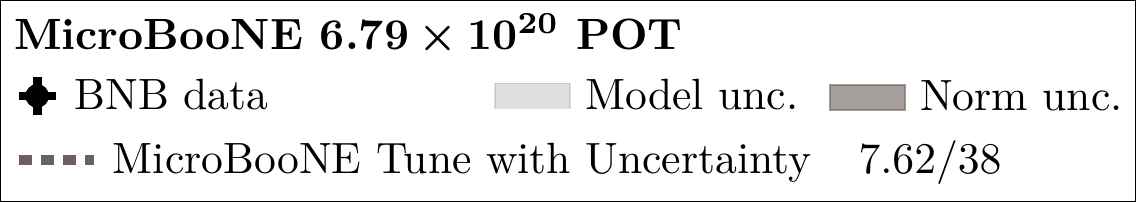}%
}
\hfill

\caption{Measured differential cross sections for block \#6
$(\delta p_{T_y}, \delta p_{T_x})$.
The overall
$\chi^2$ value includes contributions from three underflow and three overflow
$\delta p_{T_x}$ bins that are not plotted.
Statistical (shape-only systematic) uncertainties are included in the inner
(outer) error bars.
The dark (light) gray band shows the
remainder of the measurement uncertainty (theoretical uncertainty on the
prediction). Error bars in the lower panels show the full measurement
uncertainty.}
\label{fig:ResultsTuneUncblock6}
\end{figure*}

\begin{figure*}
\centering

\includegraphics[page=43, width=0.64\textwidth]{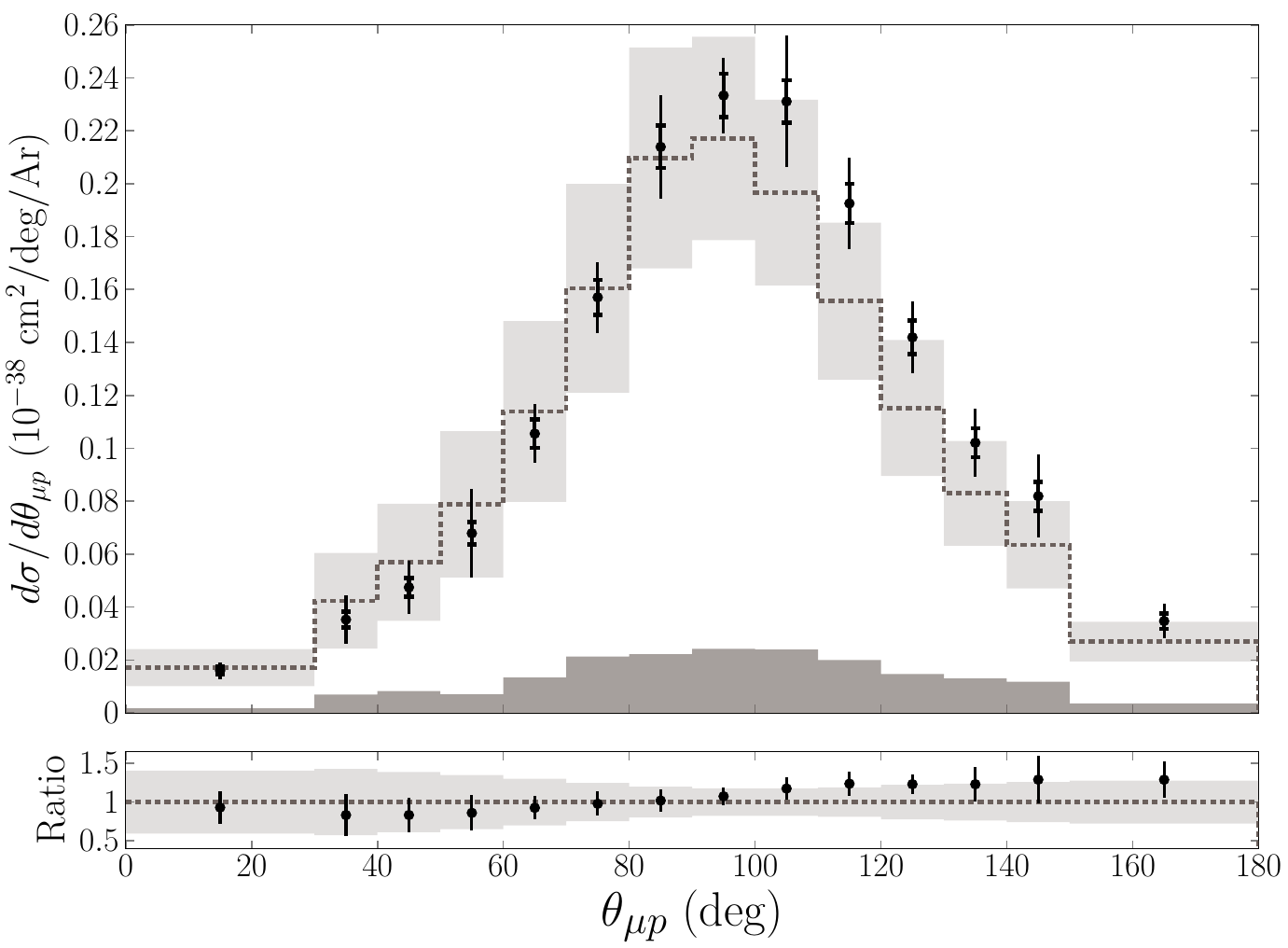}
\hfill
\belowbaseline[-0.23\textheight]{
  \includegraphics[page=44, width=0.35\textwidth]{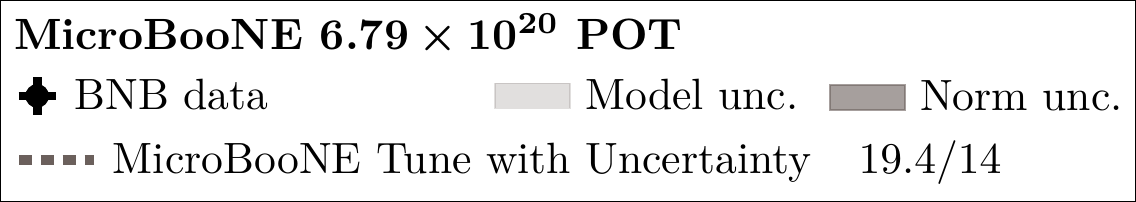}%
}
\hfill

\caption{Measured differential cross sections for block \#7
($\theta_{\mu p}$).
Statistical (shape-only systematic) uncertainties are included in the inner
(outer) error bars.
The dark (light) gray band shows the
remainder of the measurement uncertainty (theoretical uncertainty on the
prediction). Error bars in the lower panels show the full measurement
uncertainty.}
\label{fig:ResultsTuneUncblock7}
\end{figure*}

\begin{figure*}
\centering

\includegraphics[page=45, width=0.64\textwidth]{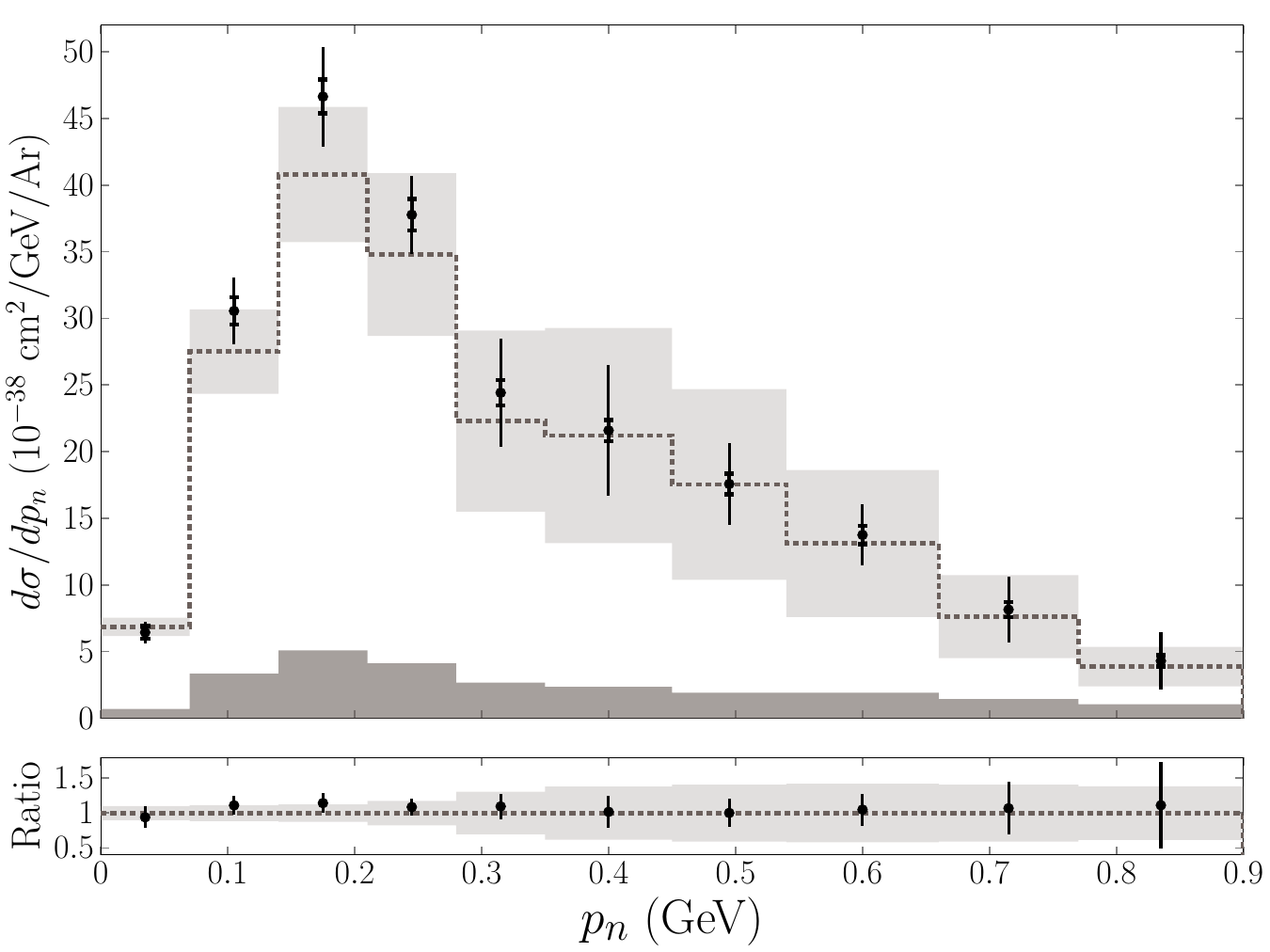}
\hfill
\belowbaseline[-0.23\textheight]{
  \includegraphics[page=46, width=0.35\textwidth]{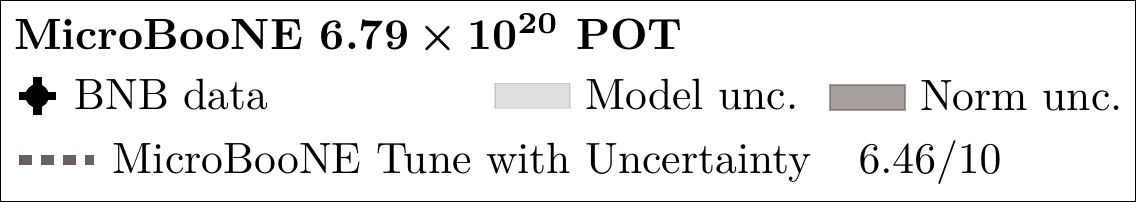}%
}
\hfill

\caption{Measured differential cross sections for block \#8 ($p_n$).
Statistical (shape-only systematic) uncertainties are included in the inner
(outer) error bars.
The dark (light) gray band shows the
remainder of the measurement uncertainty (theoretical uncertainty on the
prediction). Error bars in the lower panels show the full measurement
uncertainty.}
\label{fig:ResultsTuneUncblock8}
\end{figure*}

\begin{figure*}
\centering
\includegraphics[page=47, width=0.49\textwidth]{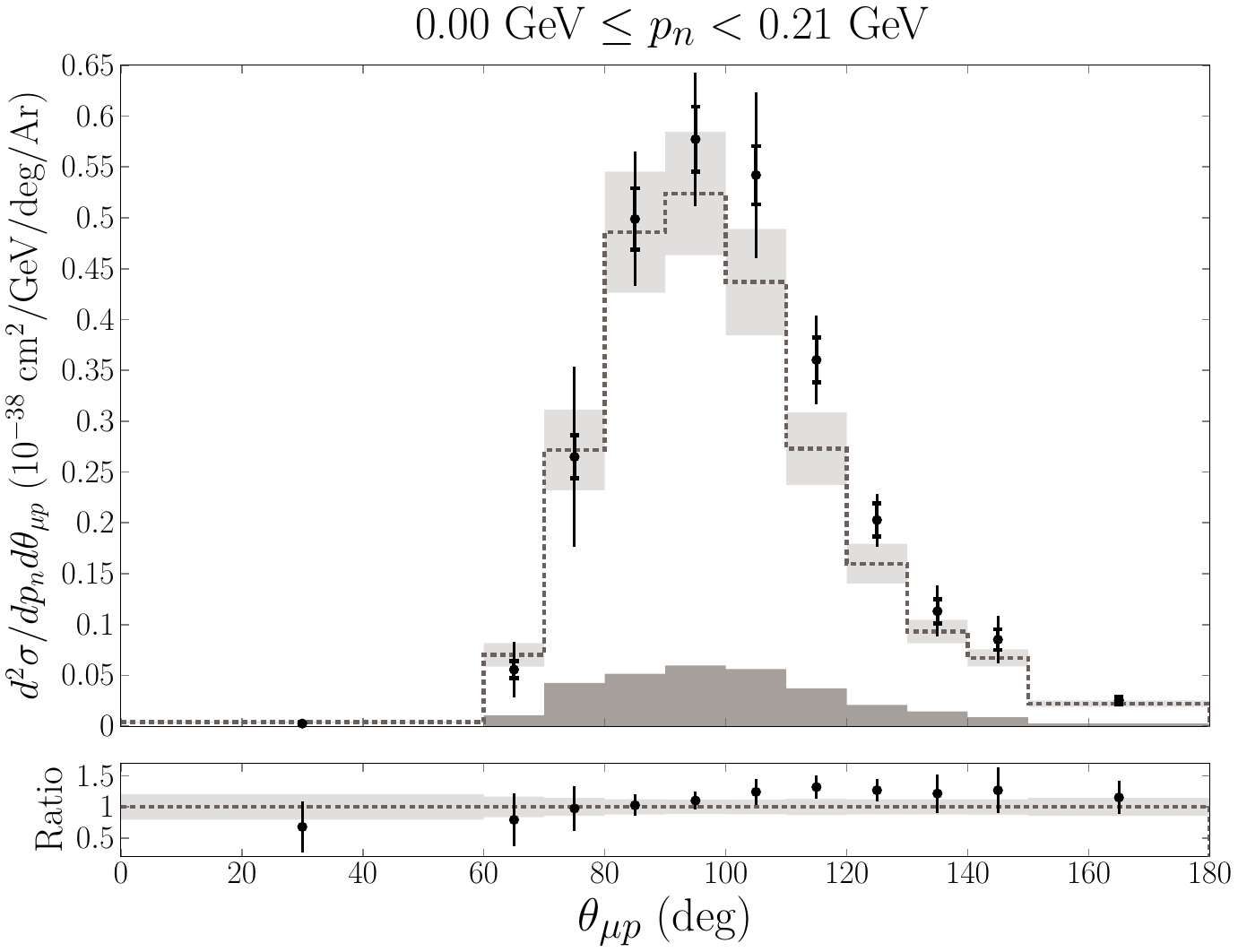}
\hfill
\includegraphics[page=48, width=0.49\textwidth]{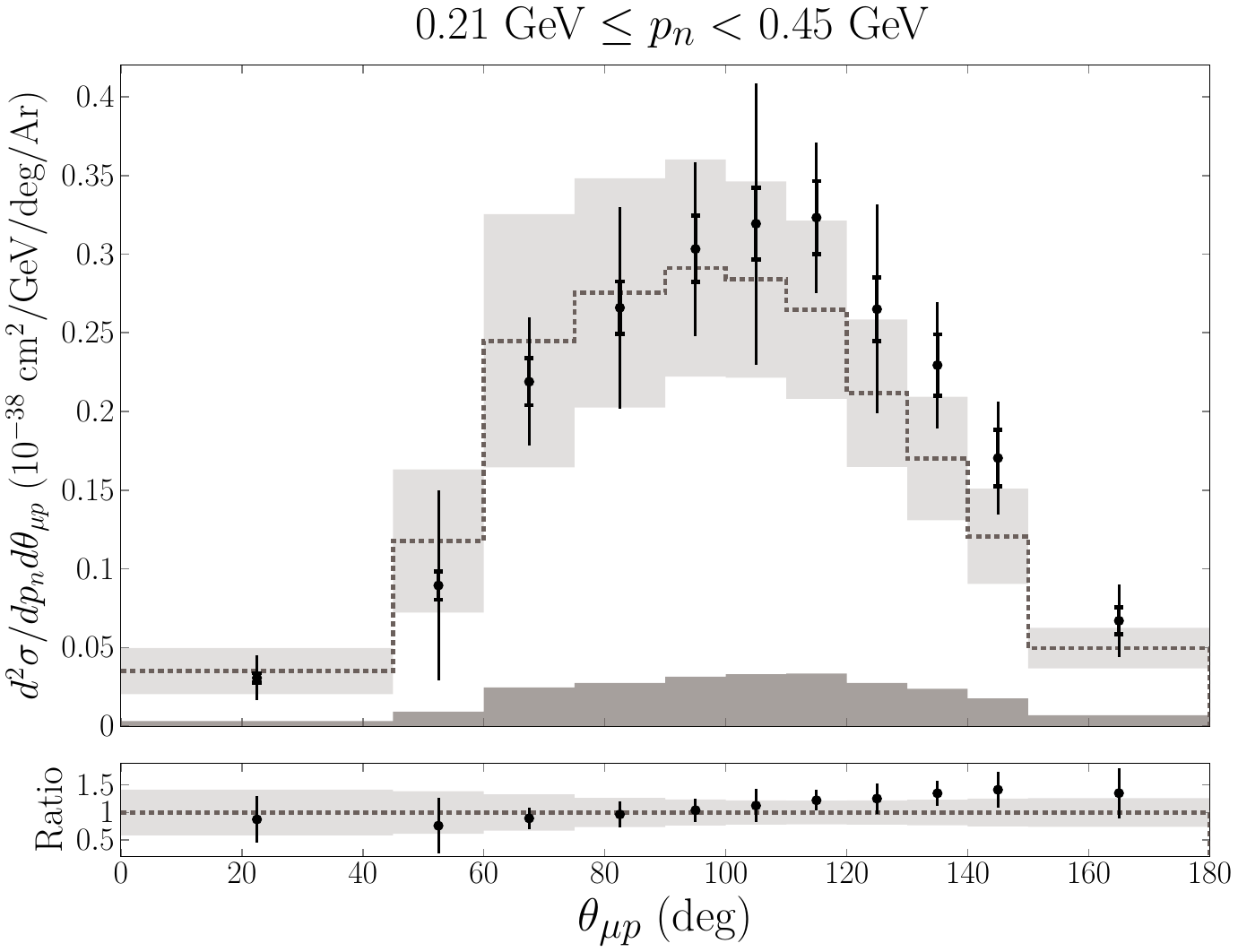}
\vspace{0.2cm}

\includegraphics[page=49, width=0.49\textwidth]{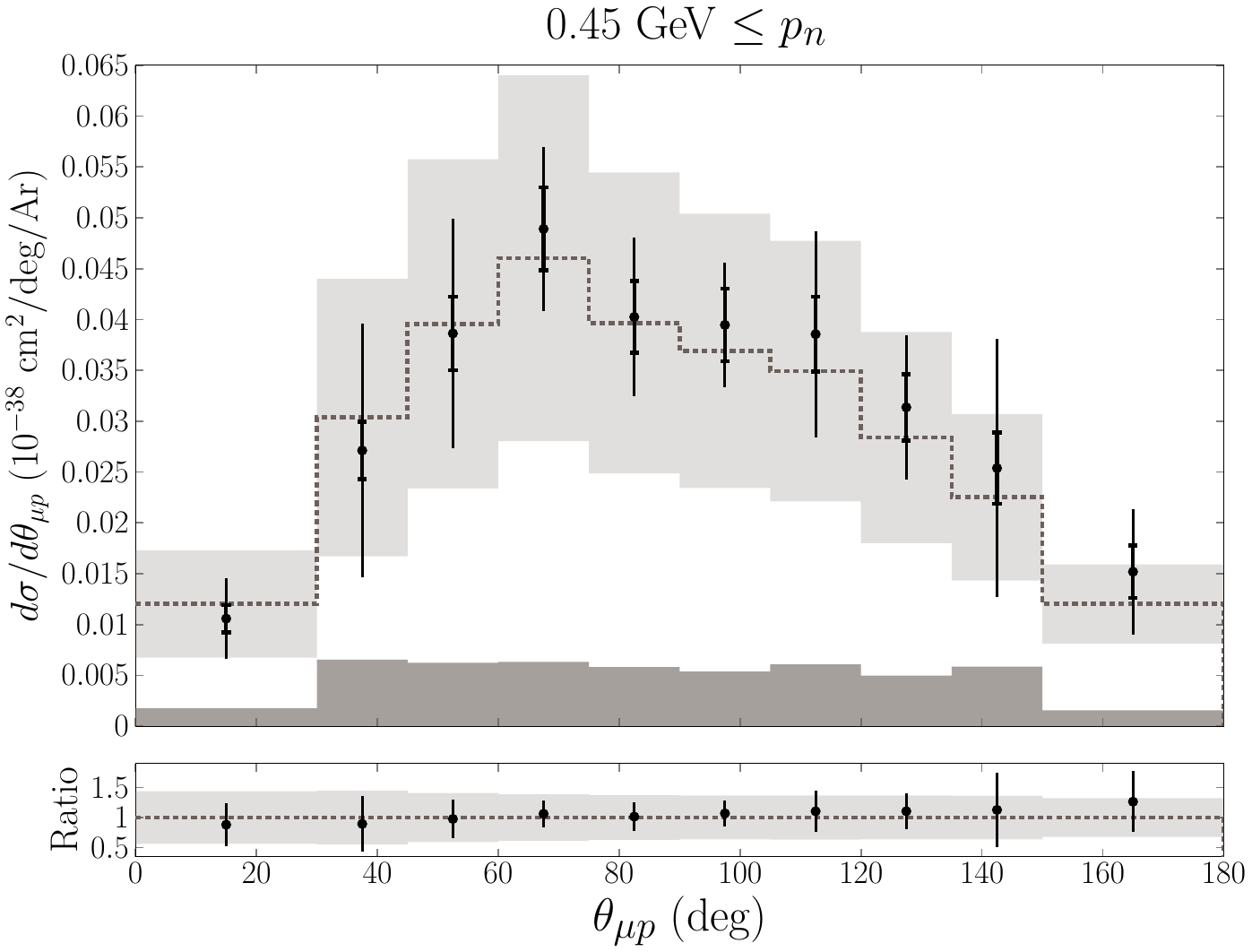}
\hfill
\belowbaseline[-0.23\textheight]{
  \includegraphics[page=51, width=0.35\textwidth]{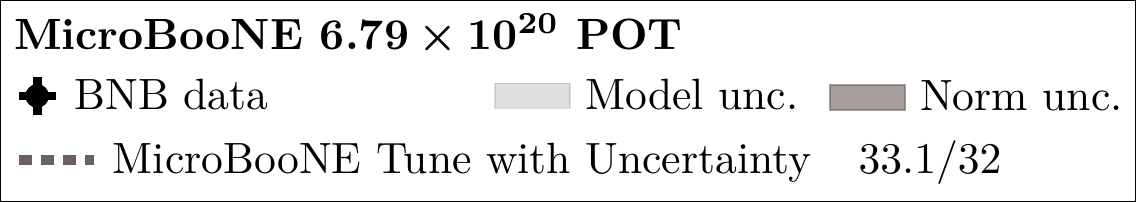}%
}
\hfill

\caption{Measured differential cross sections for block \#9
$(p_n, \theta_{\mu p})$.
Statistical (shape-only systematic) uncertainties are included in the inner
(outer) error bars.
The dark (light) gray band shows the
remainder of the measurement uncertainty (theoretical uncertainty on the
prediction). Error bars in the lower panels show the full measurement
uncertainty.}
\label{fig:ResultsTuneUncblock9}
\end{figure*}

\begin{figure*}
\centering

\includegraphics[page=10, width=0.64\textwidth]{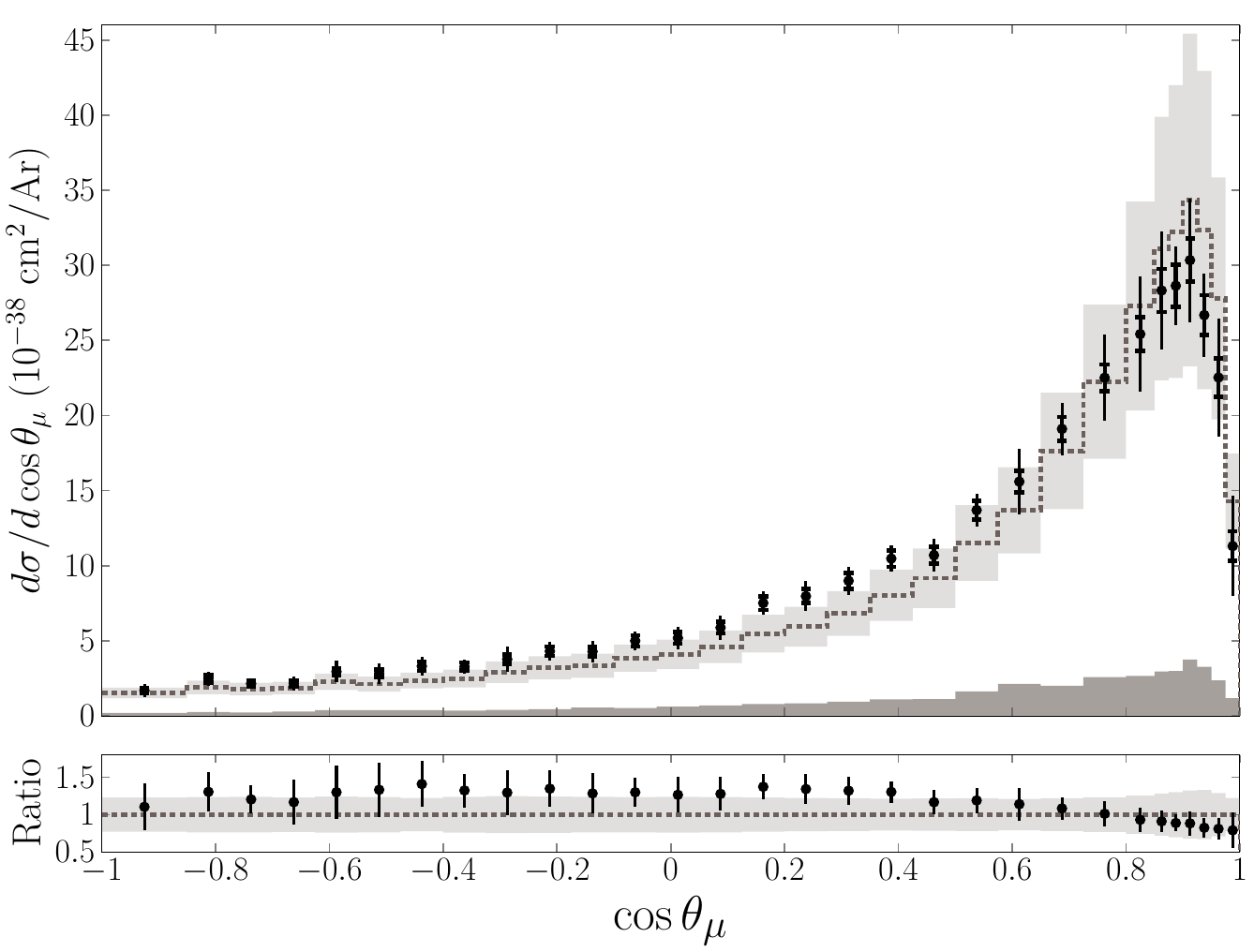}
\hfill
\belowbaseline[-0.23\textheight]{
  \includegraphics[page=11, width=0.35\textwidth]{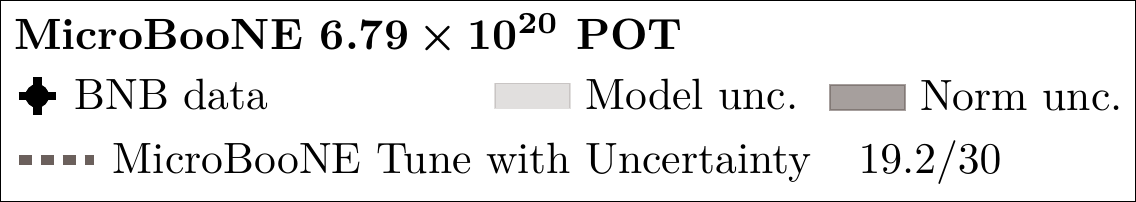}%
}
\hfill

\caption{Measured differential cross sections for block \#10
($\cos\theta_\mu$).
Statistical (shape-only systematic) uncertainties are included in the inner
(outer) error bars.
The dark (light) gray band shows the
remainder of the measurement uncertainty (theoretical uncertainty on the
prediction). Error bars in the lower panels show the full measurement
uncertainty.}
\label{fig:ResultsTuneUncblock10}
\end{figure*}

\begin{figure*}
\centering

\includegraphics[page=52, width=0.64\textwidth]{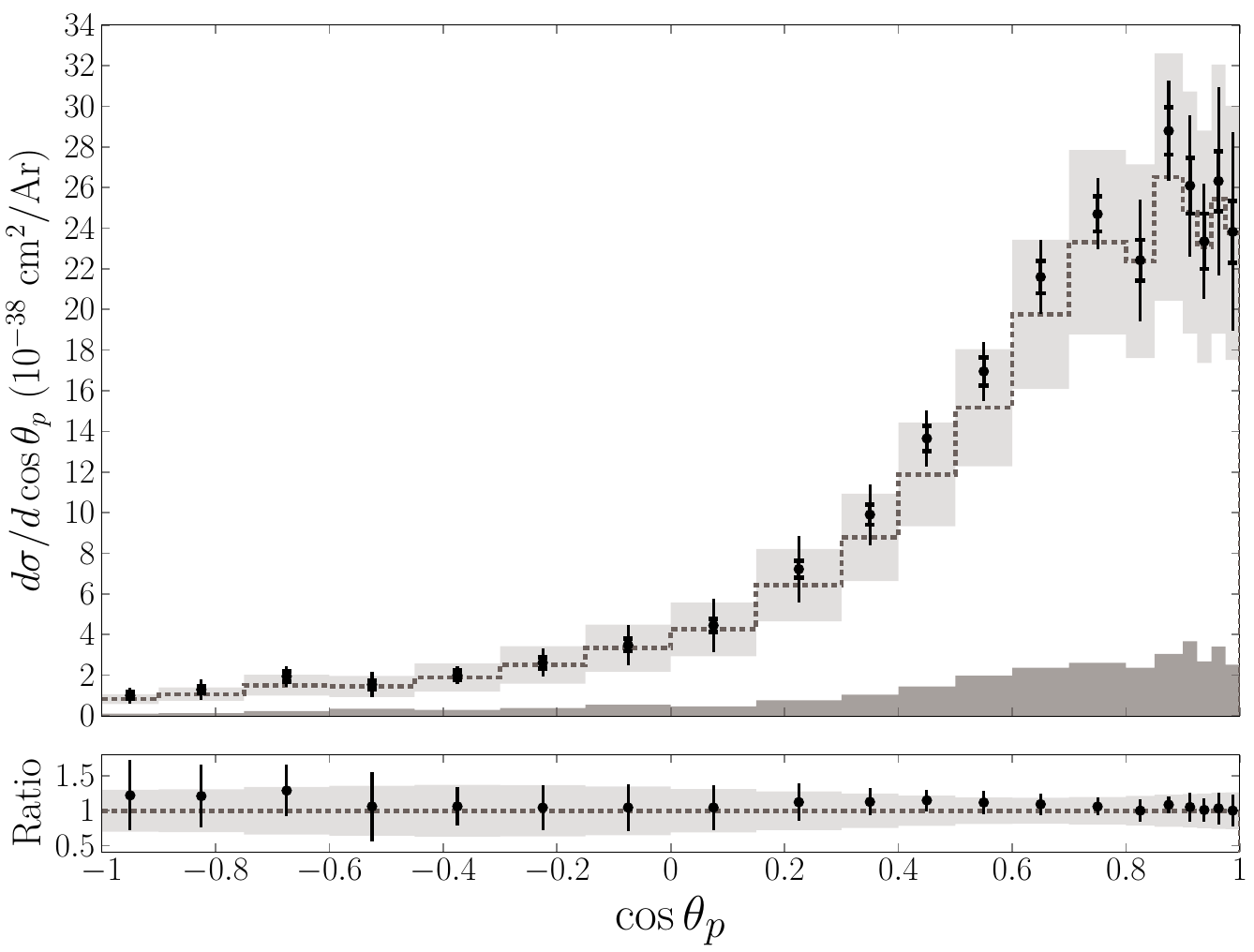}
\hfill
\belowbaseline[-0.23\textheight]{
  \includegraphics[page=53, width=0.35\textwidth]{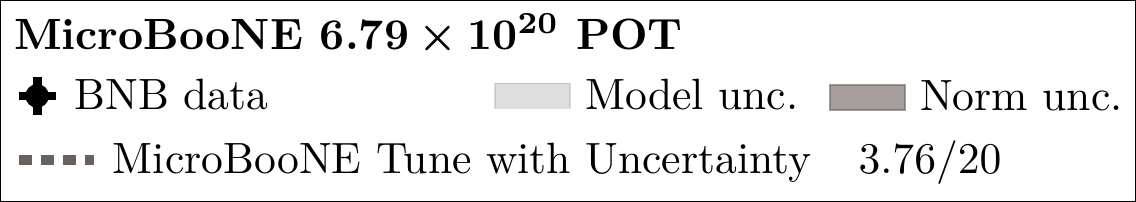}%
}
\hfill

\caption{Measured differential cross sections for block \#11
($\cos\theta_p$).
Statistical (shape-only systematic) uncertainties are included in the inner
(outer) error bars.
The dark (light) gray band shows the
remainder of the measurement uncertainty (theoretical uncertainty on the
prediction). Error bars in the lower panels show the full measurement
uncertainty.}
\label{fig:ResultsTuneUncblock11}
\end{figure*}

\begin{figure*}
\centering

\includegraphics[page=54, width=0.64\textwidth]{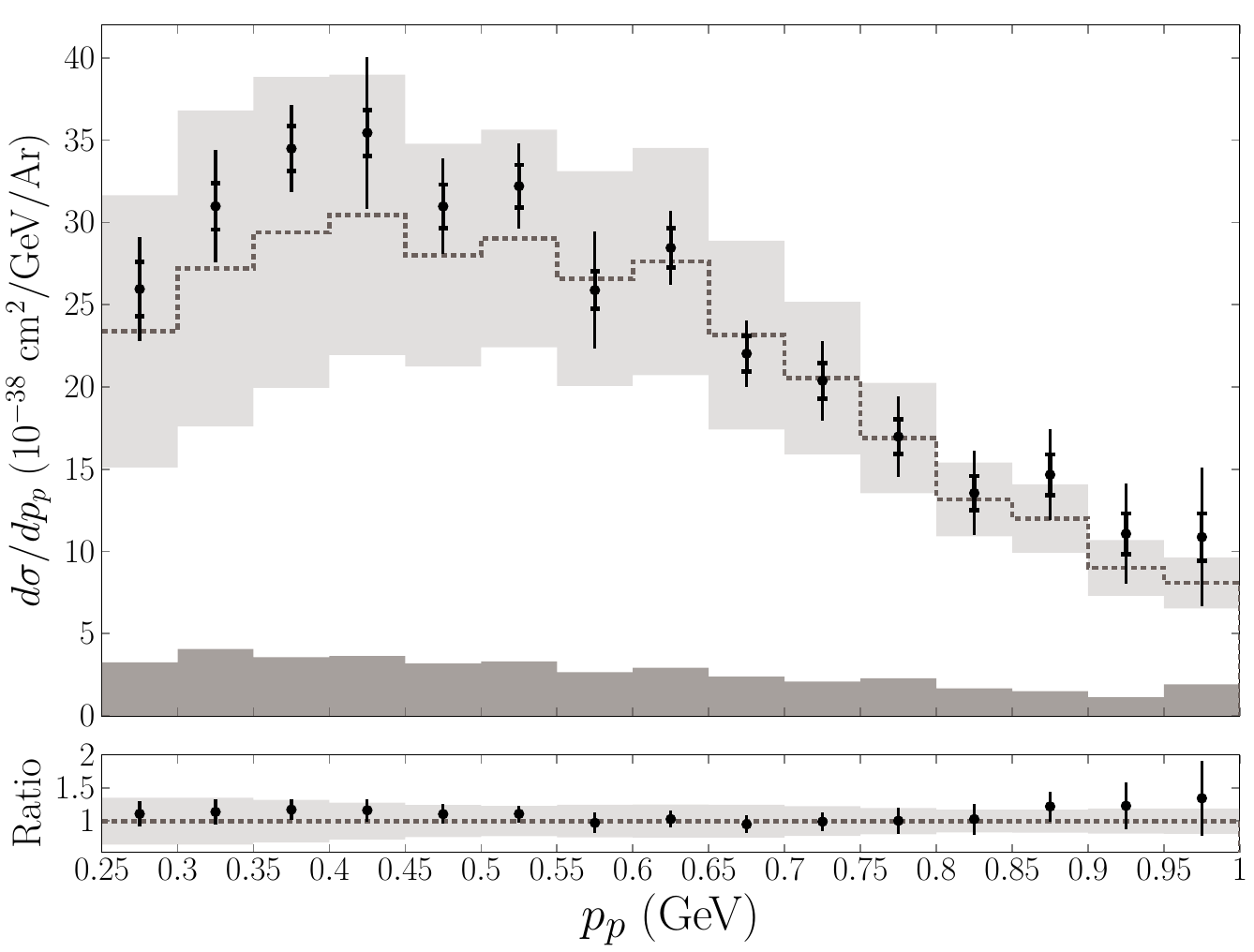}
\hfill
\belowbaseline[-0.23\textheight]{
  \includegraphics[page=55, width=0.35\textwidth]{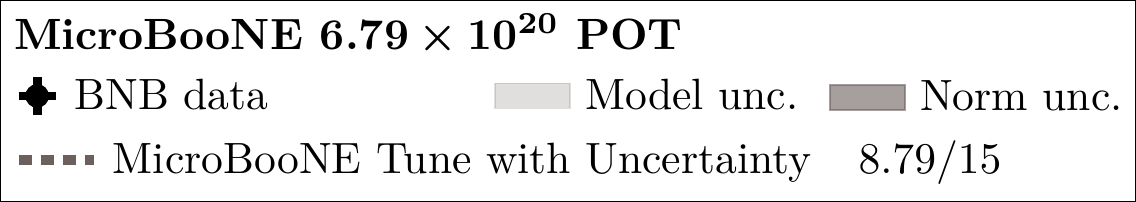}%
}
\hfill

\caption{Measured differential cross sections for block \#12
($p_p$).
Statistical (shape-only systematic) uncertainties are included in the inner
(outer) error bars.
The dark (light) gray band shows the
remainder of the measurement uncertainty (theoretical uncertainty on the
prediction). Error bars in the lower panels show the full measurement
uncertainty.}
\label{fig:ResultsTuneUncblock12}
\end{figure*}

\begin{figure*}
\centering

\includegraphics[page=56, width=0.64\textwidth]{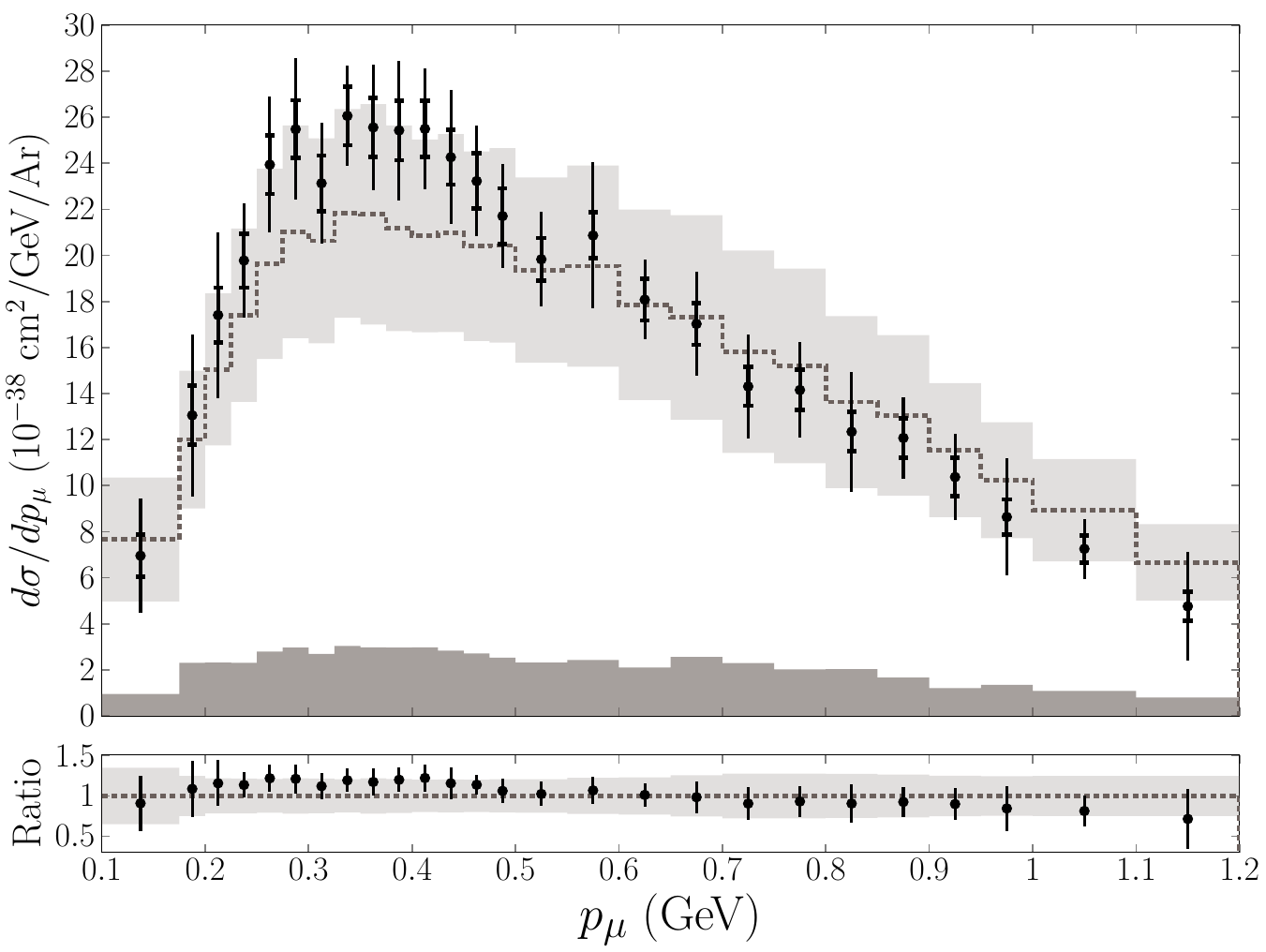}
\hfill
\belowbaseline[-0.23\textheight]{
  \includegraphics[page=57, width=0.35\textwidth]{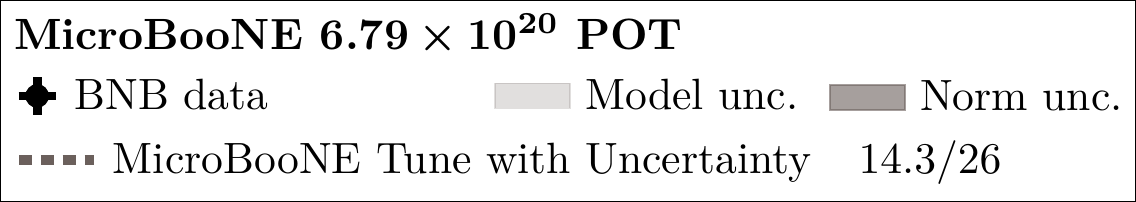}%
}
\hfill

\caption{Measured differential cross sections for block \#13
($p_\mu$).
Statistical (shape-only systematic) uncertainties are included in the inner
(outer) error bars.
The dark (light) gray band shows the
remainder of the measurement uncertainty (theoretical uncertainty on the
prediction). Error bars in the lower panels show the full measurement
uncertainty.}
\label{fig:ResultsTuneUncblock13}
\end{figure*}

Plots of the \ccnp\ differential cross-section results shown in Sec.~VII$\,$B
of the main text are reproduced here with all model predictions other than the
MicroBooNE Tune removed.
Figures~\ref{fig:ResultsTuneUncblock0}--\ref{fig:ResultsTuneUncblock13} follow
the same conventions as the main text for displaying the data points in the
main panel of each plot; the inner error bars represent the statistical
uncertainty only, while the outer error bars also include shape-only systematic
uncertainties. The remaining portion of the total measurement uncertainty is
shown by the dark gray band along the $x$-axis.

The bottom panel of each plot displays the ratio of the measured data points to
the MicroBooNE Tune prediction. The error bars on the black points in the ratio
plots display the full uncertainty on the measurement.

The light gray band that appears in all panels of
Figs.~\ref{fig:ResultsTuneUncblock0}--\ref{fig:ResultsTuneUncblock13} displays
the theoretical uncertainty on the \mbox{MicroBooNE} Tune prediction. This
uncertainty is calculated using the same variations to the neutrino interaction
model mentioned in Sec.~V$\,$A of the main text. All other sources of
uncertainty are omitted.

To calculate the theoretical uncertainty, the covariances between the expected
signal event counts in the $\trueBinIdx$-th and $\secondTrueBinIdx$-th true
bins are evaluated via
\begin{equation}
\label{eq:theory_unc}
\CovMat_{\trueBinIdx \secondTrueBinIdx} =
\frac{ 1 }{ \UnivCount } \sum_{\UnivIdx = 1}^{\UnivCount} \big(
\PredictedSignalEvtCount_{\trueBinIdx}^\text{CV} -
\PredictedSignalEvtCount_{\trueBinIdx}^{\UnivIdx} \big) \big(
\PredictedSignalEvtCount_{\secondTrueBinIdx}^\text{CV} -
\PredictedSignalEvtCount_{\secondTrueBinIdx}^{\UnivIdx} \big) \,.
\end{equation}
Here, $\PredictedSignalEvtCount_{\trueBinIdx}^\text{CV}$ is the number of
\ccnp\ events in true bin $\trueBinIdx$ predicted by the central-value
MicroBooNE simulation, while
$\PredictedSignalEvtCount_{\trueBinIdx}^{\UnivIdx}$ is a corresponding
prediction for the $\UnivIdx$-th alternative universe. The total number of
alternative universes for the systematic variation of interest is $\UnivCount$.
Here the special prescription from Sec.~V$\,$B is not followed; The signal
event counts are directly varied in each alternative universe. To make them
comparable to the data, the theoretical covariances from
Eq.~(\ref{eq:theory_unc}) have been scaled to differential cross section units
and transformed using the additional smearing matrix $A_C$.

The legends accompanying
Figs.~\ref{fig:ResultsTuneUncblock0}--\ref{fig:ResultsTuneUncblock13} list a
$\chi^2$ score for the ``MicroBooNE Tune with Uncertainty.'' The format is
identical to the one used in similar legends from the main text: Each $\chi^2$
score is separated from the number of bins for which it was calculated by a
\texttt{/} character. Unlike the $\chi^2$ scores given in the figures from the
main text, however, the ones shown here include the theoretical uncertainty
described above. The $\chi^2$ calculation also accounts for the correlations
between the signal prediction and the measured data points that were introduced
by using the MicroBooNE Tune model during the cross-section extraction
procedure. The overall MicroBooNE Tune value of $\chi^2 =
\GetTableParam{uBTune_chi2}$ for \GetTableParam{num_bins} bins (see Table~II
from the main text) improves to $\chi^2 = \GetTableUncParam{uBTuneUnc_chi2}$
when the theoretical uncertainty is included in this way.

\clearpage
\section{Extended data release}
\label{sec:extended_data_release}

The compressed tar archive file \texttt{extended\_data\_release.tar.bz2}
contains detailed information from the analysis that allows many plots from the
main text and this supplement to be reproduced. The archive file's contents may
be extracted on Unix-like operating systems by running the command
\begin{center}
\texttt{tar xvfj extended\_data\_release.tar.bz2}
\end{center}
in a terminal. All files discussed in the remainder of this section will be
made available by this procedure.

The file \texttt{universes.txt} tabulates vectors of predicted event counts
from the MicroBooNE simulation needed to reproduce all systematic covariance
matrices used in the analysis. The first line of \texttt{universes.txt}
contains a header of the form
\begin{center}
  \texttt{numBins numCV numAltTypes numFullCorr numSum}
\end{center}
where each of these variables takes an integer value. The header variables
have the definitions given below.
\begin{description}
  \item[numBins] The number of elements (1077) in each vector of predicted
event counts reported in the file. The first 359 of these contain the quantity
  \begin{equation}
  \label{eq:univ_reco_content}
  \PredictedRecoEvtCount_\recoBinIdx - \EXTCount_\recoBinIdx
    = \PredictedSignalEvtCount_\recoBinIdx + \PredictedBkgdCount_\recoBinIdx
  \end{equation}
in each reconstructed bin $0 \leq \recoBinIdx \leq 358$ used for the
cross-section measurements. The order and numbering scheme used for these bins
is the same as in Table~III of the main text. The notation used in
Eq.~(\ref{eq:univ_reco_content}) is the same as in Eq.~(13) from the main text.
The following group of 359 vector elements contain the same quantity but for
the duplicate set of reconstructed bins in which the sideband selection
discussed in Sec.~V$\,$F from the main text has been applied. The bin ordering
and definitions are otherwise the same as the first 359 elements. The final 359
elements of each vector contain the expected number of signal events
$\PredictedSignalEvtCount_\trueBinIdx$ in each true bin $\trueBinIdx$. The bin
ordering and definitions once again follow Table~III from the main text.
  \item[numCV] The number of central-value universes (3) reported in the file.
The nominal prediction of the MicroBooNE simulation is given by the first of
these, which is labeled \texttt{CV}. To mitigate Monte Carlo statistical
fluctuations when assessing detector-related systematic uncertainties, however,
two additional central-value universes, \texttt{detVarCV1} and
\texttt{detVarCV2}, were constructed with identical simulation parameters and
used as replacement central-value predictions in specific cases when computing
covariance matrix elements according to Eq.~(14) from the main text.
  \item[numAltTypes] The number of alternative-universe covariance matrices
(24) needed to compute the full set of systematic uncertainties adopted in the
analysis. Each of these corresponds to a specific type of variation to the
MicroBooNE simulation whose impact is studied with $\UnivCount \geq 1$
alternative universes.
  \item[numFullCorr] The number of fully-correlated systematic uncertainties
(2) included in the analysis. As mentioned in Sec.~V$\,$A of the main text,
these are computed using a fractional uncertainty applied to each bin
of the main central-value universe (\texttt{CV}).
  \item[numSum] The number of covariance matrices defined at the end of
\texttt{universes.txt} as sums of other previously-defined covariance matrices.
\end{description}

The following \texttt{numCV} (3) lines of the file contain the event count
vectors for each of the central-value universes. Each line starts with the name
of the central value universe followed by its \texttt{numBins} (1077) elements.

After the central-value universe definitions, \texttt{universes.txt} contains
\texttt{numAltType} definitions of alternative universes. Each of these begins
with a line of the form
\begin{center}
  \texttt{altName refCV numAltUniv}
\end{center}
in which the first two fields are whitespace-delimited strings and the third is
a positive integer. The \texttt{altName} field labels the simulation variation
of interest, and $\text{\texttt{refCV}} \in \{\text{\texttt{CV}},
\text{\texttt{detVarCV1}}, \text{\texttt{detVarCV2}}\}$ specifies which of the
central-value universes should be used when computing covariance matrix
elements according to Eq.~(14) from the main text. The integer
\texttt{numAltUniv} gives the number $\UnivCount$ of alternative universes that
are defined for the current systematic variation. The following
\texttt{numAltUniv} lines of the file each contain a vector with
\texttt{numBins} (1077) elements corresponding to one of the relevant
alternative universes. In the case of variations to the MicroBooNE Tune
neutrino interaction model (\texttt{altName} begins with \texttt{xsec\_} for
these), the reconstructed bin counts are evaluated according to the special
prescription described in Sec.~V$\,$B of the main text. The true signal event
counts $\PredictedSignalEvtCount_\trueBinIdx$ are varied directly without
special treatment for the last 359 elements of each alternative universe
vector.

Following the \texttt{numAltType} (24) definitions of alternative universes,
there are \texttt{numFullCorr} (2) definitions of fully-correlated systematic
uncertainties given in \texttt{universes.txt}. Each of these appears as a
single line of the form
\begin{center}
  \texttt{FullCorrName FracUnc}
\end{center}
in which $\text{\texttt{FullCorrName}}\in\{\text{\texttt{POT}},
\text{\texttt{numTargets}}\}$ labels the type of uncertainty (beam exposure and
number of Ar targets in the fiducial volume, respectively). The
\texttt{FracUnc} field contains the corresponding fractional uncertainty.

Finally, the last \texttt{numSum} (5) lines of \texttt{universes.txt} give
definitions of new covariance matrices in terms of sums of others. Each
definition appears on a single line. The line begins with the name
\texttt{sumName} of the new
covariance matrix followed by the number \texttt{numOther} of other covariance
matrices included in the sum. The line concludes with \texttt{numOther}
whitespace-delimited strings giving the names of the other covariance matrices.
Allowed values for these names are the strings used earlier in the file
within the \texttt{altName}, \texttt{FullCorrName}, and \texttt{sumName}
fields, as well as the additional names \texttt{MCstats}, \texttt{EXTstats},
and \texttt{BNBstats}. These last three allowed names correspond to the
precomputed statistical covariance matrices that appear in the
subfolder \texttt{cov\_matrices/}. Respectively, these label the Monte Carlo
statistical uncertainty, the statistical uncertainty on the measured
beam-off background, and the statistical uncertainty on the data measured
when the BNB was active. These covariance matrices are provided
explicitly because they cannot be computed from simple vectors of
event counts; due to the multi-block structure of the analysis,
events are shared between multiple bins, and a correct treatment of
statistical correlations requires knowledge of the overlaps between
each pair of bins.

The final line of \texttt{universes.txt} provides a definition for the summed
covariance matrix \texttt{total}. This covariance matrix describes the full
uncertainty on the measured event counts used for cross-section extraction
(bins 0 to 358), the measured event counts in the sidebands (bins 359 to 717)
and the MicroBooNE Tune prediction for the true signal event counts (bins 718
to 1077).

An example \cpp\ program that interprets the contents of \texttt{universes.txt}
is provided in the file \texttt{calc\_covariances.C}. Like the
\texttt{calc\_chi2.C} program provided in the basic data release (see
Sec.~\ref{sec:basic_data_release} of this supplement),
\texttt{calc\_covariances.C} relies on the \texttt{TMatrixD} class defined by
ROOT and must either be executed using the ROOT \cpp\ interpreter or be
compiled against the ROOT shared libraries. Executing
\texttt{calc\_covariances.C} will parse the precomputed statistical covariance
matrices and \texttt{universes.txt}. It will then create new covariance matrix
files in the \texttt{cov\_matrices/} subfolder representing all uncertainties
used in the analysis. The total covariance matrix describing all uncertainties
can be found in the file
\texttt{cov\_matrices/mat\_table\_extendedCov\_total.txt} after the program has
finished running. The format for both the precomputed statistical covariance
matrix files and the new ones created by \texttt{calc\_covariances.C} is
identical to the \texttt{mat\_table\_cov\_total.txt} file from the basic data
release except for two differences. First, the number of bins is larger (1077
rather than 359). Second, the numerical values of the covariance matrix
elements now represent uncertainties on event counts rather than flux-averaged
total cross sections, so they are dimensionless.

For the first 718 bins represented in the universe vectors tabulated in
\texttt{universes.txt}, the main central-value prediction (\texttt{CV}) is
split into 11 separate event categories in the file
\texttt{mat\_table\_mc\_categories.txt}. Each row of the matrix defined in this
file gives the contribution of a specific event category to the reconstructed
bins in the \texttt{CV} universe vector. The categories $c \in [0,10]$ are
defined below in terms of the descriptions from Sec.~III$\,$B of the main text.
\begin{description}
  \item[0] Signal \ccnp\ events in which the primary interaction mode
was QE.
  \item[1] Signal \ccnp\ events in which the primary interaction mode
was 2p2h.
  \item[2] Signal \ccnp\ events in which the primary interaction mode
was resonance production.
  \item[3] Signal \ccnp\ events with a primary interaction mode not specified
by a prior category. In Figs.~1 and 7--14 from the main text,
\textit{Signal (CC other)} is used to label the sum of categories 2 and 3.
  \item[4] Background CC$N\pi$ events
  \item[5] Background CC0$\pi$0$p$ events
  \item[6] Background Other $\nu_\mu$ CC events
  \item[7] Background $\nu_e$ CC events
  \item[8] Background NC events
  \item[9] Background Out FV events
  \item[10] Background Other events
\end{description}
As described below, the beam-off background events are tabulated separately.
Note that, as shown in Eq.~(\ref{eq:univ_reco_content}), the beam-off
background events do not contribute to the universe vectors by construction.

As implied by Eq.~(29) of the main text, the portion of the extended data
release covariance matrices representing the first 359 bins can be adjusted to
represent uncertainties on the unfolded signal event counts
$\MeasuredSignalEvtCount_\trueBinIdx$ via transformation by the error
propagation matrix $\ErrPropMatrix$. The elements of $\ErrPropMatrix$ are
tabulated in the file \texttt{mat\_table\_err\_prop.txt} for this purpose. A
similar transformation using the additional smearing matrix $\AddSmearMatrix$
can be performed on the covariance matrix elements describing the MicroBooNE
tune signal prediction (the last 359 bins) in order to obtain the theoretical
uncertainty shown in Sec.~\ref{sec:tune_unc} above. The elements of
$\AddSmearMatrix$ are provided in the basic data release as described in
Sec.~\ref{sec:basic_data_release} of this supplement. For completeness, the
elements of the unfolding matrix $\UnfoldingMatrix$ are also provided in the
file \texttt{mat\_table\_unfolding.txt}. Note that all elements of
$\ErrPropMatrix$, $\AddSmearMatrix$, and $\UnfoldingMatrix$ are dimensionless.

The file \texttt{scale\_factors.txt} contains the values of the integrated BNB
$\nu_\mu$ flux $\IntegratedFlux$ and the number of Ar targets in the fiducial
volume $\NumTargets$ needed to convert event counts to cross sections. These
are labeled in the file as \texttt{IntegratedFlux} ($\nu_\mu /
\si{\centi\meter\squared}$) and \texttt{NumArTargets} (dimensionless),
respectively. Although it is not needed for the unit conversion, the beam
exposure in \si{\pot} used for the analysis is also given in the file with the
label \texttt{BeamExposurePOT}.

The file \texttt{vec\_table\_data\_bnb.txt} tabulates the measured number of
events $\AllRecoEvents_\recoBinIdx$ in each reconstructed bin $\recoBinIdx$ for
data taken when the BNB was active. The first 359 entries give the event counts
for the selection criteria used to obtain the cross-section results. The
remaining 359 entries give the corresponding results for the sideband selection
defined in Sec.~V$\,$F from the main text. Both groups of reconstructed bins
use the same order and numbering scheme from Table~III of the main text. The
file \texttt{vec\_table\_data\_ext.txt} has the same organization, but it
tabulates the measured beam-off background event counts
$\EXTCount_\recoBinIdx$.

\FloatBarrier
\bibliography{main.bib}